\documentclass{aa}  
\usepackage{graphicx}
\usepackage{txfonts}
\usepackage[utf8]{inputenc}
\usepackage{amssymb,amsmath}
\usepackage{lscape}
\usepackage{longtable}
\usepackage{natbib}
\usepackage{color}
\begin{document} 

\title{Characterization of galactic bars from $3.6\,\mu$m S$^{4}$G imaging
\thanks{Full Tables \ref{truncated65}, \ref{truncated1} and \ref{truncated2}, 
the tabulated radial force profiles and the rotation curve decomposition model of each individual galaxy,
are only available in electronic form at the CDS via anonymous ftp to cdsarc.u-strasbg.fr (130.79.128.5) or via http://cdsweb.u-strasbg.fr/cgi-bin/qcat?J/A+A/}
}
   \author{S. Díaz-García\inst{1}
          \and        
          H. Salo\inst{1}
          \and        
          E. Laurikainen\inst{1,2}
          \and        
          M. Herrera-Endoqui\inst{1}
          }
   \institute{Astronomy and Space Physics research center, University of Oulu, FI-90014 Finland\\
              \email{simon.diazgarcia@oulu.fi}
         \and
             Finnish Centre of Astronomy with ESO (FINCA), University of Turku, Väisäläntie 20, FI-21500, Piikkiö, Finland\\
             }
   \date{Received March 23, 2015; accepted July 27, 2015}
  \abstract
   {Stellar bars play an essential role in the secular evolution of disk
galaxies because they are responsible for the redistribution of matter and
angular momentum. Dynamical models predict that bars become stronger and longer in time, while their rotation speed slows down.
}
   {We use the Spitzer Survey of Stellar Structure in Galaxies (S$^{4}$G) 3.6~$\mu$m
imaging to study the properties (length and strength) and fraction of bars at $z=0$ over a wide range of galaxy masses ($M_{\ast} \approx 10^8 - 10^{11} M_{\odot}$) and Hubble types ($-3 \le T \le 10$).}
   {We calculated gravitational forces from the 3.6~$\mu$m images for galaxies with a disk inclination lower than 65$^{\circ}$. 
We used the maximum of the tangential-to-radial force ratio in the bar region ($Q_{\rm b}$) as a measure of the bar-induced perturbation strength for a sample of $\sim 600$ barred galaxies.
We also used the maximum of the normalized m=2 Fourier density amplitude ($A_{2}^{\rm max}$) and the bar isophotal ellipticity ($\varepsilon$) to characterize the bar.
Bar sizes were estimated i) visually, ii) from ellipse fitting, iii) from the radii of the strongest torque, and iv) from the radii of the largest m=2 Fourier amplitude 
in the bar region. By combining our force calculations with the H{\sc\,i} kinematics from the literature, we estimated the ratio of the halo-to-stellar
mass ($M_{\rm h}/M_{\ast}$) within the optical disk and by further using the universal rotation curve models, we obtained a first-order model of the rotation curve decomposition of $1128$ disk galaxies.
}
{
We probe possible sources of uncertainty in our $Q_{\rm b}$ measurements:
the assumed scale height and its radial variation, the influence of the spiral arms torques,
the effect of non-stellar emission in the bar region, and the dilution of the bar forces by the dark matter halo (our models imply that only $\sim 10\%$ of the disks in our sample are maximal).
We find that for early- and intermediate-type disks ($-3 \le T < 5$), the relatively modest influence of the dark matter halo leads to a systematic reduction of the mean $Q_{\rm b}$ by about $10-15\%$,
which is of the same order as the uncertainty associated with estimating the vertical scale height.
The halo correction on $Q_{\rm b}$ becomes important for later types, implying a reduction of $\sim 20-25\%$  for $T = 7-10$.
Whether the halo correction is included or not, the mean $Q_{\rm b}$ shows an increasing trend with $T$. However, the mean $A_{2}^{\rm max}$
decreases for lower mass late-type systems. These opposing trends are most likely related to the reduced force dilution by bulges when moving towards later type galaxies.
Nevertheless, when treated separately, both the early- and late-type disk galaxies show a strong positive correlation between $Q_{\rm b}$ and $A_{2}^{\rm max}$.
For spirals the mean $\varepsilon \approx 0.5$ is nearly independent of $T$, but it drops among S0s ($\approx 0.2$). 
The $Q_{\rm b}$ and $\varepsilon$ show a relatively tight dependence, with only a slight difference between early and late disks.
For spirals, all our bar strength indicators correlate with the bar length (scaled to isophotal size).
Late-type bars are longer than previously found in the literature.
The bar fraction shows a double-humped distribution in the Hubble sequence ($\sim 75\%$
for Sab galaxies),
with a local minimum at $T=4$ ($\sim 40\%$), and it drops for $M_{\ast} \lesssim 10^{9.5-10} M_\odot$.
If we use bar identification methods based on Fourier decomposition or ellipse fitting instead of the morphological classification, the bar fraction decreases by $\sim 30-50\%$
for late-type systems with $T \ge 5$ and correlates with $M_{\rm h}/M_{\ast}$.
Our $M_{\rm h}/M_{\ast}$ ratios agree well with studies based on weak lensing analysis, abundance matching, and halo occupation distribution methods,
under the assumption that the halo inside the optical disk contributes roughly a constant fraction of the total halo mass ($\sim 4 \%$).
}
{We find possible evidence for the growth of bars within a Hubble time, as (1) bars in early-type galaxies show larger density
amplitudes and disk-relative sizes than their intermediate-type counterparts, and (2) long bars are typically strong.
We also observe two clearly distinct types of bars, between early- and intermediate-type galaxies ($T<5$) on one side, and the late-type systems on the other,
based on the differences in the bar properties. 
Most likely this distinction is connected to the higher halo-to-stellar ratio that we observe in later types, which affects the disk stability properties.
}
   \maketitle
%
%
\section{Introduction}
The lambda cold dark matter ($\Lambda$CDM) model predicts that galaxies originate in dark matter haloes. In the first stage of their evolution, galaxies
undergo a process of continuous mergers. As the redshift decreases, galaxy mergers become less frequent and the evolution becomes internally driven
(the so-called secular evolution).

Nearly two thirds of all galaxies in the nearby Universe have bars
\citep[e.g.][]{1991rc3..book.....D,2000ApJ...529...93K,2002MNRAS.336.1281W,2004ApJ...607..103L,2007ApJ...657..790M}.
Approximately one half of these are strongly barred \citep[e.g.][]{1963ApJS....8...31D,1993RPPh...56..173S}.
Although early studies suggested a nearly constant bar fraction at least up to $\rm z=1.2$ \citep[e.g.][]{2004ApJ...612..191E}, it has become clear that bar fraction increases with time and is around 20$\%$ at $\rm z=0.84$
\citep[][]{2008ApJ...675.1141S,2010ApJ...714L.260N}. From all these studies it is known that the bar fraction depends on colour, stellar mass, and bulge prominence.

Bars continuously interact dynamically with the other structure components of galaxies, such as the underlying disks, the bulges, or the
dark matter halos. This causes them to evolve over time \citep[e.g.][]{1992ApJ...400...80H,2000ApJ...543..704D,2003MNRAS.341.1179A,2004ARA&A..42..603K,2006ApJ...637..582B,2009ApJ...707..218V}.
Galaxies are not isolated entities in the Universe, they are for instance actively interacting with the surrounding extragalactic gas and with the small satellite galaxies.
Bars play an important role in the secular evolution of disk galaxies \citep[][]{2013seg..book..305A}.
Bars are known to be robust so that their effects cover a wide range of redshifts \citep[e.g.][]{2004ApJ...604..614S,2010ApJ...719.1470V,2013MNRAS.429.1949A}.
This is supported by studies showing that bars are typically composed of old stars \citep[][]{2006ApJS..163..270G,2011MNRAS.415..709S}.

Bars are responsible for the redistribution of the angular momentum of the baryonic and dark matter components of disk galaxies \citep[e.g.][]{1985MNRAS.213..451W, 2002MNRAS.330...35A,2007ApJ...659.1176M}. Bar-induced secular evolution is manifested
in many features of the disk, such as resonance rings \citep[e.g.][]{1981ApJ...247...77S,1986ApJS...61..609B,1996FCPh...17...95B,2000A&A...362..465R}, or long-lasting spiral density waves \citep[e.g.][]{1969ApJ...158..899T,1979ApJ...233..539K}.

Bars also trigger gas inflow towards the central regions of the galaxies, which leads to nuclear starbursts. They drive the secular evolution of bulges and cause the kinematic heating of the inner disk \citep[e.g.][]{1990A&A...233...82C}.
The causality between the bar-driven gas inflow and the fuelling of the active galactic nucleus (AGN) \citep[e.g.][]{1989Natur.338...45S} has been a much-debated topic during the past decades \citep[for a review, see][]{2006LNP...693..143J}.
\citet[][]{2004ApJ...607..103L} observed a higher fraction of bars in Seyfert galaxies than in the non-active counterparts;
this is consistent with other studies at near-infrared \citep[e.g.][]{2000ApJ...529...93K,2002ApJ...567...97L} and optical wavelengths \citep[e.g.][]{2009ASPC..419..402H}. 
\citet[][]{2012ApJ...750..141L} recently claimed that the connection between the bar fraction and the nuclear activity only depends on the fact that AGN-host galaxies are on average redder and more massive,
since no correlation is found when the colour or the stellar mass of the host galaxy are fixed.
Likewise, \citet[][]{2013ApJ...776...50C} did not observed any correlation between the bar strength and the degree of nuclear activity as determined from \emph{Chandra} X-ray observations,
which suggests that other mechanisms are required to fuel the AGN \citep[see also the discussion in][and references therein]{2009ASPC..419..402H}.

Bars themselves evolve, which among other things can manifest as buckling in the vertical
direction that results in vertically thick structure components \citep[][]{1981A&A....96..164C,1990A&A...233...82C,1991Natur.352..411R,2002MNRAS.330...35A,2004ApJ...604L..93D,2004ApJ...613L..29M},
also called boxy/peanut bulges. \citet[][]{2007MNRAS.381..401L}, \citet[][]{2014MNRAS.444L..80L} and \citet[][]{2014arXiv1405.6726A}
proposed that the so-called barlenses \citep[][]{2011MNRAS.418.1452L} are boxy/peanut bulges seen face-on.
For a review of the properties of boxy/peanut bulges, the reader is refereed to \citet[][]{2015arXiv150500590L} and \citet[][]{2015arXiv150304804A}.

Many studies have been devoted to characterizing bars from an observational point of view. Bars can have either flat or exponential
radial surface brightness profiles with respect to the surrounding disk \citep{1982modg.proc..113K, 1985ApJ...288..438E}, and a few of them
present blobs at the end of the bar \citep[ansae morphology;][]{1965AJ.....70..501D,2007MNRAS.381..401L,2007AJ....134.1863M}. 
Bar strengths, that is, the ratio of the tangential force to the mean axisymmetric radial force field, can be computed from the gravitational potentials
inferred from optical and infrared images \citep{1981A&A....96..164C,2001ApJ...550..243B,2002MNRAS.337.1118L,2002MNRAS.331..880L,2004ApJ...607..103L,2004MNRAS.355.1251L}.
There are alternative ways of estimating the strength of the bars photometrically,
namely by calculating their intrinsic ellipticity \citep{1995AJ....109.2428M,2002MNRAS.331..880L} and from the
normalized Fourier intensity amplitude \citep{2004MNRAS.355.1251L,2005MNRAS.362.1319L}. Another indirect method is provided by studying dust lanes, whose curvature is
inversely proportional to the bar strength according to the studies of \citet[][]{1992MNRAS.259..345A} \citep[probed in][]{2002MNRAS.337..808K,2009ApJ...706L.256C,2015MNRAS.450.2670S},
or by 2D decomposition, fitting  generalized ellipses to the bar component and using the boxyness as a proxy of the bar strength \citep{2011MNRAS.415.3308G}.
Bar lengths have also been measured by different methods in the literature \citep[for a detailed review see][hereafter E2005]{2005MNRAS.364..283E}
such as ellipse fitting to bar isophotes \citep[e.g.][]{1995A&AS..111..115W}, Fourier analysis of the images based on the bar-interbar luminosity contrast or phase
\citep[e.g.][]{1990ApJ...357...71O,2000A&A...361..841A,1994ApJ...437..162Q,2002MNRAS.337.1118L}, and they have been estimated visually \citep[e.g.][]{1979ApJ...227..714K}.

From N-body simulations it has been shown that bars form spontaneously in galactic disks
\citep[e.g.][]{1970ApJ...161..903M,1971ApJ...168..343H,1973ApJ...186..467O,1980A&A....89..296S,1981A&A....99..362S,1986MNRAS.221..213A},
provided that the disk velocity dispersion is not too high and the contribution of the disk to the total force field is considerable. 
A key process in the bar-induced secular evolution is the interaction between the bar and the dark matter halo, which is expected to take
place in resonance regions that are sufficiently filled with the baryonic matter \citep[][]{2003MNRAS.341.1179A}. Theoretical models
\citep[e.g.][]{2010ApJ...719.1470V,2013MNRAS.429.1949A} predict that when the bar evolves, it becomes longer, stronger, and slower. The last
of these predictions is not always confirmed with the bar pattern speed observations \citep[e.g.][]{2005ApJ...631L.129R,2008MNRAS.388.1803R},
or at most, contradictory observations are shown.
In fact, direct Tremaine-Weinberg measurements \citep[][]{1984MNRAS.209..729T} indicate fast bars \citep[][]{2015A&A...576A.102A},
but in this study our aim is not to solve this problem.
Instead, we estimate bar strengths or the prominence of bars using different methods for a representative sample of galaxies in the nearby Universe that cover a
wide range of galaxy masses. In a forthcoming paper these measurements will be connected to bar pattern speed estimates for the same galaxies.

Infrared (IR) observations, where the dust absorption is lower than in the visual wavelengths, are good tracers of old stellar
populations that probe the underlying mass distribution in galaxies \citep{2000AJ....119..536E}, which makes this spectral range well suited 
for the study of stellar structures. For this reason, the mid-IR imaging of the Spitzer Survey of Stellar Structure in Galaxies
\citep[S$^{4}$G,][]{2010PASP..122.1397S}, which includes 2352 galaxies in the nearby Universe of different masses and morphological stages ($T$)
from which we take our sample, is ideal for studying the properties of stellar bars.

In addition to ellipse fitting, we use Fourier decomposition as a method for estimating bar strengths.
We evaluated the gravitational potential from the flux in the 3.6~$\mu$m images and calculated tangential forces normalized to the axisymmetric force field. 
In these calculations we took into account what is learned from the previous studies using a 
similar method \citep[e.g.][]{2001ApJ...550..243B,2002MNRAS.337.1118L,2002MNRAS.331..880L,2004ApJ...607..103L,2004MNRAS.355.1251L,2005MNRAS.362.1319L,2010ApJ...715L..56S}:
we included the effects of the bulge while de-projecting the images and considered the effects of spiral arms.
We also estimated the effect of the dark matter halo and the possible effects of the vertical thickening of the inner part of the bar associated with a boxy/peanut bulge.
We repeated the calculations with mass maps in which the non-stellar contaminants are eliminated, taken from \citet[][]{2015ApJS..219....5Q}.

The paper is organized as follows: in Sect. 2, we present the data used in this study and the sample selection.
In Sect. 3, we provide a thorough description of the way in which we calculated bar strengths and the stellar contributions to the circular velocity,
as well as the different tests that we carried out to probe possible sources of uncertainty in our methodology.
In Sect. 4, we explain the different bar size measurements. 
In Sects. 5 and 6 we analyse the bar fraction and the various bar length and strength parameters as a function of Hubble stage and galaxy mass.
In Sect. 7 we discuss the evidence provided by the different measurements for the growth of bars over a Hubble time. 
Finally, in Sect. 8 we summarize the main results and the implications for the secular evolution of disk galaxies.

\section{Data and sample selection}
%
\subsection{S$^{4}$G sample}
The 2352 galaxies of the S$^{4}$G survey \citep[][]{2010PASP..122.1397S} were observed in the 3.6~$\mu$m and 4.5~$\mu$m filters with the InfraRed Array Camera \citep[IRAC;][]{2004ApJS..154...10F},
installed on-board the \emph{Spitzer} Space Telescope \citep{2004ApJS..154....1W}. 
The sample was defined based on data gathered from HyperLEDA\footnote{We acknowledge the usage of the database (http://leda.univ-lyon1.fr).}\citep{2003A&A...412...45P}
with the following criteria:
\begin{itemize}
 \item Radial velocity $\nu_{\rm radio}$ $\textless$ 3000 km s$^{\rm -1}$, which corresponds to a distance of $d \lesssim 40$ Mpc.
 \item Galactic latitude $|b|\textgreater 30^{\circ}$.
 \item Total blue magnitude corrected for internal extinction, $m_{B_{\rm corr}} \textless 15.5$ mag.
 \item Blue light isophotal angular diameter $D_{25} \textgreater 1^{\prime}$.
\end{itemize}

All Hubble types ($T$) are included, although the requirement of HyperLEDA $v_{{\rm radio}}$ measurement means that many gas-poor
early-type galaxies are missing from the original S$^{4}$G-sample 
\citep[the observations are currently extended to include the missing early types, see][]{2013sptz.prop10043S}.
The S$^{4}$G sample covers a wide range of stellar masses that span over several orders of magnitude.

The surface brightness depth is $\mu_{\rm 3.6 \mu \rm m}$(AB)(1$\sigma$) $\sim$27 mag arcsec$^{-2}$. 
As the mid-infrared images are barely affected by dust absorption or polluted by star formation \citep[little contamination from hot gas and PAHs;][]{2012ApJ...744...17M, 2011MNRAS.417..812Z},
they trace the old stellar population in galaxies well. Possible contaminants are controlled in the mass maps in which they are eliminated \citep[][]{2015ApJS..219....5Q}.
\subsubsection{Selection of sub-samples}
The selection of barred galaxies is based on the classification in
\citet[][B2015 hereafter]{2015ApJS..217...32B}, made for the complete S$^4$G.
We first selected disk galaxies (ellipticals in B2015 were excluded) with an inclination $i\le 65^{\circ}$, which leaves 1345 galaxies, for which the gravitational potentials and the stellar component of the circular velocity were obtained.
This inclination upper limit is very similar to that ($i=60^{\circ}$) used in \citet[][]{2014A&A...562A.121C} and recommended in \citet{2014ApJ...791...11Z}.
Of these non-highly inclined galaxies, 860 are barred according to B2015, forming our sample of barred galaxies for which bar lengths are measured visually.
This means that we did not use any intrinsic bar ellipticity cut-off for the sample selection \citep[as in e.g.][]{1999MNRAS.308..569A},
which means that oval-like structures are also included in our analysis as if they were normal bars.

To calculate the bar force and determine bar lengths and shapes through ellipse fitting, we tried to cover the largest possible number of barred galaxies within
S$^{4}$G. However, some faint dwarf, irregular and/or gas-rich late-type galaxies have peculiar bars that are complicated to
measure. Another reason to exclude some galaxies are offset bars, that is,{\em } bars whose centres are displaced with
respect to the centre of the galaxy (NGC$\,$1345 is a good example). These are typical among Magellanic and irregular galaxies ($8\le T \le 10$).
Of the 860 barred galaxies in B2015, 654 have bar sizes, axial ratios, and position angles estimated from the isophotal profiles; bar force calculations based on Fourier decomposition and gravitational torques are made for 599 systems.
\begin{figure*}
   \centering
      \includegraphics[width=16.4cm,clip]{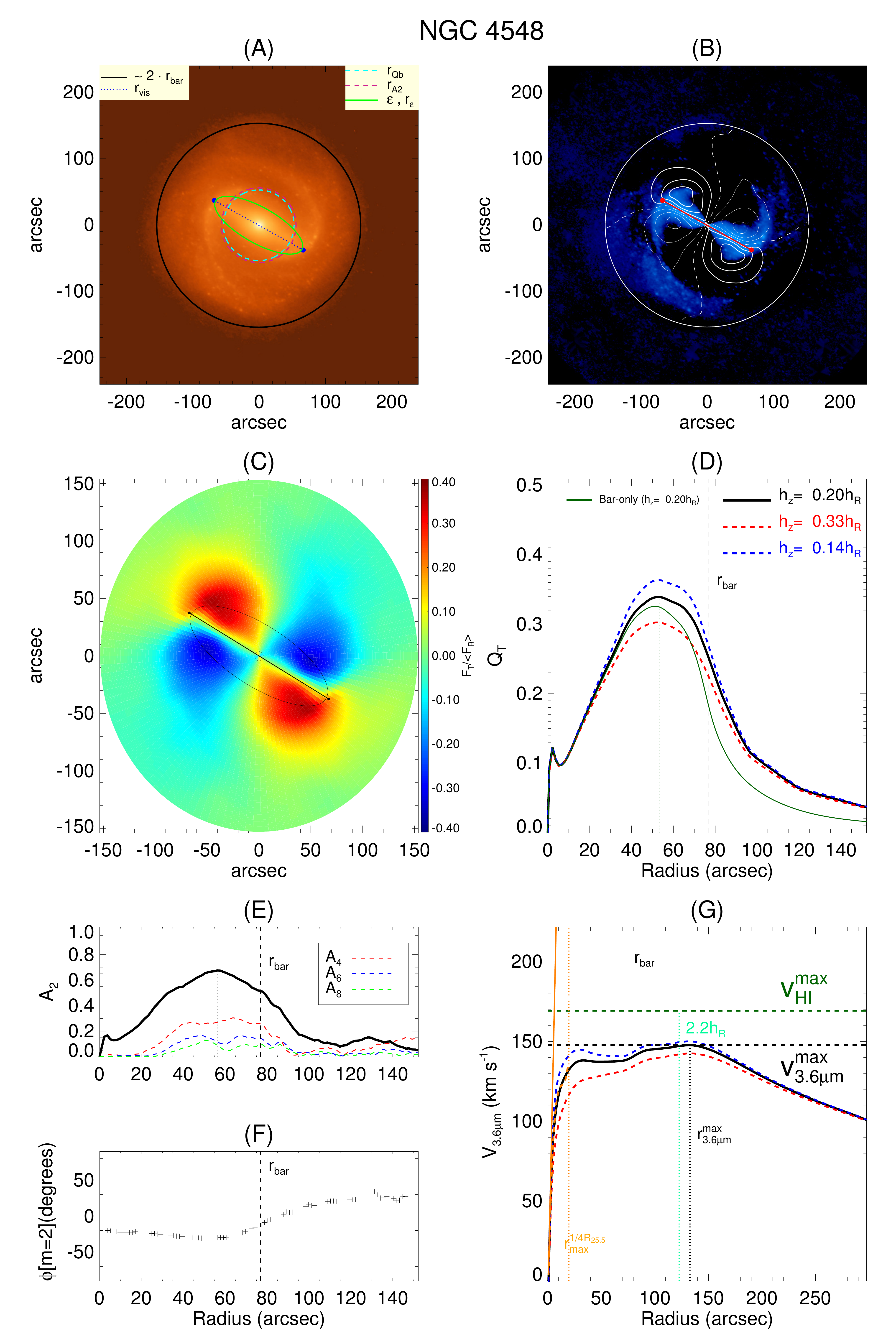} 
\caption{See next page for caption.}
\label{NGC4548_torque_maps}
\end{figure*}  
\begin{figure*}
\addtocounter{figure}{-1}
\caption{Example of force calculation for the barred galaxy NGC$\,$4548.
\emph{Panel A}: The de-projected image of NGC$\,$4548 in magnitude
scale with a range of 17-25 $\mu_{3.6\mu \rm m}(AB)$. The blue dotted line indicates the visually estimated length and
position angle of the bar. The light green ellipse corresponds to the
bar radius from the isophotal fit (maximum ellipticity). 
The purple and light blue circles have radii
equal  to $r_{\rm A2}$ and $r_{\rm Qb}$, respectively. 
\emph{Panel B}:
Image of the galaxy after subtracting the axisymmetric m=0 component, together
with contours tracing the force ratio $F_{\rm T}/\left<F_{\rm R}\right>$ (separated by 0.1 intervals). 
The dashed lines indicate the regions where the tangential forces change sign. 
The outer circle delimits a region of radius twice the size of the bar (the same as in Panel A).
\emph{Panel C}: The $F_{\rm T}/\left<F_{\rm R}\right>$ force map (butterfly pattern) of
the galaxy. The length and ellipticity of the bar are traced with black lines. The inner dotted circle corresponds to the bulge effective
radii from P4 decompositions.
\emph{Panel D}: The thick solid curve indicates the normalized
tangential force amplitude ($Q_{\rm T}$ in Eq.~\ref{eqqb}) calculated from
the force maps, using the nominal vertical scale height estimated from
the radial scale length. The dashed lines correspond to the assumed
upper (red) and lower bounds (blue) for the scale height. The thin green 
line corresponds to the bar-only force profile (see the text).
\emph{Panel E}: The normalized $m=2$ Fourier
density amplitude vs. radius. The dashed lines show
higher order even Fourier amplitudes.
\emph{Panel F}: The phase of m=2 Fourier amplitude vs. radius.
\emph{Panel G}: The solid line indicates the circular velocity curve
calculated from the image (dashed lines correspond to the same upper and lower 
bounds as in panel D). The horizontal dotted lines indicate the maxima of the observed H{\sc\,i} velocity amplitude (green)
and that of our calculated stellar component of the velocity curve.  
The solid orange line indicates the slope inferred from the linear term of the third degree polynomial fit of the inner part of the rotation curve (dashed orange line).
The dotted vertical orange line delimits the region where this fit is made, which is taken between the galactic centre and the radius of the maximum rotation within one fourth of $R_{25.5}$.
The dark green vertical dotted line corresponds to a radius of $2.2$ times the disk scale length \citep[radius of the maximum
velocity of an exponential disk;][]{1970ApJ...160..811F},
while the vertical dashed line traces the bar length.
}
\end{figure*}
\subsection{S$^{4}$G pipelines}
Five pipelines (P1-5) are behind the S$^{4}$G data release and subsequent scientific products:
\begin{itemize}
\item P1 produced science-ready mosaics from the raw data. The pixel scale of the 3.6~$\mu$m mosaics after the image processing and reduction is $0.75\arcsec$pixel$^{-1}$
\citep[for further details see][]{2010PASP..122.1397S}. The FWHM of the images is $\sim2.1\arcsec$.
\item P2 created object masks automatically using SExtractor \citep{1996A&AS..117..393B}, which later went through a human-supervised verification and edition by hand (see also P4).
\item P3 \citep[][describing also P1 and P2]{2015ApJS..219....3M} provides RC3-type parameters, such as the total magnitudes at the two S$^{4}$G bands
and the isophotal radii at the surface brightness 25.5~mag~arcsec$^{-2}$ ($R_{25.5}$) obtained from the $3.6\,\mu$m images. Based on the calibrations from \citet[][E2012 hereafter]{2012AJ....143..139E},
total stellar masses were derived from the $3.6$ and $4.5\,\mu$m fluxes in the following manner:
\begin{equation}   
M_{\ast}/M_{\odot}=10^{5.65}F_{\rm 3.6\mu m}^{2.85}F_{4.5\mu \rm m}^{-1.85}(D/0.05)^{2}, 
\label{stellar-mass}
\end{equation} 
where $D$ is the distance to the galaxy in Mpc and $F$ is the galaxy flux in MJy.
\item P4 \citep[][]{2015ApJS..219....4S} is dedicated to the decomposition of the two-dimensional light distribution into different structure components such as bulges, disks, bars, nuclear point sources, and various other disk components.
The decompositions were carried out using GALFIT \citep{2010AJ....139.2097P} with the help of IDL-based visualization procedures. 
In P4 ellipse fitting was also made using the 3.6~$\mu$m images, and the disk orientation parameters were estimated based on the ellipticity and position angle of the outer isophotes 
($PA_{\rm outer}$, $\epsilon_{\rm outer}=1-(b/a)_{\rm outer}$, where a and b refer to the semi-major and semi-minor axes, respectively).
The inclination of each galaxy was obtained under the assumption of an infinitesimally thin and intrinsically circular disk: $i=\rm cos^{-1}(b/a)_{\rm outer}$. 
The inclinations of the galaxies were visually checked by de-projecting the galaxy images and ensuring that no stellar structures appeared artificially stretched in face-on view.
These P4 orientation parameters were used to de-project the galaxy images and also to convert ellipse fit parameters measured in the sky plane to the intrinsic disk plane values.
\item P5 \citep[][]{2015ApJS..219....5Q} separates the old stellar population from the contaminant non-stellar emission.
Using the 3.6~$\mu$m and 4.5~$\mu$m images \citep[see][]{2012ApJ...744...17M}, a statistical independent component analysis (ICA) technique was used to decompose the galaxy emission into old stellar population light
and non-stellar emission associated either with hot dust, PAHs, or asymptotic giant and red super-giant stars.
\end{itemize}
We used the 3.6 $\mu$m science-ready P1 images.
Compared to the $4.5\,\mu$m band, $3.6\,\mu$m images have a greater depth that reaches stellar mass surface densities as low as one solar mass per square parsec \citep[][]{2010PASP..122.1397S}.
They also have a smaller FWHM than $4.5\,\mu$m images.

Before the force calculations, the sky background was subtracted, and the images were cleaned by filling the masked regions (with foreground objects or image defects) with values
obtained by linear interpolation from the adjacent good-image pixels. Distances and $R_{25.5}$ values are from P3. 
The total stellar masses are also from P3, computed using Eq.~\ref{stellar-mass}. For the parameters of the bulge and the disk scale lengths ($h_{\rm R}$) and orientations, we used the P4-values.

When converting the 3.6~$\mu$m flux to mass in the force calculations, we used the formula given in \citet{2013ApJ...771...59M}
to estimate the stellar mass from the raw 3.6~$\mu$m absolute magnitude ($M_{\rm 3.6 AB}$):
\begin{equation}   
{\rm log} (M_{\ast}/M_{\odot})=-0.4 M_{\rm 3.6 AB} + 2.13,
\label{munoz1}
\end{equation} 
assuming a mass-to-light ratio $M/L=\Upsilon_{3.6 \rm \mu m}=0.53$ (E2012). 
In practice, Eqs.~\ref{stellar-mass} and \ref{munoz1} are very similar, given the narrow range of $[3.6]-[4.5]$ colours \citep{2013ApJ...771...59M}.
\subsection{Other sources of information} 
Absolute magnitudes in the $I$ and $B$-bands (needed in Sect.~\ref{halo_section}) were calculated from the apparent magnitudes available in HyperLEDA, 
which were corrected for Galactic \citep[$AG$;][]{1998ApJ...500..525S} and internal extinction \citep[$AI$;][]{1995A&A...296...64B} and K-correction \citep[$AK$;][]{1976rcbg.book.....D} as follows:
\begin{equation}
m_{\rm B} = m_{\rm B}^{\rm leda}-AG-AI-AK(T)\cdot v/10000,
\end{equation}
\begin{equation}
m_{\rm I} = m_{\rm I}^{\rm leda} + 0.44(m_{\rm B}-m_{\rm B}^{\rm leda}),
\end{equation}
where $v$ is the heliocentric radial velocity (see HyperLEDA documentation for further details).
When $I$-band photometry was not available, we derived M$_{I}$ from the $B$-band absolute magnitude \citep[e.g.][]{1996MNRAS.281...27P}: $M_{I}=1.087\cdot(M_{B}+0.38)$.

We identified the bars based on the morphological classifications by B2015, 
which were made visually using the 3.6~$\mu$m images and following the comprehensive de Vaucouleurs-revised Hubble-Sandage system \citep[][]{1959HDP....53..275D,2007dvag.book.....B}.
This catalogue uses the following notation for the prominence of the bar: $\rm SB$, $\rm SA\underline{B}$, $\rm SAB,$ and $\rm S\underline{A}B$.
The underline notation means that the galaxy is more likely barred or non-barred, hosting a strong or a weak bar.
Magellanic and irregular barred galaxies (I) are also included in our analysis (Hubble type T=10). 
The maximum ellipticities and visual measurements of bars used in this study are taken from the catalogue of structures by \citet[][hereafter HE2015]{2015A&A...582A..86H}.

The Cosmic Flows project provides a complete database of H{\sc\,i} line profiles obtained at Green Bank in the USA and at Parkes in Australia
\citep{2009AJ....138.1938C,2011MNRAS.414.2005C}, from which we obtained 
the maximum circular velocity of our galaxies in a similar manner as for instance in \citet{2014AJ....147..134Z}:
\begin{equation}   
V_{\rm HI}^{\rm max}=W_{\rm mx}^{\rm av}/(2\,\rm sin\,i),
\label{gas_flows}
\end{equation} 
where $W_{\rm mx}^{\rm av}$ is the line width and $i$ is the P4 inclination.
For the galaxies for which Cosmic Flows data were not available ($\sim 40\,\%$), we obtained the gas velocity amplitude from HyperLEDA, and corrected it to correspond to P4 inclination. 
\section{Gravitational potential and calculation of bar forcing}
%
%
\subsection{Method for calculating bar-induced tangential forces}
We inferred the gravity potentials from the 3.6 $\mu$m images and constructed maps of the tangential forces ($F_{\rm T}$) normalized to the azimuthally averaged radial force ($F_{\rm R}$).
The calculations were made with the NIRQB-code \citep[][hereafter LS2002]{2002MNRAS.337.1118L}, which is based on the polar method of \cite{1999AJ....117..792S}.
Figure \ref{NGC4548_torque_maps} shows a typical example of force calculations in a barred galaxy, with a well-defined \emph{\textup{butterfly pattern}} \citep{2001ApJ...550..243B} in the torque map, 
which is roughly symmetric with respect to the bar major axis (with some twists caused by the spiral arms).
From the torque maps we constructed the radial profile of the normalized tangential force amplitude \citep[][]{1981A&A....96..164C}:
\begin{equation}
Q_{\rm T}(r)=\frac{{\rm max}\left( |F_{\rm T}(r,\phi)| \right)}{\langle |F_{\rm R}(r,\phi)|\rangle}.
\label{eqqb}
\end{equation}
\noindent The maximum of $Q_{\rm T}$ in the  bar region characterizes the bar strength, called $Q_{\rm b}$ in the following. The radial distance where the maximum occurs is denoted as $r_{\rm Qb}$.

Instead of computing the gravitational potential directly from the image pixels, as is done in the Cartesian method \citep[][]{1994ApJ...437..162Q}, the de-projected image was first decomposed into polar coordinates
\citep[][LS2002]{1999AJ....117..792S,2010ApJ...715L..56S}. This is based on the azimuthal Fourier decomposition of the surface densities in different radial zones:
\begin{equation}
\Sigma(r,\phi)=I_{0}(r)\left[1+\sum_{m=1}^{m=\infty} A_{m}(r)cos[m\big(\phi-\phi_{m}(r)\big)]\right].
\end{equation}
The amplitude of each Fourier component, $A_m$, was tabulated separately as a function of radius (we note that amplitudes are normalized with respect to the azimuthally averaged $m=0$ component).
Altogether, the summation was made over the even Fourier modes $m$=0-20. The main modes in the bar region are $m$=2, 4, 6, and 8 \citep[][LS2002]{1996ASPC...91...37O}. 
The maximum of the $m=2$ Fourier amplitude in the bar region was used as another proxy of the prominence of the bar (denoted by $A_{2}^{\rm max}$).
The radius where this maximum is found is called $r_{\rm A2}$.

The advantage of the polar method is that it effectively suppresses possible spurious maxima that may arise in the noisy outer parts of the images \citep[][]{2004ASSL..319..673S}.
The potential at the equatorial plane is obtained by 
\begin{equation}
\Phi_m(r,\phi,z=0)=-G\int^{\infty}_{0}r^{\prime}dr^{\prime}\int^{2\pi}_{0}\Sigma_m(r^{\prime},\phi^{\prime})g(\Delta r)d\phi^{\prime},
\end{equation}
\noindent where $\Delta r^{2}=r^{\prime^{2}}+r^{2}-2rr^{\prime}{\rm cos}(\phi^{\prime}-\phi)$, G is the gravitational constant,
and g($\Delta$r) is a convolution function including the pre-tabulated integration over the vertical direction (see LS2002).
We note that the azimuthal integration can be made with FFT for each $m$ while integration over radius is done by direct summation.
The assumptions made in our nominal force calculations are that 
(1) the mass-to-light ratio (M/L) is constant, 
(2) the vertical profile follows an exponential law,
(3) the scale height is constant over radius, 
(4) the scale height is tied to the disk scale length. 
The effect of varying the assumptions (1) and (3) are studied below (Sects. \ref{contaminants} and \ref{peanut}).
The role of the functional form of the density distribution (2) was explored in detail in LS2002, where it was concluded that $Q_{\rm b}$
is not sensitive to the detailed vertical profile as long as the same dispersion is assumed (for the exponential law $\sqrt{<z^2>}= 2h_z$).
\subsubsection{Bulge stretching correction}\label{bpdepro}
\begin{figure}
\centering    
\begin{tabular}{c c}
   \includegraphics[width=0.5\textwidth]{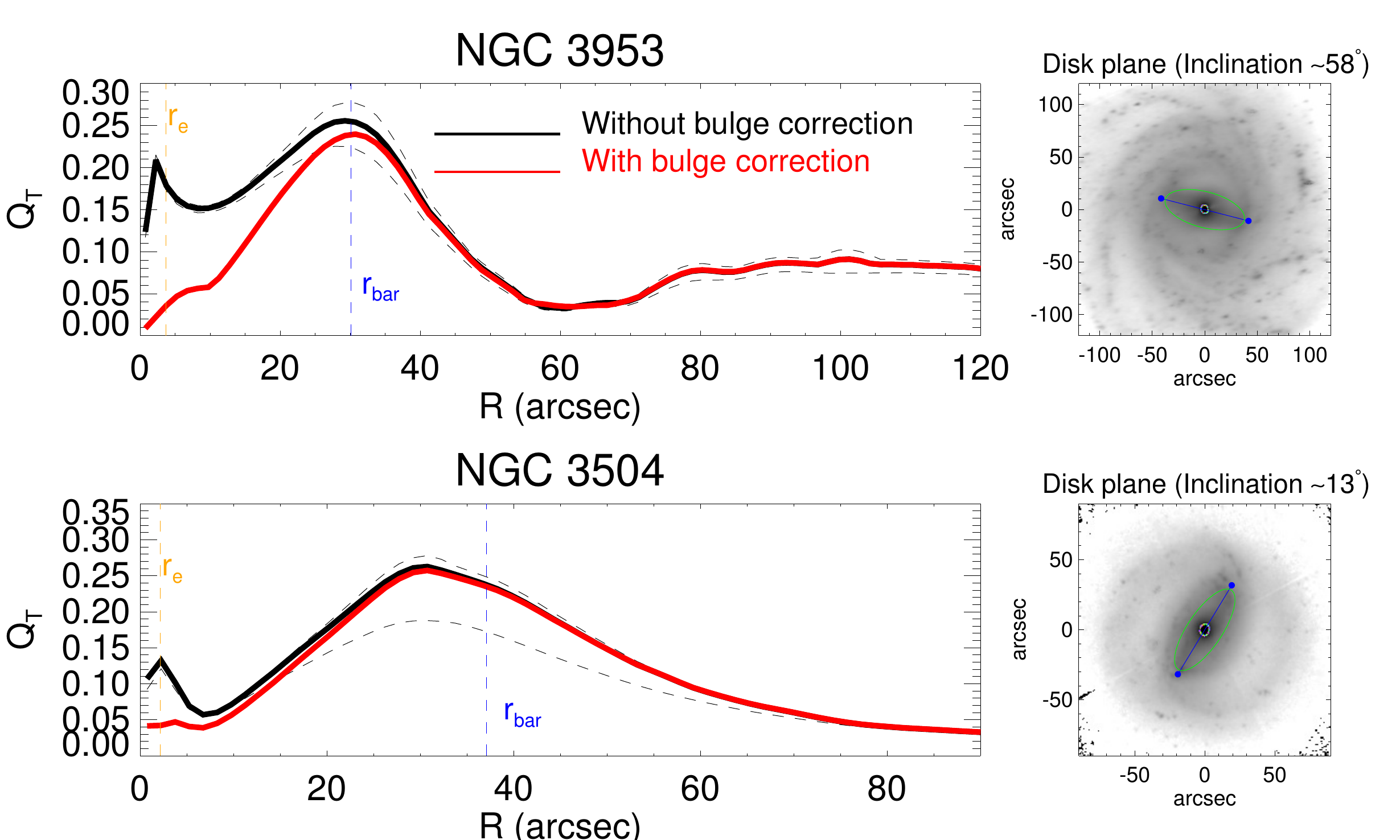}\\
   \includegraphics[width=0.5\textwidth]{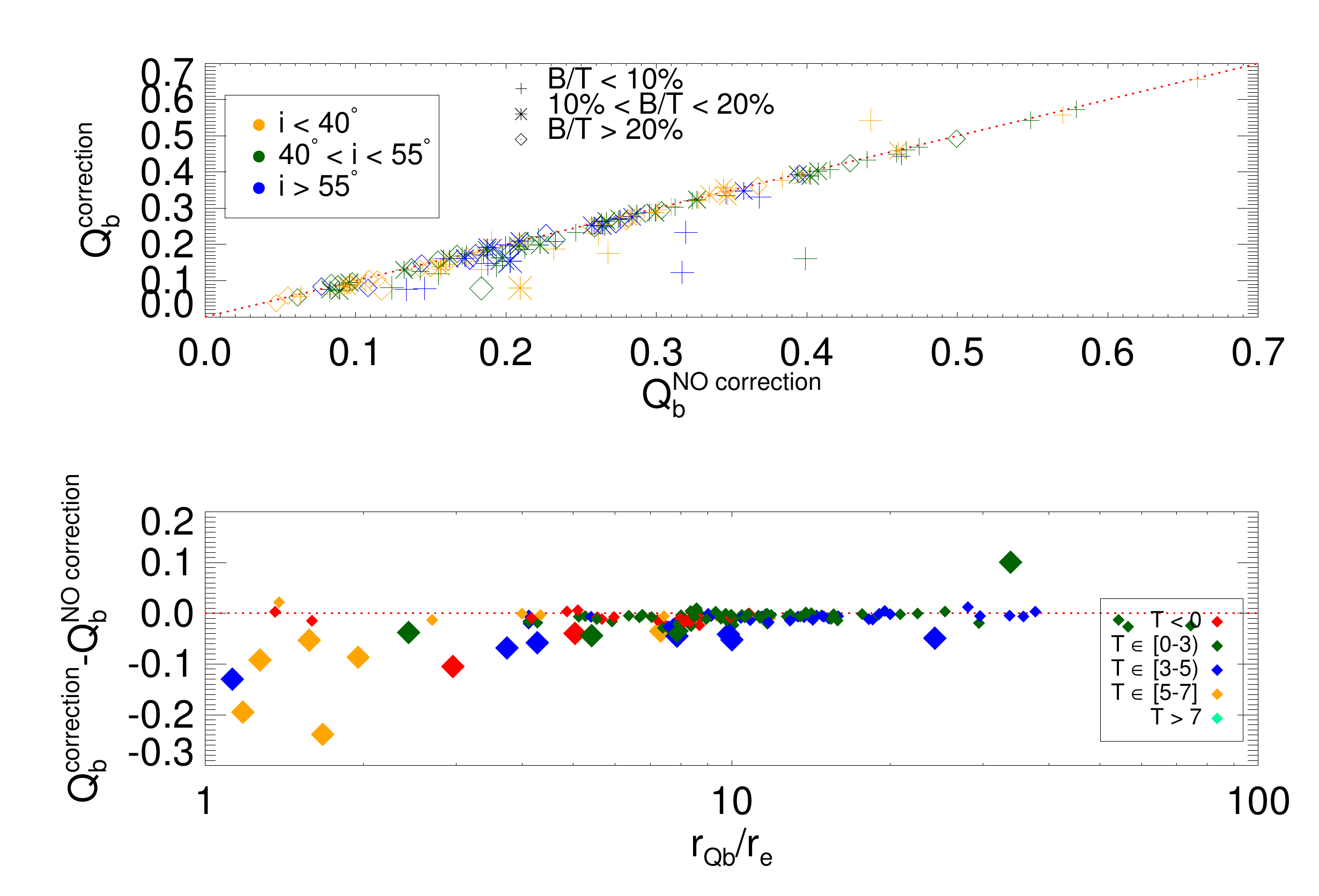}
\end{tabular}
\caption{
\emph{First and second rows:}
The radial force profile of NGC$\,$3953 and NGC$\,$3504 with and without spherical bulge correction.
The former (red line) represents the $Q_{\rm T}$ profile when the bulge flux is subtracted from the image before de-projection of the image to face-on orientation and added back afterwards for the potential calculation.
For the first galaxy, the global maximum of $Q_{\rm T}$, which corresponds to $Q_{\rm b}$, is slightly overestimated if no correction is considered, but remains within the limits determined by the disk thickness uncertainty (black dashed lines).
The visual estimate of the bar length and the effective radius of the bulge are indicated with vertical dashed lines.
\emph{Third and fourth rows:}
Comparison of $Q_{\rm b}$ measurements with and without bulge stretching correction for the galaxies with bulge in P4, 
with the data binned based on the disk inclination and the bulge-to-total ratio (P4). 
The strength of the correction is also studied in terms of the location of $r_{\rm Qb}$ with respect to the effective radius of the bulge from P4 ($r_{\rm e}$), separating the galaxies according to their morphological class.
The outliers in the third row are also indicated with larger symbols in row 4.
}
\label{bulge_corr}
\end{figure}
When necessary, a bulge stretching correction was also applied in the force calculations to prevent obtaining artificially strong tangential
forces caused by the bulge appearing elongated after de-projection \citep[as in e.g.][]{2004MNRAS.355.1251L}. 
Thus for a prominent bulge, its flux is subtracted from the image before de-projection of the image to face-on orientation. The obtained force
then represents the disk-only contribution. The contribution of the bulge force is added, calculated by assuming that the bulge has a
spherically symmetric intrinsic light distribution (in which case it is unambiguously determined by its projected light distribution). 
To obtain the flux of the bulge, the model parameters were taken from P4-decompositions (the fitted bulge model also corrects for the smearing due to the seeing) 
in which the model assumed elliptical bulge isophotes. We used an equivalent spherical radius that kept the fitted bulge flux constant.
The typical result of the bulge correction is to remove the artificial inner peaks of the
$Q_{\rm T}$ profile in the bulge region (see the examples in the two upper frames of Fig.~\ref{bulge_corr}). Naturally, the assumption of spherical bulges does
not work well for non-classical disky bulges: the force calculations were thus also re-run without any bulge correction,
that is,\emph{} assuming that the bulge has the same flattening as the disk.  A third possibility, that the bulge represents the inner
boxy/peanut part of the bar, is briefly addressed in Sect.~\ref{peanut}. To avoid over-correcting the images, the P4-decompositions were visually inspected to exclude highly extended bulge fits.

Altogether, the bulge correction and its uncertainties are expected to affect only galaxies that have relatively extended bulges compared to
the bar size \citep[][]{2004MNRAS.355.1251L}. Indeed, as shown in Fig.~\ref{bulge_corr}, the strength of the bulge correction is not
very relevant in the statistical sense. The few outliers in the plot correspond to galaxies with very small bars, with $r_{\rm Qb}/r_{\rm e}$ close
to unity, where $r_{\rm e}$ is the effective radius of the bulge. In total, 251 of the galaxies in our sample of barred galaxies have a bulge
fitted in P4, of which 145 galaxies present central peaks in the force profiles that are due to bulge de-projection.  However, only in 65 cases did bulge
stretching change the $Q_{\rm b}$ value noticeably. For these we obtain median $Q_{\rm b}^{\rm corrected}/Q_{\rm b}$ of $0.9 \pm 0.02$
and a mean $|\Delta Q_{\rm b}|$ of $0.03$. Overall, the role of bulge correction is small, and the different treatments used are expected to delimit the uncertainty in bulge contribution to $Q_{\rm b}$ quite well.
\subsubsection{Vertical scale height}\label{thickmethods}
We estimated the vertical scale height based on the empirical relation from \citet[][]{1998MNRAS.299..595D}, in which it is shown that the
ratio of the disk vertical thickness to the disk scale length depends on the Hubble type. The morphological stage binning and the
corresponding mean values (and range) are the same as in \citet[][]{2004MNRAS.355.1251L}: $h_{\rm R}/ h_{\rm z} = 4 \, (1-5)$ if $T\le1$, $ 5 \, (3-7)$ if $T\in[2,4]$  and  $9 \, (5-12)$ if $T\ge5$.
To estimate the largest uncertainties on the forces, 
the gravitational potentials were also calculated using the highest and lowest values for each bin (the number listed in the ranges inside the parenthesis).

When $h_{\rm R}$ values were not reliable (129 cases in the sample of 1345 non-highly inclined disk galaxies have quality flags lower than 5 in P4), we assumed that the vertical scale height scales
with the disk size as $h_{\rm z}=0.1r_{\rm k_{20}}$,
where $r_{\rm k_{20}}$ is the K-band isophote radius at 20 mag arcsec$^{-2}$ from 2MASS\footnote{2MASS is a joint project of the University of Massachusetts and the Infrared Processing and Analysis 
Center/California Institute of Technology, funded by the National Aeronautics and Space Administration and the National Science Foundation.}. 
This empirical relation, also used in \citet{2010ApJ...715L..56S}, was found in \citet[][]{2008MNRAS.383..317S} to approximate the de Grijs relation in an adequate manner 
\footnote{The bar torque parameter ($Q_{\rm b}$) was also computed based on this relation \citep[][]{2008MNRAS.383..317S} for all the barred galaxies in our sample with an identifiable maximum torque. It is not used in the discussion 
except for Figs.~\ref{vert_thick} and~\ref{halo-qb}. Both approaches for the disk thickness determination give very similar $Q_{\rm b}$ values.
All the measurements are listed in Table~\ref{truncated1}.}.
Finally, we used $a_{25}$/1.5 in substitution of $r_{\rm k_{20}}$ when neither $r_{\rm k_{20}}$ nor $h_{\rm R}$ values were available (17 galaxies).
Here a$_{25}$ denotes the radius of the $B$-band surface brightness isophote of 25 mag arcsec$^{-2}$, taken from HyperLEDA. This relation between these isophotal B- and K-band radii is purely empirical.

The uncertainties of our force calculation method are summarized by \citet[][]{2004MNRAS.355.1251L}, where it was shown that the largest
expected contribution (10-15 \% in the value of $Q_{\rm b}$) comes from the uncertain vertical thickness. 
On the other hand, the functional form of the vertical density distribution (LS2002), superposition of spiral arms
\citep[][]{2003AJ....126.1148B}, the position angle of the bar relative to the line of nodes \citep[][BLS2004
  hereafter]{2004AJ....127..279B}, possible contribution of the dark matter halo (BLS2004), or the radial variations in the vertical scale
height (LS2002) were estimated to affect $Q_{\rm b}$ by only about 5$\%$.
However, since our current sample extends to galaxies of much later type, a more thorough estimation of the effects of the dark matter halos is needed (see Sect.~$\ref{halo_section}$).
\subsection{Circular velocity curves}
We used the mass-to-light ratio at 3.6 $\mu$m by E2012 ($\Upsilon_{3.6 \rm \mu m}=0.53$)
and assumed it to be constant throughout the galaxy to obtain the stellar contribution to the circular velocity curve:
\begin{equation}
V_{\rm 3.6\mu m}(r)=\sqrt{\Upsilon_{3.6 \rm \mu m}\left<F_{\rm R}(r)\right> r}, 
\end{equation}
where $r$ is the galactocentric radius and $\rm F_{\rm R}$ is the radial force calculated for $M/L=1$.
We calculated the maximum of these rotation curves ($V^{\rm max}_{\rm 3.6\mu m}$) and the radius where $V_{\rm 3.6\mu m}$ peaks ($r^{\rm max}_{\rm 3.6\mu m}$) for all the 1345 disk galaxies with inclinations lower than $65^{\circ}$,
regardless of whether they host a stellar bar or not.
Uncertainties in the determination of $V_{\rm 3.6\mu m}$ are partly influenced by the disk scale height uncertainty (see Fig. \ref{NGC4548_torque_maps}).
However, unlike in the calculation of normalized tangential forces, the uncertainty in the mass-to-light ratio ($\sim30\%$ according to E2012) is dominant.
Variations in $\Upsilon_{3.6 \rm \mu m}$ are associated with the $[3.6]-[4.5]$ colour \citep[E2012,][]{2014ApJ...788..144M},
but the colour differences in the S$^{4}$G sample are small enough {\citep[average of $-0.41\pm0.08$ mag (AB), according to][]{2013ApJ...771...59M},} to safely apply the same mass-to-light ratio to all Hubble types without any systematic error.

Furthermore, we obtained the inner stellar velocity gradient of $V_{\rm 3.6\mu m}$, d$_{\rm R}V_{\rm 3.6\mu m}(0)$, by fitting the inner rotation curve with a polynomial function of order $m=3$
and taking the linear term as an estimate of the inner slope \citep[in a similar manner as in][]{2013MNRAS.433L..30L}.
\subsection{$Q_{\rm b}$ from 3.6 $\mu$m images and from P5 mass maps}\label{contaminants}
%
%
\begin{figure}
     \centering
     \begin{tabular}{c c}
     \includegraphics[width=0.44\textwidth]{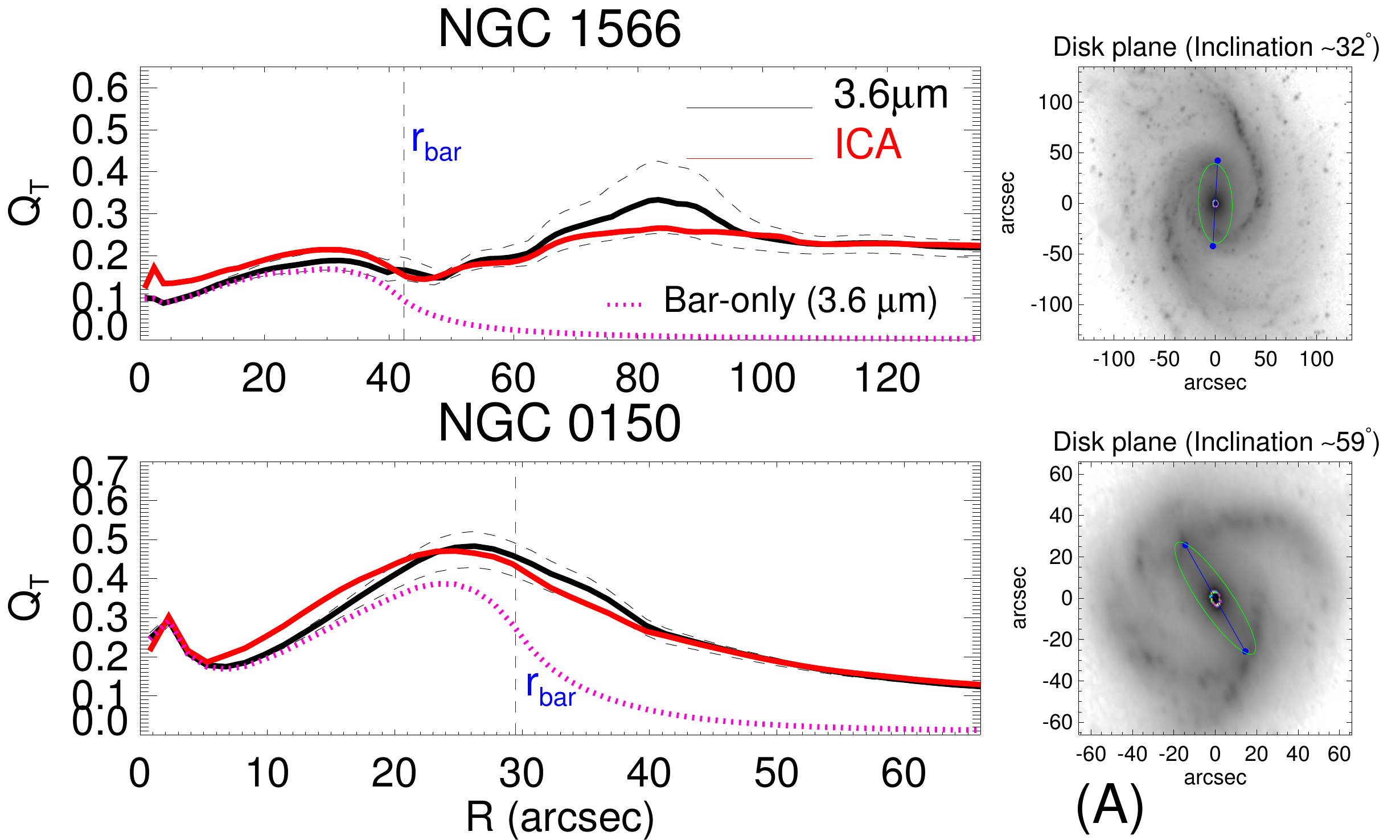}\\
     \includegraphics[width=0.44\textwidth]{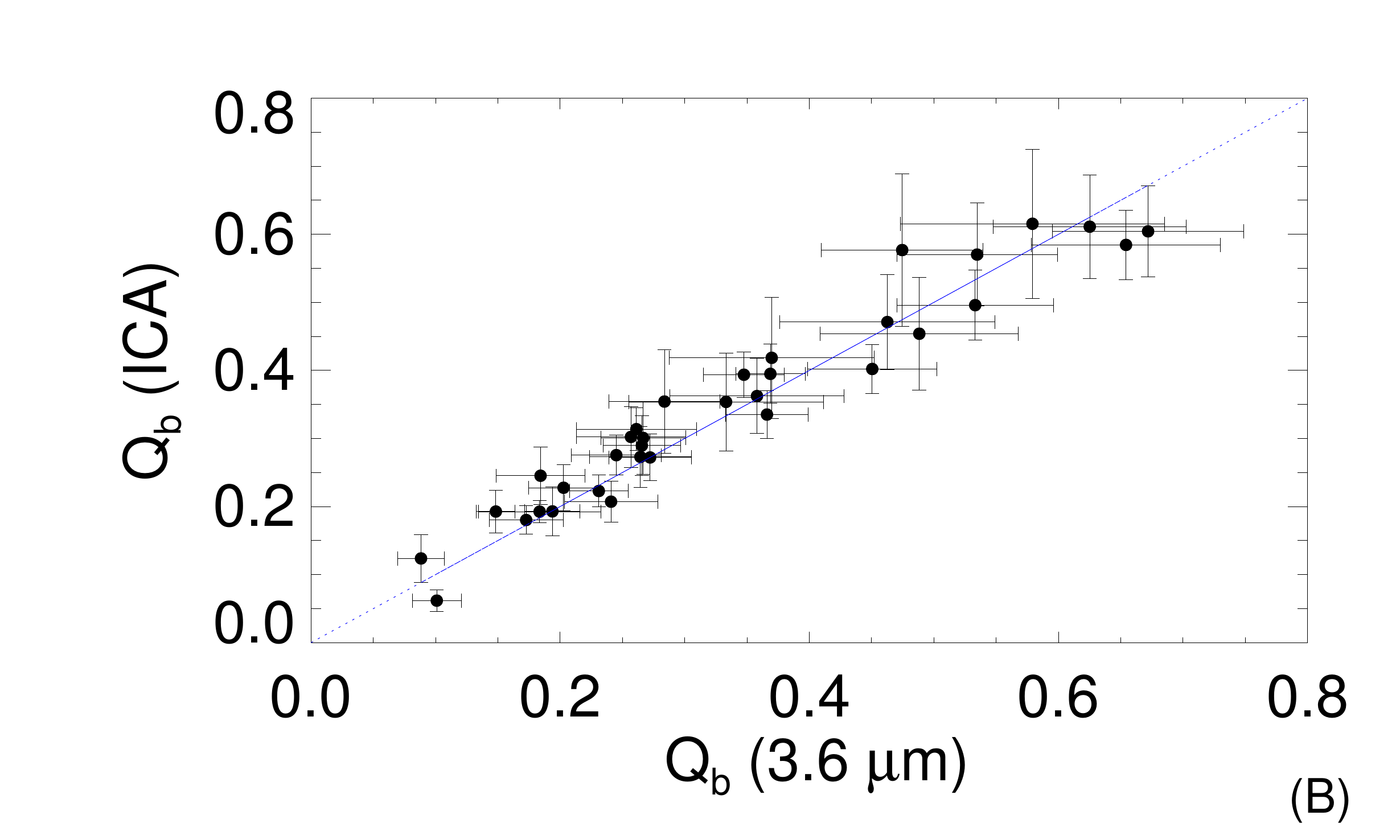}
     \end{tabular}
\caption{
\emph{Panel A}: Radial force profiles derived from $3.6\,\mu$m imaging (solid black line) and from P5 mass maps (red solid line) for the barred galaxies NGC$\,1566$ and NGC$\,0150$.
For NGC$\,1566$, which is representative of the galaxies with the largest difference in the force profiles due to non-stellar contaminants,
the $Q_{\rm T}$ derived with the P5 mass map slightly differs from the raw $Q_{\rm T}$, but the deviation is lower than the uncertainty associated with the disk thickness, whereas for NGC$\,0150$ the $Q_{\rm b}$ is roughly the same.
We also show the ``bar-only'' $Q_{\rm T}$ profiles using $3.6\,\mu$m images for these two galaxies.
\emph{Panel B}: 
For a sample of 34 barred galaxies, we compare the gravitational torque parameter as measured from 3.6~$\mu$m and ICA corrected maps.
The error bars are determined by the disk thickness uncertainty.
No systematic deviation is observed.
}
\label{parameter_comparisons_mass}
\end{figure}
In spite of the azimuthal smoothing in our polar method and the use of near-IR imaging, non-stellar emission in H{\sc\,ii} regions,
which also appears along the bar, may affect the $Q_{\rm T}$ radial profiles. This is the case particularly for the low-luminosity late-type galaxies,
in which this clumpiness induces local maxima that can be mixed with the bar-induced amplitude maxima.

To test the impact of non-stellar contaminants in our estimates, the force calculations made for the direct 3.6~$\mu$m images
were repeated for a random sample of 72 barred galaxies using the P5 mass maps \citep[][]{2015ApJS..219....5Q}, including only galaxies with reliable ICA analysis.
For $34$ of these galaxies, it was possible to reliably identify $Q_{\rm b}$ .
Two examples of force profile comparisons are shown in Fig. \ref{parameter_comparisons_mass}: P5 indicates a large contribution of non-stellar contaminants for NGC 1566,
in particular associated with the spiral arms, whilst for NGC$\,$0150, the role of contaminants in the bar strength calculation seems less important.
Figure \ref{parameter_comparisons_mass} also shows the comparison of bar strengths derived from the direct images and from the mass maps for our sub-sample. 
The deviation in this relation is $\sim$~10-15~$\%$ (see also Fig.~\ref{ICA_36_statistics}).
The median $Q_{\rm b}(\rm ICA)/Q_{\rm b}(3.6\,\mu \rm m)$ is $1.06\pm0.03$, and the mean $|\Delta Q_{\rm b}|=|Q_{\rm b}(\rm ICA)-Q_{\rm b}(3.6\,\mu m)|$ is $0.03$,
which is smaller than the estimated uncertainty related to the thickness of the disk.
Although some individual galaxies can be affected by the non-stellar contaminants, there is no systematic difference in the two $Q_{\rm b}$ values. 

It might naively be expected that using mass maps would increase the number of low-luminosity galaxies for which bar strengths can be reliably estimated.
However, even though the ICA method eliminates non-stellar components that are associated with star-forming regions, for most of the low-luminosity galaxies in
our sub-sample the $Q_{\rm T}$ profiles also remain noisy when mass maps are used.

In Appendix~\ref{appendix4} we also discuss the effect of non-stellar contaminants on the Fourier amplitudes of our bars and on the calculation of the stellar component of the circular velocity curves.
%
%
\subsection{Contribution of spiral arms}
The $Q_{\rm T}$ profile of NGC$\,$1566 (see the upper plot in Fig.~\ref{parameter_comparisons_mass}) is representative of a barred galaxy with prominent spiral arms, presenting a local maximum
due to the bar perturbation at a radius of $\sim$~30 arcsec, while the global maximum at $\sim$~85 arcsec corresponds to the spiral arms.  

We wish to determine the contribution of the spiral arms to our bar force measurements. \citet{2003AJ....126.1148B} developed a
Fourier-based method for filtering the spiral arm contribution to the bar strength measurements in near-infrared images. 
They assumed that the $A_2(r)$ Fourier amplitude profile due to the bar alone is symmetric with respect to its maximum, $r_{\rm A2}$,
if the effect of the spiral arms were suppressed, and used this symmetry assumption to eliminate the spiral contributions from $A_2$ (and likewise from $A_4$).
However, this method is fairly time-consuming to apply and it is unclear how well it applies for the galaxies presenting inner non-axisymmetric structures (e.g. double barred galaxies)
and for the early-type galaxies that have an ansae morphology. Moreover, using measurements from the OSUBSG sample, \citet{2007MNRAS.381..401L} showed that the trend of $Q_{\rm b}$ in the
Hubble sequence was not affected by whether this spiral arm correction was applied or not.

\citet[][]{2010ApJ...715L..56S} found a statistically significant correlation between the local bar-only forcing and the
local spiral density amplitude. In this study the spiral contribution to forcing was eliminated simply by setting the $m>0$ Fourier density
amplitudes to zero beyond a certain radius ($r_{\rm cut}$) (the polar method facilitates this separation). We followed a similar method here and chose $r_{\rm cut}$ to equal the bar length 
(see Fig.~1 in \citet[][]{2010ApJ...715L..56S} and Fig.~\ref{parameter_comparisons_mass} in this work for an illustration of force profiles with and without spiral contribution).

We denote as $Q_{\rm b}^{\rm bar-only}$ the value of the bar torque parameter after excluding the spiral arm contribution to the force maps in the bar region. 
In the statistical sense, bar-only force measurements do not deviate significantly from the raw $Q_{\rm b}$ values ($\left<Q_{\rm b}^{\rm bar-only}/Q_{\rm b}\right>=0.89$; $\sigma=0.12$),
in agreement with \citet{2007MNRAS.381..401L}. 
To test the sensitivity of $Q_{\rm b}^{\rm bar-only}$ to the assumed $r_{\rm cut}$, we also computed the bar-only profiles after setting $r_{\rm cut}=1.2 r_{\rm bar}$, obtaining a mean $Q_{\rm b}^{\rm bar-only}/Q_{\rm b}$ of $0.97$.
\subsection{Radial profile of the vertical disk thickness}\label{peanut}
\begin{figure}
   \centering
   \includegraphics[width=0.5\textwidth]{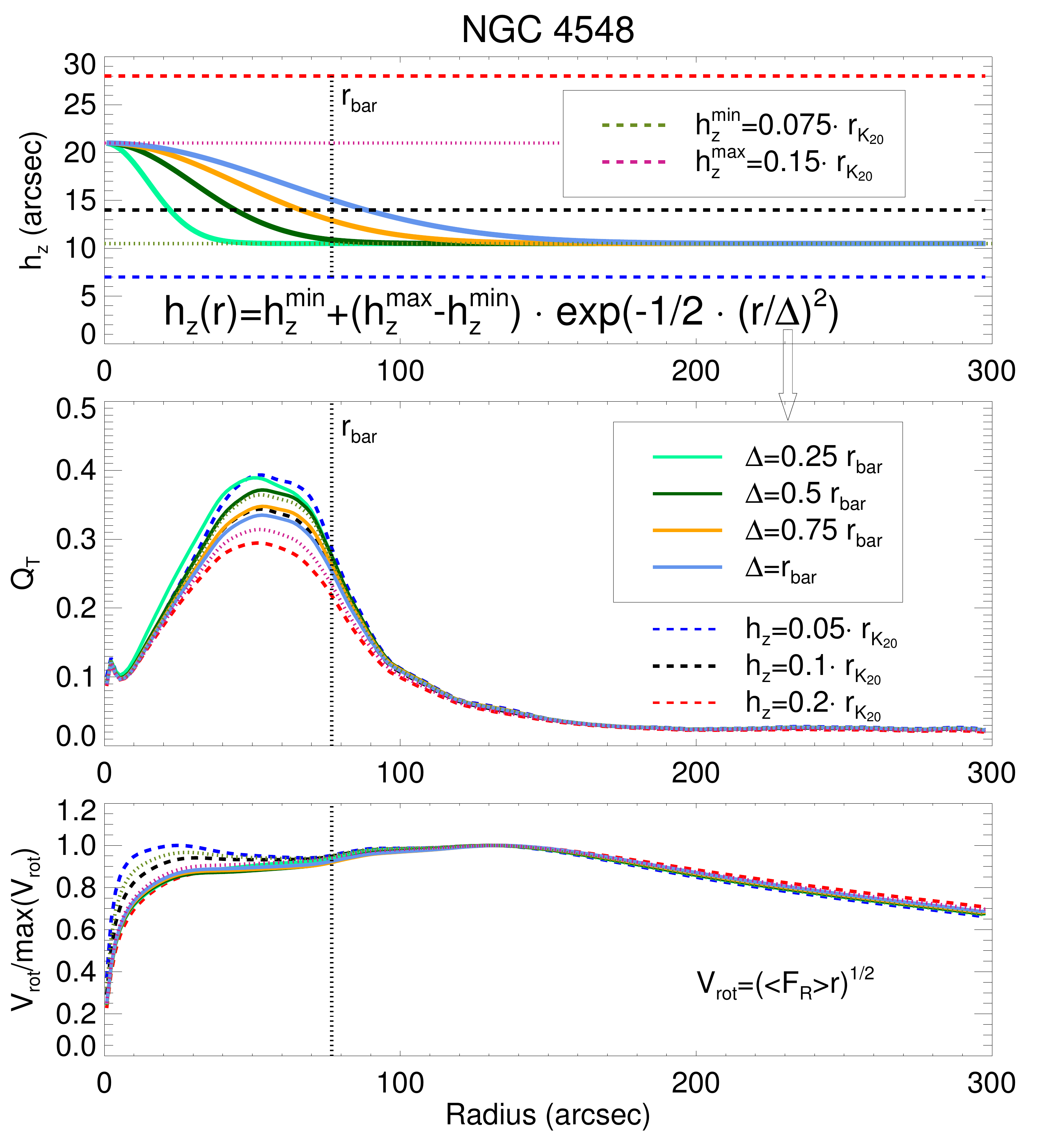}
\caption{
\emph{First panel:}
We compare different models of a disk scale hight profile for NGC$\,4548$.
The profiles assume a Gaussian function, with the spread $\Delta$ defined in terms of the bar size.
The black horizontal dashed line represents the nominal thickness to this particular galaxy, while the uppermost and lowermost lines correspond to the limits for the disk thickness as determined in \citet[][]{2008MNRAS.383..317S}.
\emph{Second and third panels:}
Radial force profiles and rotation curves inferred from the potential associated with the distance-dependent exponential scale heights defined above (solid lines)
and under the assumption of radially constant disk thickness (dashed lines).
}
\label{vert_thick}
\end{figure}
\begin{figure}
   \centering  
   \begin{tabular}{c c}
   \includegraphics[width=0.5\textwidth]{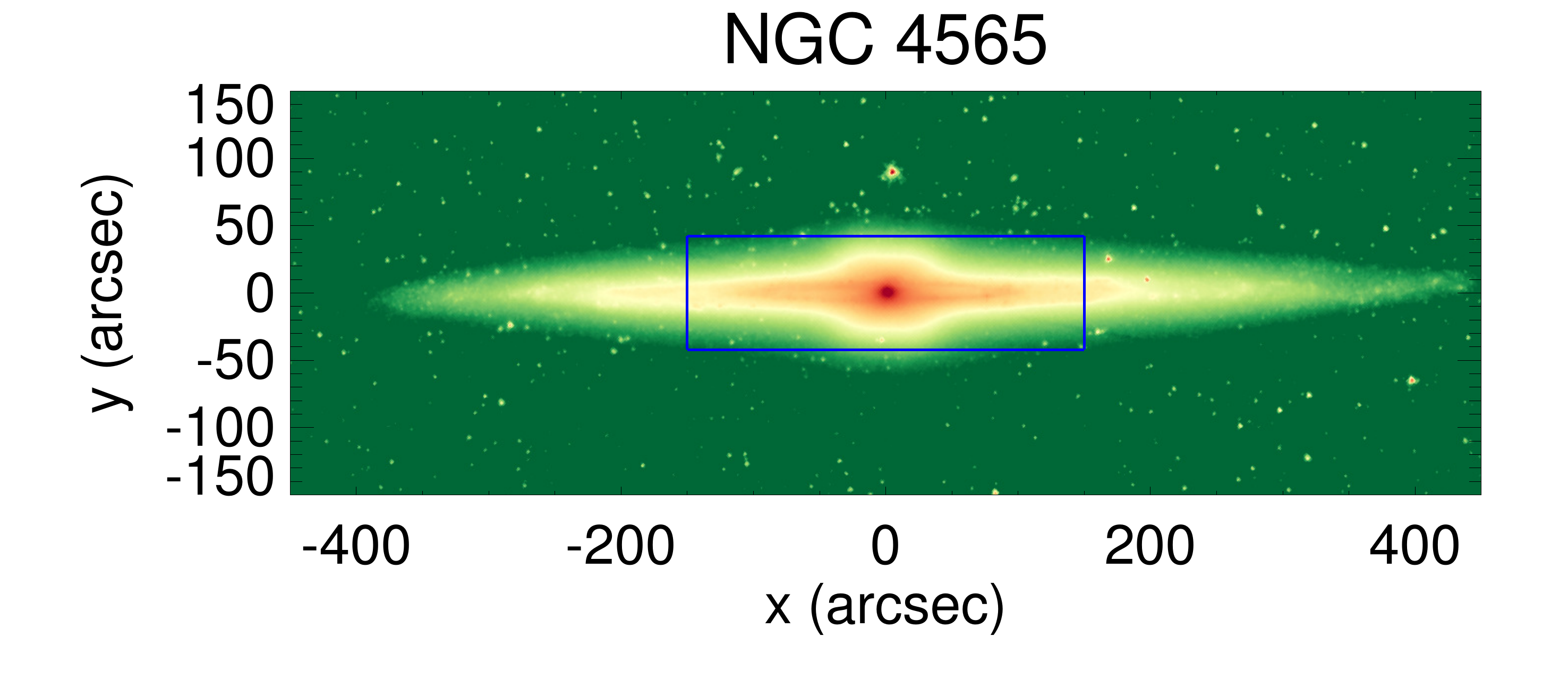}\\[-5ex]
   \includegraphics[width=0.5\textwidth]{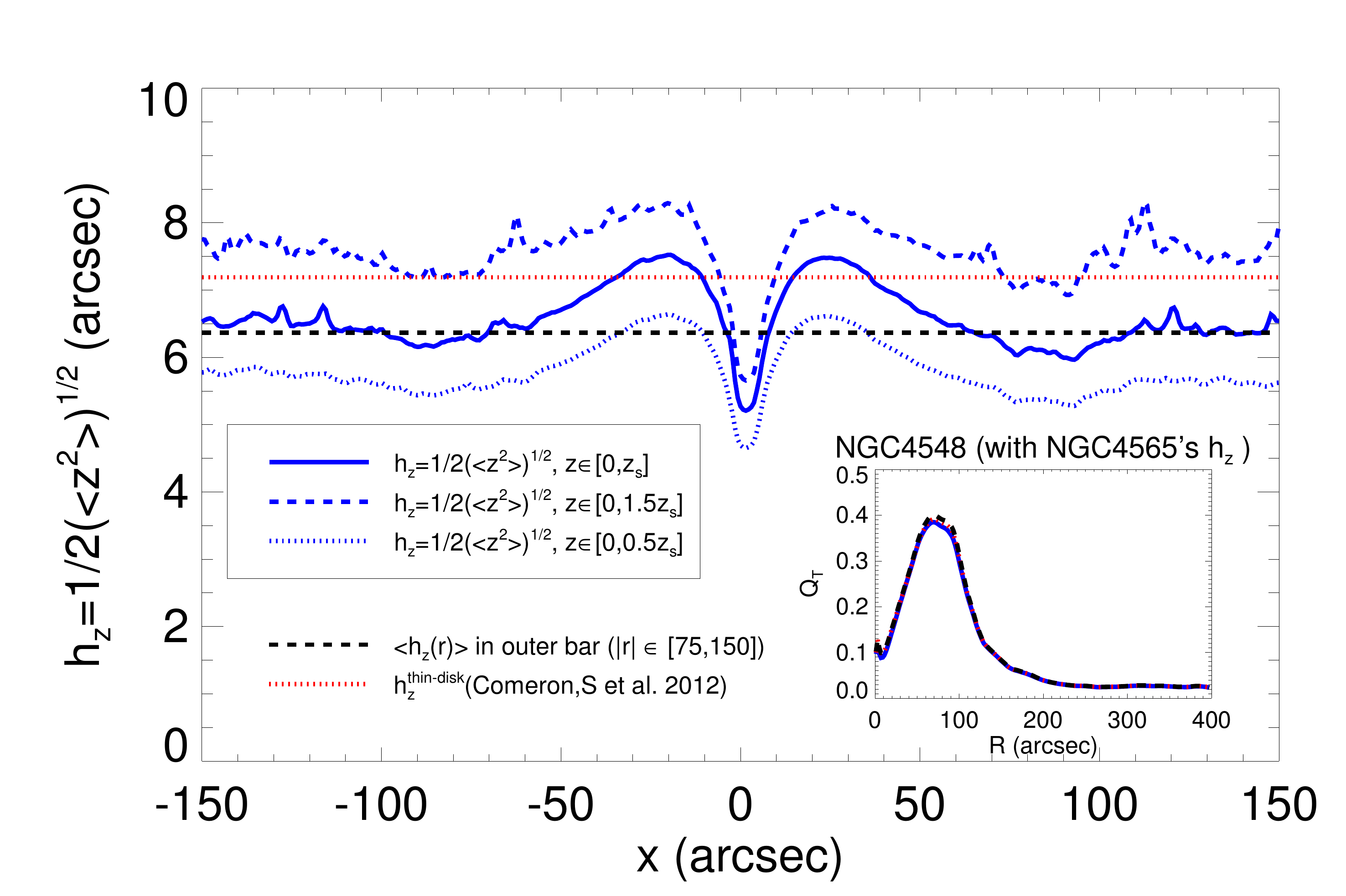}
   \end{tabular}
\caption{
\emph{First panel:}
Image of the edge-on galaxy NGC$\,$4565, which hosts a prominent peanut-shaped bar, in magnitude scale with range 17-24 $\mu_{3.6\mu \rm m}(AB)$.
The galaxy has been rotated so that the major axis appears horizontal in the image.
The blue box roughly covers the bar region and extends vertically as far as $z_{\rm s}$,
which is the height above which $90\%$ of the light emission comes from the thick disk component.
\emph{Second panel:}
Disk scale height estimated from the line-of-sight vertical dispersion.
The continuous, dashed, and dotted blue lines correspond to the resulting $h_{\rm z}$ integrating up to $z_{\rm s}$, $1.5z_{\rm s}$ , and $0.5z_{\rm s}$, respectively.
For the first case, the black dashed horizontal line shows the mean $h_{\rm z}$ outside the range of the peanut-shaped bar.
The horizontal lines trace the thin disk $h_{\rm z}$ estimated from the fits to the vertical luminosity profiles (red).
The resulting $Q_{\rm T}$ profiles from the different disk scale
heights (solid blue, dashed black, and dotted red lines; with the meaning indicated in the legends) are shown in the lower right plot.
}
\label{vert_thick_edge_on}
\end{figure}
\noindent 
Our standard force calculations assumed a constant scale height throughout the disk. 
As mentioned above, LS2002 showed that shallow gradients in vertical thickness do not have a significant effect on $Q_{\rm b}$
compared to that obtained by assuming a constant average $h_{\rm z}$.

Here we determine the change in $Q_{\rm T}$  for a more strongly
varying disk thickness, for example in the presence of inner thicker
structures such as the boxy/peanut bulges. These are vertically thick inner regions of bars that have gone through a buckling episode
\citep[see simulations of][]{1981A&A....96..164C,1990A&A...233...82C,1991Natur.352..411R}
and are readily identifiable in observations of nearly edge-on galaxies \citep[][]{1986AJ.....91...65J,2000A&AS..145..405L};
they appear as X-shapes in unsharp mask images \citep[][]{2006MNRAS.370..753B,2014MNRAS.444L..80L}.
Moreover, it has been suggested by \citet[][]{2007MNRAS.381..401L} that the round inner parts of bars seen in many more face-on galaxies, termed barlenses \citep[][]{2011MNRAS.418.1452L}, might in fact be the same phenomenon.
Recent support for this speculation has been given in \citet[][]{2014MNRAS.444L..80L} and \citet[][]{2014arXiv1405.6726A}.

To obtain a rough estimate of the possible effect of a thickened central bar component, we calculated the gravitational forces by assuming that the disk scale height declines radially from 
$h_{\rm z}^{\rm max}=0.15\cdot \rm r_{\rm K_{20}}$ to $h_{\rm z}^{\rm min}=0.075\cdot\rm r_{\rm K_{20}}$ following a function
\begin{equation}
h_{\rm z}(r)=h_{\rm z}^{\rm min}+(h_{\rm z}^{\rm max}-h_{\rm z}^{\rm min}) \cdot {\rm exp}\big(-1/2 \cdot (r/\Delta)^{2}\big),
\label{eq_thick}
\end{equation}
with the range $\Delta$ varying from $0.25r_{\rm bar}$ to $r_{\rm bar}$.
In Fig.~\ref{vert_thick} we assess the effect of the disk scale
height variations on the $Q_{\rm b}$ measurement for the particular case of NGC$\,$4548.
For the studied range of $\Delta$s, we observe that the differences in $Q_{\rm b}$ are of the same order as the uncertainty arising
from the unknown disk thickness itself (for the model with the smallest $\Delta$, the force is enhanced by $\sim 15\%$).
The rotation curves derived from the azimuthally averaged radial forces were also compared and led to a similar conclusion: 
the resulting curves show very similar values for the four models (they drop by less than $10\%$ at the thick bar region 
and they closely approach the rotation curve associated with the thickest disk under the constant $h_{\rm z}$ approximation).

Our selection of $h_{\rm z}^{\rm max}$ and $h_{\rm z}^{\rm min}$ implies a difference of a factor of two between the inner and outer bar thicknesses, which is taken as an upper bound.
To verify whether this limit is realistic, we examined the vertical profile of the edge-on galaxy NGC$\,$4565 (see Fig.~\ref{vert_thick_edge_on}),
which is known to host a prominent peanut-shaped bulge \citep[][]{1986AJ.....91...65J}.
For this particular case, we estimated the disk scale height from the line-of-sight vertical dispersion:
\begin{equation}
h_{\rm z}(x)=\frac{1}{2}\sqrt{<z^2>}=\frac{1}{2}\sqrt{\frac{\int_{-z_{\rm s}}^{z_{\rm s}}S(x,z)z^2dz}{\int_{-z_{\rm s}}^{z_{\rm s}}S(x,z)dz}},
\end{equation}
where $x$ and $z$ are the axial and vertical distances, $S$ is the flux density, and $z_{\rm s}$ is the height above which $90\%$ of the light emission comes from the thick disk \citep[taken from][]{2012ApJ...759...98C}.
Varying $z_{\rm s}$ by a factor 2 has no significant effect on the final $h_{\rm z}$ radial profile, as shown in the same figure.
We obtain a mean $h_{\rm z}$ value in the bar region ($\sim 7$ arcsec) consistent with \citet[][]{2012ApJ...759...98C}, who performed
fits to the vertical luminosity profiles in the disk region outside the peanut-shaped bar area.
More importantly, we observe a difference between the maximum $h_{\rm z}$ at the peanut-shaped bulge and that
of the surrounding thin bar/disk ($<50\%$) that is much lower than the above tested values.
The same most likely holds for the true $h_{\rm z}$ as a function of the galactocentric distance.

In the same figure we also show the force profile of NGC$\,$4548 assuming it has the same scale height as NGC$\,$4565, 
under the assumption that the peanut-shaped part covers one half of the bar size \citep[][]{2000A&A...362..435L}.
These two galaxies supposedly have similar bar lengths, and both are classified as Sa in B2015.
The resulting $Q_{\rm T}$ barely deviates from the force profile under the assumption of constant thickness.

Using an image taken from an N-body+SPH simulation, \citet[][]{2015MNRAS.450..229F} concluded that $Q_{\rm b}$ can be strongly overestimated by not taking into account the boxy/peanut bulge geometry (regardless of measurement uncertainties). 
Determining the scale height in the peanut-shaped region from observed non-highly inclined galaxy images is complex (as discussed also in this work).
We have analysed different models of a thick inner bar (twice the thickness of the surrounding thin disk), with the functional form of the scale height radial profiles defined in terms of the bar length.
It is probably safe to conclude that the uncertainty on $Q_{\rm b}$ coming from the assumption of a constant $h_{\rm z}$
is not larger than the uncertainty arising from the disk thickness determination itself.
In addition, for a complete view of the possible effect of the boxy/peanut bulge, projection effects on the analysis
also need to be included \citep[in a similar manner as in][]{2004ASSL..319..673S},
which can be particularly strong in the inner regions (as discussed in Sect.~\ref{bpdepro} without taking into account the detailed bulge geometry).
%
%
\subsection{Dark matter halos}\label{halo_section}
\begin{figure}
   \centering  
   \includegraphics[width=0.5\textwidth]{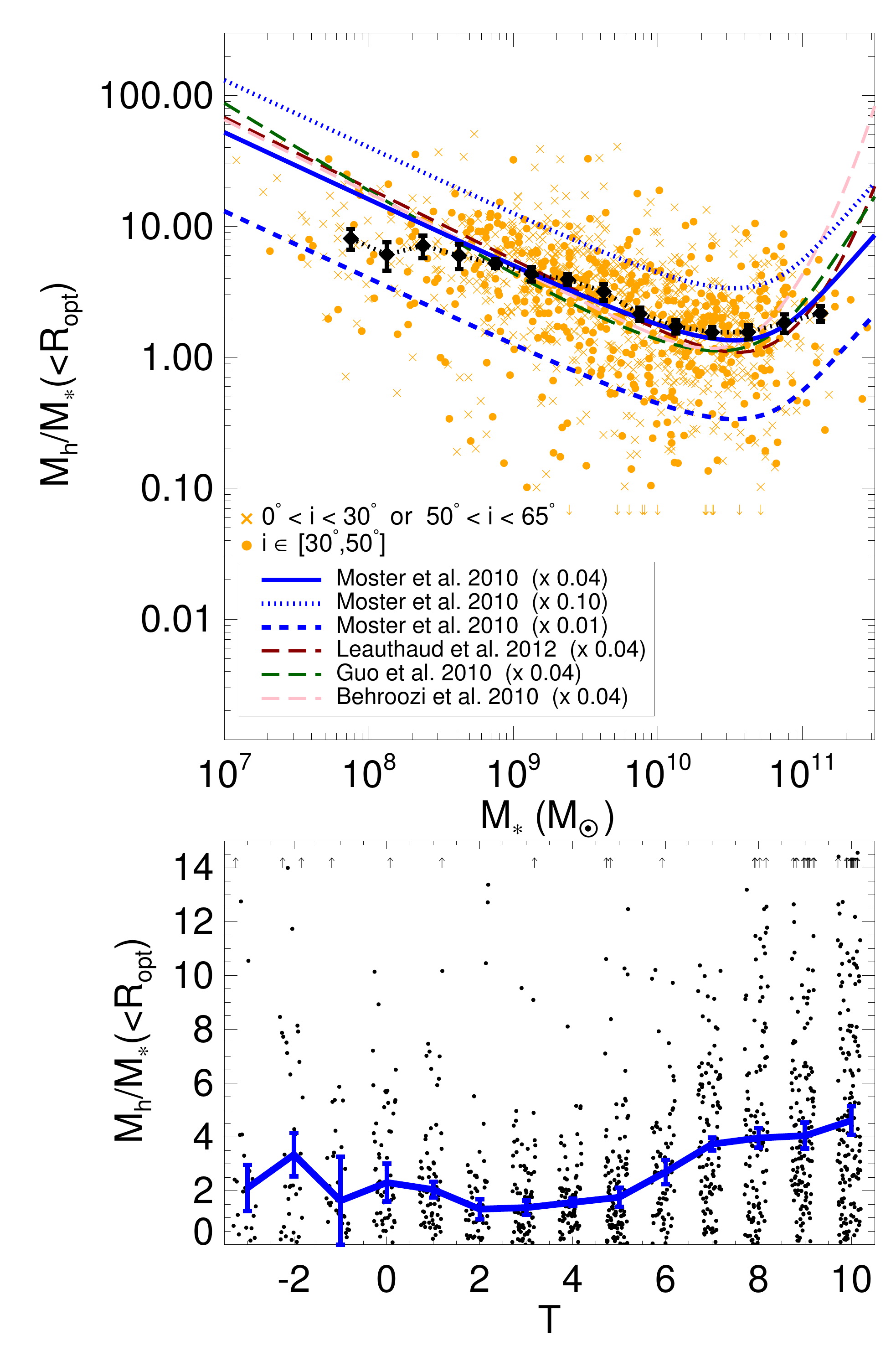}
\caption{
\emph{Upper frame:}
The halo-to-stellar mass ratio within the optical radius for the non-highly inclined spiral galaxies ($i<65^{\circ}$) in the S$^{4}$G sample with usable H{\sc\,i} maximum velocities (1227 systems).
Galaxies with $V_{\rm HI}^{\rm max}>400 $ km s$^{-1}$  (gas-poor galaxies with unreliable velocity measurements) are excluded from our statistics.
Both barred and non-barred systems form part of this sample.
Moderately inclined galaxies ($30^{\circ}<i<50^{\circ}$) are plotted with orange filled circles, while more inclined and face-on galaxies are displayed with orange $X$ symbols:
 for both sub-samples the statistical trend is practically the same, which means that uncertainties in the velocity measurements due to inclination are not a major issue here.
The black dots represent the total running median of our data and the error bars are determined by applying the bootstrapping statistical method.
The different lines show the estimates in the literature for the \textup{\textup{{\em \textup{total}}}} halo-to-stellar mass ratio once these have been scaled down by the factor specified in the labels.
\emph{Lower frame:}
$M_{\rm h}/M_{\ast}(<R_{\rm opt})$ as a function of Hubble type.
The running median is overplotted in blue (error bars obtained via bootstrapping).
}
\label{stellar-halo-relation}
\end{figure}
Our goal is to estimate the amount of dark matter halo in the disk regions of our sample galaxies and to quantify the dark halo contribution to the mean axisymmetric radial force field, and thereby to the torque parameter $Q_{\rm b}$.
To do this accurately would require a sophisticated kinematic decomposition of observed rotation curves, but clearly such a study is not feasible given the size of the sample and the lack of rotation curve measurements for all of the galaxies. 
\subsubsection{Halo-to-stellar mass ratio}\label{SHMR}
We obtained a first-order estimate of the halo-to-stellar mass ratio ($M_{\rm h}/M_{\ast}$) inside the optical disk by comparing the
circular velocity curve calculated from the 3.6 $\mu$m images with the inclination-corrected H{\sc\,i} velocity amplitude ($V_{\rm HI}^{\rm max}$). We assumed that at the optical radius $R_{\rm opt}$ (the radius enclosing $83\%$ of the light in the blue
band; $R_{\rm opt}\approx3.2h_{\rm R}$ for an exponential disk) the circular rotation velocity is close to the observed maximum velocity ($V_{\rm HI}^{\rm max}$):
\begin{equation}
(V_{\rm HI}^{\rm max})^{2}\approx V_{\rm 3.6\mu m}^{2}(R_{\rm opt})+V_{\rm halo}^{2}(R_{\rm opt})
.\end{equation}
\noindent From this we can infer
\begin{equation}
M_{\rm h}/M_{\ast}(<R_{\rm opt})\approx F\cdot\bigg(\frac{(V_{\rm HI}^{\rm max})^{2}}{V_{3.6 \mu \rm m}^{2}(R_{\rm opt})}-1\bigg),
\label{halo-to-stellar-eq}
\end{equation}
where the factor $F=1.34$ is the ratio between the mass contained by a spherical mass distribution and that enclosed by an exponential disk yielding a similar radial force at $R_{\rm opt}$ \citep[][]{1987gady.book.....B}.
The gas contribution to the total rotation at the optical radius is here assumed to be negligible \citep[e.g.][]{1996A&AS..115..407R,1997PhDT........13V}.

In Fig.~\ref{stellar-halo-relation} we show the estimated halo-to-stellar mass ratio within $R_{\rm opt}$ as a function of stellar mass and compare this to various estimates in the literature for the
total halo-to-stellar ratio.
These estimates, based on abundance matching, weak lensing analysis in galaxy clusters, and halo occupation distribution methods \citep[][]{2010ApJ...710..903M,2010ApJ...717..379B,2010MNRAS.404.1111G,2012ApJ...746...95L},
all predict a minimum of low-redshift $M_{\rm h}/M_{\ast}$ at $M_{\ast} \sim 10^{10.6}M_{\odot}$.
Our estimate for the halo mass within $R_{\rm opt}$ agrees well with the predictions if these are scaled down by a constant factor $\sim 0.04$.
For the interval $10^{10}\,M_{\odot}\lesssim M_{\ast} \lesssim 10^{11}\,M_{\odot}$ we obtain a mean $M_{\rm h}/M_{\ast}(<R_{\rm opt})\sim2$ and disks that are clearly less dominated by dark matter than the fainter counterparts
($M_{\rm h}/M_{\ast}(<R_{\rm opt})\sim8$ for $M_{\ast}\lesssim 10^{9}\,M_{\odot}$).

The distribution of $M_{\rm h}/M_{\ast}(<R_{\rm opt})$ as a function of the revised Hubble stage is shown in the lower panel of Fig.~\ref{stellar-halo-relation}.
Extreme late-type systems (Scd-Im) show a high dark matter relative content within the optical disk (median of $\sim 3.6\pm0.2$),
which is twice the median $M_{\rm h}/M_{\ast}(<R_{\rm opt})$ of earlier-type systems ($\sim 1.8 \pm 0.2$ if $T\le5$).
Although measurement uncertainties are larger among early-type S0s because of their gas-poor nature,
lenticulars and early-type systems show a $\sim 15 \%$ higher relative dark matter content compared to intermediate-type galaxies.
This higher dark matter fraction among latest-type systems seems consistent with the reported trend
in \citet[][see their Fig.~3]{2015IAUS..311...78F} that was based on the CALIFA survey.
\begin{figure*}
   \centering 
   \includegraphics[width=18.6cm,clip]{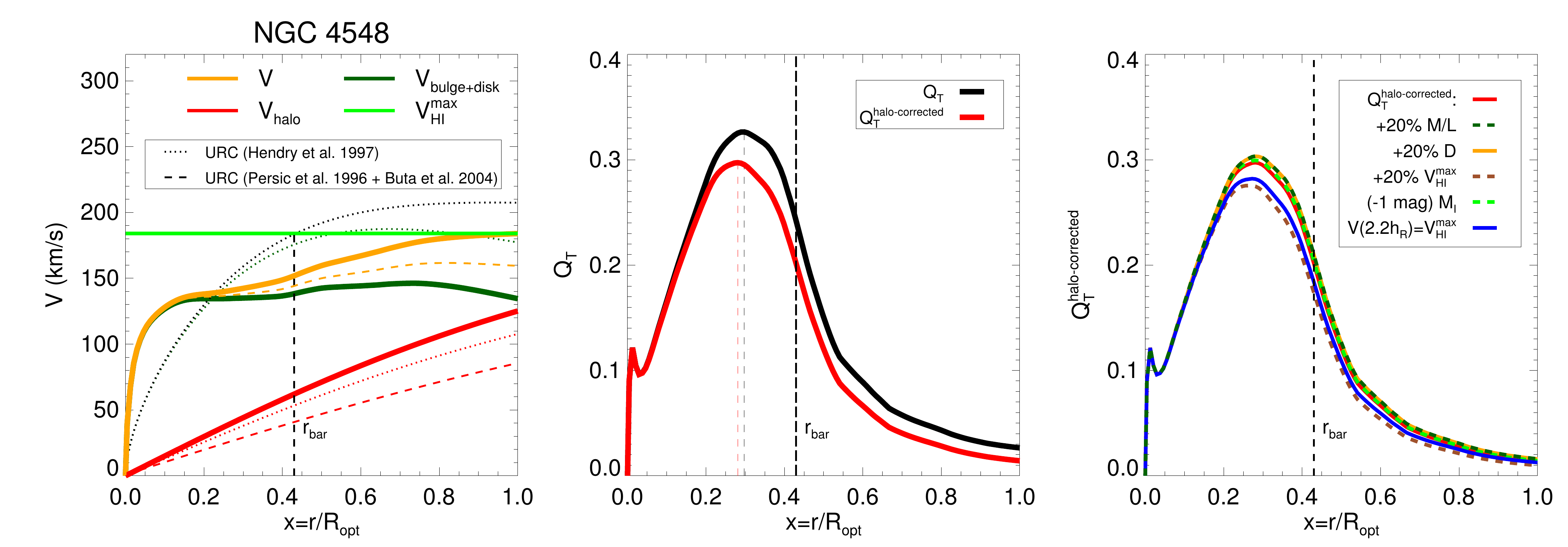}
\caption{
Ad hoc RC model for the barred galaxy NGC$\,$4548 and the effect of the dark matter halo on the force profile.
\emph{Left panel:}
Solid curves indicate the RC model based on the stellar component of the rotation curve, inferred from the 3.6 $\mu$m photometry.
The amplitude of the URC halo component (isothermal sphere) has been corrected so that the total rotation curve matches the H{\sc\,i} maximum velocity
at the optical radius ($R_{\rm opt}=3.2h_{\rm R}$). The dotted line displays the URC model (exponential disk + isothermal sphere), while the dashed lines correspond to the models used in B2004.
\emph{Central panel:}
$Q_{\rm T}$ profiles, with and without halo correction.
\emph{Right panel:}
Halo-corrected $Q_{\rm T}$ profiles resulting from a measurement error in input parameters.
A $20~\%$ overestimate on the input parameters is assumed (in the mass-to-light ratio, the distance to the galaxy, and the observed velocity amplitude).
We also assess the impact of reducing by one mag the $I$-band total luminosity (affecting the core radius in the URC models).
The effect of matching the observed maximum velocity at $2.2h_{\rm R}$ instead of at $3.2h_{\rm R}$ is tested as well.
}
\label{halo-galaxy}
\end{figure*}
\subsubsection{Halo correction on the force profiles}\label{h-corrr}
Figure~\ref{stellar-halo-relation} indicates that the dark matter contribution inside the optical radius is not negligible, in particular for galaxies with stellar masses below $10^{10}M_{\odot}$.
Assuming that the halo distribution is spherically symmetric, the dilution of the normalized tangential force is
\begin{equation}
Q_{\rm T}^{\rm halo-corr}(r)=Q_{\rm T}(r) \cdot \frac{F_{\rm R}(r)}{F_{\rm R}(r)+F_{\rm halo}(r)}.
\label{halo_dilution}
\end{equation}
\noindent To study the influence of the dark halo in our bar strength measurements, we need to estimate its force profile inside the
visible disk. In what follows we use several approximations, based on the so-called universal rotation curve models (URC)
\citet[][PSS hereafter]{1996MNRAS.281...27P}, combined with our S$^{4}$G measurements and H{\sc\,i} amplitudes from literature.
In Appendix~\ref{URC_appendix} we compare in more detail how the URC fits the kinematic measurements for the S$^{4}$G sample galaxies, 
and we provide further details of our halo density profile determination.

The URC models are based on a fit of exponential disk + isothermal halo model on measured velocity profiles (see Appendix~\ref{URC_appendix} for details). 
The isothermal halo implies
\begin{equation}
V^{2}_{\rm halo}(x)=V_{\infty}^2 \cdot \frac{x^2}{(x^2+a^2)},
\label{urc_iso}
\end{equation}
where $x=r/R_{\rm opt}$ is the radius normalized to the optical radius
and $V_{\infty}$ (halo velocity amplitude) and $a$ (halo core radius) are given in terms of total luminosity in URC.
Several corrections of the bar force accounting for the halo contribution are contemplated:

i) The first estimate for halo dilution on bar forces would be to replace the total radial force in the denominator of Eq.~\ref{halo_dilution} with
the force predicted by the URC (Eq.~\ref{urc_tot}). According to Fig.~\ref{fig_urc_comparison}, this should 
be more or less acceptable in a statistical sense.
However, it would completely discard the information about the individual stellar contribution to the velocity curve calculated from the S$^{4}$G image.

ii) A better estimate is to use the disk velocity curve as calculated from the 3.6~$\mu$m images and use URC solely to estimate the halo
contribution. This was also the approach in \citet[][B2004 hereafter]{2004AJ....127..279B}, where the halo core radius was taken from the URC model, based on the galaxy's total blue band
luminosity, while the halo amplitude $V_{\infty}$ was obtained by fitting the relative contributions of the disk and halo at $R_{\rm opt}$
to those in the URC-models. Halo parameters were inferred from the URC halo-to-stellar mass relation.

iii) However, compared to B2004, where no velocity measurements were available for the galaxies, we now have H{\sc\,i} velocity amplitudes for
most of the sample galaxies. We may use this by fitting $V_{\infty}$ from the requirement that the modelled maximum velocity within $R_{\rm opt}$
matches the H{\sc\,i} velocity amplitude. In addition, we estimate the halo core from the $I$-band luminosity based on the URC parameterization from \citet{1997ASPC..117..198H}.
Altogether, such an improved estimate can be made for 484 galaxies with usable H{\sc\,i} data.

Our final halo models imply that $\sim10\%$ of the disks in our sample are maximal 
according to the criterion of \citet{1985ApJ...294..494C} and \citet{1985ApJ...295..305V},
based on the velocity of the stellar component relative to the total rotation evaluated at $2.2h_{\rm R}$ (for maximal disks $V_{\rm disk}/V_{\rm total}(2.2h_{\rm R})>0.85$).

In the left panel of Fig.~\ref{halo-galaxy} we show the RC model for the barred galaxy NGC$\,$4548. This is based on the combination of stellar
disk rotation curve combined with the estimated halo profile based on approximation (iii) above (see also Fig.~\ref{all_plot_summary} for an example with a more prominent dark matter halo).
The middle panel then indicates the corresponding halo-corrected force profile.
For this particular early-type spiral galaxy, the effect of the halo is weak at $r_{\rm Qb}$ and the
deviation from the non-corrected value is smaller than the uncertainty related to the unknown vertical thickness.
However, for late-type galaxies, which typically host a higher relative amount of dark matter,
the halo correction will have somewhat larger contributions, as discussed below in Sect. \ref{qb-uncertainty}.
In addition, in the right panel of Fig.~\ref{halo-galaxy} we check the robustness of the halo model (see Appendix~\ref{URC_appendix} for a more detailed explanation) by
assessing the uncertainties on the used input parameters, namely the mass-to-light ratio, the distance to the galaxy, and the observed velocity.
In addition, we check the effect of matching the observed maximum velocity at a closer radius ($2.2h_{\rm R}$ instead of $3.2h_{\rm R}$).
None of these changes result in a strong deviation in the $Q_{\rm b}^{\rm halo-corr}$ value ($<10 \%$).
%
%
\section{Estimation of bar lengths and ellipticities}
%
%
\begin{figure}
    \centering   
    \includegraphics[width=0.4\textwidth]{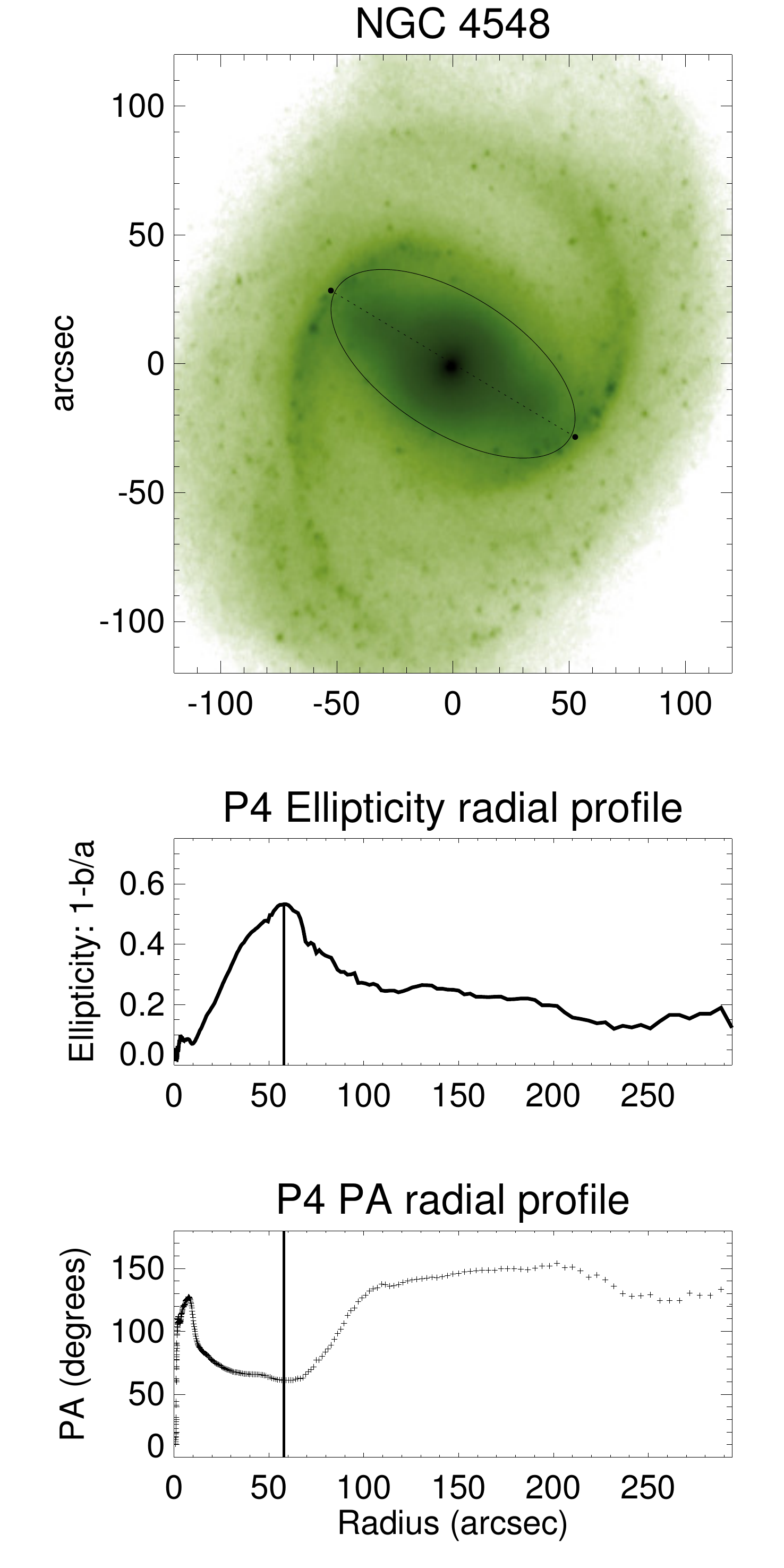}  
\caption{NGC$\,$4548 bar characterization based on P4 isophotal ellipticity profiles.
  \emph{Top panel:} Image of NGC$\,$4548, in magnitude scale with range 17-25 $\mu_{3.6\mu \rm m}(AB)$. 
  The frame is aligned in a way that the \emph{y-axis} points north and east is to the left.  Axes are in units of arcsec.
  The ellipse corresponds to the maximum ellipticity isophote, while the straight line shows the visual estimate of the bar length and PA.
  \emph{Middle panel:} ellipticity profile, with the maximum highlighted with a vertical line.
  \emph{Bottom panel:} PA profile, with $r_{\varepsilon}$ displayed with a vertical line.
}
\label{bar_ellipse_NGC4548}
\end{figure}
\begin{figure}
    \centering   
    \begin{tabular}{c c c}
    \includegraphics[width=0.5\textwidth]{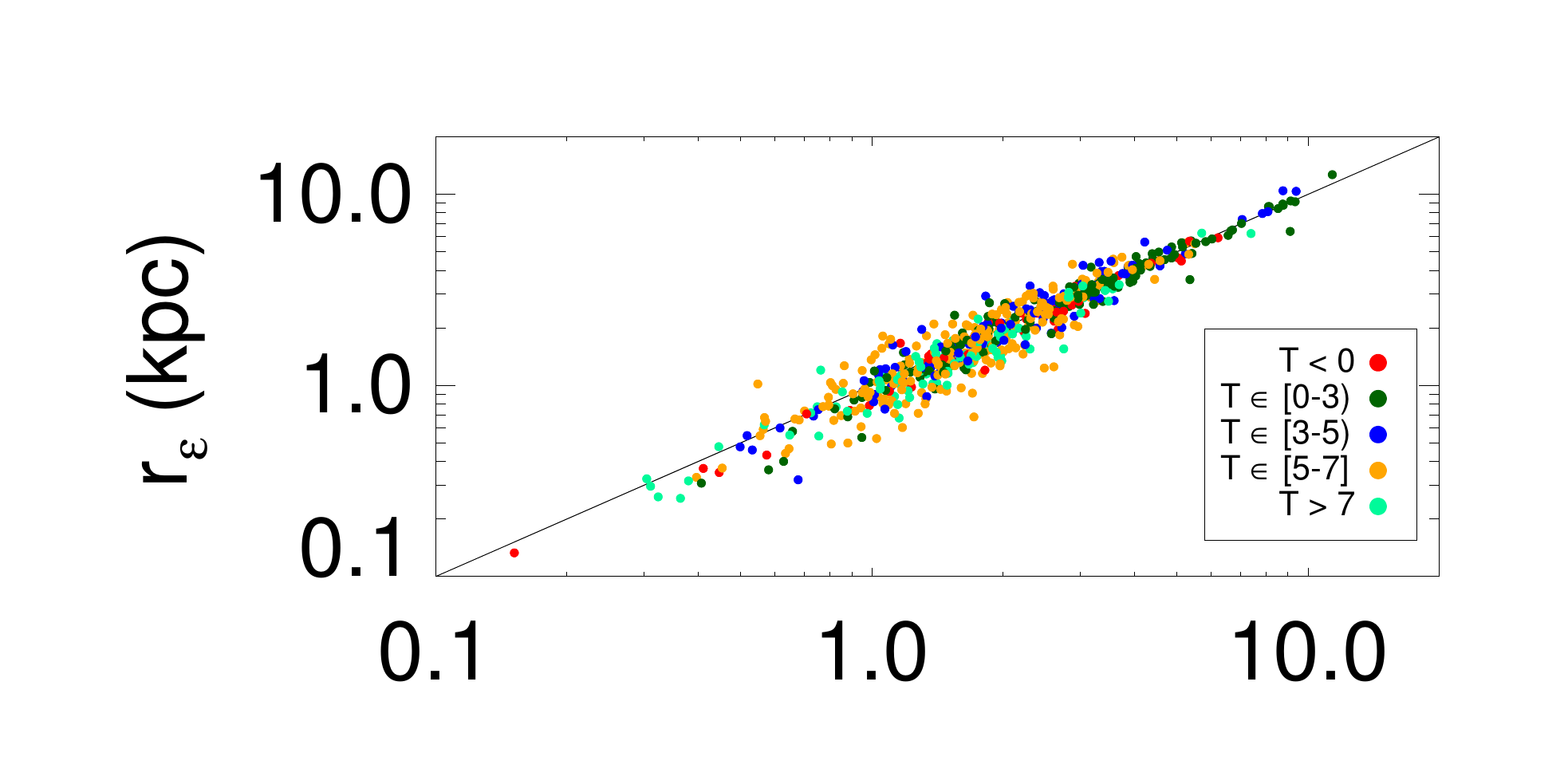}\\[-5ex]
    \includegraphics[width=0.5\textwidth]{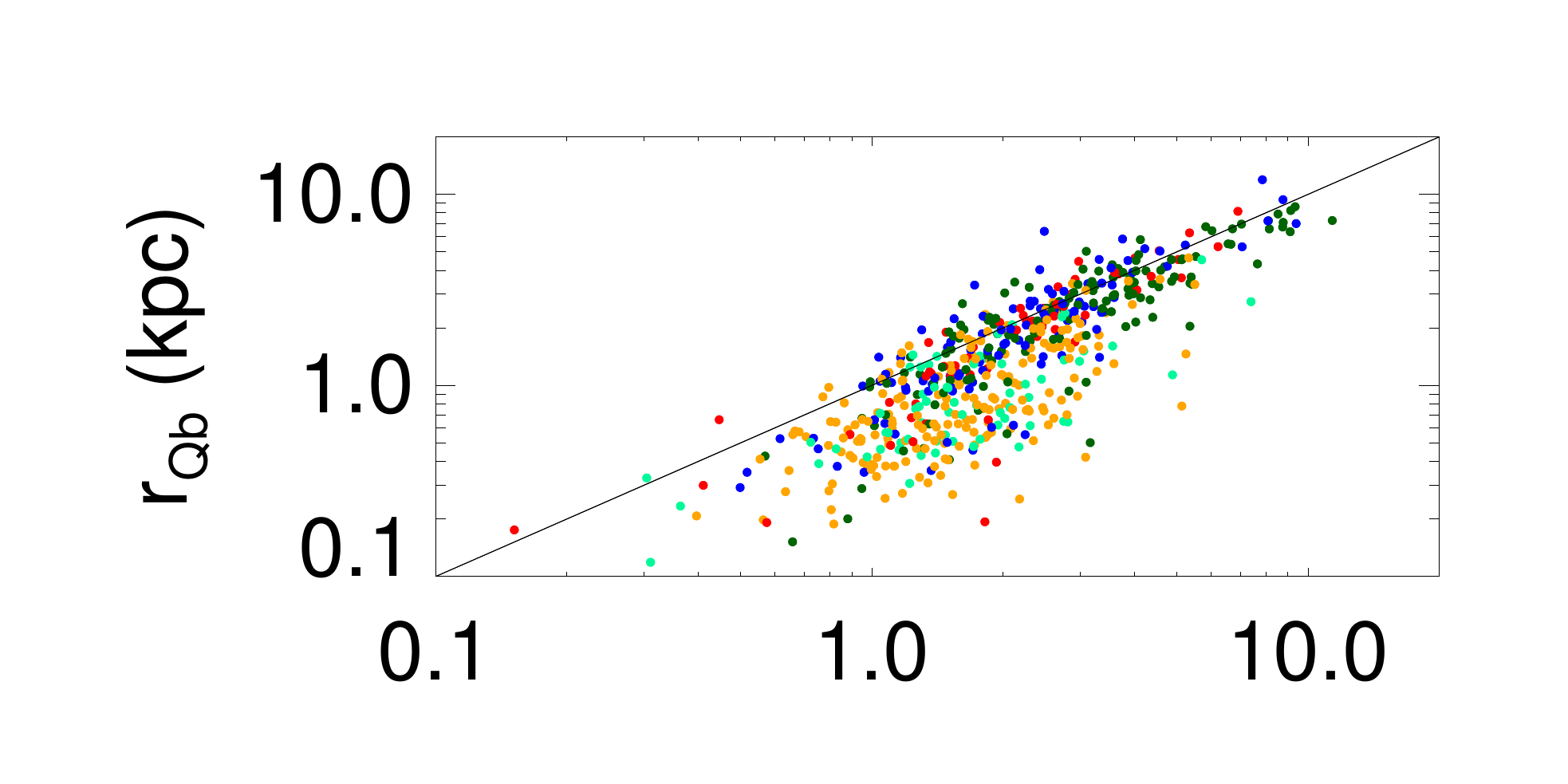}\\
    \includegraphics[width=0.5\textwidth]{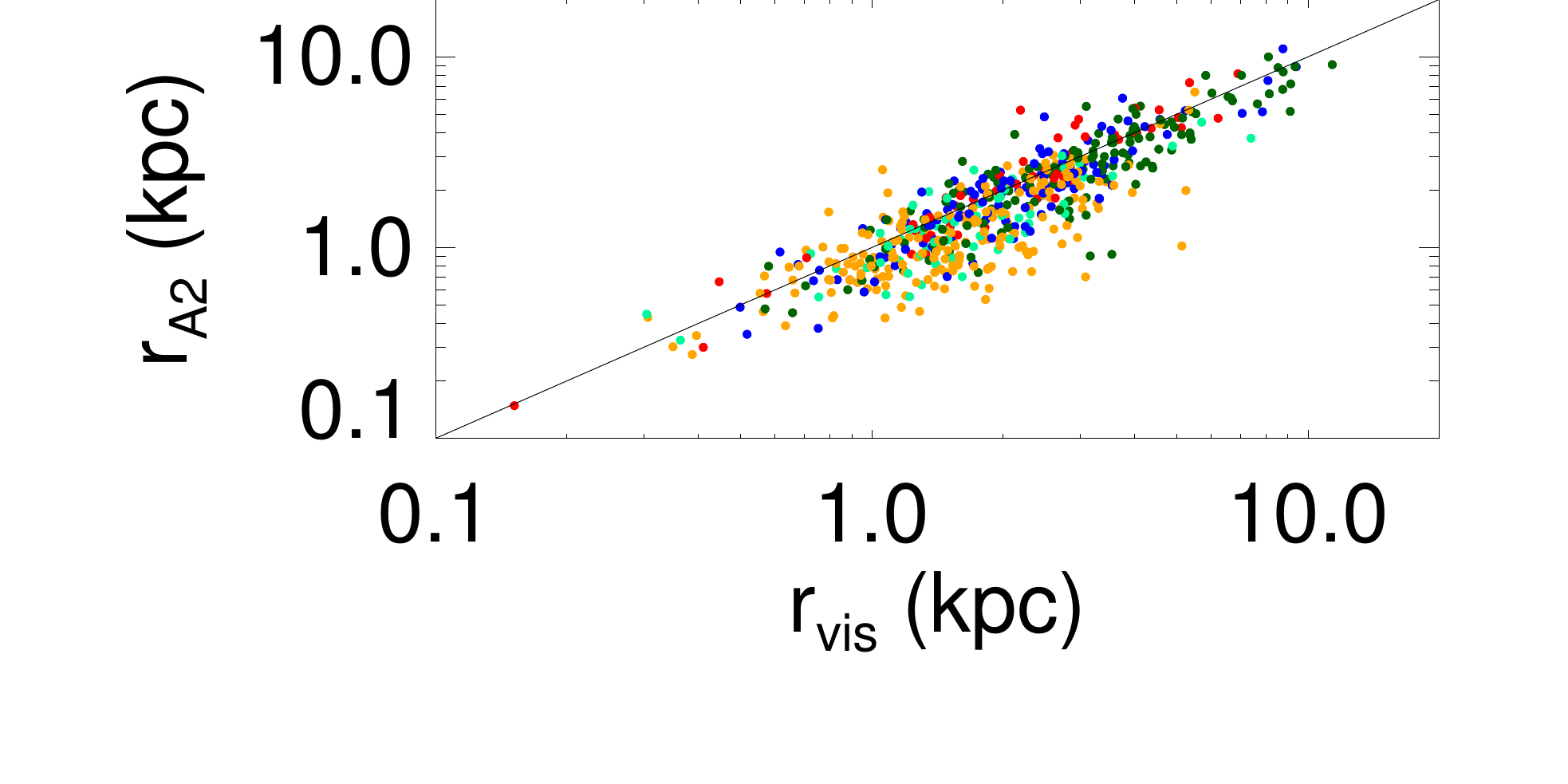}\\
    \end{tabular}
\caption{Comparisons of different measurements of bar sizes, with the symbols colour-coded according to the different Hubble types.
\emph{Top panel:} Bar length calculated from the radius of maximum ellipticity and visual estimate (both converted to intrinsic disk plane values). 
\emph{Central panel:} Visual bar length compared to the radius of maximum gravitational torque.
\emph{Bottom panel:} Visual bar length compared to the radius of maximum $A_{2}$ Fourier amplitude.
}
\label{bar_visual_ellipse_flat}
\end{figure}
We used bar length measurements based on ellipse fitting
($r_{\varepsilon}$) and on visual inspection of the 3.6$\mu$m images ($r_{\rm vis}$), taken from HE2015.  We also used $r_{\rm A2}$ and $r_{\rm Qb}$ values obtained from our force calculation.

The analysis of ellipticity profiles in HE2015 was made for all those galaxies for which bars were identified in the morphological
classification of B2015 for the complete S$^4$G sample.  Based on P4 ellipse profiles, using the ellipticity maxima in the bar region and
the constancy of the position angle in that region \citep{1991A&AS...88..325W,1995A&AS..111..115W}, 654 bars were
identified in the 860 barred galaxies in B2015.
Bar parameters in HE2015 were measured with two different methods: in the first method
the bar length, position angle, and ellipticity were taken from the isophote fit corresponding to the ellipticity maximum in the bar
region. In the second method, the length of the bar and its position angle were visually marked on top of the image. In both cases the bar
was measured on the original sky image. The bar parameters (length, PA, $\epsilon$) were then converted to the disk plane, using the
2D analytical de-projection code in \citet[][]{1999AJ....117..792S} and LS2002 \citep[the formulae are the same as in][]{2007MNRAS.381..943G}.
Since the visual measurement of the bar only gave its length and position angle, the maximum of isophotal ellipticity was used in the conversion to the disk plane.

An example for obtaining bar lengths from ellipse fitting is illustrated in Fig.~\ref{bar_ellipse_NGC4548} for NGC$\,$4548. 
The length is the radius of the maximum ellipticity, $r_{\varepsilon}$. 
As an upper limit of bar length, \citet[][]{2003ApJS..146..299E} and \citet[][E2005 hereafter]{2005MNRAS.364..283E}
have used the first minimum in the ellipticity profile after $r_{\varepsilon}$, 
named as $L_{\rm bar}$.  However, as discussed in \citet[][]{2006A&A...452...97M}, the location and value of such a
minimum depends on the galaxy inclination and/or the real non-axisymmetric shape of the disk and the type of bar profiles, 
which is obvious also from the S$^4$G images. Therefore $L_{\rm bar}$ is not used in this study. 
If not otherwise mentioned, in the following the visually estimated bar lengths, after conversion to the disk plane, are used and referred as $r_{\rm bar}$.
Furthermore, $r_{\rm bar}$ always refers to the bar radius and not to the full length. 

Studies based on both observations \citep{1991A&AS...88..325W,2002MNRAS.337.1118L,2003ApJS..146..299E} and on N-body simulations \citep{1999A&A...348..737R,2002MNRAS.330...35A,2006A&A...452...97M} 
show that $r_{\varepsilon}$ tends to underestimate the real size of bars. Using synthetic images, \citet{2009A&A...495..491A} 
demonstrated that the length actually depends on the model describing the surface brightness profile of the bar (e.g. larger underestimation for Ferrers bars compared to Freeman and flat bars).

A statistical comparison between the bar sizes calculated visually and the different independent systematic measurements of bar lengths
($r_\varepsilon$, $r_{\rm A2}$, and $r_{\rm Qb}$) is presented in Fig.~\ref{bar_visual_ellipse_flat}.  The $r_\varepsilon$
measurements correlate tightly with $r_{\rm vis}$. We also confirm the tight correlation between the visual estimation of the bar size and
both $r_{\rm A2}$ \citep{2002MNRAS.330...35A} and $r_{\rm Qb}$ \citep{2002MNRAS.331..880L}.  However, the later relation is much more
scattered among the late-type galaxies.  We obtain mean values $<r_{\varepsilon}/r_{\rm vis}>=0.97$ $(\sigma=0.18)$, $<r_{\rm Qb}/r_{\rm
  vis}>=0.62$ $(\sigma=0.25)$ and $<r_{\rm A2}/r_{\rm vis}>=0.75$ $(\sigma=0.25)$.
Thus, subject to natural measurement uncertainties, all the bar size proxies underestimate the bar length compared to the visual estimate, in agreement with what is reported in the literature. 
However, the difference between $r_{\varepsilon}$ and $r_{\rm vis}$ is not as large as in previous studies.

Finally, a small fraction of our galaxies show $r_{\rm A2}$ and $r_{\rm Qb}$ distances that fall beyond the bar length. Cases
like this are seen for all Hubble types. 
These are cases in which the spiral arms strongly influence the potential in the bar region, where both maxima are
determined by the combined contribution of the two non-axisymmetric stellar structures and occasionally reach beyond the bar limit.
In the case of the ellipticity, bars with $r_{\varepsilon}$/$r_{\rm vis}>1$ are explained by the presence of spiral arms and rings,
as discussed in E2005 and \citet[][]{2014A&A...562A.121C}.
%
%
\section{ Characterization of bars as a function of Hubble stage and family}\label{barlen_ttype}
%
In Table~\ref{bar_strength_statistics_structures} we present the mean values of the bar strengths ($Q_{\rm b}$, $A_{2}^{\rm max}$, $\epsilon$) 
and bar sizes (in physical units and normalized to $R_{25.5}$ and $h_{\rm R}$) for the different morphological classes and bar families.

From now on, we refer to the galaxies in our sample as S0s if $T<0$, early-type spirals
if $T \in [0,3)$, intermediate-type spirals when $T \in [3,5)$, late-type spirals for $T \in [5,7]$, and Magellanics and
irregulars for those with $T > 7$.  With regard to the study of the bar properties in the Hubble sequence, we exclude ellipticals
($T<-3$) and also those galaxies that have a double morphological classification in B2015: this is the case of some lenticular and
early-type spiral galaxies with an embedded disk-like inner structure with spiral(s) and a bar. However, there are only 
six such cases in our sample, and the measurements of bars are still provided, but are not included in any analysis.
\subsection{Bar fraction in the Hubble sequence}\label{barfrachubble}
\begin{figure}
    \centering   
    \begin{tabular}{l r}
    \includegraphics[width=0.5\textwidth]{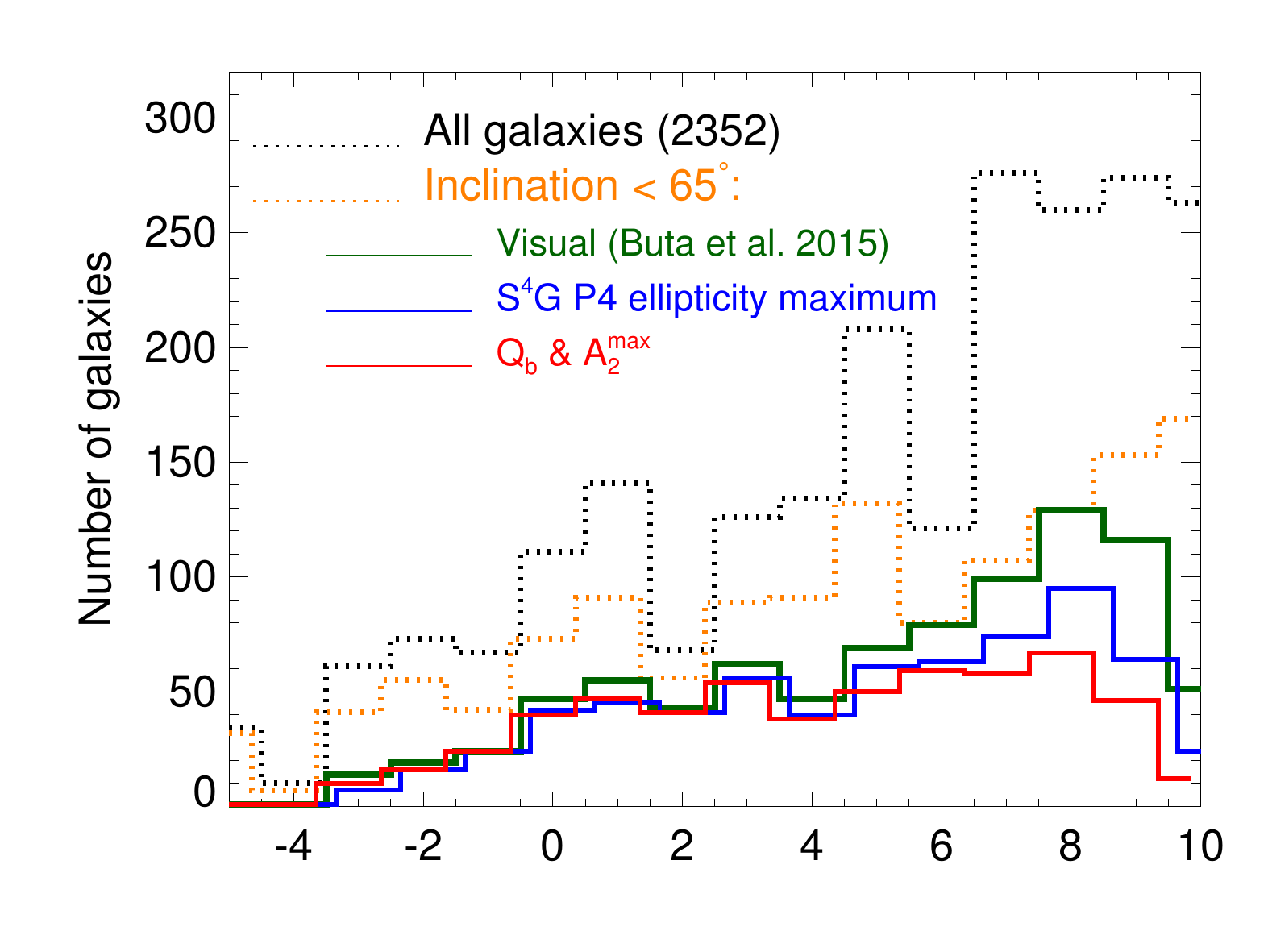}\\[-6ex]
    \includegraphics[width=0.5\textwidth]{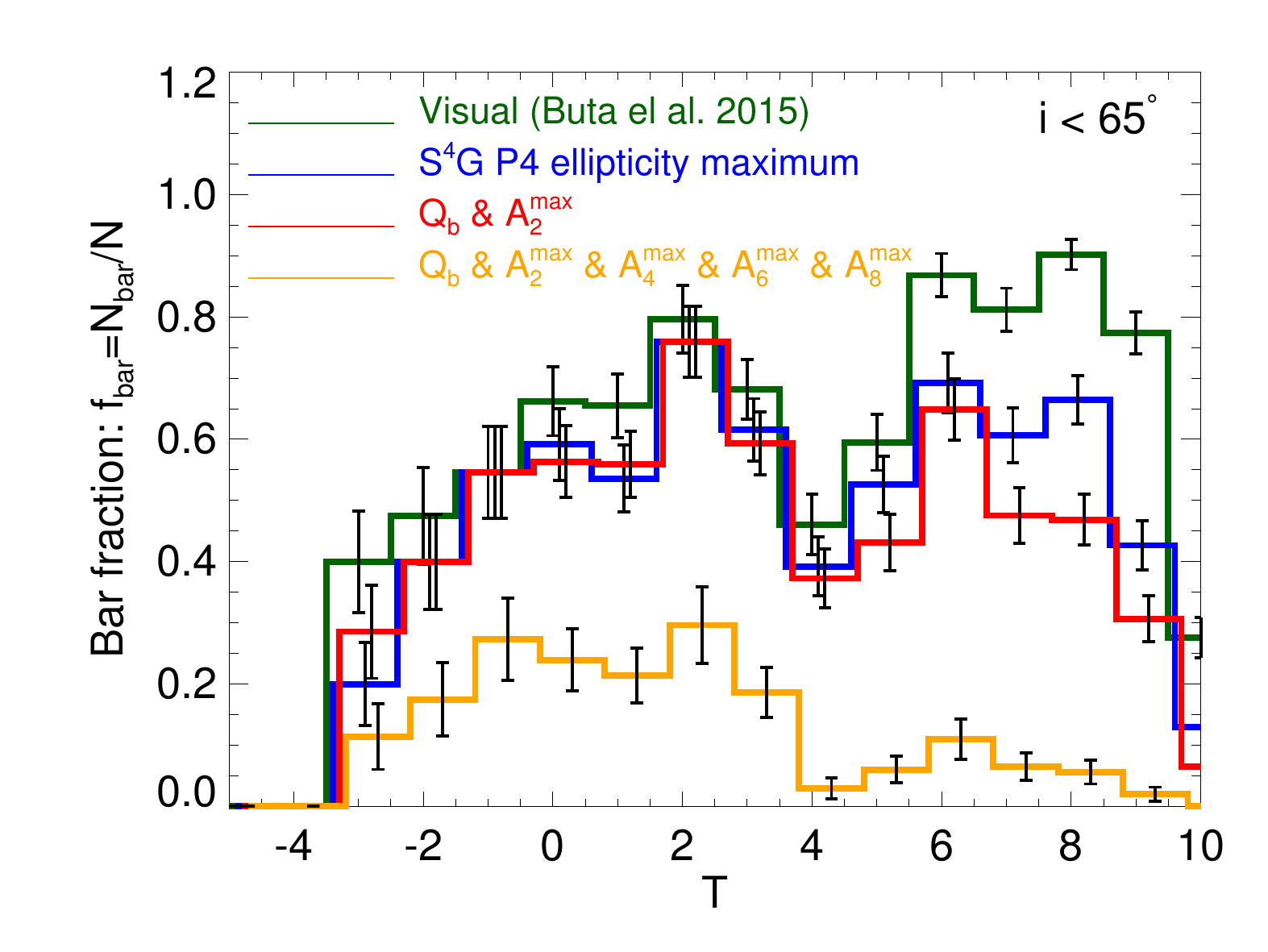}
    \end{tabular}
\caption{
\emph{Upper panel:}
Distribution of galaxies in our sample vs. their Hubble type.  
The black and orange dotted lines correspond to the histogram of all the S$^{4}$G galaxies and those with inclination lower than 65$^{\circ}$, respectively.
For the galaxies with $i<65^{\circ}$, we display the histogram of barred galaxies in the Hubble sequence (solid lines),
according to the various criteria explained in the text.
\emph{Lower panel:} Bar fraction vs. Hubble stage for a sample of face-on and moderately inclined galaxies ($i<65^{\circ}$).
The fraction of galaxies with an identifiable maximum in the amplitude of their higher order even Fourier components is also
shown.
Uncertainties are calculated assuming a binomial distribution.
}
\label{bar_dist}
\end{figure}
We studied the bar fraction ($f_{\rm bar}$) for our sample of face-on and moderately inclined S$^{4}$G galaxies ($i<65^{\circ}$). 
Based on the methods explained above, three different criteria were used to determine whether or not a galaxy has a bar: (1) from the visual
classification in B2015, (2) by inspection of the radial ellipticity profiles, and (3) from the $A_2$ Fourier amplitude profiles and the presence of a well-defined four-quadrant butterfly pattern in the torque map.

The upper panel of Fig.~\ref{bar_dist} shows the histogram of all S$^4$G galaxies as a function of Hubble type and that of the barred
galaxies, based on the three different criteria. It indicates that the sample is dominated by late-type systems, among which the frequency of bars is clearly decreased when criteria 2 and 3 are used instead of visual classifications.
The lower panel shows the same, normalized to the total number of galaxies in the bin.

A double-humped distribution of the bar fraction stands out regardless of the bar detection criterion, with a local minimum at $T=4$.
The physical reason for a lower bar fraction of these transitional Sb/c systems is not obvious, since they seem to be as massive and 
have similar dark matter fractions as earlier types of   spiral galaxies in the S$^{4}$G sample.
Interestingly, no similar dip is present when the bar fraction as a function of $M_{\ast}$ is studied (see Sect.~\ref{barfrachubble}).

For later types, the bar fraction rises again.
However, it appears that more bars are identified visually than with the other methods.
Indeed, the bar fraction is reduced by $\sim 30 \%$ and $\sim 50 \%$ for $T>4$ if $\varepsilon$ or $A_{2}$ and $Q_{\rm T}$ profiles are used (criteria 2 and 3 above, respectively).
This is natural, taking into account that very weak bars that
are still visible in the images can be lost in ellipse fitting analysis where clumpy star-forming regions can induce stronger ellipticity maxima than the bars. 
Similarly, dust lanes can distort the shape of the isophotes. Dust lanes and H{\sc\,ii} regions can also make $A_{2}$ profiles noisy.
Many of these late-type bars would possibly be overlooked if they were observed at higher redshifts, given their faint disks.

The bar fraction is remarkably high for Sab ($T=2$) galaxies, and there is a drop in the bar fraction at Hubble types earlier than $T=0$,
consistent with \citet[][]{2013MNRAS.430.3489L}. 

\citet[][]{1990ApJ...357...71O} showed that for early-type galaxies, $m=4$ and $m=6$ components are not negligible in the azimuthal density profiles of barred galaxies.
In the lower panel of Fig.~\ref{bar_dist} we also show the bars for which the maximum of the higher order component amplitudes could be detected in the bar region. 
Of the early-type systems ($T<4$) for which $Q_{\rm b}$ and $A_{2}^{\rm max}$ were calculated, approximately one half of the galaxies present reliably identifiable $A_{4}^{\rm max}$, $A_{6}^{\rm max}$ and $A_{8}^{\rm max}$.
In contrast, very few galaxies with $T>4$ have a significant contribution from the higher-order even components to the Fourier decompositions of bars.

The bar fractions based on different bar detection criteria and for different morphological types are listed in Table~\ref{bar_frac_ttype}.
For complementary details about the visual bar fraction in the S$^4$G, see B2015.
\begin{table}
\centering
\small
\begin{tabular}{|| c| c| c| c| c|| c||}
\hline
\multicolumn{1}{||c}{\centering Bar fraction:} & \multicolumn{1}{|c}{\centering Visual} &
\multicolumn{1}{|c}{\centering Ellipse fitting} & \multicolumn{1}{|c||}{\centering $\exists$ $A_{2}^{\rm max}$ \& $Q_{\rm b}$} & \multicolumn{1}{|c||}{\centering $N_{\rm total}$} \tabularnewline
\hline
\hline
$T\in[-3,0)$&  49.1 $\pm$   4.6 $\%$&  40.5 $\pm$   4.6 $\%$&  43.1 $\pm$   4.6 $\%$&116 \\
$T\in[0,3)$&  69.8 $\pm$   3.2 $\%$&  61.8 $\pm$   3.3 $\%$&  61.8 $\pm$   3.3$ \%$&212 \\
$T\in[3,5)$&  55.8 $\pm$   3.6 $\%$&  48.9 $\pm$   3.6 $\%$&  46.8 $\pm$   3.6$ \%$&190 \\
$T\in[5,7]$&  75.1 $\pm$   2.4 $\%$&  60.2 $\pm$   2.7 $\%$&  50.8 $\pm$   2.8$ \%$&329 \\
$T>7$&  60.6 $\pm$   2.2 $\%$&  37.1 $\pm$   2.2 $\%$&  25.5 $\pm$   2.0$ \%$&498 \\
\hline
\end{tabular}
\caption{
The mean bar fractions based on different bar identification criteria and for different morphological types. Binomial counting errors and the total number of galaxies within the bins are indicated.}
\label{bar_frac_ttype}
\end{table}

%
%
\subsection{Bar lengths in the Hubble sequence}
\begin{figure}
   \centering   
   \includegraphics[width=0.475\textwidth]{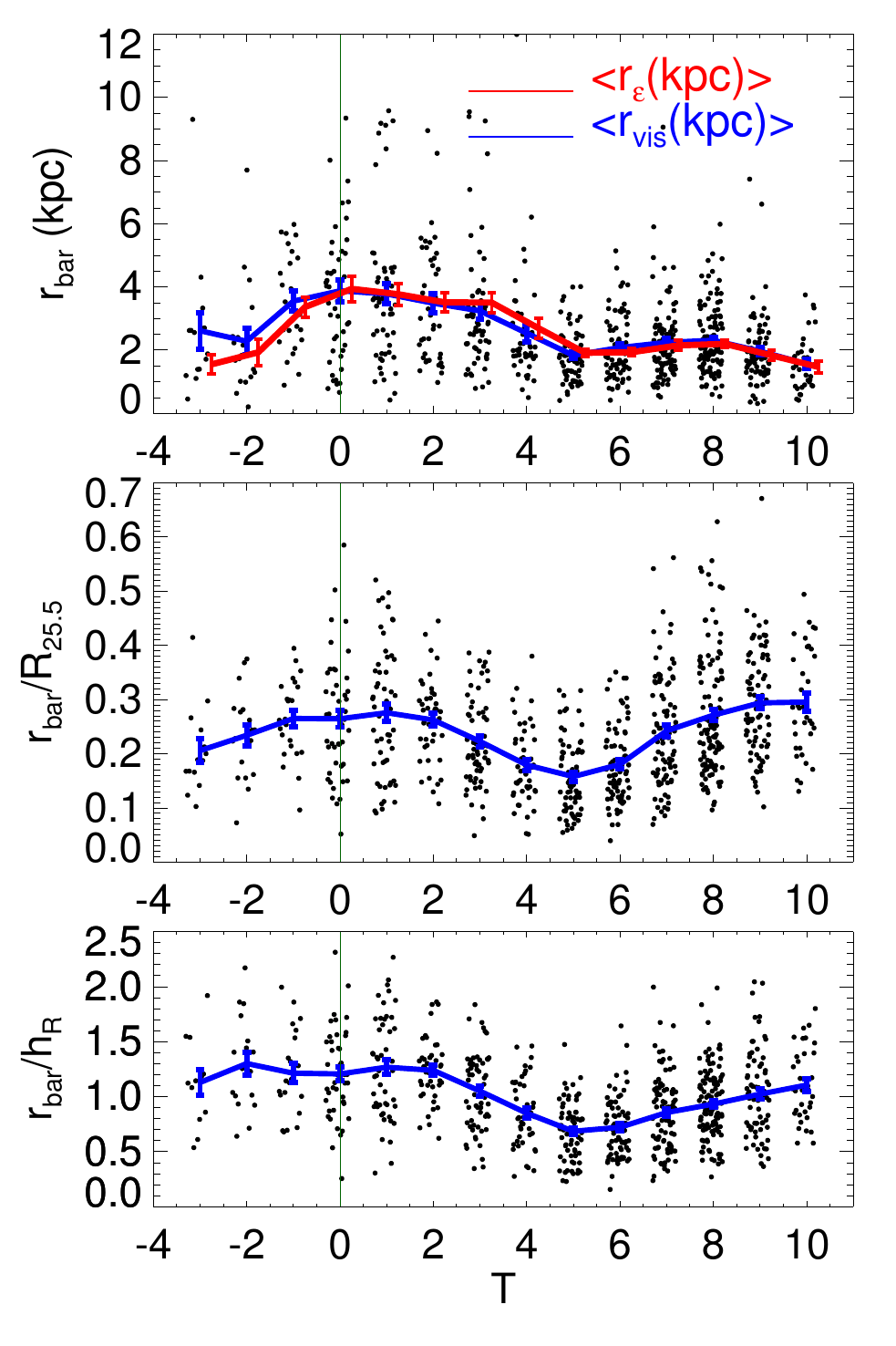}
\caption{
Bar length distribution in terms of the galaxy Hubble stage. 
The blue line corresponds to the running mean, with the standard error of the mean indicated with an error bar.
The green vertical line delimits the region of S0 galaxies.
The \emph{first row} displays the bar length (in units of kpc) vs. $T$.
The red line shows the running mean of the radius of maximum ellipticity, with an offset of $+0.25$ in the x-axis. 
Shown in the \emph{second and third rows} are the distributions of bar length (visual estimate), normalized by $R_{25.5}$ and $h_{\rm R}$.
To avoid overlapping of points, the $T$ values in the $x$-axis have been randomly displaced from the integer.
}
\label{barlen_ttype}
\end{figure}
Bar lengths are displayed as a function of the galaxy Hubble stage in Fig.~\ref{barlen_ttype}.
In the upper panel the distribution of bar lengths in kpc is displayed, and in the second and third panels we normalized them by the galaxy size 
(as measured by $R_{25.5}$ and $h_{\rm R}$).
We confirm the result from \citet[][]{1985ApJ...288..438E}, who showed that the bars in early-type spiral galaxies ($T=0-2$) are typically longer than the bars hosted by late-type spirals ($T=3-7$).

More specifically, we observe that bar sizes tend to increase from $T=5$ towards the Sa and S0/a galaxies
\citep[in agreement with e.g.][]{1995AJ....109.2428M}, and then drop among the S0s \citep[as found earlier in E2005 and][]{2007MNRAS.381..401L}. 
This is roughly the same for the absolute and relative bar sizes 
and is independent of the normalization used (either by $R_{25.5}$ or $h_{\rm R}$), but the trend flattens among S0s when normalizing by the disk scale length.
The difference between the disk-relative bar sizes of early-type and late-type spirals is not as pronounced as the
contrasts of a factor $\sim 3$ and $\sim 2.5$  reported in \citet[][]{1995AJ....109.2428M} and E2005, respectively. 
Indeed, the strongest difference is found by comparing Sa to Sc galaxies, the former hosting bars that are $\sim 1.75$ larger on average.
On the other hand, Sd galaxies also typically have longer bars than Sc galaxies.
The mean values for the different morphological types and bar classes are summarized in Table~\ref{bar_strength_statistics_structures}.

The distribution of bar lengths in the Hubble sequence that we obtain is more or less consistent with the statistics in E2005 (we have a sample larger by a
factor $\sim 6$ ), who found a mean bar size of $\sim 3.3$ kpc for S0-Sb galaxies (with the bar sizes distributed in a
similar range as we observe, although we find bars shorter than $1$ kpc among early-type galaxies). Nevertheless, his late-type
galaxies (Sc-Sd) had bars with a mean size of $\sim 1.5$ kpc (ranging from $0.5$ to $3.5$ kpc), while in their $T$ range ours are typically
$\sim 2.5$ kpc, spanning in a wider interval ($0.5-6$ kpc). Altogether, our estimates give a mean bar size of $\sim 2.5$ kpc,
early-type systems (S0$^{-}$-Sb) hosting bars that are a factor $\sim
1.5$ longer in physical size than those in late-type galaxies (Sbc-Im).

For scaled bar sizes, the measurements in E2005 are made in intervals $0.2-0.8 R_{25}$ and $0.5-2.5h_{\rm R}$ for S0-Sab galaxies, with mean values 
$\sim 0.38 R_{25}$ and $1.4h_{\rm R}$.  In their $T$ range we also observe shorter bars (as short as $\sim 0.1R_{25.5}$ and a few cases with $r_{\rm bar}/h_{\rm R}<0.5$), and 
we obtain lower mean bar sizes ($\sim 0.27R_{25.5}$ and $\sim 1.25h_{R}$).
The disk-relative bar sizes of late-type systems in E2005 are found in the ranges $0.05-0.35 R_{25}$ and $0.2-1.5h_{\rm R}$,
with mean values of $\sim 0.14 R_{25}$ and $\sim 0.6h_{\rm R}$.
We also measure bars among Sc-Sd galaxies that can be as large as $\sim 0.6 R_{25.5}$ and $\sim 2 h_{\rm R}$,
and in general we observe slightly longer bars in these systems
than E2005.

\citet[][]{2005A&A...434..109A} found a mean relative bar length $\left<r_{\rm bar}/h_{\rm R}\right>=1.21\pm0.08$ (as measured from the photometric decomposition including lenses).
In addition, numerical simulations made by \citet[][]{2003MNRAS.346..251O} and \citet[][]{2003MNRAS.345..406V} predicted $\left<r_{\rm bar}/h_{\rm R}\right>=1.1\pm0.33$ and $1.0-1.2$, respectively. 
In this work, S0s and early-type spirals are the only systems with $<r_{\rm bar}/h_{\rm R}>\,\ge1$, while for galaxies with $T\ge3$ bar sizes relative to $h_{\rm R}$ are on average below the predicted values.
This is natural considering that early-type bars resemble the type of bars resulting from numerical simulations, as stated in E2005.

The lower mass systems at the end of the Hubble sequence are interesting galaxies.  As expected, the absolute bar sizes are smaller than
in the bright galaxies (S0$^{-}$-Sbc galaxies have on average the same size), but the normalized sizes are on average even larger (compared
to intermediate-type spirals). In fact, there is a tendency of increasing bar length from $T=6$ towards the irregular galaxies.

It seems that among the late-type systems, the more clumpy the galaxy, the longer the relative size of the elongated structure embedded in the underlying disk. However, it is worth to mention
that even among the smallest galaxies, the bars are long enough ($97\%$ of them are longer than 10 pixels, with the shortest bar being 6 pixels long) so that they are not likely to arise solely from the joining
of a pair or group of star-forming clumps, that is, they are not visual artifacts (as can happen with measurements of bars at high redshifts). This trend of increased relative size among the faint galaxies, which was also
present in the observations of \citet[][]{2007MNRAS.381..401L} in spite of the poorer sampling of $T$-types $\ge7$, is maintained regardless of the bar size estimate that we use (see also Table~\ref{barlen_tabulate}).
In addition, we have divided our sample into two bins in distance and checked that the trends for the bar sizes are exactly the same, confirming that the angular resolution is not a problem here.
\subsection{$\varepsilon$, $Q_{\rm b}$, and $A_2$ in the Hubble sequence}\label{bar_force_HS}
\begin{figure*}
    \centering   
    \begin{tabular}{ l c r}
    \includegraphics[width=0.3\textwidth]{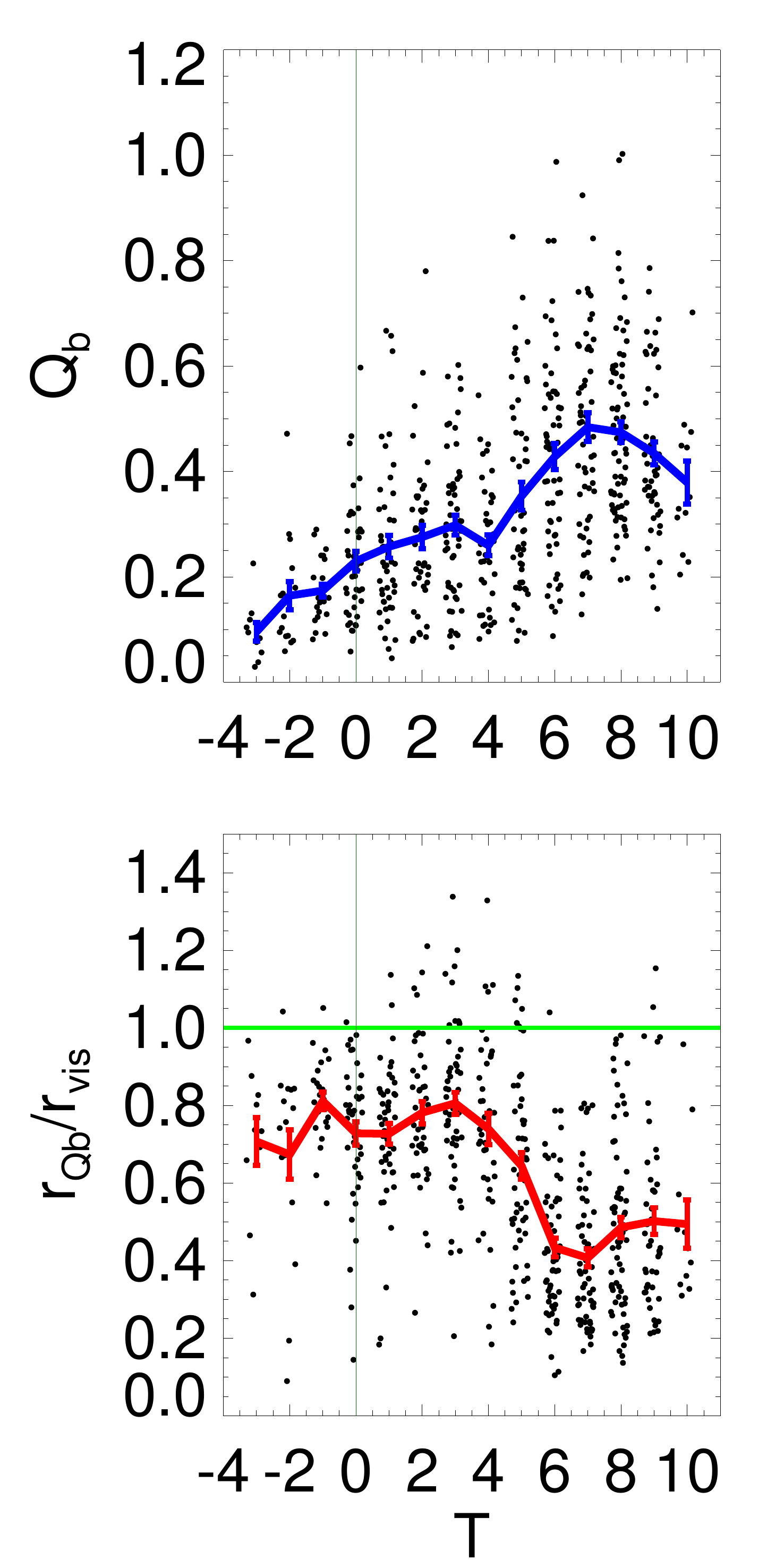}
    \includegraphics[width=0.3\textwidth]{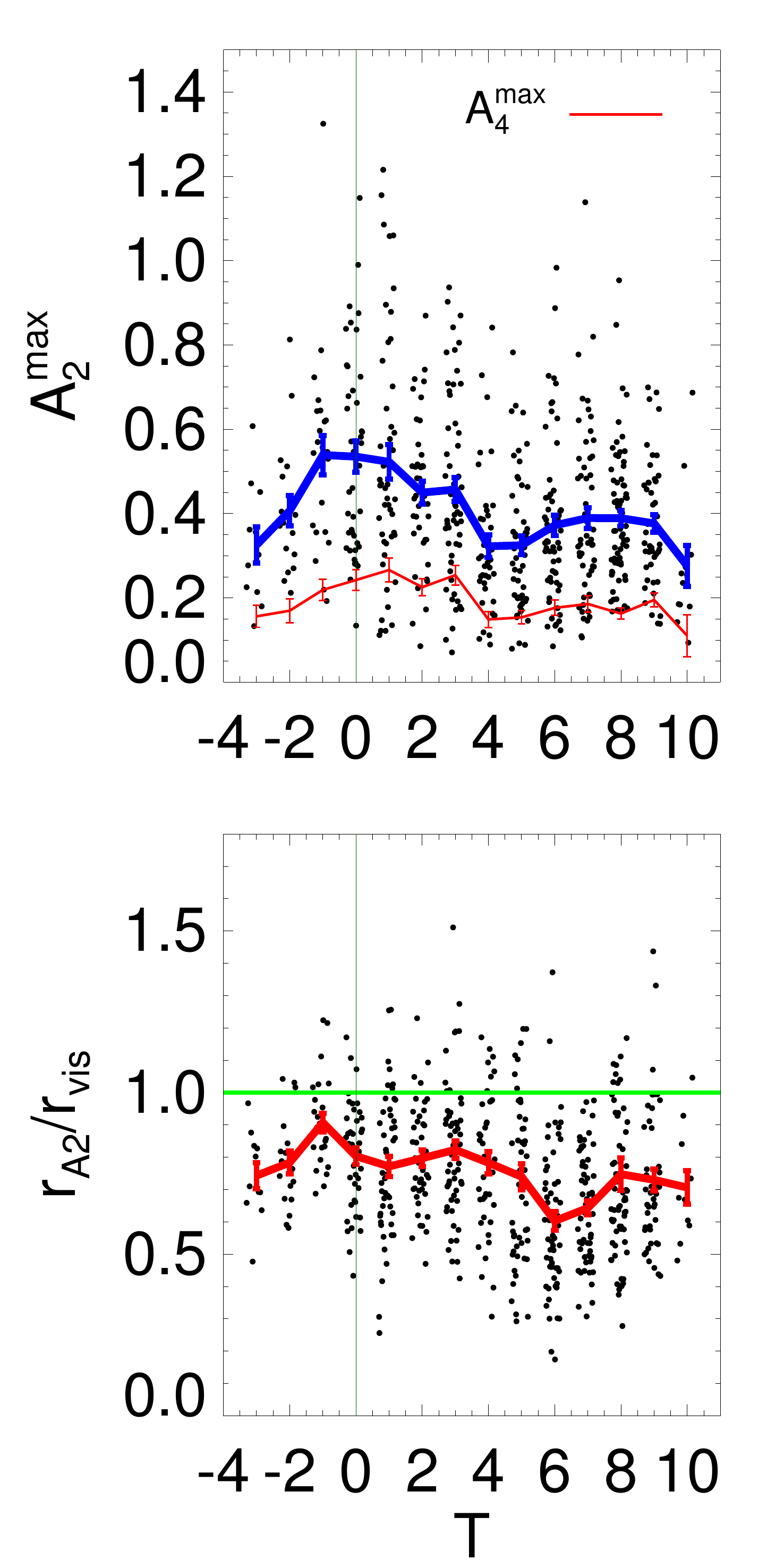}
    \includegraphics[width=0.3\textwidth]{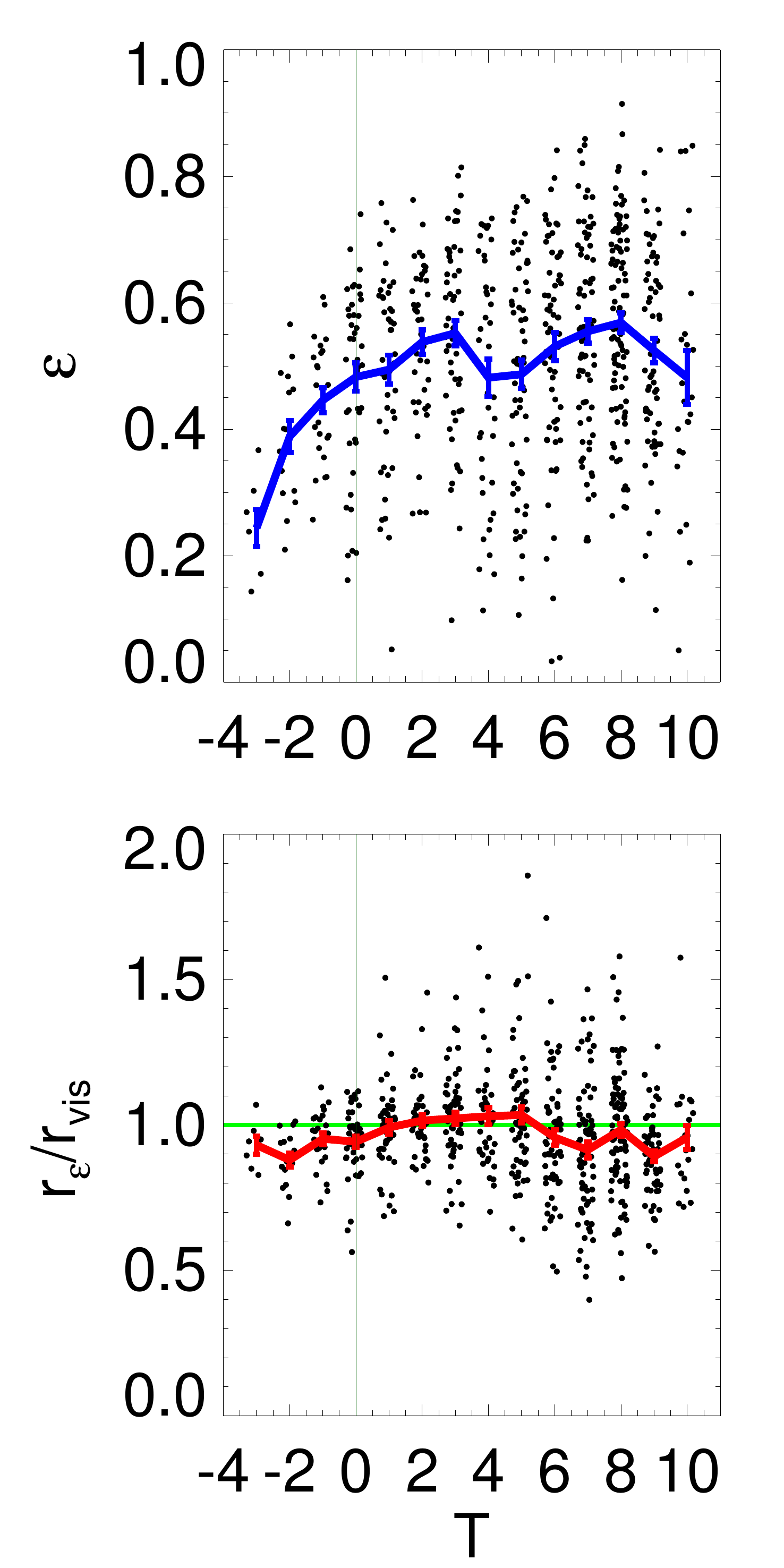}
    \end{tabular}  
\caption{
Different bar strength indicators as a function of the Hubble Type.
\emph{Upper row:}
The $Q_{\rm b}$, $A_{2}^{\rm max}$, $A_{4}^{\rm max}$ and $\varepsilon$ are plotted vs. the integer value of the revised numerical Hubble stage (from left to right, respectively).
\emph{Lower row:} Radius of maximum gravitational torque, Fourier amplitude and ellipticity, normalized to the visual estimate of the bar length as a function of the Hubble stage.
The running mean is overplotted with a solid line and the standard error of the mean is indicated with a vertical error bar. 
$T$ values in the $x$-axis (integers) have been randomly displaced for the sake of avoiding point overlapping.
For all the plots, the green vertical line demarcates the region of S0 galaxies.
In the central upper panel, the moving average of bar maximum amplitude of the $m=4$ Fourier component ($A_{4}^{\rm max}$) is also displayed.
}       
\label{ttype_qb_a2_ellip}
\end{figure*}
\begin{figure*}
    \centering   
    \begin{tabular}{l c r}
    \includegraphics[width=0.32\textwidth]{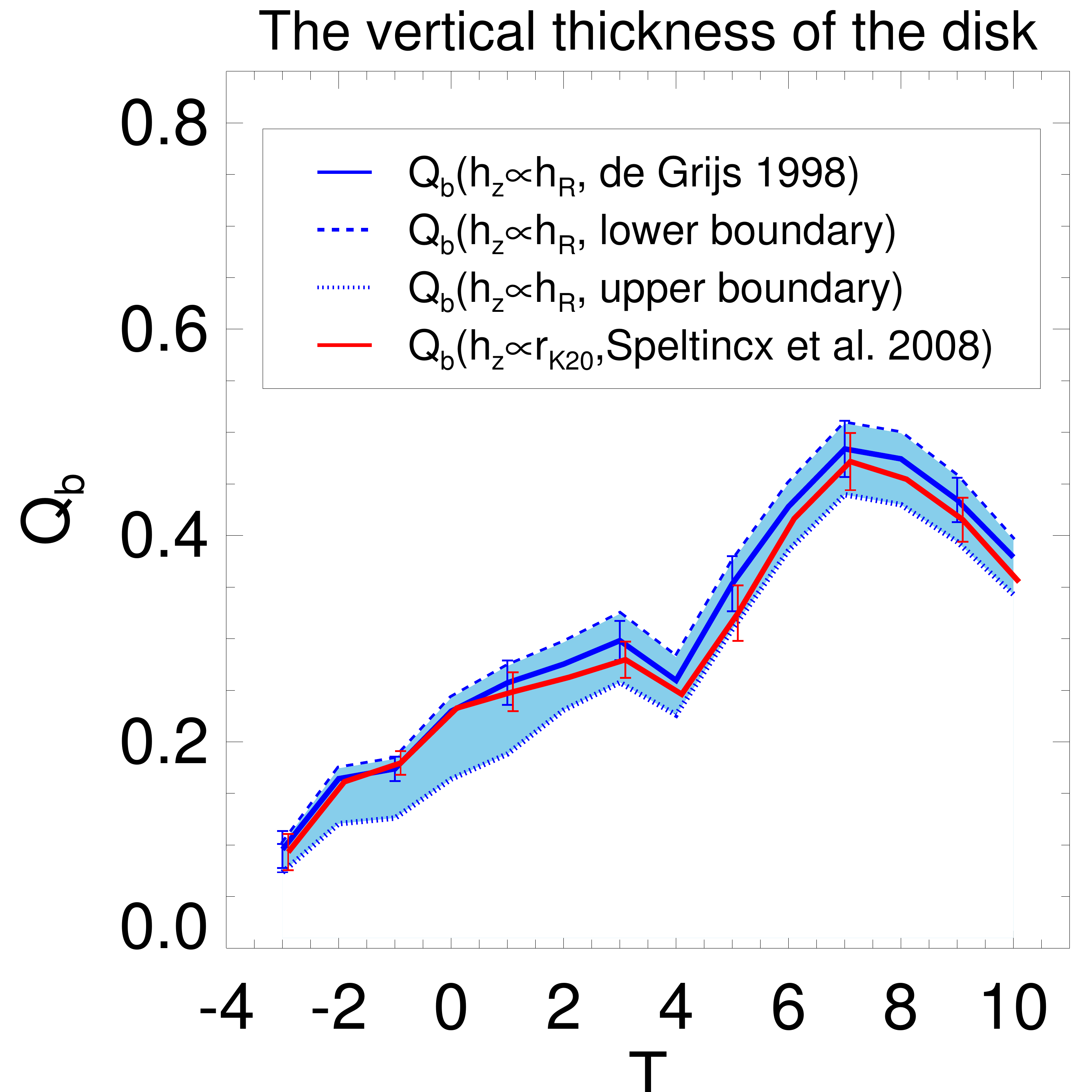}
    \includegraphics[width=0.32\textwidth]{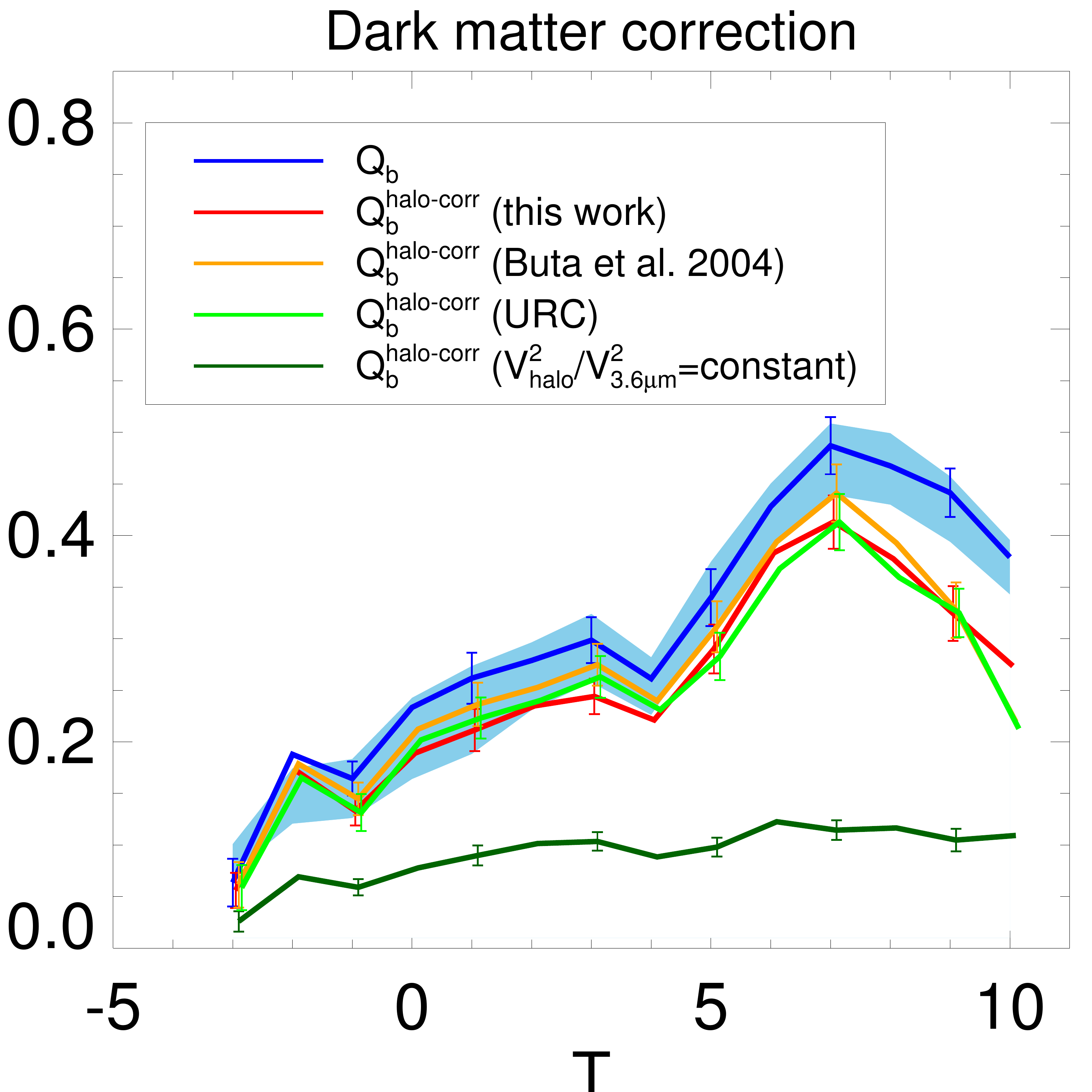}
    \includegraphics[width=0.32\textwidth]{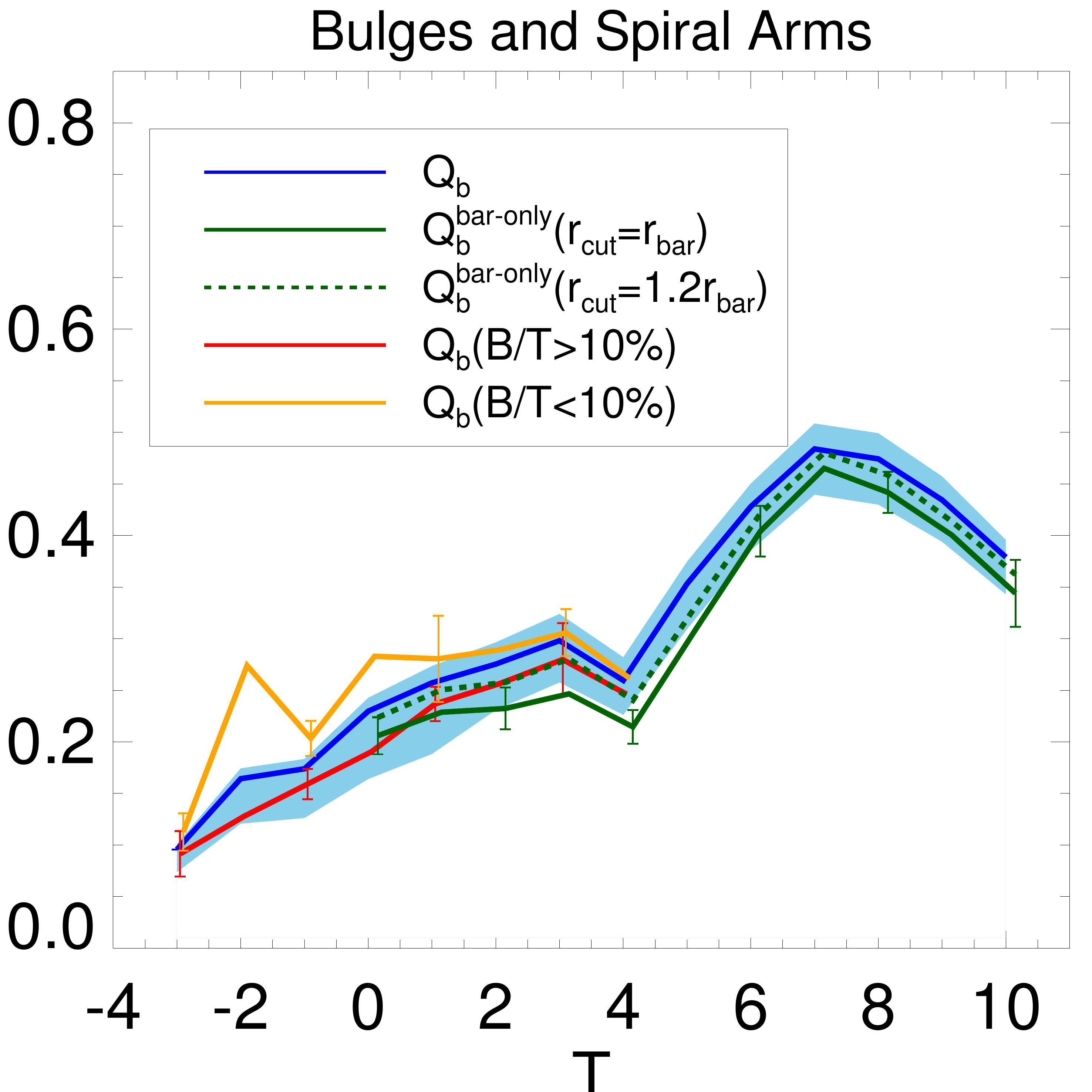}
    \end{tabular}
\caption{
Running mean of $Q_{\rm b}$ in the Hubble sequence, with and without spiral arms and dark halo corrections.
\emph{Left panel:}
Mean bar torque parameter obtained using the two disk thickness estimates applied in this study (see Sect.~\ref{thickmethods}): \citet[][]{1998MNRAS.299..595D} and \citet[][]{2008MNRAS.383..317S} (blue and red solid lines, respectively.).
The lower and upper limits for the disk thickness are calculated using the de Grijs' relation,
from which the corresponding mean bar torque parameter delimits the uncertainty in the estimate of $Q_{\rm b}$ (blue shaded area).
\emph{Central panel:}
The mean torque after the halo correction, following the different approaches described in Sect.~\ref{h-corrr}:
(i) using the universal rotation curve, (ii) following the methodology of B2004, and (iii) adding $V_{3.6 \mu \rm m}$+URC halo,
and correcting the URC halo amplitude to match the observed velocities. 
The unrealistic extreme case (iv) of assuming $V^{2}_{\rm halo}/V^{2}_{3.6 \mu\rm m}$ to be constant at all radii is also shown.
Corrections (i), (ii), (iii), and (iv) correspond to the light green, orange, red, and dark green solid lines, respectively.
The shaded area is computed from all the reliable bar strength measurements presented in this paper,
while the remaining moving averages (including that one of the raw $Q_{\rm b}$) are computed using only the galaxies with available halo correction.
\emph{Right panel:}
The moving average of $Q_{\rm b}$ after the spiral arm correction.
Galaxies with prominent bulges are studied separately, showing the role they play in diluting the bar forces.
We obtain $Q_{\rm b}^{\rm bar-only}$ setting the $m>0$ Fourier components to zero (1) beyond the bar length and 
(2) beyond 1.2 times the bar length in the potential calculation.
}
\label{halo-qb}
\end{figure*}
\begin{table}
\centering
\scalebox{0.875}{
\small
\begin{tabular}{|| c| c| c| c| c| c||}
\hline
\multicolumn{1}{||c}{\centering } & \multicolumn{1}{|c}{\centering $T < 0$} & \multicolumn{1}{|c}{\centering $T\in[0,3)$} &
\multicolumn{1}{|c}{\centering $T\in[3,5)$} & \multicolumn{1}{|c}{\centering $T\in[5,7]$} & \multicolumn{1}{|c||}{\centering $T\in(7,10]$} \tabularnewline
\hline
$r_{\rm Qb}/R_{\rm 25.5}$&0.18$\pm$0.01&0.21$\pm$0.01&0.17$\pm$0.01&0.09$\pm$0.00&0.13$\pm$0.01 \\
$r_{\rm Qb}/h_{\rm R}$&0.87$\pm$0.06&0.95$\pm$0.03&0.77$\pm$0.04&0.35$\pm$0.02&0.43$\pm$0.02 \\
\hline
$r_{\rm A2}/R_{\rm 25.5}$&0.20$\pm$0.01&0.22$\pm$0.01&0.17$\pm$0.01&0.12$\pm$0.00&0.19$\pm$0.01 \\
$r_{\rm A2}/h_{\rm R}$&0.97$\pm$0.06&0.99$\pm$0.04&0.79$\pm$0.04&0.47$\pm$0.02&0.70$\pm$0.04 \\
\hline
$r_{\epsilon}/R_{\rm 25.5}$&0.22$\pm$0.01&0.26$\pm$0.01&0.21$\pm$0.01&0.19$\pm$0.01&0.27$\pm$0.01 \\
$r_{\epsilon}/h_{\rm R}$&1.07$\pm$0.06&1.21$\pm$0.04&1.00$\pm$0.04&0.72$\pm$0.02&0.95$\pm$0.03 \\
\hline
$r_{\rm Qb}/r_{\rm vis}$&0.75$\pm$0.03&0.75$\pm$0.02&0.77$\pm$0.02&0.49$\pm$0.02&0.50$\pm$0.02 \\
\hline
$r_{\rm A2}/r_{\rm vis}$&0.83$\pm$0.02&0.79$\pm$0.02&0.80$\pm$0.02&0.66$\pm$0.02&0.74$\pm$0.03 \\
\hline
$r_{\epsilon}/r_{\rm vis}$&0.91$\pm$0.01&0.98$\pm$0.01&1.03$\pm$0.02&0.96$\pm$0.02&0.95$\pm$0.01 \\
\hline
\end{tabular}}
\caption{
Statistics of the scaled bar sizes as measured from $r_{\rm Qb}$, $r_{\rm A2}$ and $r_{\epsilon}$, and comparison of the visual measurements with the rest of the estimates. It is presented the mean and standard deviation of the mean.
}
\label{barlen_tabulate}
\end{table}

The prominence of a bar is displayed as a function of Hubble type in Fig.~\ref{ttype_qb_a2_ellip}.
In addition to $Q_{\rm b}$ and $A_2^{\rm max}$ discussed above, we also used the ellipticity maximum in the bar region, $\varepsilon$, as a proxy of the bar strength.

We confirm the tendencies from the previous studies, which showed that for the bright spiral galaxies with Hubble types T=0-6, the mean $Q_{\rm b}$ increases towards the later types, while the mean $A_2^{\rm max}$ decreases (likewise $A_4^{\rm max}$).
Our results agree with those of \citet[][]{2010ApJ...721..259B}, who showed that the frequency of strong bars, as estimated with $Q_{\rm b}$, is lower in S0 systems.
Early-type S0s are also characterized by a lower $A_2^{\rm max}$.
On the other hand, the maximum ellipticity is maintained nearly constant in each Hubble type bin, in agreement with \citet[][]{2007ApJ...659.1176M}. 
Here we additionally show that $\varepsilon$ decreases from T=0 towards the early-type S0s, as found earlier in \citet[][]{2007MNRAS.381..401L}.

The behaviour of the torque parameter among early-type galaxies is mostly explained by the dilution of $Q_{\rm b}$ by the underlying radial force field, primarily generated by bulges \citep[so-called \textup{\emph{\textup{bulge dilution,}}}][LS2002]{2001A&A...375..761B}.
The high $Q_{\rm b}$ values in the low-mass galaxies can be understood for the weak underlying disks (hence low $F_{\rm R}$, which is the denominator in the equation of $Q_{\rm b}$).
Furthermore, we have already given additional evidence that dark halos dominate the mass distribution of these systems,
whose effect on $Q_{\rm b}$ is discussed in Sect.~\ref{qb-uncertainty}.

We further investigate the loci of $r_{\rm A2}$, $r_{\rm Qb}$ , and $r_{\varepsilon}$ with respect to the visual bar length in the lower three panels of Fig.~\ref{ttype_qb_a2_ellip} and also in
Table~\ref{barlen_tabulate} with galaxies separated in morphological classes (see also Fig.~\ref{bar_visual_ellipse_flat}). 
As we have already discussed, these three proxies of the bar size tend to underestimate the bar length as compared to the visual measurement.
In addition, both $r_{\rm A2}$/$r_{\rm vis}$ and $r_{\rm Qb}$/$r_{\rm vis}$ ratios drop for late-type galaxies. 
This drop is remarkably large for the maximum gravitational torque and is related to the reduced central concentration towards later-type systems,
causing $r_{\rm Qb}$ to move inwards \citep[see][]{2002MNRAS.331..880L}.
In spite of this, all of these bar length proxies have the same behaviour in the Hubble sequence.  

How much the above tendencies depend on the mass of the galaxy or on the masses of the different structure components in galaxies
is discussed in Sect.~\ref{barlen_mass_sec}.
\subsubsection{Sources of uncertainty on the $Q_{\rm b}$ estimate}\label{qb-uncertainty}
We have carried out tests to probe possible sources of uncertainty in our $Q_{\rm b}$ estimates: (1) the assumption of a constant
scale height, (2) the impact of non-stellar emission in the bar region, (3) the contribution of the spiral arms torques to the overall
$Q_{\rm b}$ value, and (4) the effect of dark matter halos on the bar forcing.  

Here, we check whether these uncertainties can be large enough to change the observed 
trend of the bar torque parameter in the Hubble sequence (see Fig.~\ref{halo-qb}).

We have already concluded that uncertainties associated to the constant $h_{\rm z}$ assumption and the effect of non-stellar
contaminants are similar in size as the uncertainty related to the thickness of the disk ($\sim10-15\%$). However, as the effect of the
dark matter halo on the bar force depends on whether our disks are maximal or sub-maximal, the halo correction is likely to increase with $T$.
This is indeed the case for $T>4$.
Interestingly, earlier-type systems also have a strong halo correction, which is fairly constant in the T-type range
$[-3,4]$: the amplitude of the halo correction on $Q_{\rm T}$ increases with radius and the loci of $r_{\rm Qb}$ with respect to the
galaxy centre moves outwards for more concentrated galaxies, explaining the lack of difference on the mean $Q_{\rm b}^{\rm halo-corr}$
between intermediate-type spirals and early-type spirals and S0s. We obtain $\left<Q_{\rm b}^{\rm halo-corr}/Q_{\rm b}\right>=0.83\,(\sigma=0.15)$.

Here, we also compare the approaches described in Sect.~\ref{h-corrr} for the halo correction.
As shown in Fig.~\ref{halo-qb}, the halo correction made in B2004 (URC halo + circular velocity from the potential) and the direct correction using the URC halo+disk yield quite similar mean values.
Compared to the improved method applied in this work, we obtain a slightly larger correction for the earliest types.
The smaller $Q_{\rm b}$ correction present in the data of B2004 ($6\%$) might depend on the use of a sample with brighter galaxies and also on the methodology.

In the statistical sense the bar-only force measurements do not deviate considerably from the raw $Q_{\rm b}$ measurements ($\sim 10 \%$ on average),
in agreement with \citet{2007MNRAS.381..401L}. 
Only Sb-Sc galaxies show $Q_{\rm b}^{\rm bar-only}$ values that are slightly higher than the uncertainty in the raw estimate of $Q_{\rm b}$.
Setting the $m>0$ Fourier components to zero beyond $1.2$ times the bar size for the potential calculation 
(instead of using $r_{\rm cut}=r_{\rm bar}$) gives practically the same results as using the raw $Q_{\rm b}$.

In conclusion, none of the analysed sources of uncertainty alter the statistical trend of $Q_{\rm b}$ in the Hubble sequence. 
The difference of a factor $\sim 2$ in $Q_{\rm b}$ between early- and late-type
systems is slightly reduced when taking into account the dark halo correction, but this is not sufficient to change the monotonically 
raising trend of $Q_{\rm b}$ in the Hubble sequence. 
In spite of this, for individual galaxies dark halos might change the bar force measurements substantially.

\subsubsection{Gravitational torque at the end of the bar}

\begin{figure}
    \centering   
    \includegraphics[width=0.5\textwidth]{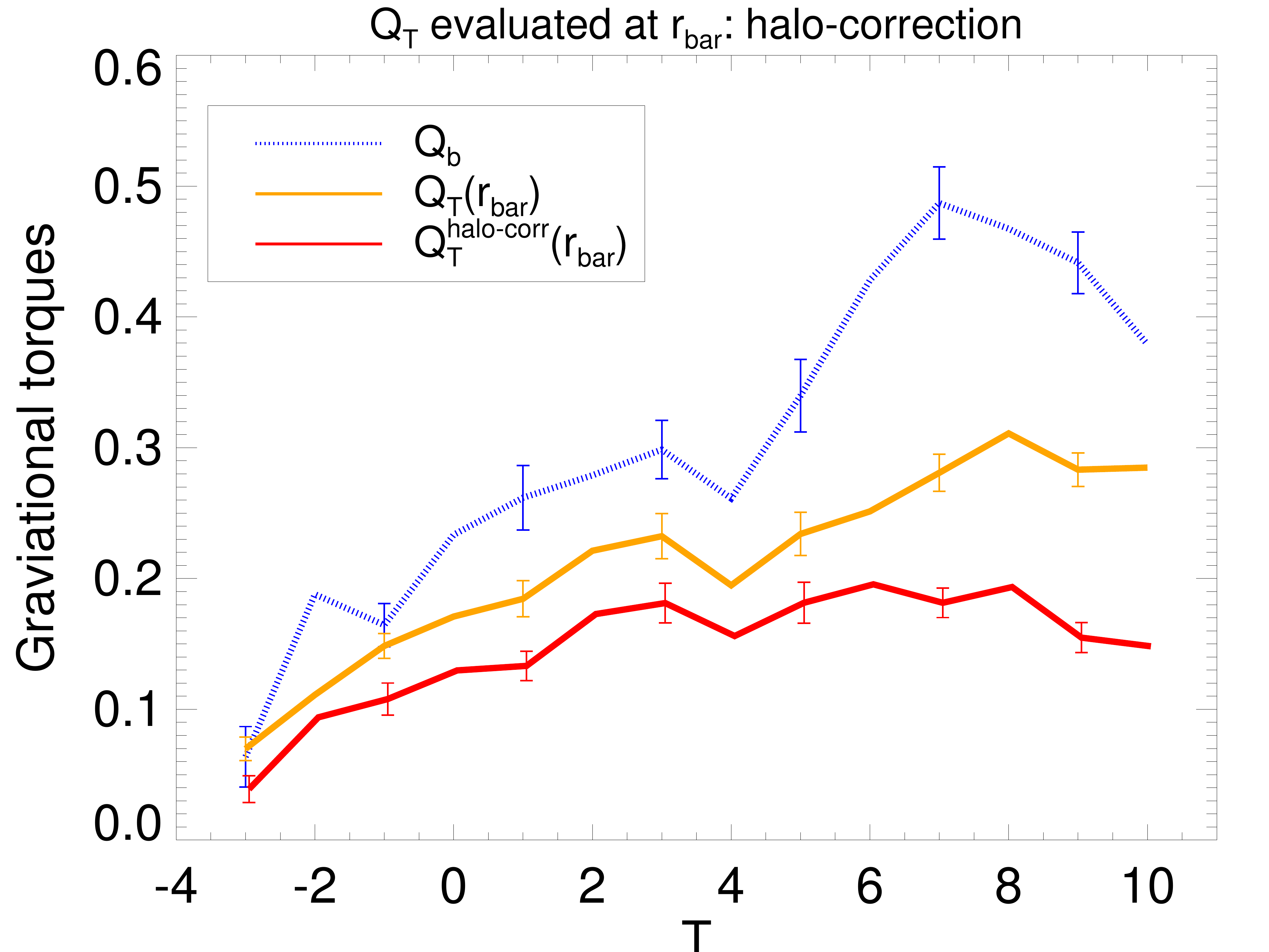}
\caption{
Moving average of the gravitational force evaluated at the end of the bar - $Q_{\rm T}(r_{\rm bar})$ - with and without halo correction.
}
\label{halo-qb-barend}
\end{figure}

In Fig.~\ref{halo-qb-barend} we assess how strong the halo correction would be if we were to evaluate the gravitational force at a radius larger than $r_{\rm Qb}$.
For instance, this is of vital importance to shed light on the bar-spiral interplay based on comparisons between their strengths, an idea which has remained controversial 
\citep[examples either supporting or refuting a connection in the coupling of these structures can be found in e.g.][]{2003MNRAS.342....1S,2004AJ....128..183B,2005AJ....130..506B,2009MNRAS.397.1756D},
although the recent analysis carried out in \citet{2009AJ....137.4487B} and \citet{2010ApJ...715L..56S} found evidence of a correlation between local bar forcing and local spiral amplitude.

We observe that the behaviour of $Q_{\rm T}(r_{\rm bar})$ in the Hubble sequence is similar to the distribution of $Q_{\rm b}$ shown in Fig.~\ref{halo-qb-barend}, with certain flattening among the later-type systems.
This is explained by the aforementioned tendency of $r_{\rm Qb}$ to move inwards for the faintest disks,
which causes $Q_{\rm T}$ to experience a more pronounced drop at $r_{\rm bar}$ relative to $Q_{\rm b}$.

The halo correction seems to be of the same order as in $Q_{\rm b}$ for $T\le 5$ and larger for later types ($\sim 30-50\%$),
making the average $Q_{\rm T}^{\rm halo-corr}(r_{\rm bar})$ value roughly constant ($\sim~0.2$) for intermediate-type spirals and later-type systems, and following a trend that is fairly similar to the trend of the bar intrinsic ellipticity.
\subsection{Bar parameters and the family class}
\begin{table*}
\footnotesize
\caption{Statistics of the measured bar strengths and lengths (in kpc and normalized to $R_{25.5}$ and $h_{\rm R}$), for all barred galaxies and also separated into morphological types and families.
For each of the different measurements and subsamples, the mean value, the standard deviation of the mean and the number of galaxies within the bin are given.
}
\label{bar_strength_statistics_structures}
\centering
\begin{tabular}{ c c c c c c c}
\hline
\hline
\multicolumn{1}{c}{\centering Buta et al. 2015} & \multicolumn{1}{c}{\centering $\rm Q_{\rm b}$} & \multicolumn{1}{c}{\centering $\rm A_{2}^{\rm max}$} &
\multicolumn{1}{c}{\centering $\rm \epsilon$} & \multicolumn{1}{c}{\centering $\rm r_{\rm bar}/R_{25.5}$} & \multicolumn{1}{c}{\centering $\rm r_{\rm bar}/h_{R}$} & \multicolumn{1}{c}{\centering $\rm r_{\rm bar} (kpc)$} \tabularnewline
\hline
\hline
{\tiny } & {\tiny } & {\tiny } & {\tiny } & {\tiny } & {\tiny } & {\tiny }\\
$\rm Barred$ &  0.35 $\pm$  0.01 $\emph{(576)}$ &   0.41 $\pm$  0.01 $\emph{(587)}$ &   0.52 $\pm$  0.01 $\emph{(654)}$ &   0.24 $\pm$  0.00 $\emph{(782)}$ &   0.98 $\pm$  0.01 $\emph{(742)}$ &   2.55 $\pm$  0.06 $\emph{(784)}$\\
{\tiny } & {\tiny } & {\tiny } & {\tiny } & {\tiny } & {\tiny } & {\tiny }\\
\multicolumn{7}{c}{Morphological type}  \vspace{0.1cm} \\
\hline
{\tiny } & {\tiny } & {\tiny } & {\tiny } & {\tiny } & {\tiny } & {\tiny }\\
$T<$ 0 &  0.16 $\pm$  0.01 $\emph{(50)}$ &   0.45 $\pm$  0.03 $\emph{(53)}$ &   0.40 $\pm$  0.02 $\emph{(47)}$ &   0.24 $\pm$  0.01 $\emph{(56)}$ &   1.22 $\pm$  0.06 $\emph{(50)}$ &   2.91 $\pm$  0.25 $\emph{(56)}$\\
$T\in$ [0,3) &  0.26 $\pm$  0.01 $\emph{(132)}$ &   0.51 $\pm$  0.02 $\emph{(134)}$ &   0.51 $\pm$  0.01 $\emph{(131)}$ &   0.27 $\pm$  0.01 $\emph{(146)}$ &   1.24 $\pm$  0.03 $\emph{(137)}$ &   3.76 $\pm$  0.19 $\emph{(146)}$\\
$T\in$ [3,5) &  0.28 $\pm$  0.01 $\emph{(89)}$ &   0.39 $\pm$  0.02 $\emph{(95)}$ &   0.52 $\pm$  0.02 $\emph{(93)}$ &   0.20 $\pm$  0.01 $\emph{(105)}$ &   0.96 $\pm$  0.03 $\emph{(102)}$ &   2.87 $\pm$  0.18 $\emph{(105)}$\\
$T\in$ [5,7] &  0.42 $\pm$  0.02 $\emph{(168)}$ &   0.36 $\pm$  0.01 $\emph{(174)}$ &   0.53 $\pm$  0.01 $\emph{(198)}$ &   0.20 $\pm$  0.01 $\emph{(234)}$ &   0.76 $\pm$  0.02 $\emph{(230)}$ &   2.07 $\pm$  0.07 $\emph{(235)}$\\
$T>$ 7 &  0.45 $\pm$  0.01 $\emph{(137)}$ &   0.37 $\pm$  0.01 $\emph{(131)}$ &   0.54 $\pm$  0.01 $\emph{(185)}$ &   0.28 $\pm$  0.01 $\emph{(241)}$ &   0.99 $\pm$  0.02 $\emph{(223)}$ &   2.06 $\pm$  0.07 $\emph{(242)}$\\
{\tiny } & {\tiny } & {\tiny } & {\tiny } & {\tiny } & {\tiny } & {\tiny }\\
$\rm S0^{-}$ &  0.10 $\pm$  0.02 $\emph{(10)}$ &   0.33 $\pm$  0.04 $\emph{(11)}$ &   0.24 $\pm$  0.03 $\emph{(7)}$ &   0.21 $\pm$  0.02 $\emph{(14)}$ &   1.13 $\pm$  0.12 $\emph{(12)}$ &   2.61 $\pm$  0.58 $\emph{(14)}$\\
$\rm S0^{0}$ &  0.16 $\pm$  0.03 $\emph{(16)}$ &   0.41 $\pm$  0.04 $\emph{(18)}$ &   0.39 $\pm$  0.03 $\emph{(16)}$ &   0.23 $\pm$  0.02 $\emph{(18)}$ &   1.30 $\pm$  0.10 $\emph{(17)}$ &   2.28 $\pm$  0.42 $\emph{(18)}$\\
$\rm S0^{+}$ &  0.17 $\pm$  0.01 $\emph{(24)}$ &   0.54 $\pm$  0.05 $\emph{(24)}$ &   0.45 $\pm$  0.02 $\emph{(24)}$ &   0.27 $\pm$  0.01 $\emph{(24)}$ &   1.21 $\pm$  0.09 $\emph{(21)}$ &   3.55 $\pm$  0.32 $\emph{(24)}$\\
$\rm S0/a$ &  0.23 $\pm$  0.02 $\emph{(40)}$ &   0.54 $\pm$  0.04 $\emph{(41)}$ &   0.48 $\pm$  0.02 $\emph{(42)}$ &   0.26 $\pm$  0.02 $\emph{(47)}$ &   1.21 $\pm$  0.06 $\emph{(43)}$ &   3.88 $\pm$  0.35 $\emph{(47)}$\\
$\rm Sa$ &  0.26 $\pm$  0.02 $\emph{(47)}$ &   0.52 $\pm$  0.04 $\emph{(49)}$ &   0.49 $\pm$  0.02 $\emph{(45)}$ &   0.28 $\pm$  0.02 $\emph{(54)}$ &   1.27 $\pm$  0.07 $\emph{(49)}$ &   3.78 $\pm$  0.32 $\emph{(54)}$\\
$\rm Sab$ &  0.27 $\pm$  0.02 $\emph{(41)}$ &   0.44 $\pm$  0.03 $\emph{(40)}$ &   0.54 $\pm$  0.02 $\emph{(40)}$ &   0.26 $\pm$  0.01 $\emph{(41)}$ &   1.24 $\pm$  0.05 $\emph{(41)}$ &   3.47 $\pm$  0.30 $\emph{(41)}$\\
$\rm Sb$ &  0.30 $\pm$  0.02 $\emph{(50)}$ &   0.44 $\pm$  0.03 $\emph{(54)}$ &   0.55 $\pm$  0.02 $\emph{(52)}$ &   0.22 $\pm$  0.01 $\emph{(58)}$ &   1.05 $\pm$  0.05 $\emph{(56)}$ &   3.13 $\pm$  0.24 $\emph{(58)}$\\
$\rm Sbc$ &  0.26 $\pm$  0.02 $\emph{(36)}$ &   0.31 $\pm$  0.03 $\emph{(38)}$ &   0.49 $\pm$  0.03 $\emph{(38)}$ &   0.18 $\pm$  0.01 $\emph{(43)}$ &   0.84 $\pm$  0.05 $\emph{(42)}$ &   2.50 $\pm$  0.28 $\emph{(43)}$\\
$\rm Sc$ &  0.35 $\pm$  0.03 $\emph{(51)}$ &   0.32 $\pm$  0.02 $\emph{(52)}$ &   0.49 $\pm$  0.02 $\emph{(61)}$ &   0.16 $\pm$  0.01 $\emph{(68)}$ &   0.69 $\pm$  0.03 $\emph{(67)}$ &   1.84 $\pm$  0.10 $\emph{(68)}$\\
$\rm Scd$ &  0.43 $\pm$  0.02 $\emph{(59)}$ &   0.37 $\pm$  0.02 $\emph{(61)}$ &   0.53 $\pm$  0.02 $\emph{(63)}$ &   0.18 $\pm$  0.01 $\emph{(74)}$ &   0.72 $\pm$  0.03 $\emph{(73)}$ &   2.06 $\pm$  0.12 $\emph{(75)}$\\
$\rm Sd$ &  0.48 $\pm$  0.03 $\emph{(57)}$ &   0.39 $\pm$  0.03 $\emph{(60)}$ &   0.55 $\pm$  0.02 $\emph{(73)}$ &   0.24 $\pm$  0.01 $\emph{(91)}$ &   0.86 $\pm$  0.04 $\emph{(89)}$ &   2.26 $\pm$  0.13 $\emph{(91)}$\\
$\rm Sdm$ &  0.47 $\pm$  0.02 $\emph{(74)}$ &   0.39 $\pm$  0.02 $\emph{(70)}$ &   0.57 $\pm$  0.02 $\emph{(95)}$ &   0.27 $\pm$  0.01 $\emph{(117)}$ &   0.93 $\pm$  0.03 $\emph{(114)}$ &   2.30 $\pm$  0.10 $\emph{(118)}$\\
$\rm Sm$ &  0.43 $\pm$  0.02 $\emph{(49)}$ &   0.38 $\pm$  0.02 $\emph{(47)}$ &   0.52 $\pm$  0.02 $\emph{(64)}$ &   0.29 $\pm$  0.01 $\emph{(84)}$ &   1.02 $\pm$  0.05 $\emph{(72)}$ &   1.95 $\pm$  0.12 $\emph{(84)}$\\
$\rm Im$ &  0.38 $\pm$  0.04 $\emph{(12)}$ &   0.28 $\pm$  0.05 $\emph{(12)}$ &   0.48 $\pm$  0.04 $\emph{(24)}$ &   0.30 $\pm$  0.02 $\emph{(35)}$ &   1.11 $\pm$  0.06 $\emph{(34)}$ &   1.57 $\pm$  0.15 $\emph{(35)}$\\
{\tiny } & {\tiny } & {\tiny } & {\tiny } & {\tiny } & {\tiny } & {\tiny }\\
\multicolumn{7}{c}{Galaxy family}  \vspace{0.1cm} \\
\hline
{\tiny } & {\tiny } & {\tiny } & {\tiny } & {\tiny } & {\tiny } & {\tiny }\\
$\rm SB$ &  0.45 $\pm$  0.01 $\emph{(260)}$ &   0.48 $\pm$  0.01 $\emph{(261)}$ &   0.59 $\pm$  0.01 $\emph{(277)}$ &   0.25 $\pm$  0.01 $\emph{(323)}$ &   1.02 $\pm$  0.02 $\emph{(308)}$ &   2.73 $\pm$  0.10 $\emph{(323)}$\\
$\rm SA\underline{B}$ &  0.34 $\pm$  0.02 $\emph{(67)}$ &   0.39 $\pm$  0.02 $\emph{(69)}$ &   0.55 $\pm$  0.02 $\emph{(76)}$ &   0.27 $\pm$  0.01 $\emph{(96)}$ &   1.03 $\pm$  0.04 $\emph{(89)}$ &   2.38 $\pm$  1.44 $\emph{(96)}$\\
$\rm SAB$ &  0.26 $\pm$  0.01 $\emph{(193)}$ &   0.38 $\pm$  0.02 $\emph{(193)}$ &   0.46 $\pm$  0.01 $\emph{(226)}$ &   0.24 $\pm$  0.01 $\emph{(264)}$ &   0.97 $\pm$  0.02 $\emph{(253)}$ &   2.54 $\pm$  0.10 $\emph{(266)}$\\
$\rm S\underline{A}B$ &  0.20 $\pm$  0.02 $\emph{(33)}$ &   0.26 $\pm$  0.02 $\emph{(37)}$ &   0.36 $\pm$  0.02 $\emph{(46)}$ &   0.18 $\pm$  0.01 $\emph{(63)}$ &   0.78 $\pm$  0.06 $\emph{(61)}$ &   2.06 $\pm$  0.19 $\emph{(63)}$\\
{\tiny } & {\tiny } & {\tiny } & {\tiny } & {\tiny } & {\tiny } & {\tiny }\\
{\tiny } & {\tiny } & {\tiny } & {\tiny } & {\tiny } & {\tiny } & {\tiny }\\
\multicolumn{7}{c}{S0 galaxies: $T<$ 0}  \vspace{0.1cm} \\
$\rm SB$ &  0.19 $\pm$  0.02 $\emph{(17)}$ &   0.50 $\pm$  0.04 $\emph{(17)}$ &   0.44 $\pm$  0.02 $\emph{(18)}$ &   0.25 $\pm$  0.02 $\emph{(18)}$ &   1.31 $\pm$  0.07 $\emph{(17)}$ &   2.88 $\pm$  0.43 $\emph{(18)}$\\
$\rm SA\underline{B}$ &  0.17 $\pm$  0.03 $\emph{(7)}$ &   0.42 $\pm$  0.05 $\emph{(8)}$ &   0.39 $\pm$  0.04 $\emph{(7)}$ &   0.18 $\pm$  0.02 $\emph{(9)}$ &   1.13 $\pm$  0.10 $\emph{(8)}$ &   2.14 $\pm$  1.42 $\emph{(9)}$\\
$\rm SAB$ &  0.13 $\pm$  0.01 $\emph{(17)}$ &   0.46 $\pm$  0.07 $\emph{(17)}$ &   0.39 $\pm$  0.03 $\emph{(17)}$ &   0.24 $\pm$  0.02 $\emph{(18)}$ &   1.10 $\pm$  0.11 $\emph{(16)}$ &   2.79 $\pm$  0.39 $\emph{(18)}$\\
$\rm S\underline{A}B$ &  0.13 $\pm$  0.03 $\emph{(8)}$ &   0.39 $\pm$  0.07 $\emph{(8)}$ &   0.32 $\pm$  0.09 $\emph{(4)}$ &   0.29 $\pm$  0.03 $\emph{(8)}$ &   1.45 $\pm$  0.23 $\emph{(7)}$ &   3.98 $\pm$  1.00 $\emph{(8)}$\\
{\tiny } & {\tiny } & {\tiny } & {\tiny } & {\tiny } & {\tiny } & {\tiny }\\
\multicolumn{7}{c}{Early-type spirals [S0/a-Sb): $T\in$ [0,3)}  \vspace{0.1cm} \\
$\rm SB$ &  0.33 $\pm$  0.02 $\emph{(54)}$ &   0.59 $\pm$  0.03 $\emph{(55)}$ &   0.59 $\pm$  0.01 $\emph{(51)}$ &   0.29 $\pm$  0.01 $\emph{(59)}$ &   1.31 $\pm$  0.05 $\emph{(56)}$ &   4.15 $\pm$  0.31 $\emph{(59)}$\\
$\rm SA\underline{B}$ &  0.28 $\pm$  0.05 $\emph{(8)}$ &   0.57 $\pm$  0.08 $\emph{(8)}$ &   0.56 $\pm$  0.03 $\emph{(8)}$ &   0.30 $\pm$  0.02 $\emph{(8)}$ &   1.36 $\pm$  0.14 $\emph{(8)}$ &   4.23 $\pm$  0.86 $\emph{(8)}$\\
$\rm SAB$ &  0.20 $\pm$  0.01 $\emph{(58)}$ &   0.47 $\pm$  0.03 $\emph{(57)}$ &   0.46 $\pm$  0.02 $\emph{(56)}$ &   0.26 $\pm$  0.01 $\emph{(61)}$ &   1.22 $\pm$  0.05 $\emph{(56)}$ &   3.63 $\pm$  0.30 $\emph{(61)}$\\
$\rm S\underline{A}B$ &  0.19 $\pm$  0.08 $\emph{(3)}$ &   0.23 $\pm$  0.07 $\emph{(4)}$ &   0.34 $\pm$  0.09 $\emph{(5)}$ &   0.23 $\pm$  0.03 $\emph{(7)}$ &   1.05 $\pm$  0.17 $\emph{(7)}$ &   2.90 $\pm$  0.62 $\emph{(7)}$\\
{\tiny } & {\tiny } & {\tiny } & {\tiny } & {\tiny } & {\tiny } & {\tiny }\\
\multicolumn{7}{c}{Intermediate-type spirals [Sb-Sc): $T\in$ [3,5)}  \vspace{0.1cm} \\
$\rm SB$ &  0.43 $\pm$  0.03 $\emph{(39)}$ &   0.49 $\pm$  0.03 $\emph{(38)}$ &   0.61 $\pm$  0.02 $\emph{(39)}$ &   0.23 $\pm$  0.01 $\emph{(41)}$ &   1.02 $\pm$  0.05 $\emph{(41)}$ &   3.30 $\pm$  0.36 $\emph{(41)}$\\
$\rm SA\underline{B}$ &  0.28 $\pm$  0.02 $\emph{(17)}$ &   0.37 $\pm$  0.05 $\emph{(17)}$ &   0.58 $\pm$  0.03 $\emph{(16)}$ &   0.22 $\pm$  0.02 $\emph{(17)}$ &   1.02 $\pm$  0.08 $\emph{(17)}$ &   2.88 $\pm$  1.96 $\emph{(17)}$\\
$\rm SAB$ &  0.27 $\pm$  0.02 $\emph{(59)}$ &   0.36 $\pm$  0.02 $\emph{(64)}$ &   0.49 $\pm$  0.02 $\emph{(67)}$ &   0.18 $\pm$  0.01 $\emph{(78)}$ &   0.83 $\pm$  0.03 $\emph{(77)}$ &   2.29 $\pm$  0.13 $\emph{(78)}$\\
$\rm S\underline{A}B$ &  0.23 $\pm$  0.04 $\emph{(14)}$ &   0.25 $\pm$  0.02 $\emph{(16)}$ &   0.39 $\pm$  0.04 $\emph{(19)}$ &   0.12 $\pm$  0.01 $\emph{(23)}$ &   0.63 $\pm$  0.07 $\emph{(23)}$ &   1.59 $\pm$  0.12 $\emph{(23)}$\\
{\tiny } & {\tiny } & {\tiny } & {\tiny } & {\tiny } & {\tiny } & {\tiny }\\
\multicolumn{7}{c}{Late-type spirals [Sc-Sd]: $T\in$ [5,7]}  \vspace{0.1cm} \\
$\rm SB$ &  0.52 $\pm$  0.02 $\emph{(92)}$ &   0.44 $\pm$  0.02 $\emph{(93)}$ &   0.61 $\pm$  0.01 $\emph{(94)}$ &   0.21 $\pm$  0.01 $\emph{(108)}$ &   0.80 $\pm$  0.03 $\emph{(107)}$ &   2.26 $\pm$  0.12 $\emph{(108)}$\\
$\rm SA\underline{B}$ &  0.41 $\pm$  0.04 $\emph{(12)}$ &   0.34 $\pm$  0.02 $\emph{(13)}$ &   0.62 $\pm$  0.02 $\emph{(15)}$ &   0.22 $\pm$  0.03 $\emph{(15)}$ &   0.72 $\pm$  0.06 $\emph{(14)}$ &   2.14 $\pm$  0.91 $\emph{(15)}$\\
$\rm SAB$ &  0.32 $\pm$  0.02 $\emph{(46)}$ &   0.29 $\pm$  0.02 $\emph{(49)}$ &   0.47 $\pm$  0.02 $\emph{(60)}$ &   0.20 $\pm$  0.01 $\emph{(74)}$ &   0.79 $\pm$  0.04 $\emph{(73)}$ &   1.97 $\pm$  0.11 $\emph{(75)}$\\
$\rm S\underline{A}B$ &  0.25 $\pm$  0.06 $\emph{(8)}$ &   0.17 $\pm$  0.02 $\emph{(9)}$ &   0.32 $\pm$  0.03 $\emph{(18)}$ &   0.13 $\pm$  0.01 $\emph{(24)}$ &   0.60 $\pm$  0.06 $\emph{(24)}$ &   1.46 $\pm$  0.12 $\emph{(24)}$\\
{\tiny } & {\tiny } & {\tiny } & {\tiny } & {\tiny } & {\tiny } & {\tiny }\\
\multicolumn{7}{c}{Magellanic and Irregular galaxies [Sdm,Sm,Im]: $T > $ 7}  \vspace{0.1cm} \\
$\rm (S/I)B$ &  0.52 $\pm$  0.02 $\emph{(74)}$ &   0.42 $\pm$  0.02 $\emph{(73)}$ &   0.60 $\pm$  0.02 $\emph{(89)}$ &   0.27 $\pm$  0.01 $\emph{(113)}$ &   1.00 $\pm$  0.04 $\emph{(103)}$ &   2.12 $\pm$  0.10 $\emph{(113)}$\\
$\rm (S/I)A\underline{B}$ &  0.42 $\pm$  0.02 $\emph{(25)}$ &   0.37 $\pm$  0.03 $\emph{(25)}$ &   0.54 $\pm$  0.03 $\emph{(32)}$ &   0.31 $\pm$  0.01 $\emph{(49)}$ &   1.04 $\pm$  0.05 $\emph{(44)}$ &   2.02 $\pm$  1.16 $\emph{(49)}$\\
$\rm (S/I)AB$ &  0.35 $\pm$  0.01 $\emph{(33)}$ &   0.29 $\pm$  0.02 $\emph{(28)}$ &   0.47 $\pm$  0.02 $\emph{(52)}$ &   0.30 $\pm$  0.01 $\emph{(62)}$ &   0.99 $\pm$  0.04 $\emph{(60)}$ &   2.05 $\pm$  0.13 $\emph{(63)}$\\
$\rm (S/I)\underline{A}B$ &  0.28 $\pm$  0.04 $\emph{(4)}$ &   0.19 $\pm$  0.02 $\emph{(4)}$ &   0.39 $\pm$  0.06 $\emph{(9)}$ &   0.23 $\pm$  0.02 $\emph{(12)}$ &   0.70 $\pm$  0.08 $\emph{(11)}$ &   1.70 $\pm$  0.23 $\emph{(12)}$\\
{\tiny } & {\tiny } & {\tiny } & {\tiny } & {\tiny } & {\tiny } & {\tiny }\\
\hline
\hline
\end{tabular}
\end{table*}

\begin{figure*}
   \centering   
   \includegraphics[width=1.0\textwidth]{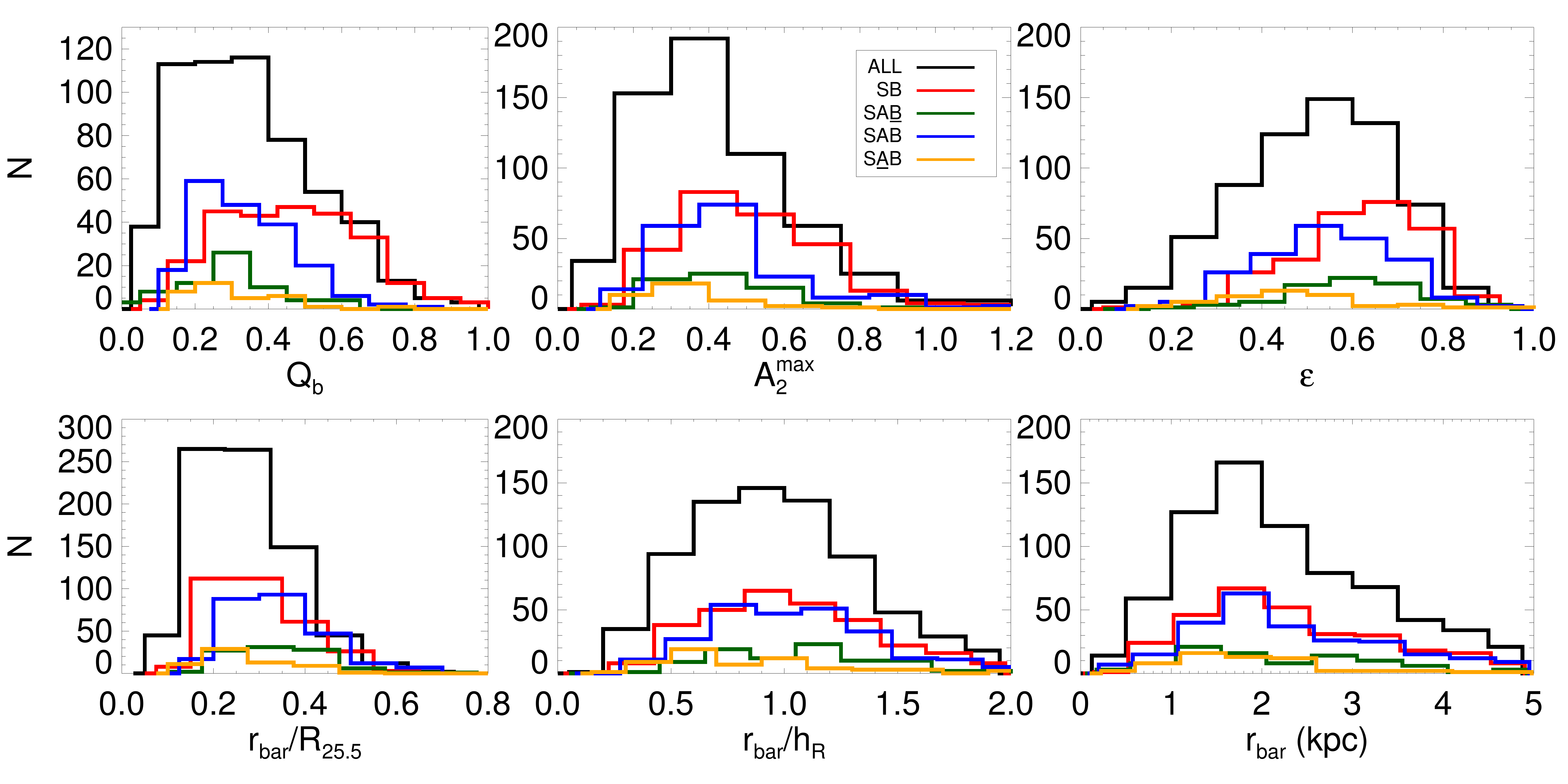}
\caption{
From left to right and top to bottom, histograms of the distributions of the bar gravitational torque parameter $Q_{\rm b}$, bar density amplitude  $A_{2}^{\rm max}$, bar intrinsic ellipticity $\varepsilon$,
$R_{25.5}$-relative bar size, $h_{\rm R}$-relative bar size, and bar length in physical units.
}
\label{histo_barfamily_ttype}
\end{figure*}
\begin{figure}
   \centering   
   \includegraphics[width=0.5\textwidth]{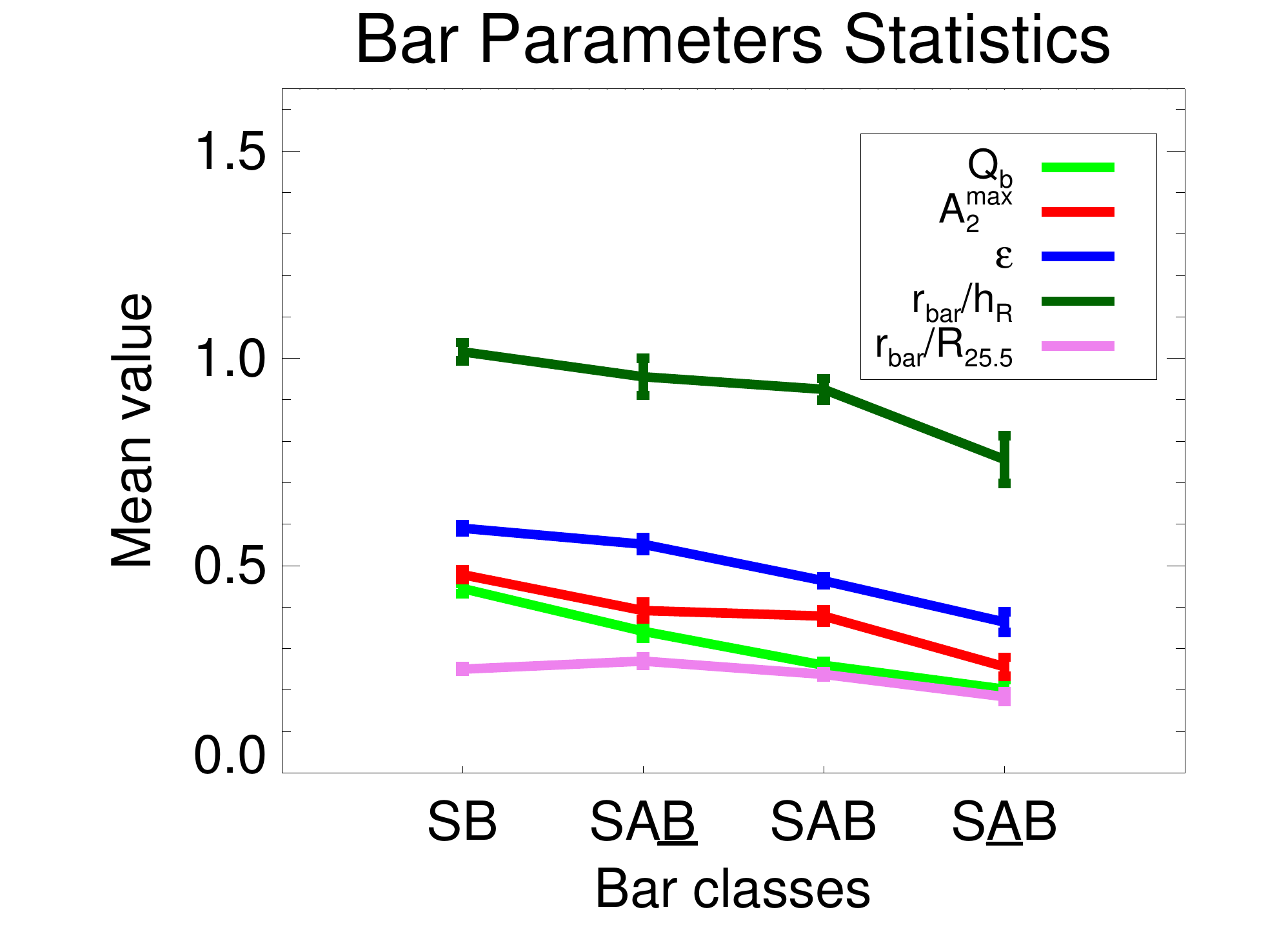} 
\caption{
Mean bar strengths and disk-relative bar sizes for all the different bar families. Error bars are the standard deviation of the mean.
Both irregular and spiral galaxies are considered in the analysis.
}
\label{histo_barfamily_mean}
\end{figure}
While comparing the visual bar classifications in B2015 (using the de Vauculeurs family classification nomenclature from RC3) with the measures of the prominence of the bar obtained in this study,
it is evident that the quantitative estimates are well consistent with the visual ones. 
This is shown in Figs.~\ref{histo_barfamily_ttype} and \ref{histo_barfamily_mean} (see also Table~\ref{bar_strength_statistics_structures}),
where the distributions of bar strengths and lengths are shown separated for the different bar families. 

\citet[][]{2005MNRAS.364..283E} concluded that strong (SB) and weak (SAB) bars hosted by early-type galaxies differ primarily in ellipticity, being very similar in size,
while SB bars in late-type spiral galaxies are twice as large as SAB bars.
We observe that SB, SA$\underline{\rm B}$, SAB, and S$\underline{\rm A}$B bars differ in ellipticity as well as in $A_{2}^{\rm max}$ and $Q_{\rm b}$ for all morphological types, as shown in Table~\ref{bar_strength_statistics_structures}.
The decreasing tendency from SB towards the S$\underline{\rm A}$B family is particularly clear for the bar torque parameter.
In the statistical sense, there is also a difference between weak and strong bars for the bar sizes, as visually categorized: 
SB and SA$\underline{\rm B}$ galaxies are $\sim 10-15\%$ longer relative to the disk size.
We already find a similar difference in disk-relative bar sizes between weak and strong bars among early-type spirals (0 $\le T$ < 3).
However, for S0s weak and strong bars have similar mean bar sizes.
On the other hand, although our bars are on average slightly longer for SB and SA$\underline{\rm B}$ families in the late-type systems,
we do not obtain such a huge difference between weak and strong bars (factor $\sim 2$ in E2005).
%
%
\subsection{Comparison of bar strength measurements}
\begin{figure*}
   \includegraphics[width=1.0\textwidth]{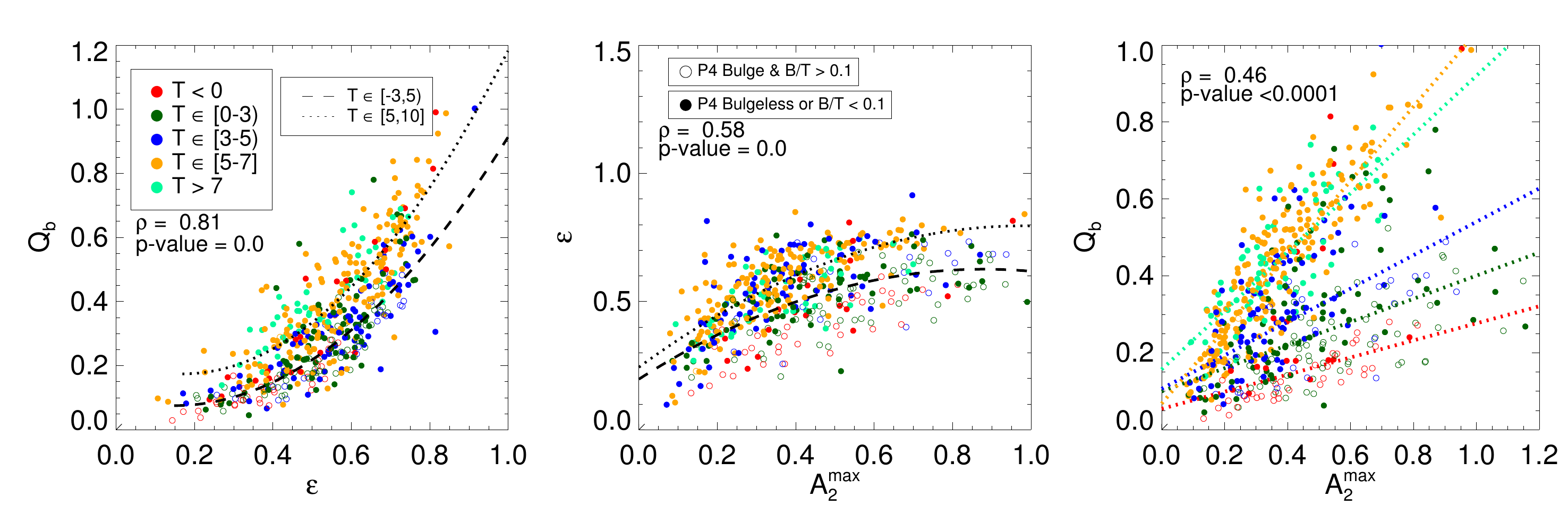}
\caption{
Comparison of bar strength indexes.
Galaxies are separated into $T$ bins.
Spearman's rank correlation coefficients (together with the significance) are indicated in all the plots for the total sample (see Table~\ref{barstrength_correlations} for the different morphological types).
Systems with relatively massive (faint) bulges are plotted with filled (empty) circles.
In the left and central panels, the dotted and dashed lines display a second-order polynomial fit to the data separately for $T<5$ and $T\ge5$, respectively.
In the right panel we show the linear fits for different morphological classes.
}
\label{forces_correlation}
\end{figure*}
\begin{table*}
\scalebox{0.7}{
\small
\centering
\begin{tabular}{|| c| c| c| c|| c| c| c|| c| c| c|}
\hline
\multicolumn{1}{||c}{\centering Correlations:} & \multicolumn{1}{|c}{\centering $Q_{\rm b}$ vs. $\varepsilon$} & \multicolumn{1}{|c}{\centering $\varepsilon$ vs. $A_{2}^{\rm max}$} &
\multicolumn{1}{|c||}{\centering $Q_{\rm b}$ vs. $A_{2}^{\rm max}$} &
\multicolumn{1}{|c}{\centering $Q_{\rm b}^{\rm halo-corr}$ vs. $\varepsilon$} & \multicolumn{1}{|c}{\centering $Q_{\rm T}(r_{\varepsilon})$ vs. $\varepsilon$} &
\multicolumn{1}{|c||}{\centering $Q_{\rm T}^{\rm halo-corr}(r_{\varepsilon})$ vs. $\varepsilon$} &
\multicolumn{1}{|c}{\centering $Q_{\rm b}^{\rm halo-corr}$ vs. $A_{2}^{\rm max}$} & \multicolumn{1}{|c}{\centering $Q_{\rm T}(r_{\rm bar})$ vs. $A_{2}(r_{\rm bar})$} &
\multicolumn{1}{|c||}{\centering $Q_{\rm T}^{\rm halo-corr}(r_{\rm bar})$ vs. $A_{2}(r_{\rm bar})$} \tabularnewline
\hline
\hline
$T < 0$& $  0.80\:(  6.5\cdot 10^{-11})$&$  0.84\:(  5.2\cdot 10^{-13})$&$  0.72\:(  3.9\cdot 10^{-9})$&$  0.79\:(  8.1\cdot 10^{-6})$&$  0.78\:(  9.0\cdot 10^{-11})$&$  0.78\:(  9.0\cdot 10^{-11})$&$  0.73\:(  3.2\cdot 10^{-5})$&$  0.40\:(   0.\cdot 10^{-3})$&$  0.44\:(    0\cdot 10^{-2})$\\
\hline
$T\in [0,3)$& $  0.88\:(  3.1\cdot 10^{-41})$&$  0.58\:(  1.3\cdot 10^{-12})$&$  0.62\:(  3.4\cdot 10^{-15})$&$  0.80\:(  3.2\cdot 10^{-23})$&$  0.85\:(  5.9\cdot 10^{-37})$&$  0.85\:(  5.9\cdot 10^{-37})$&$  0.64\:(  9.8\cdot 10^{-13})$&$  0.43\:(  3.7\cdot 10^{-8})$&$  0.42\:(  4.7\cdot 10^{-6})$\\
\hline
$T\in [3,5)$& $  0.88\:(  1.6\cdot 10^{-27})$&$  0.65\:(  1.9\cdot 10^{-11})$&$  0.70\:(  3.9\cdot 10^{-14})$&$  0.84\:(  4.8\cdot 10^{-20})$&$  0.68\:(  7.3\cdot 10^{-14})$&$  0.68\:(  7.3\cdot 10^{-14})$&$  0.67\:(  2.4\cdot 10^{-11})$&$  0.42\:(  6.3\cdot 10^{-6})$&$  0.44\:(  1.3\cdot 10^{-5})$\\
\hline
$T\in [5,7]$& $  0.86\:(  7.5\cdot 10^{-44})$&$  0.81\:(  7.9\cdot 10^{-36})$&$  0.88\:(0.0)$&$  0.81\:(  3.0\cdot 10^{-33})$&$  0.70\:(  1.2\cdot 10^{-30})$&$  0.70\:(  1.2\cdot 10^{-30})$&$  0.83\:(  1.0\cdot 10^{-40})$&$  0.58\:(  8.2\cdot 10^{-24})$&$  0.48\:(  1.4\cdot 10^{-14})$\\
\hline
$T > 7$& $  0.84\:(  3.5\cdot 10^{-29})$&$  0.69\:(  2.8\cdot 10^{-15})$&$  0.78\:(  2.0\cdot 10^{-27})$&$  0.76\:(  5.0\cdot 10^{-18})$&$  0.74\:(  2.4\cdot 10^{-33})$&$  0.74\:(  2.4\cdot 10^{-33})$&$  0.66\:(  4.2\cdot 10^{-15})$&$  0.69\:(  1.2\cdot 10^{-43})$&$  0.50\:(  1.6\cdot 10^{-18})$\\
\hline
\end{tabular}}
\caption{
Spearman's rank correlation coeficient and significance (in parenthesis) for the relation between the various bar strength proxies, with and without halo correction, in different morphological class bins.}
\label{barstrength_correlations}
\end{table*}

\begin{figure*}
   \centering   
   \begin{tabular}{ l c r}
   \includegraphics[width=0.32\textwidth]{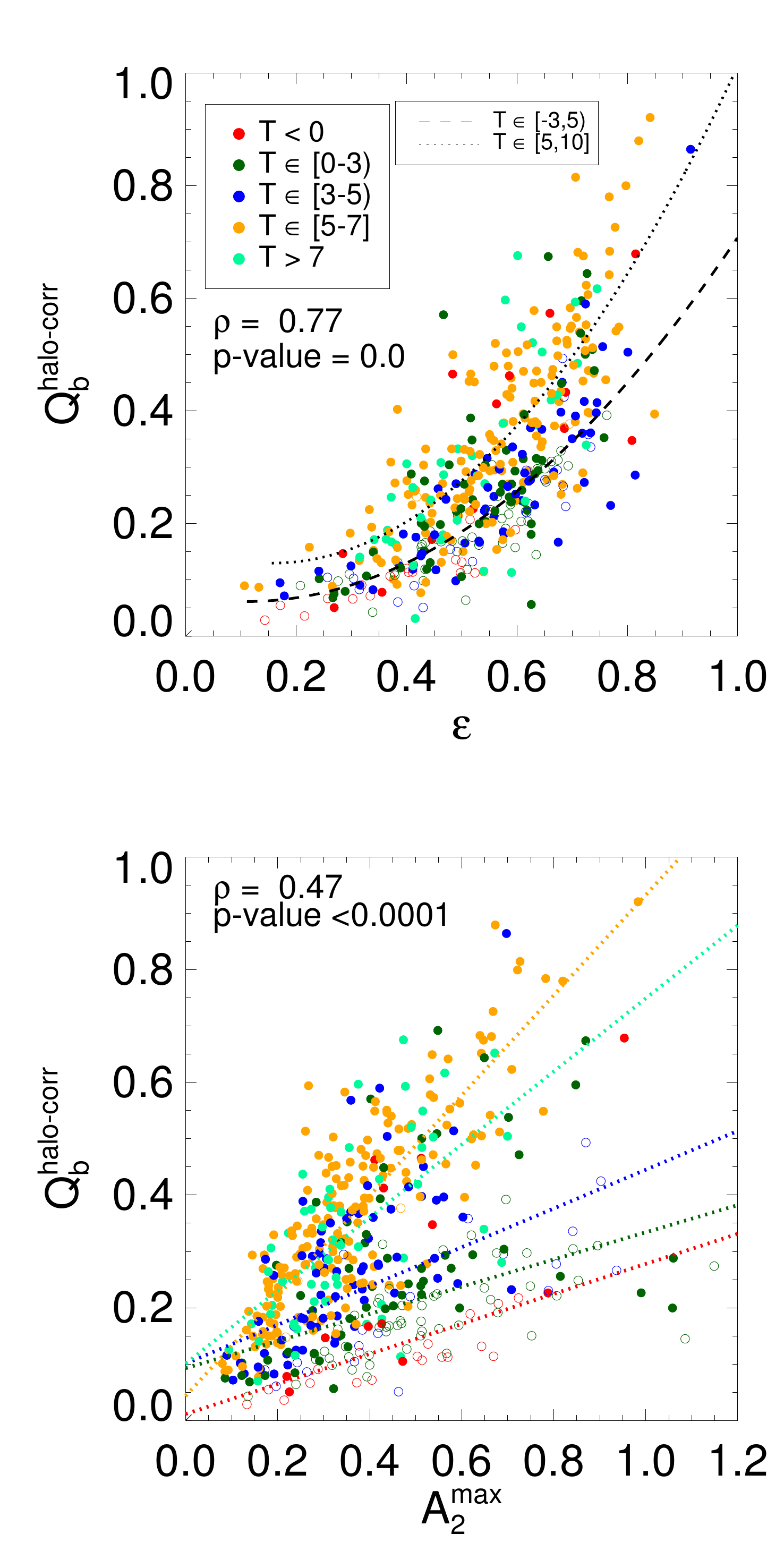}
   \includegraphics[width=0.32\textwidth]{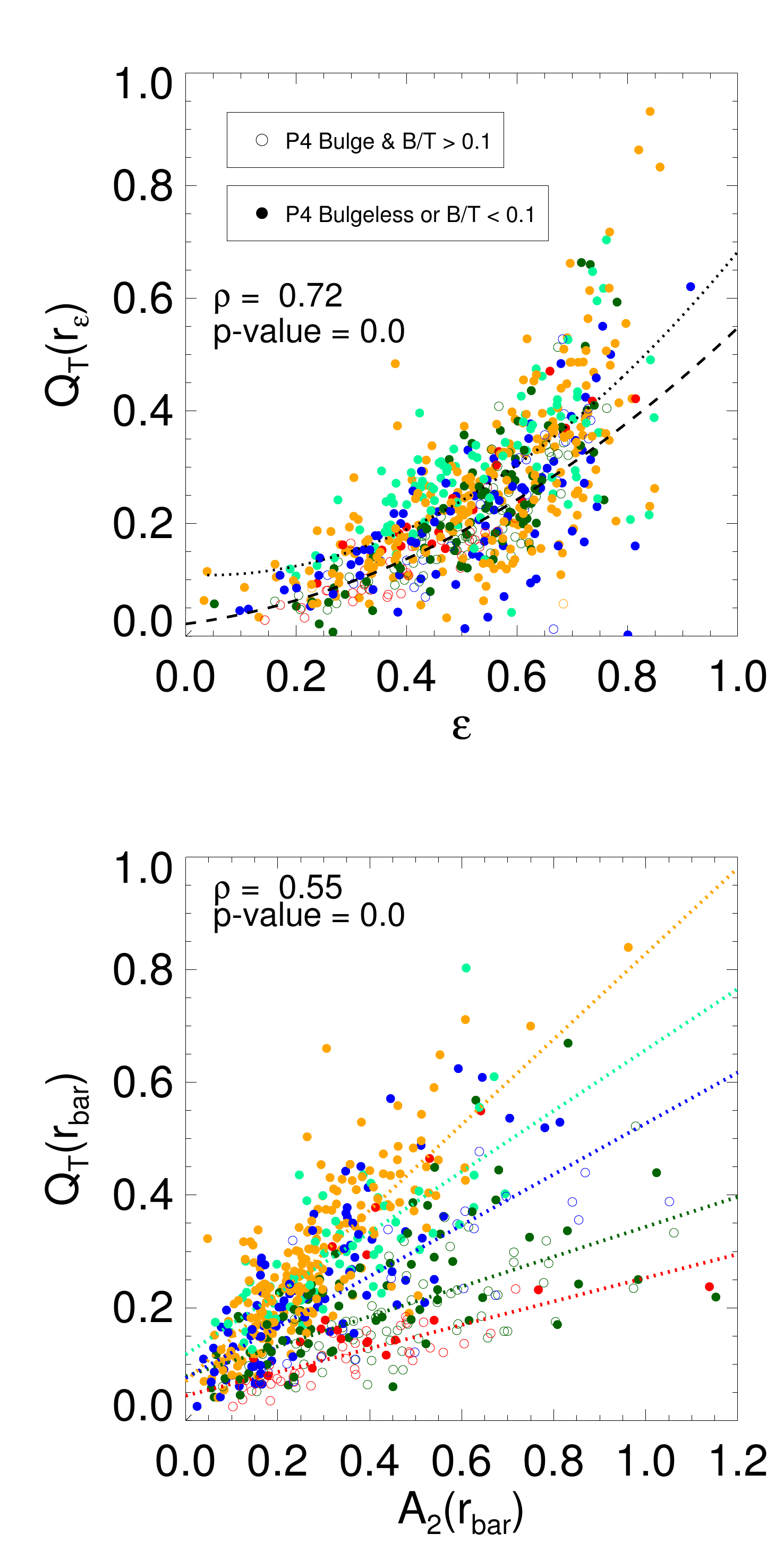}
   \includegraphics[width=0.32\textwidth]{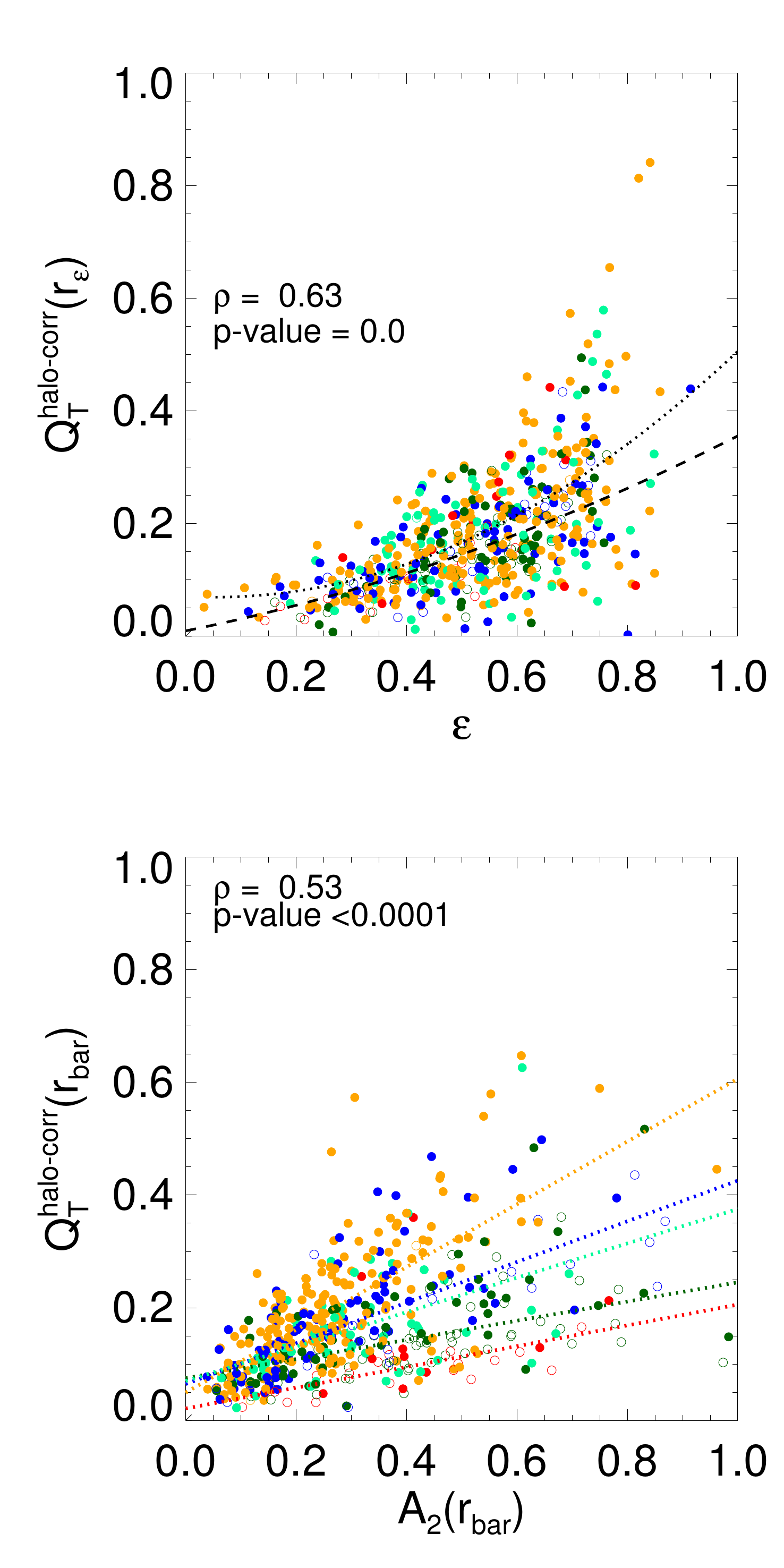}
   \end{tabular}
\caption{
Same as in Fig.~\ref{forces_correlation}, but using the halo-corrected values of $Q_{\rm b}$ and also the forces evaluated at the bar end ($r_{\rm bar}$).
The same colour palette and symbols are used as in the previous plots.
The Spearman's rank correlation coefficient and significance of the relation for different $T$ bins are displayed in Table~\ref{barstrength_correlations}.
}
\label{forces_correlation_dm_barend}
\end{figure*}
In Fig.~\ref{forces_correlation} we show a comparison of the different bar strength indexes, confirming some of the trends earlier reported in the literature.
We observe a correlation between the intrinsic ellipticity of the bar and the bar gravitational torque \citep[e.g.][LS2002]{2001A&A...375..761B}, which
shows that underlying potential tends to adjust the orbit of the stars that constitute the bar.
Galaxies of all Hubble types present the same pattern, although two families of bars can be identified based on the trends in the plot:
S0s, early-type, and intermediate-type spirals occupy the lower part, and late-type spirals, Magellanics, and irregulars are
located in the upper area.
For these families the data are fitted separately with a polynomial of order 2:
\begin{equation}
\begin{split}
\hat{Q_{\rm b}}=0.1-0.34\varepsilon+1.16\varepsilon^{2} \:\: \text{, if} \:T<5 \: \text{and} \: \varepsilon\in(0.15,0.8),\\
\hat{Q_{\rm b}}=0.21-0.49\varepsilon+1.46\varepsilon^{2} \:\: \text{, if} \:T\ge5 \: \text{and} \: \varepsilon\in(0.25,0.9).\\
\end{split}
\end{equation}
For the galaxies with $T<5$, the deviation with respect to the fitting curve is small ($\sim 5\%$), while for later types the scatter is slightly larger ($\sim 7.5\%$).
The bar ellipticity is fairly easy to obtain from photometry,
therefore these equations provide an empirical proxy to the bar torque parameter.

As discussed in \citet[][]{1995AJ....109.2428M}, $\varepsilon$ is insensitive to uniform changes in the relative mass of a bar of fixed shape and size.
This constitutes one of the weaknesses of the bar strength proxies that exclusively
rely on the bar shape (e.g. ellipticity and boxyness), and it partly explains the shape of the $Q_{\rm b}-\varepsilon$
relation and its dependence on the bar-to-total ratio (as illustrated in LS2002 for analytical Ferrers bar models).
Both $Q_{\rm b}$ and $A_{2}^{\rm max}$ account for the relative mass of the bar and for other stellar structures.

When comparing $Q_{\rm b}$ with $A_{2}^{\rm max}$ (right plot of Fig.~\ref{forces_correlation}), a bimodality is evident.
A similar trend was present in the measurements of \citet[][]{2004ApJ...607..103L}, who separated galaxies into early- and late-type disks.
The lower part of the plot is mostly occupied by S0 and early-type spirals, and the upper arm is formed by late-type systems.
Sb and Sc galaxies are more or less evenly distributed in the two branches of this relation.
As we argue in Sect.~\ref{bar_bulges}, this trend is probably
mostly determined by the influence of the degree of galaxy central mass concentration in the bar force calculation,
which increases this effect more strongly on $Q_{\rm b}$ than on $A_{2}^{\rm max}$.

Finally, we show a monotonic rising trend between $A_{2}^{\rm max}$ and $\varepsilon$ (central panel of Fig.~\ref{forces_correlation})
that resembles the $Q_{\rm b}-\varepsilon$ relation in shape, but with a convex curve.
As for the bar torque parameter, for a narrow ellipticity interval, $A_{2}^{\rm max}$ is larger in this branch for earlier types.
Because the $m=2$ Fourier amplitude is a good proxy of the bar-to-total mass ratio, this trend could again be partially explained by the insensitivity of $\varepsilon$ to the relative mass of the bar of a certain shape.

In Fig.~\ref{forces_correlation_dm_barend} we study the $Q_{\rm b}-\varepsilon$ and $A_{2}^{\rm max}-Q_{\rm b}$ relations taking into account the halo correction and also considering an evaluation
of the forces at the end of the bar.
The observed dichotomy between late-type and early-type systems in these relations is maintained regardless of whether we use $Q_{\rm b}^{\rm halo-corr}$ values or not.
The relation between $\varepsilon$ and $Q_{\rm T}$ evaluated at the same radius resembles the same trend as obtained for $Q_{\rm b}$,
Likewise, the bimodality in the comparison with $A_{2}$  weakens, but is undoubtedly present.
Finally, we observe that considering the effect of halos for the $Q_{\rm T}$ evaluation at the bar radius obscures the segregation in the $\varepsilon-Q_{\rm T}^{\rm halo-corr}(r_{\varepsilon})$ relation;
it converges to a common trend for all Hubble types, but slightly reduces the tightness of the correlation.
For the $A_{2}(r_{\rm bar})-Q_{\rm T}^{\rm halo-corr}(r_{\rm bar})$ relation, early- and late-type systems still occupy different regions of the plot, but there is no longer an empty area separating them.

Altogether, it appears that the bar orbits are, to a major extent, controlled by the underlying disk potential, but the role of halos cannot be ignored for the relation between bar strength and shape.
%
\section{Characterization of bars as a function of galaxy mass}\label{barlen_mass_sec}
%
The tendencies of the bar parameters in the Hubble sequence as shown in the previous section can be better understood when they
are considered as a function of the galaxy mass.
\subsection{Bar fraction dependence on the parent galaxy mass}\label{barfracmass}
\begin{figure}
    \centering   
    \begin{tabular}{l r}
    \includegraphics[width=0.49\textwidth]{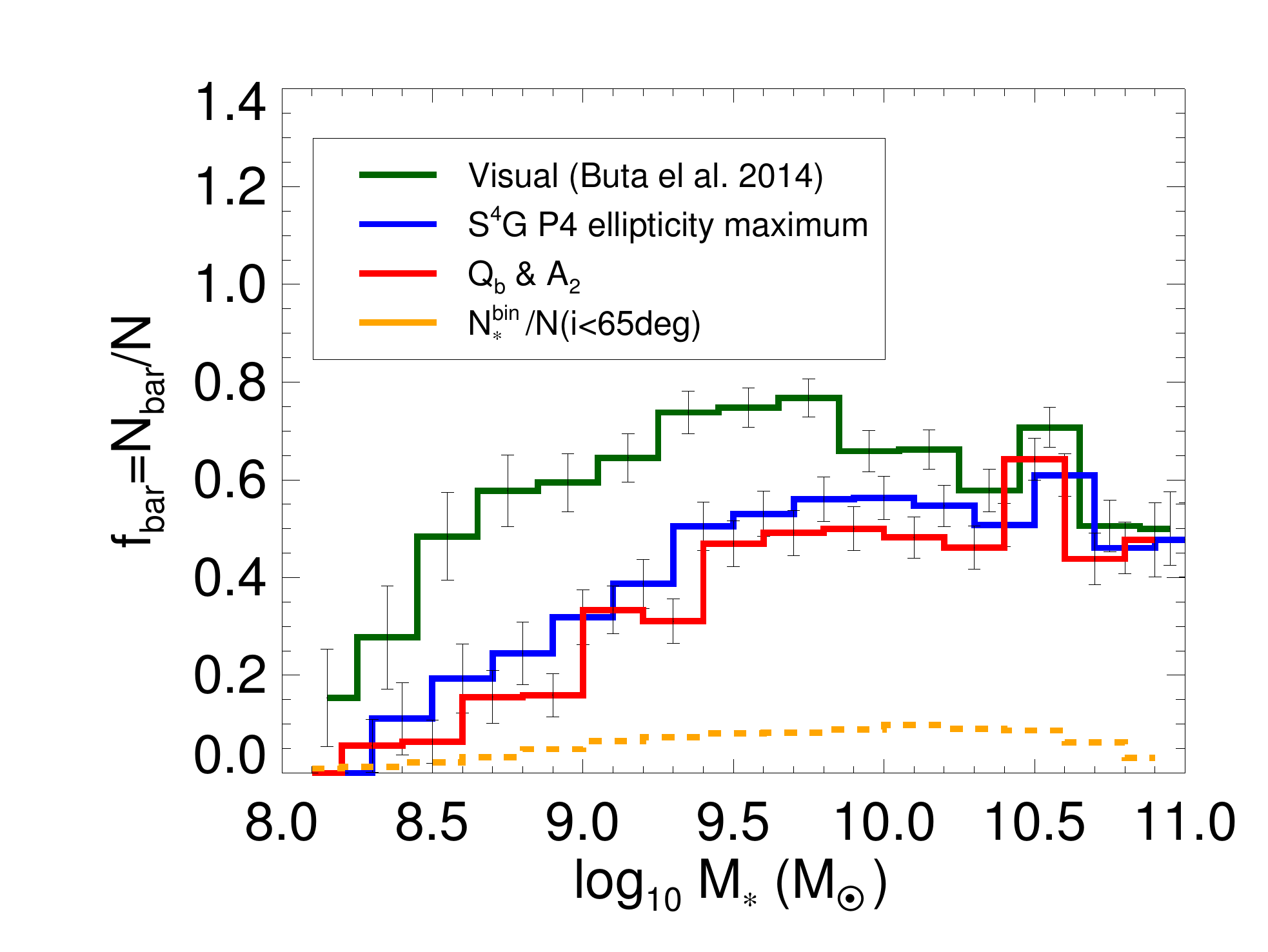}\\[-5ex]
    \includegraphics[width=0.49\textwidth]{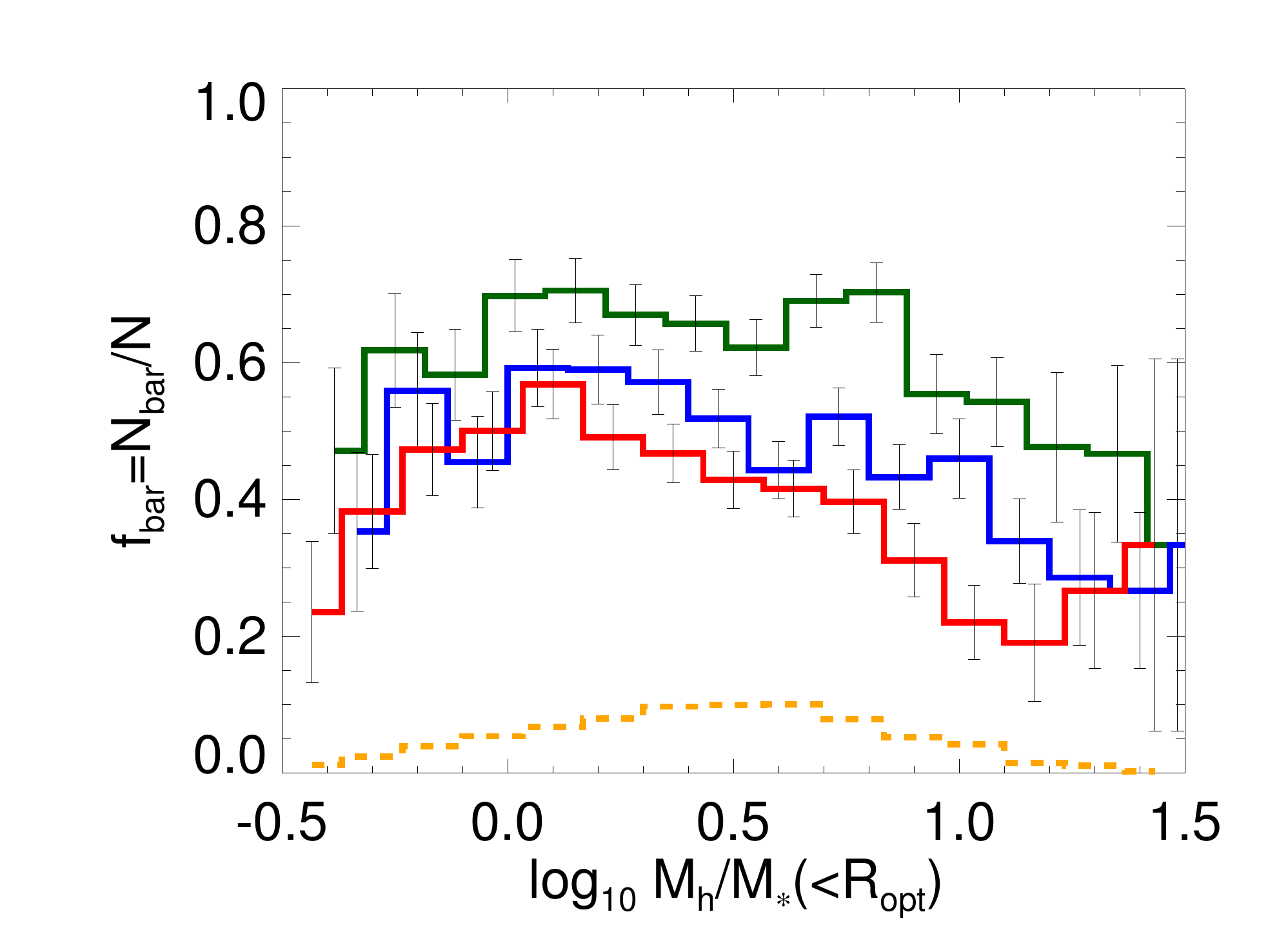}
    \end{tabular}
\caption{
Bar fraction as a function of stellar mass (\emph{upper panel}) and halo-to-stellar mass ratio (\emph{bottom panel})
for the different bar detection criteria contemplated in this work. The error bars are binomial errors.
}
\label{bar_frac_mass}
\end{figure}

As discussed in B2015, the double-peaked trend of the bar fraction in the Hubble sequence discussed previously might be linked to the bimodality found at intermediate redshifts $z\in(0.01,0.1)$ 
by \citet[][]{2010ApJ...714L.260N}. Consistent with \citet[][]{2008ApJ...675.1194B},
these authors found a bimodal dependence of the bar fraction on the central concentration and the total stellar mass with a minimum $f_{\rm bar}$ at $\rm log_{10}(M_{\ast}/M_{\odot})\sim 10.2$.
According to the authors, this might imply that the creation of the red and blue sequences are linked to the origin and evolution of bars,
since that is the characteristic mass at which a bimodality in the stellar population and fundamental properties of galaxies is observed \citep[e.g.][]{2003MNRAS.343..978S,2003MNRAS.341...54K,2004ApJ...600..681B}.

Here we searched for a possible dependence between $f_{\rm bar}$ and the total stellar mass (upper panel of Fig.~\ref{bar_frac_mass}),
but found no obvious trend from the visual classification of bars for galaxies of stellar masses $\gtrsim 10^{9} M_{\odot}$. 
$f_{\rm bar}$ clearly drops for the faintest systems \citep[][]{2014AAS...22320502S}, which is consistent with studies of the bar fraction at higher redshifts \citep[e.g.][]{2008ApJ...675.1141S,2010ApJ...711L..61M}.

Nevertheless, methods based on Fourier decomposition and ellipse fitting for the detection of bars reveal that $f_{\rm bar}$ increases with stellar mass up to $\sim 10^{9.5-10}M_{\odot}$,
where it reaches a constant level of $\sim 50\%$ in the range $10^{10}-10^{11}M_{\odot}$ with an isolated peak ($\sim 60\%$) at $\sim 10^{10.5}M_{\odot}$.
Curiously enough, this is also the mass where galaxies show, statistically, the smallest amount of dark matter relative to the stellar mass (see again Fig.~{\ref{stellar-halo-relation}}),
and this mass is probably linked to the maximum $f_{\rm bar}$ observed for the Sab galaxies in Sect.~\ref{barfrachubble}.

\citet[][]{2015ApJ...807..111C} found a relationship between the bar fraction and the halo-to-stellar mass ratio for relatively long bars,
using the galaxy group catalogue of \citet[][]{2007ApJ...671..153Y} and photometry from SDSS DR7 (stellar masses ranging from $10^{9.5}$ to $10^{11.5}$ $M_{\odot}$): $f_{\rm bar}$ drops with increasing $M_{\rm h}/M_{\ast}$.
This tendency is weak when visually identifying bars (lower panel of Fig.~\ref{bar_frac_mass}).
However, it becomes clearer when relying on the Fourier and ellipse fitting detection of bars. This could be interpreted
as caused by the dark matter halos stabilizing the stellar disk  against bar formation \citep[][]{1976AJ.....81...30H,1997ApJ...477L..79M,2012MNRAS.426..983D}, 
which in turn would create the dependence between bar fraction and stellar mass (or Hubble stage) given the coupling between luminous and dark matter (Fig.~\ref{stellar-halo-relation}).
However, we did not find any clear dependence of $f_{\rm bar}$ on the halo-to-stellar mass ratio
at a fixed stellar mass with any bar detection criterion (see Appendix~\ref{DM_appendix}), although this contrast could be due to our smaller sample size.
\subsection{Dependence of the bar parameters on galaxy mass}\label{barlen_mass_sec_param}
\begin{table}
\scalebox{0.85}{
\tiny
\centering
\begin{tabular}{|| c| c| c| c||}
\hline
\multicolumn{1}{||c}{\centering Correlations:} & \multicolumn{1}{|c}{\centering $r_{\rm bar}$(kpc) vs. $R_{25.5}$(kpc)} & \multicolumn{1}{|c}{\centering $r_{\rm bar}$(kpc) vs. $h_{\rm R}$(kpc)} &
\multicolumn{1}{|c||}{\centering $r_{\rm bar}$(kpc) vs. $M_{\ast}$($M_{\odot}$)} \tabularnewline
\hline
\hline
$T < 0$& $  0.85\:(  1.6\cdot 10^{-16})$&$  0.87\:(  2.3\cdot 10^{-16})$&$  0.78\:(  1.5\cdot 10^{-12})$\\
\hline
$T\in [0,3)$& $  0.72\:(  7.1\cdot 10^{-25})$&$  0.81\:(  1.1\cdot 10^{-32})$&$  0.62\:(  5.3\cdot 10^{-17})$\\
\hline
$T\in [3,5)$& $  0.60\:(  1.6\cdot 10^{-11})$&$  0.65\:(  1.9\cdot 10^{-13})$&$  0.54\:(  3.1\cdot 10^{-9})$\\
\hline
$T\in [5,7]$& $  0.39\:(  5.5\cdot 10^{-10})$&$  0.53\:(  9.0\cdot 10^{-18})$&$  0.12\:(0.0778)$\\
\hline
$T > 7$& $  0.68\:(  2.5\cdot 10^{-34})$&$  0.70\:(  2.2\cdot 10^{-34})$&$  0.40\:(  1.4\cdot 10^{-10})$\\
\hline
\end{tabular}}
\caption{
Spearman's rank correlation coeficient and significance for the relation between the bar absolute size (in kpc), the disk size ($h_{\rm R}$ and $R_{25.5}$) and total stellar mass. Different Hubble stage bins are considered.}
\label{barlen_correlations}
\end{table}

It is well known that the scatter among the bar parameters is very high within the different Hubble type bins, and therefore it is natural to study the same parameters also as a function of galaxy mass.

\begin{figure}
    \centering   
    \includegraphics[width=0.49\textwidth]{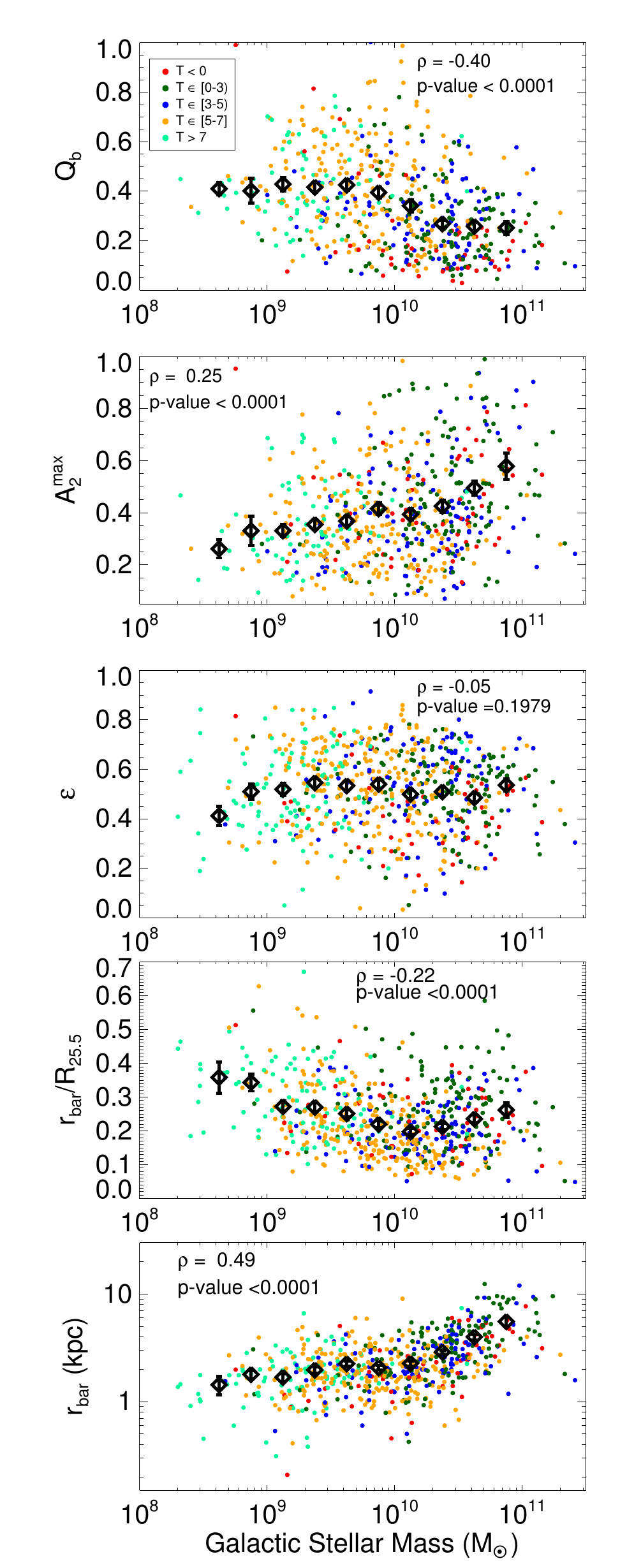}
\caption{
Gravitational torque, bar density amplitude, bar intrinsic ellipticity, and $R_{25.5}$-relative and absolute bar size (in physical units) vs. stellar mass.
Spearman's rank correlation coefficient and significance are displayed in the plots.
Different morphological types are separated following the colour palette shown in the legend of the first panel.
The black symbols correspond to the running mean, and the error bars correspond to the standard deviation of the mean.
}
\label{mass_barlen}
\end{figure}

It appears that the ellipticity of a bar is independent of the parent galaxy mass (third panel in Fig.~\ref{mass_barlen}).
The S0s are simply lacking the highest bar ellipticities,
which might be related to the bar morphology in these galaxies: for example, barlenses and ansae are typical of early-type galaxies \citep[][]{2011MNRAS.418.1452L}.
For the bar torque parameter, for which the trend was obvious in the Hubble sequence,
we observe that $Q_{\rm b}$ increases towards lower galaxy masses. For $A_2^{\rm max}$ an opposite correlation is found, which is expected because the trend with $T$ was also reversed.
None of the bar strength indexes studied as a function of the stellar mass is clearly segregated in the different Hubble types for a given galaxy mass.

The bar sizes in physical units show a correlation with the total stellar mass (see the lowest panel of Fig.~\ref{mass_barlen}),
which is quite natural (massive galaxies have larger stellar structures in general).
E2005 observed that late-type spiral galaxies seem to have no correlation between bar size and either the disk scale
length or the absolute magnitude of the host galaxies
(with the galaxy parameters measured in the $B  $ band).
We have reassessed this result (see Table~\ref{barlen_correlations}) and found that even though trends between $r_{\rm bar}$ and $h_{\rm R}$ and $R_{25.5}$ are weaker for late-type systems,
the correlations hold in general, regardless of $T$-type (but for the particular case of Sc and Sd galaxies we do not observe any correlation between $r_{\rm bar}$ and $M_{\ast}$ either, as noted in E2005).
Early-type spirals and especially S0s are the galaxies with the strongest scaling relations between bar and disk sizes.
Furthermore, as shown in the lower panels of Fig.~\ref{mass_barlen}, bar-disk size relations show distinct trends for the systems with revised Hubble stages $T\ge5$ on one hand and $T<5$ on the other.

Moreover, E2005 (see their Fig. 14) concluded that there is no dependence between disk-relative bar size and the parent galaxy absolute blue magnitude.
This is more or less consistent with our results (see Fig. \ref{mass_barlen}), assuming a similar behaviour between blue and near-IR absolute magnitudes and limiting the analysis to their $T$ range. 
However, the author speculates that fainter galaxies should have a similar range of relative bar sizes, which we confirm do not.
We observe a decreasing trend of the disk-relative bar sizes with increasing stellar masses for the late-type systems.
On the other hand, the relative size increases with the stellar mass for early types.
In conclusion, we find a dichotomy in the behaviour of both the absolute and relative bar sizes
with respect to the stellar mass of the host galaxy. 

Based on the analysis of the bar parameters, both as a function of Hubble type and galaxy mass,
it appears that bars in the S0 galaxies ($T<1$) are slightly different from other types.
Bars in these galaxies are rounder and have lower $Q_{\rm b}$ (which most probably cannot be explained solely by bulge dilution effects). 
In addition, S0$^{-}$ and S0$^{0}$ are typically weaker based on $A_2^{\rm max}$.

Another conclusion is that the bars in the low-luminosity galaxies at the end of the Hubble sequence have similar ellipticities as the rest of the spirals: they do not have more oval shapes,
for instance.
In addition, while the $A_2^{\rm max}$ amplitudes are weaker, the $Q_{\rm b}$ are higher because of the weak underlying disks of the parent galaxies.
In the simulations by \citet[][]{2015MNRAS.451..936S}, higher values of $Q_{\rm b}$ are obtained for the systems with a lower disk-to-total fraction,
which may support the observed difference between Sb/c and S0/a galaxies, under the assumption that the former are more dominated
by dark matter than the early-type systems 
\citep[see][for a review of the dark matter problem]{1998CeMDA..72...69B}. 
%
\subsection{Bar strength dependence on the bulge-to-total mass ratio}\label{bar_bulges}
%
\begin{figure}
   \centering   
   \includegraphics[width=0.5\textwidth]{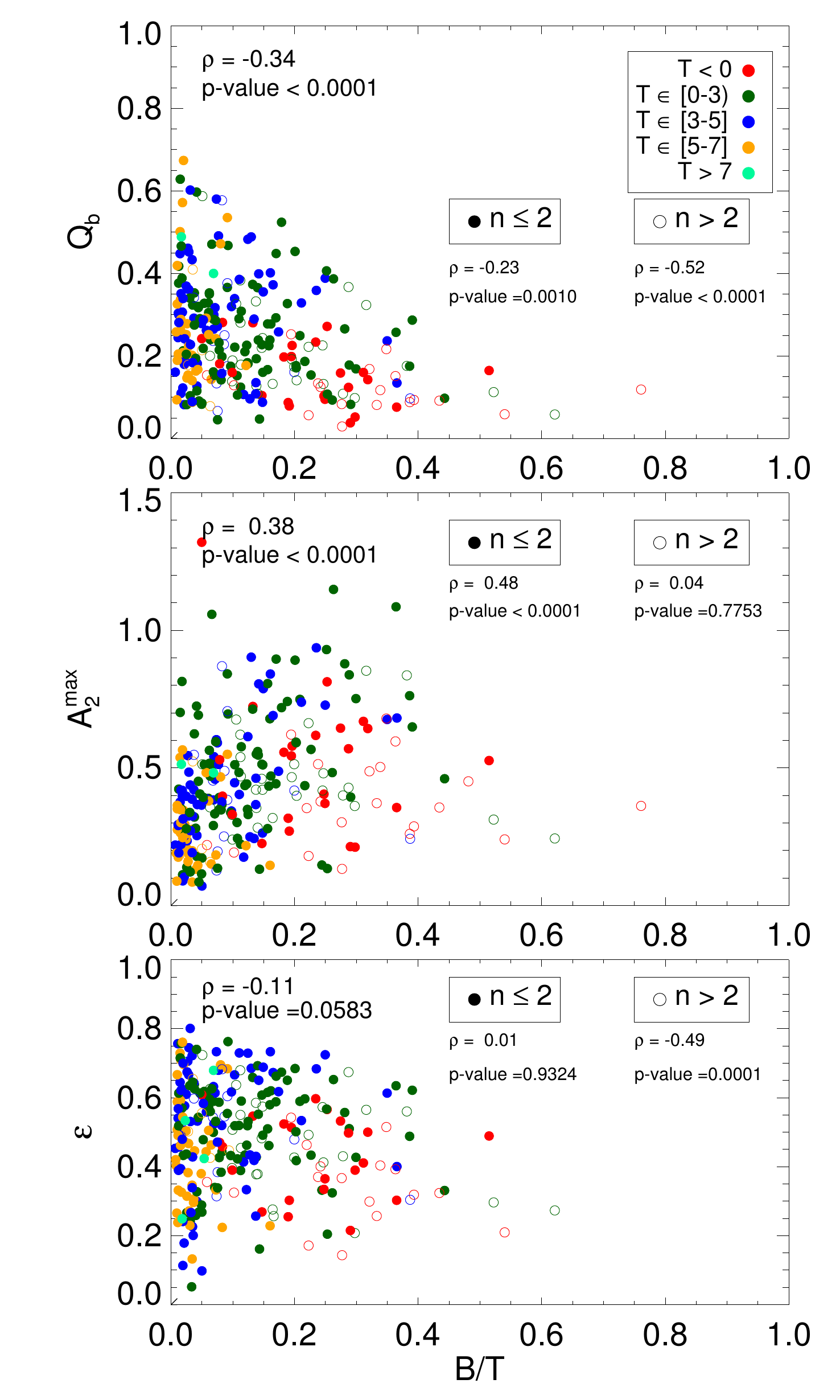}
\caption{
Bar strength indicators dependence on P4 bulge parameters.
Spearman's rank correlation coefficient and significance are displayed in each plot.
The $Q_{\rm b}$, $A_{2}^{\rm max}$ , and $\varepsilon$ as a function of the P4 bulge-to-total ratio are plotted in \emph{rows 1-3}.
Colour indicates the galaxy morphological type.
Open symbols indicate bulges with a Sérsic index $n>2$, while filled circles correspond to $n\le2$.
}
\label{qb_a2_ellip_bulge}
\end{figure}
As the axisymmetric stellar structures are known to influence the measurement of bar strengths \citep[e.g.][]{2004ApJ...607..103L}
and the formation and evolution of bars and bulges might be connected \citep[e.g.][]{2008ApJ...675.1141S}, 
we next studied in which way the above-derived bar strength parameters are associated with the properties of bulges, specifically the $B/T$.

We plot in Figure~\ref{qb_a2_ellip_bulge} the indicators of the prominence of the bar ($Q_{\rm b}$, $A_{2}^{\rm max}$, $\varepsilon$) as a function of $B/T$.
The upper panel shows that the maximum $Q_{\rm b}$ decreases with increasing $B/T$.
We confirm the previous result by \citet[][]{2004ApJ...607..103L}, where such an upper boundary was also found and interpreted as dilution of $Q_{\rm b}$ by massive bulges.
The figure also separately shows bulges with a Sérsic index $n<2$ and $n>2$, as this limit is often used to divide them into pseudo and classical bulges, respectively {\citep{2004ARA&A..42..603K,2008AJ....136..773F}}.
The fact that this upper limit exists for all kinds of bulges is consistent with this interpretation (only the relative mass matters).
In addition, galaxies with bulges have a lower mean bar torque for all Hubble stages, as shown in Fig.~\ref{halo-qb} (overall means of $0.23\pm0.13$ and $0.42\pm0.18$ for systems with and without bulge, respectively).
However, the discussed anti-correlation between $Q_{\rm b}$ and $B/T$ is stronger for bulges with $n>2$ and completely disappears for galaxies with $T\ge 3$ (when the relative mass of the bulge, if present, decreases).

The ellipticity maximum as a function of $B/T$ shows a similar distribution as $Q_{\rm b}$. 
Both the upper bound and the anti-correlation only hold for bulges with high Sérsic indexes, 
which can be interpreted as an effect of classical bulges making the bar isophotes rounder.
Furthermore, a correlation is found between $A_2^{\rm max}$ and the relative mass of the bulge if we take only galaxies with $n \le 2$.
For Sérsic indexes higher than $2$ the correlation vanishes completely.
The trend is likewise strenghtend when S0 galaxies are excluded; for these the bulge-to-total ratio and the Sérsic index in P4 might be overestimated (e.g. not all disk components such as lenses are included).

The reported trends of the different bar strength indexes as a function of stellar mass can be partially explained by the degree of influence of the bulge on the measurements,
since the more prominent bulges tend to reside in more massive galaxies (idem for other inner stellar structures that contribute to the overall radial force field):
$A_{\rm 2}^{\rm max}$ ($Q_{\rm b}$) increases (decreases) with mass and $B/T$. With the same reasoning,
$B/T$ also determines the tendencies of $A_{\rm 2}^{\rm max}$ and $Q_{\rm b}$ in the Hubble sequence.
In addition, the observed behaviour when comparing bar force measurements to each other (Fig. \ref{forces_correlation}) might
also be explained by bulges.
Clearly, the effect of the bulge dilution is more pronounced in $Q_{\rm b}$, which can explain the $T$-dependent bimodality between $A_{2}^{\rm max}$ and $Q_{\rm b}$ and also the $Q_{\rm b}-\varepsilon$ relation.

That the $Q_{\rm b}$ value is sensitive to the bulge is one of its advantages for the quantification of gravitational torques, 
since all components affect the underlying potential in the bar region and therefore the dynamics of the orbits making up the bar.
The usefulness of $Q_{\rm b}$ is supported by the results reported
in \citet[][]{2015MNRAS.451..936S} (Fig. 12), where the photometric bar torque (as obtained in this work)
is compared with the so-called kinematic bar torque ($Q_{\rm kin}$), which directly measures the kinematic perturbation related to the bar.
$Q_{\rm kin}$ is derived from the radial and tangential velocities extracted from the stellar velocity fields \citep[mapped with the IFU spectrograph SAURON,][]{2001MNRAS.326...23B} following the recipe by \citet[][]{2012MNRAS.427.3427M}.
A tight correlation between the two independent bar strength measurements is found and is also supported by N-body simulations.
In \citet[][]{2015MNRAS.451..936S} a sub-sample mainly composed of S0s and early-type S$^{4}$G galaxies was used,
which implies relatively prominent bulges and demonstrates their role in the dynamics of stars in the bar region.

The formation of disky bulges in early-type galaxies can be driven by the evolution of bars \citep[e.g.][]{2007MNRAS.381..401L}, 
but scenarios in which pseudo-bulges form without any bar-dependent mechanism (e.g. stripped spirals) are also possible \citep[][]{2010MNRAS.405.1089L}.
The observed bar weakness (as measured from $Q_{\rm b}$ and $\varepsilon$) that accompanies prominent bulges is most likely explained by the bulge dilution.
Another explanation could be the exchange of angular momentum with gas that the bar funneled inwards and eventually formed
or increased the pseudo-bulge, causing its present-day weakness.
However, the correlations between $Q_{\rm b}$ or $\varepsilon$ and $B/T$ are weaker for low Sérsic indexes, which does not support this interpretation.
On the other hand, bulges formed through mergers or by the coalescence of giant clumps at high redshift (higher Sérsix indexes) could have prevented bars from growing,
and hence bars in the presence of relatively massive bulges barely evolved since they formed and remain weak at $z=0$.
A different perspective is given by considering $A_{2}^{\rm max}$ vs. $B/T$: the positive correlation between these 
two values seems to point in the direction of bars evolving in parallel with disky bulges.
This link between bars and bulges is not observed among S0s, where a more thorough analysis of the relationship between bars and bulges in the Hubble sequence
can be achieved by considering the presence of barlenses when performing the 2D decomposition of the surface brightness profiles \citep[][]{Lauri2015}.
\section{Growth of bars}\label{bar-growth}
\begin{table}
\small
\centering
\begin{tabular}{|| c| c| c| c||}
\hline
\multicolumn{1}{||c}{\centering Correlations:} & \multicolumn{1}{|c}{\centering $r_{\rm bar}/R_{25.5}$ vs. $Q_{\rm b}$} & \multicolumn{1}{|c}{\centering $r_{\rm bar}/R_{25.5}$ vs. $A_{2}^{\rm max}$} &
\multicolumn{1}{|c||}{\centering $r_{\rm bar}/R_{25.5}$ vs. $\epsilon$} \tabularnewline
\hline
\hline
$T < 0$& $  0.15\:(0.2961)$&$  0.37\:(0.0110)$&$  0.32\:(0.0297)$\\
\hline
$T\in [0,3)$& $  0.36\:(  2.3\cdot 10^{-5})$&$  0.41\:(  2.3\cdot 10^{-6})$&$  0.40\:(  2.3\cdot 10^{-6})$\\
\hline
$T\in [3,5)$& $  0.55\:(  2.8\cdot 10^{-8})$&$  0.53\:(  4.8\cdot 10^{-8})$&$  0.43\:(  1.9\cdot 10^{-5})$\\
\hline
$T\in [5,7]$& $  0.37\:(  6.8\cdot 10^{-7})$&$  0.37\:(  7.0\cdot 10^{-7})$&$  0.17\:(0.0153)$\\
\hline
$T > 7$& $  0.09\:(0.3109)$&$  0.18\:(0.0490)$&$ -0.13\:(0.0876)$\\
\hline
\end{tabular}
\caption{
Spearman's rank correlation coeficient and significance for the relation between the $R_{25.5}$-relative bar size and the different indexes of the bar strength. Different Hubble stage bins are considered.}
\label{barlen_strength_correlations}
\end{table}

\begin{figure}
   \centering   
   \includegraphics[width=0.5\textwidth]{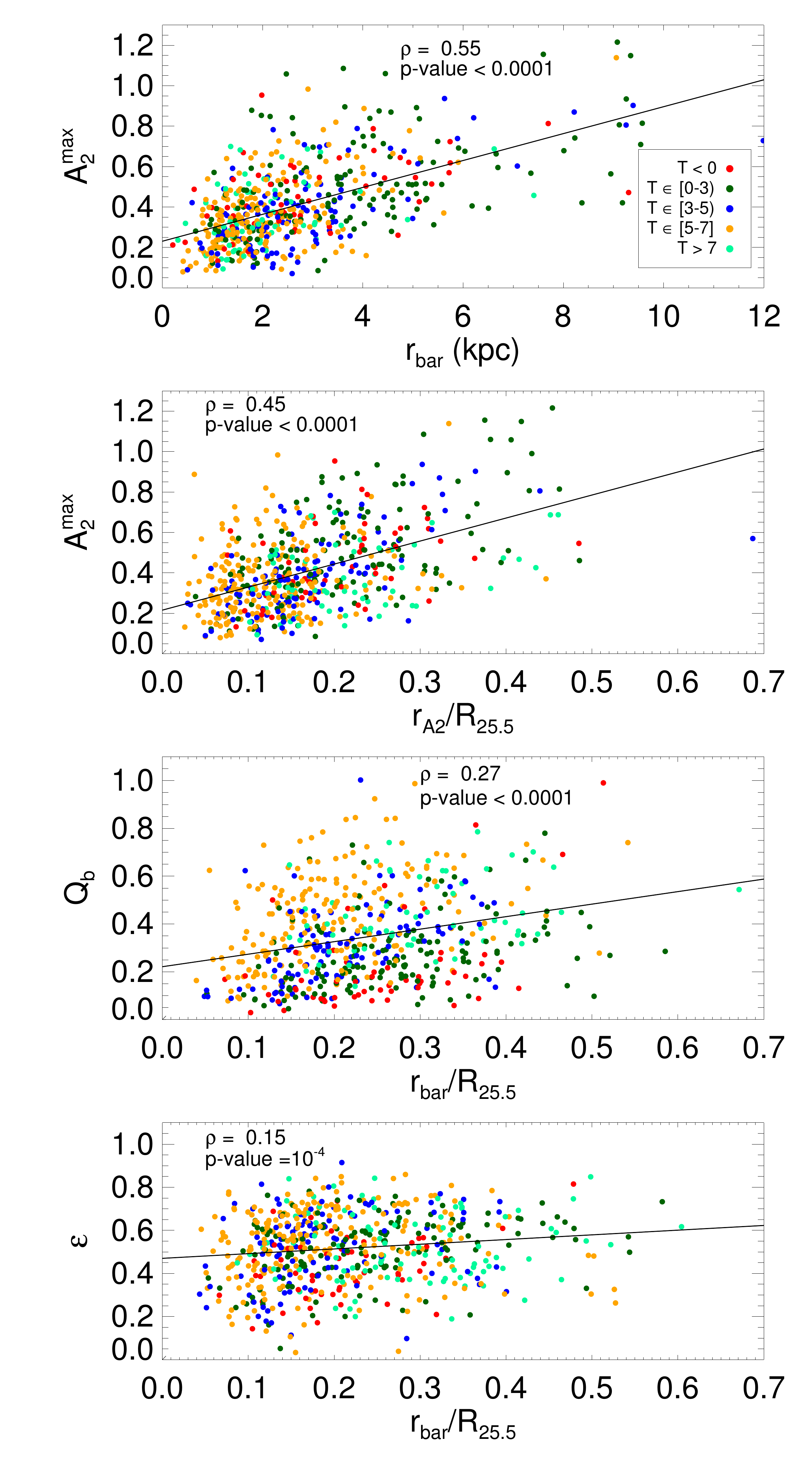} 
\caption{
Relation between the bar length (in kpc and scaled to the disk size) and the different bar strength indexes.
Spearman's rank correlation coefficient and significance are indicated in each plot.
Colour selection depends on the galaxy Hubble stage.
The straight line displays the linear fit of the cloud of points.
}
\label{barlen_evol}
\end{figure}
\begin{figure}
   \centering   
   \begin{tabular}{l c r}
   \includegraphics[width=0.5\textwidth]{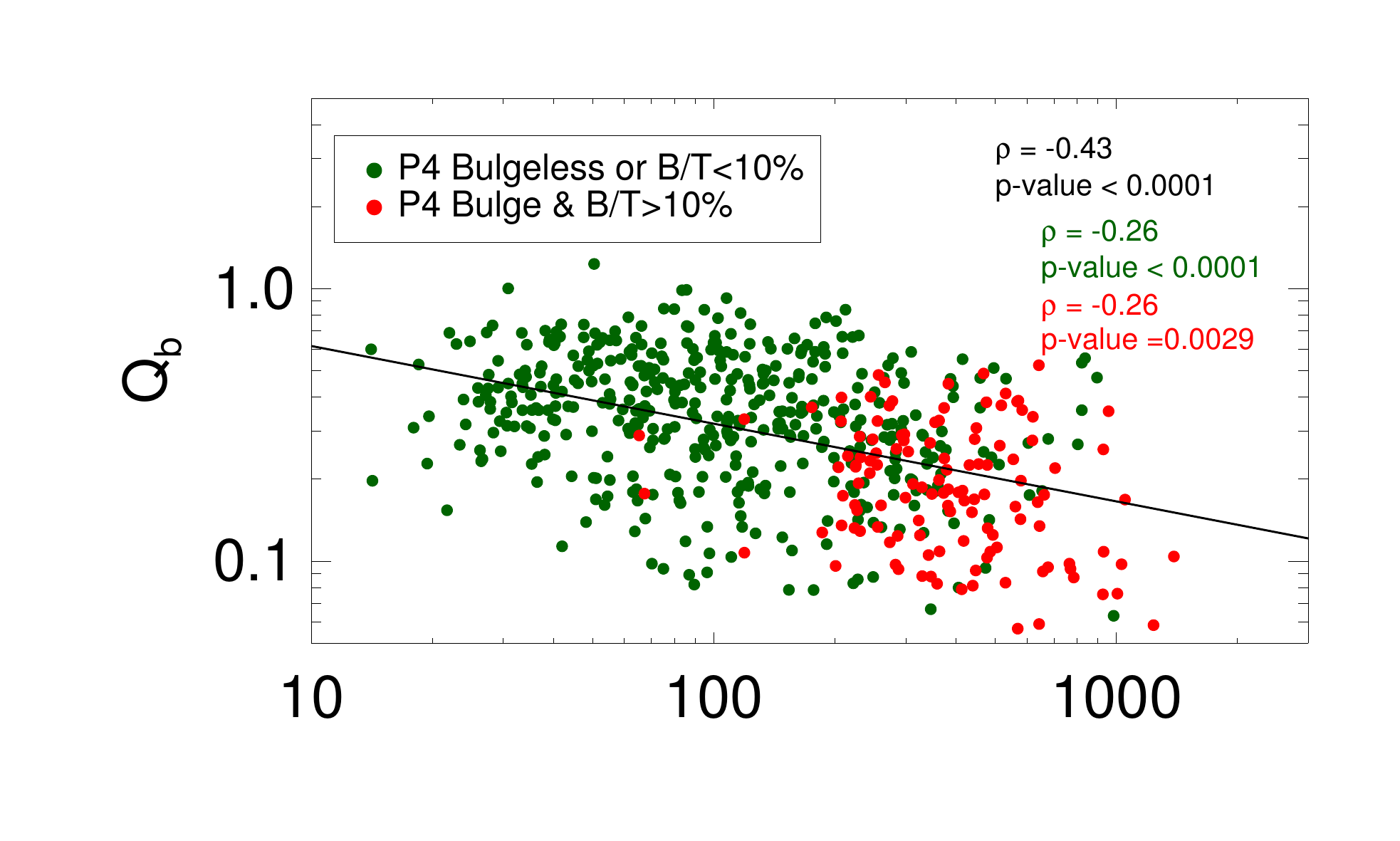}\\[-8ex]
   \includegraphics[width=0.5\textwidth]{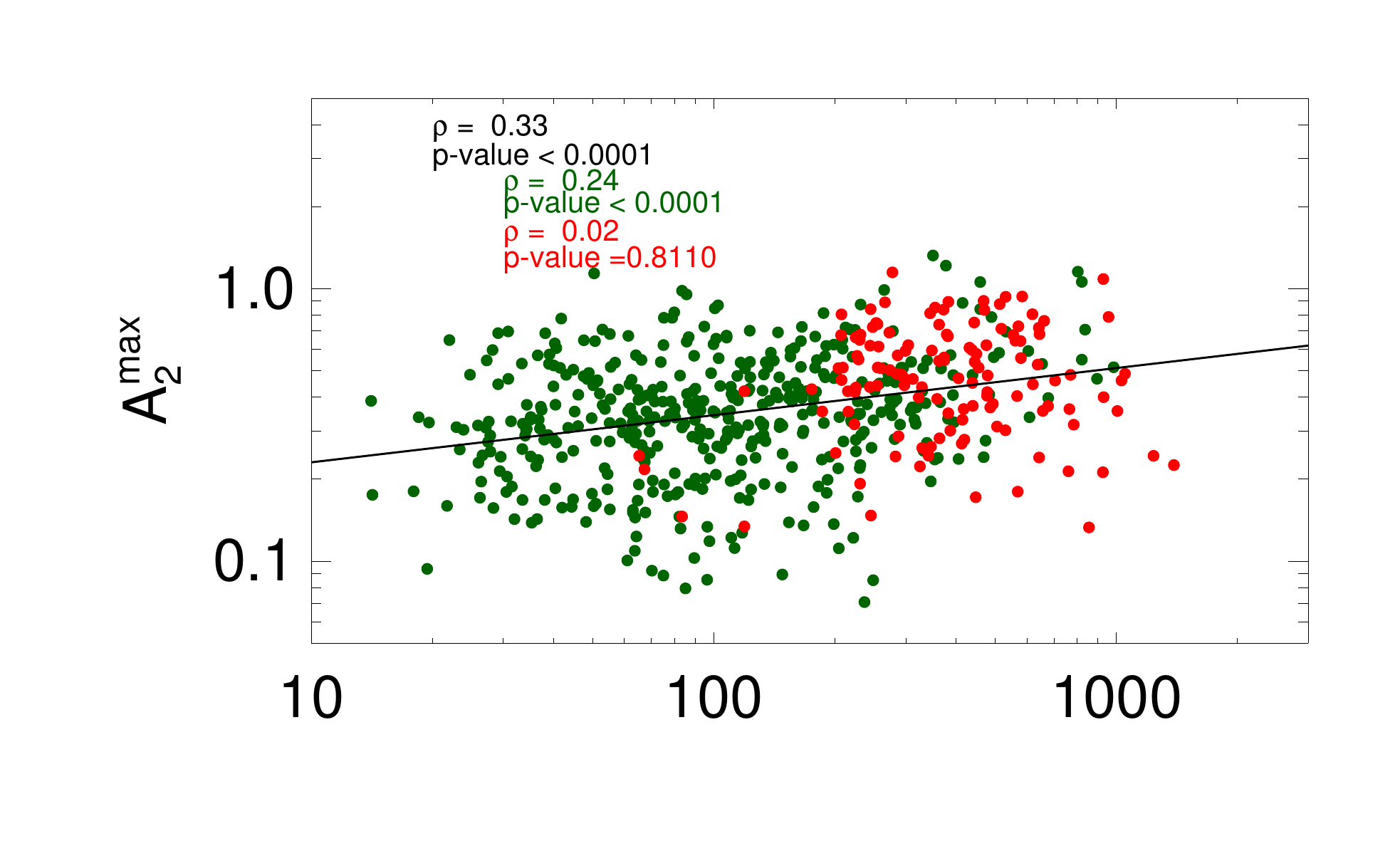}\\[-8ex]
   \includegraphics[width=0.5\textwidth]{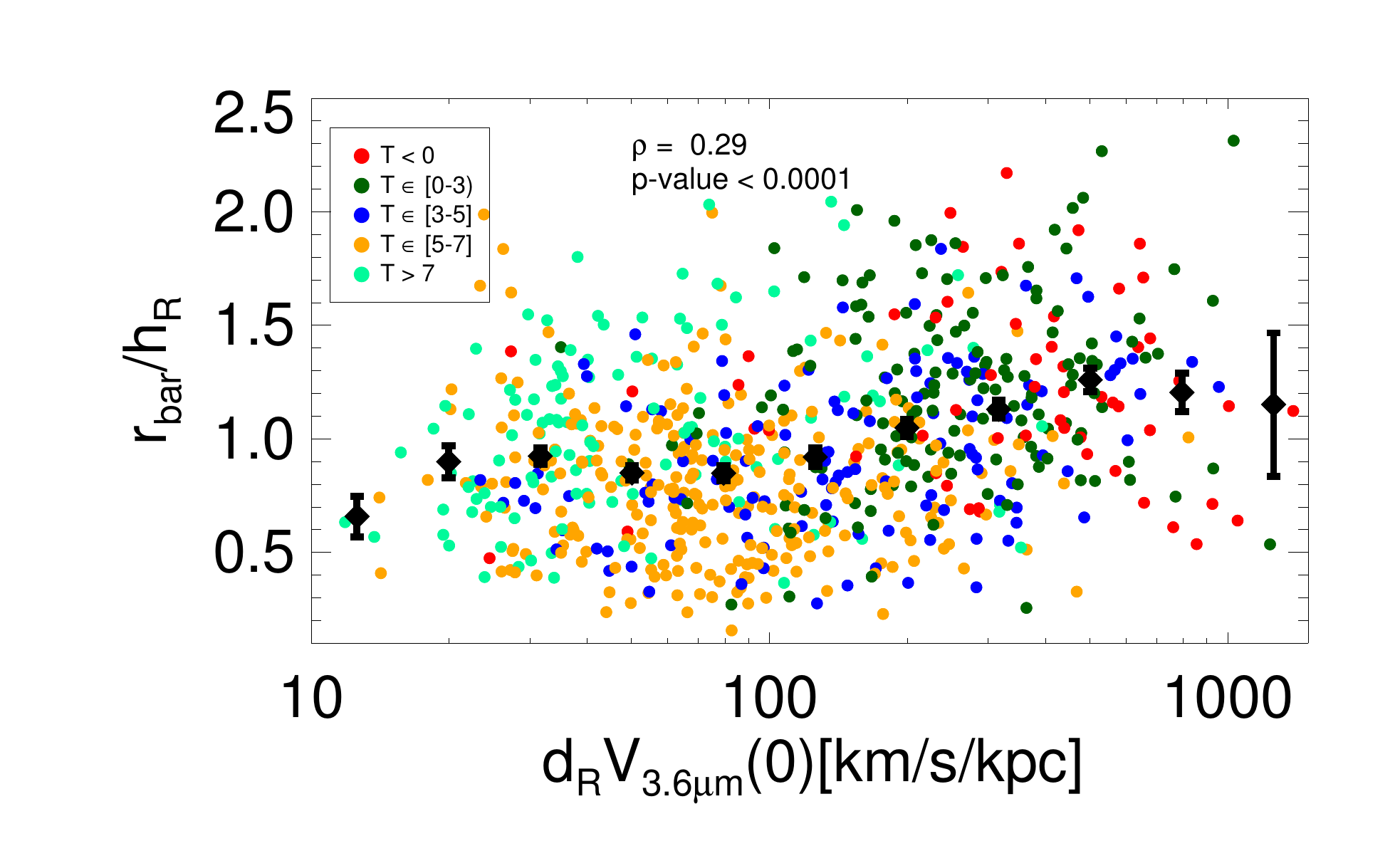}
   \end{tabular}
\caption{
Inner slope of the stellar component of the rotation curve as a function of $A_{2}^{\rm max}$, $Q_{\rm b}$ and the normalized bar length.
The Spearman's rank correlation coefficient and significance are displayed together with the linear fit of the cloud of points.
For the first two panels, galaxies with and without massive bulges are studied separately, while for the lower plot colours are chosen based on $T$, indicating the moving average with black symbols.
}
\label{barlen_evol_slope}
\end{figure}
A fundamental research question is whether galactic stellar bars become stronger, longer, thinner, and slow down as they evolve
while exchanging angular momentum with the disk and the halo components \citep[e.g.][]{1979ApJ...227..714K,1984MNRAS.209..729T}. This evolution is expected from the simulation models of various independent groups
\citep[e.g.][]{1991MNRAS.250..161L,1998ApJ...493L...5D,2002MNRAS.330...35A,2003MNRAS.341.1179A,2003MNRAS.346..251O,2006ApJ...645..209D,2003MNRAS.345..406V,2006ApJ...637..214M}.
Observational evidence of this evolution has been investigated by
different groups
\citep[e.g.][]{2007ApJ...670L..97E, 2007MNRAS.381..401L,2011MNRAS.415.3308G}.

\citet{2007ApJ...670L..97E} reported a correlation between $r_{\rm A_{2}}$ (normalized to $R_{25}$) and $A_{2}^{\rm max}$,
present for early and intermediate-type spiral galaxies based on the $H$-band and $K_{\rm s}$-band measurements by \citet{2006AJ....132.2634L}
and \citet{2004MNRAS.355.1251L}.
We also observe a definite trend between the bar size (both in kpc and relative to the disk size) and $A_{2}^{\rm max}$,
which gives evidence that longer bars are also stronger (see Fig.~\ref{barlen_evol}).
A similar statistical trend persists for $Q_{\rm b}$ or $\varepsilon$, but the correlation is weaker \citep[consistent with][]{2002MNRAS.331..880L}.
We proceed by studying the relation between the bar sizes normalized to $R_{25.5}$ and the different bar strength indexes in bins of $T$
(see the results of the statistical tests in Table~\ref{barlen_strength_correlations}). We observe 
that early- and intermediate-type spirals, $T \in [0,5)$, are indeed the galaxies with the tightest
monotonic correlation between bar size and strength, independently of the bar index used.
A moderate trend is also found among late-type spirals, which trend disappears for Magellanic and irregulars.
There is also a moderate positive correlation for S0s when considering $\varepsilon$ or $A_{2}^{\rm max}$ that
practically vanishes when taking $Q_{\rm b}$ as the bar strength proxy (for which bulges and barl enses may substantially change the bar strength).

\citet{2007ApJ...670L..97E} also claimed that both the bar length and
the relative amplitude of m=2 Fourier component correlate with the
central density, which can be understood in terms of galaxy evolution
because galaxies with higher central rotation rates evolve more quickly.
In Fig.~\ref{barlen_evol_slope} we confirm a similar trend by plotting $A_{2}^{\rm max}$ against the inner slope of the
stellar contribution to the rotation curves, which is related to the correlation between $A_{2}^{\rm max}$ and $B/T$.
On the other hand, high inner slope values are associated with low $Q_{\rm b}$, 
which can be understood because the small bar torques parameters are typical for galaxies with prominent stellar bulges.
The anti-correlation between $Q_{\rm b}$ and $\rm d_{\rm R}V_{\rm 3.6\mu m}(0)$ appears even for galaxies without bulges.
Finally, we reassessed the possible dependence between $\rm d_{\rm R}V_{\rm 3.6\mu m}(0)$ and the bar size
normalized to the scale length of the disk and found a weak positive relationship. This trend is strengthened when the very late-type systems (i.e. $T>7$) are excluded.

Intermediate-type spiral galaxies (e.g. Sbc galaxies) are known to be the most favourable candidates to evolve secularly (presence of a bar,
oval disk or a spiral structure; they have small bulges and the disk spreading is energetically favourable); they strip off their gas and move
in the Hubble sequence towards earlier types \citep[][]{2013seg..book....1K}.
The bar relative sizes and density amplitudes increase
from galaxies with Sc morphology ($T=5$) towards Sa and Sa/0 types (as
shown in Figs. \ref{barlen_ttype} and \ref{ttype_qb_a2_ellip} and in
Table~\ref{bar_strength_statistics_structures}); this is consistent with this scenario.
The tightest correlation between the relative sizes of bars and their strengths,
which can be understood as possible evidence of the growth of bars over a Hubble time, 
occurs for galaxies of morphological types Sa/0-Sc.

For the angular momentum transfer within disk galaxies to be efficient, the mass of the halo near resonances is a crucial factor \citep[][]{2013MNRAS.429.1949A}. 
In addition, the exchange of angular momentum can be hampered by dynamically hot disks and halos.  The
influence of the halo triaxiality and the dark matter profile is beyond the scope of our observational analysis for obvious limitations.
We have searched for a possible connection between $M_{\rm h}/M_{\ast}$
with $Q_{\rm b}$ and $A_{2}^{\rm max}$ (both depend on stellar
mass), but found no correlation, or if one is present, it is very weak
(Spearman's correlation coefficient and significance of $\rho=0.18, p=5\cdot10^{-5}$
and $\rho=-0.09, p=0.048$, respectively).  However, the effect of cold gas on
the evolution of bars is not considered in this study.  The role of
halos is expected to also be important for the evolution of bars among
intermediate- and early-type spiral galaxies.

The discrepant statistical behaviour of $A_{2}^{\rm max}$ and $Q_{\rm b}$ as a function of $T$, $M_{\ast}$ , or $B/T$ does not mean that one
index is better than the other, as they are rather complementary. 
While ${A_2}^{\rm max}$ accounts for the relative mass of the bar, $Q_{\rm b}$ directly measures the bar-induced tangential 
forces. As the tangential force depends on the underlying gravitation
potential, it is also sensitive to the presence of massive spheroidals (or other vertically thick inner structures such as the boxy/peanut
bulges) that dilute the $F_{\rm T}/<F_{\rm R}>$ force ratio. The ellipticity of the bar, $\varepsilon$, is also determined by a response of the stellar
orbits to the underlying gravitational potential, and it therefore is a proxy for $Q_{\rm b}$. However, the maximum ellipticity also depends
on the morphology of the bar and therefore does not directly measure the bar-induced tangential forces.
\section{Summary and conclusions}
We analysed the bar properties of a sample of 860 face-on
and moderately inclined barred galaxies ($i<65^{\circ}$) drawn from
the S$^{4}$G survey \citep[][]{2010PASP..122.1397S}, which covers a
wide range of galaxy masses in the local Universe.  We have performed
a statistical analysis of the lengths and strengths of stellar bars in the
Hubble sequence and also studied the bar fractions.

For a sub-sample of $\sim 600$ barred galaxies and using the 3.6~$\mu$m
images, we have inferred the gravitational potentials with the polar
method \citep[][]{1999AJ....117..792S,2002MNRAS.337.1118L,2010ApJ...715L..56S} and calculated gravitational torques, that is, the ratio of the maximal
tangential force to the mean axisymmetric radial force, to obtain an
estimate of the bar strength ($Q_{\rm b}$). In addition, we measured the bar strenght by the m=2 normalized Fourier density
amplitudes ($A_{2}^{\rm max}$) and used the bar isophotal ellipticities
($\varepsilon$) taken from \citet[][]{2015A&A...582A..86H}.

We showed that the three proxies of the bar strength positively
correlate with each other (Fig.~\ref{forces_correlation}), confirming earlier studies by
\citet[][]{2001A&A...375..761B} and \citet[][]{2004ApJ...607..103L},
among others.
The $\varepsilon-Q_{\rm b}$ relation follows a curved upwards trend, which shows that the underlying disk potential tends to 
control the shape of the orbits making the bar.  We found a segregation of bars determined by the Hubble type of the host galaxy.
This might be explained by the insensitivity of $\varepsilon$ to bar mass changes, and this weakness of $\varepsilon$ as a bar strength index is also reflected in the
$\varepsilon-A_{2}^{\rm max}$ correlation: for a fixed bar ellipticity, $A_{2}^{\rm max}$ increases with $T$-type.
In addition, we confirmed a strikingly clear bimodality in the relation between $A_{2}^{\rm max}$ and $Q_{\rm b}$,
which could be explained by the enhancement of the radial force field induced by axisymmetric inner structures such as bulges (bulge dilution) that reduces the 
$Q_{\rm b}$ value particularly in early-type galaxies {(Fig.~\ref{qb_a2_ellip_bulge}).
Indeed, both for the $\varepsilon-Q_{\rm b}$ and $\varepsilon-A_{2}^{\rm max}$ relations there seem to be two clearly distinct 
correlations, one associated with early- and intermediate-type galaxies ($T\lesssim 5$) and the other with late-type systems.
We provided an estimator of the torque parameter in terms of $\varepsilon$ for these two families of bars.
In addition, we reassessed these relations by evaluating the force amplitudes ($A_{2}$ and $Q_{\rm T}$) at the end of the bar,
finding more scattered but similar trends (Fig.~\ref{forces_correlation_dm_barend}): this means that bulges themselves cannot explain these two $T$-dependent behaviours without invoking other physical 
parameters, which confirms the peculiar nature of late-type bars.

We analysed the bar sizes (in kpc, as well as normalized to the scale length of the disk to and $R_{25.5}$) as a function of
the host galaxy Hubble stage and total stellar mass (Figs.~\ref{barlen_ttype}~and~\ref{mass_barlen}). We used bar lengths estimated visually, based on ellipse fitting, on the radii of
maximum torque ($r_{\rm Qb}$), and on the radii of maximum m=2 normalized Fourier amplitude
($r_{\rm A2}$). All bar size estimates, except for that based on ellipse fitting, tend to underestimate the visual measurements. As in previous studies, we
observed that bars in early-type spirals (S0/a-Sa) are typically longer than in the intermediate-type spirals ($\sim$Sc), and there is a decline in the
bar size from late-type S0s towards S0$^{-}$ and S0$^{0}$ types. However, the contrast between early- and late-types is not as high as reported before \citep[e.g.][]{1995AJ....109.2428M,2005MNRAS.364..283E}.

The richness of the S$^{4}$G sample in low-mass galaxies has allowed
us to probe the bar properties of faint and blue systems at $z=0$ on an unprecedented level of detail and sample size, even though these bars in the late-type galaxies are fairly complicated to measure.
This is the range where we found one of the most conspicuous results:
statistically, bars in galaxies of Hubble types $>5$ start to become larger in size relative to the disk. This result is found with all of the four
independent measurements of the bar sizes (even with $r_{\rm Qb}$, which tends to underestimate the real bar length in late-type
galaxies, as $r_{\rm Qb}$ moves inwards in galaxies with low central concentration), and regardless of the disk size normalization that we
used. We also showed that a similar increase in the relative bar size occurs for galaxy masses
lower than $\sim10^{10}M_{\odot}$.

We checked that our quantitative estimates of bar parameters are
consistent with the families as classified by \citet[][]{2015ApJS..217...32B}.
We found that SB, SA$\underline{\rm B}$, SAB, and S$\underline{\rm A}$B families differ primarily in bar
strength for all morphological types (Figs.~\ref{histo_barfamily_ttype}~and~\ref{histo_barfamily_mean}).
We also showed that strong bars are longer than weak bars.

According to the models of \citet[][]{2003MNRAS.341.1179A} and \citet[][]{2006ApJ...637..214M}, bars lose their angular momentum and become thinner and longer in time by trapping
particles from the outer disk, which elongates their orbits or decreases their rotational velocity.  In this scenario, the angular
momentum left at the inner Lindblad resonance is absorbed by halo particles in the surroundings of corotation and the outer Lindblad
resonance. We found observational evidence for the evolution of bars, as longer bars (relative to the disk size) are also stronger for all
the indexes (Figs.~\ref{barlen_evol}). This correlation is particularly tight for $A_{2}^{\rm max}$, which is the bar strength index used in the above simulations.  

Furthermore, our observations showed that bars have increasing $A_{2}^{\rm max}$ amplitudes and lengths from $T=5$
to $T=0$ (Fig.~\ref{ttype_qb_a2_ellip}), which is consistent with the expectations of the above simulation models in a scenario in which late-type galaxies evolve secularly towards
earlier types. In the same Hubble-type range $Q_{\rm b}$ decreases towards the early-type systems, which can be explained by the 
more massive central concentrations (bulges) in these galaxies, which dilute the bar-induced tangential forces.
The ellipticity of the bar is maintained nearly constant, which is due to the insensitivity of $\varepsilon$ to changes in the
relative bar mass and the effect of bulges and barlenses that cause the isophotes to become rounder 
\citep[][]{1995AJ....109.2428M,2002MNRAS.337.1118L,2004ApJ...607..103L,2007MNRAS.381..401L}.
All of our bar strength parameters decrease simultaneously only among the Hubble stages of $T<-1$ , most probably indicating that bars in these galaxies are less prominent 
than in the later Hubble stages.

We probed possible sources of uncertainty in our $Q_{\rm b}$ measurements (Figs.~\ref{bulge_corr},~\ref{parameter_comparisons_mass},~\ref{vert_thick},~and~\ref{vert_thick_edge_on}):\\  
(1) The influence of the spiral arm torques on the $Q_{\rm b}$ value.\\
(2) The assumption of a constant scale height.\\
(3) The effect of non-stellar emission in the bar region.\\
(4) The dilution of the bar forces by the dark matter halo.

The uncertainties associated with the first three aspects ($\sim10-15\%$) are similar in size as the estimated uncertainty related to the thickness of the disk,
although the mean $Q_{\rm b}$ of Sb and Sc galaxies can be slightly more affected by the effect of the spiral arms.
The effect of the dark matter halo is highest for the latest-type systems ($\gtrsim20\%$).
The dark matter halos can reduce the $Q_{\rm T}$ value at the end of the bar by as much as $30-50\%$ (Fig.~\ref{halo-qb-barend}).
Moreover, when using ICA mass maps for the Fourier dissection of bars, no significant deviation is observed in the calculation of the torque parameter 
(Fig.~\ref{parameter_comparisons_mass_a2_rot}).
None of the sources of uncertainty alter the statistical trend of $Q_{\rm b}$ in the Hubble sequence (Fig.~\ref{halo-qb}),
although the raw bar torque parameter of individual galaxies as derived in this work may deviate from the intrinsic bar strength for specific galaxies 
(e.g. those with very strong boxy/peanuts or strong spiral arms, massive cuspy halos, or high star formation rates along the bar dust lanes).

The tendencies of the different proxies of the bar strength in the Hubble sequence and the impact of axisymmetric stellar structures on
the measurements are better understood when studied in terms of the stellar mass of the host galaxies (Fig.~\ref{mass_barlen}): $Q_{\rm b}$ ($A_{2}^{\rm max}$) is
observed to decrease (increase) with $M_{\ast}$, while $\varepsilon$ is nearly independent of $M_{\ast}$.
The behaviour of $Q_{\rm b}$ and the normalized bar sizes of late-type spirals, Magellanic, and irregular galaxies when studied as a
function of $M_{\ast}$ or $T$ is counter-intuitive: in spite of their strong $Q_{\rm b}$ , they are not expected to be highly evolved bars. 
Their host galaxies are rich in gas, and some of them might be the result of galaxy interactions or mergers, especially the offset bars, 
but this is not sufficient to explain the whole picture.

In addition, we calculated the stellar contribution to the
rotation curves and studied the bar properties as a function
of its inner slope (Fig.~\ref{barlen_evol_slope}), which traces the central stellar mass
concentration of our galaxies.  We found a positive (negative)
monotonic trend between the central rotation gradient and $A_{2}$
($Q_{\rm b}$).  Interestingly, galaxies with higher central
concentrations (which are probably more evolved) tend to host longer bars
relative to the underlying disk size (late- and early-type system
present again different behaviours in this relation).

From the stellar components of the rotation curves and the H{\sc\,i}
maximum velocities available in the literature, we obtained a
first-order estimate of the halo contribution to the rotation curve and of the halo-to-stellar mass ratio within the
optical disk of our galaxies, $M_{\rm h}/M_{\ast}(<R_{\rm opt})$.
Assuming that the halo inside the disk radius forms roughly a constant
fraction of the total halo mass, our ratios are consistent with
studies based on abundance matching, weak lensing analysis in galaxy
clusters, and halo occupation distribution methods (Fig.~\ref{stellar-halo-relation}).
Galaxies of Hubble type $T>5$ have a larger $M_{\rm h}/M_{\ast}(<R_{\rm opt})$.
No clear dependence of the bar parameters on $M_{\rm h}/M_{\ast}(<R_{\rm opt})$ was found.

We studied the bar fractions in the S$^{4}$G sample taking
into account the different criteria for the identification of bars: \\ (1)
From visual classification.\\ (2) By inspection of the P4
ellipticity radial profiles, with the detection criterion based on the
presence of an ellipticity local maximum associated with the bar and the
constancy of the position angle at that region.\\ (3) From the
identification of $A_{2}^{\rm max}$, together with the constancy the
$m=2$ phase, and the presence of a well-defined fourth quadrant
 butterfly pattern in the torque map.

The bar fraction shows a double-peaked distribution in the Hubble
sequence, with a local minimum at $T=4$, regardless of the bar
detection criteria (Fig.~\ref{bar_dist}). 
In comparison to the visual identification of
bars, ellipse fitting and Fourier decomposition methods reduce the
bar fraction by $\sim 30-50 \%$ for galaxies with $T\ge5$.
This is natural because these bars are too faint to appear as maxima in the ellipticity or Fourier amplitude profiles. Furthermore, these
two methods reveal a clear monotonic trend between the bar fraction and both the total stellar mass and the halo-to-stellar mass ratio of the host galaxy,
confirming the result by \citet[][]{2015ApJ...807..111C} (Fig.~\ref{bar_frac_mass}). However, this is not as clear from the
visual classification.  The fact that the bar fraction (estimated with criteria 2 and 3 above) decreases with $M_{\rm h}/M_{\ast}(<R_{\rm opt})$ could be interpreted as a result of dark
matter halos stabilizing the stellar disk against bar formation \citep[][]{1976AJ.....81...30H,1997ApJ...477L..79M,2012MNRAS.426..983D}
and might cause the dependence between bar fraction and stellar mass (or Hubble stage) given the coupling between luminous and dark matter.
However, in contrast to \citet[][]{2015ApJ...807..111C}, we did not find a
statistically significant difference in $M_{\rm h}/M_{\ast}(<R_{\rm opt})$ between barred and and non-barred systems when limiting
our search to individual mass bins 
(Fig.~\ref{bar_frac_mass_ks}).

Numerous N-body simulations suggest a scenario for the evolution of disk galaxies in which bars are expected to slow down as they evolve
secularly, which is difficult to reconcile with the fact that most observed bars are fast
\citep[e.g.][]{1996AJ....111.2233E,2011MSAIS..18...23C}.  Recent work using direct Tremaine-Weingberg measurements also indicates fast bars
\citep[][]{2015A&A...576A.102A}.  In addition, \citet[][]{2008MNRAS.388.1803R} showed that bars in early-type
galaxies are typically fast rotators, while later-type galaxies present both fast and slow bars.  In future papers we will reassess
this topic by studying the bar pattern speed from the potential derived in this work, using the 2D sticky particle simulation method of
\citet[][]{1999AJ....117..792S} and linking the analysis to the bar properties obtained here.
\begin{acknowledgements}
We thank the anonymous referee for comments that improved this paper. 
We acknowledge financial support to the DAGAL network from the People Programme (Marie Curie Actions)
of the European Union's Seventh Framework Programme FP7/2007- 2013/ under REA grant agreement number PITN-GA-2011-289313.
EL and HS acknowledge financial support from the Academy of Finland.
We thank Sebastien Comerón for very useful conversation, valuable comments on the manuscript, and for technical support with different programming languages.
We thank Ryan Leaman for an insightful discussion on the dark matter properties of the S$^{4}$G sample and the interpretation of Fig.~\ref{stellar-halo-relation}
in terms of previous work carried out in the literature.
We thank Ronald Buta for making available his morphological classification before publication.
We thank Miguel Querejeta for useful discussions and for providing us the ICA-corrected mass maps of a subset of S$^{4}$G galaxies.
We thank Jarkko Laine for support with IDL scientific programming language.
We thank W.J.G de Blok, Claude Carignan, and Olivier Daigle for providing the rotation curves used in this work.
We thank in general the S$^{4}$G team for their work on the different pipelines, whose products were fundamental for the discussion of this paper.
{\it Facilities}: Spitzer (IRAC).
\end{acknowledgements}
\bibliographystyle{aa}
\bibliography{bibliography}
\clearpage
%

\onecolumn
\begin{appendix}
\section{Summary of parameters}

The following data are listed in Table~\ref{truncated65} (the complete version can be found at www.oulu.fi/astronomy/S4G$\_$BARFORCE/),
for all the 1345 disk galaxies with inclinations lower than 65$^{\circ}$ in P4 (ellipticals are excluded):

\begin{itemize}
\item Column 1: Galaxy identification.
\item Column 2: Stellar mass from P3 ($M_{\ast}$).
\item Column 3: Distance to the galaxy from P3 ($ D$).
\item Column 4: Disk scale height ($ h_{\rm z}$) estimated from the P4 disk scale length using \citet[][]{1998MNRAS.299..595D}'s relation \citep[or from 2MASS $r_{\rm k_{20}}$,][when $h_{\rm R}$ values were not available]{2008MNRAS.383..317S}.
\item Column 5: Optical radius ($ R_{\rm opt}$) calculated using the P4 scale length ($R_{\rm opt}=3.2 h_{\rm R}$), or assuming $R_{\rm opt}=R_{25.5}$ when $h_{\rm R}$ values were not available.
\item Column 6: H{\sc\,i}  maximum velocities ($ V_{\rm HI}^{\rm max}$) taken from the literature (Cosmic Flows or HyperLEDA), corrected for inclination. 
\item Column 7: Maximum of the stellar contribution to the rotation curve ($ V_{\rm 3.6\mu m}^{\rm max}$).
\item Column 8: Radius of maximum velocity of the stellar component of the rotation curve ($ r_{\rm 3.6\mu m}^{\rm max}$).
\item Column 9: The stellar contribution to the rotation curve evaluated at the optical radius $\Big(V_{\rm 3.6\mu m}(R_{\rm opt})\Big)$.
\item Column 10: Inner slope of the stellar component of the rotation curve $\Big(\rm d_{\rm R}V_{\rm 3.6\mu m}( 0)\Big)$.
\item Column 11: Halo core radius (a) computed with Eq.~\ref{urc_iso}.
\item Column 12: Halo velocity amplitude ($V_{\infty}$) computed with Eq. \ref{urc_iso}, corrected to match the H{\sc\,i} maximum velocity.
\item Column 13: Stellar-to-halo mass ratio within the optical radius $\Big( M_{\rm h}/M_{\ast}(< R_{\rm opt})\Big)$ estimated with Eq. \ref{halo-to-stellar-eq}.
\end{itemize}
\begin{table}[h]
\scalebox{0.95}{\parbox{\linewidth}{%
\begin{small}
\begin{center}
\caption[]{Stellar masses, distances, scale-heights, optical radii, maximum observed velocities, parameters of the stellar and halo components of the rotation curve, and halo-to-stellar mass ratios (i $< 65^{\bf \circ}$).}
\label{truncated65}
\begin{tabular}{|c c c c c c c c c c c c c|}
\hline
Galaxy & $M_{\ast}$ & $D$ & $h_{\rm z}$ & $R_{\rm opt}$ & $V_{\rm HI}^{\rm max}$ & $V_{\rm 3.6\mu m}^{\rm max}$ & $r_{\rm 3.6\mu m}^{\rm max}$ & $V_{\rm 3.6\mu m}(R_{\rm opt})$ & $\rm d_{\rm R}V_{\rm 3.6\mu m}(0)$ & a & $V_{\infty}$ & $M_{\rm h}/M_{\ast}(<R_{\rm opt})$ \\
 & ($M_{\odot})$ & (Mpc) & (") & (") & (km/s) & (km/s) & (") & (km/s) & (km/s/kpc) & (kpc) & (km/s) & \\
 (1) & (2) & (3) & (4) & (5) & (6) & (7) & (8) & (9) & (10) & (11) & (12) & (13) \\
\hline
ESO$\,$012-010&4.29$\cdot 10^{9}$& 32.75&   3.6$^{ -   0.9}_{+   2.8}$& 108.7& 119.6&  46.7$^{+   0.5}_{-   1.5}$& 65.25$_{-  1.50}$&  42.2$^{+   0.2}_{-   0.7}$&  24.8$^{+   2.2}_{-   4.3}$&  17.9& 160.9&   9.4\\
ESO$\,$013-016&3.66$\cdot 10^{9}$& 23.02&   2.2$^{ -   0.6}_{+   1.8}$&  67.4& 100.4&  63.9$^{+   0.5}_{-   1.6}$& 50.25&  60.8$^{+   0.3}_{-   1.0}$&  65.0$^{+   5.6}_{-  11.3}$&   8.0& 116.7&   2.3\\
ESO$\,$026-001&2.36$\cdot 10^{9}$& 19.16&   2.0$^{ -   0.5}_{+   1.6}$&  62.1& 133.7&  61.4$^{+   0.6}_{-   1.8}$& 44.25&  53.3$^{+   0.2}_{-   0.7}$& 164.6$^{+  15.9}_{-  31.4}$&   5.4& 168.2&   7.1\\
ESO$\,$027-001&9.98$\cdot 10^{9}$& 18.31&   4.5$^{ -   1.3}_{+   3.0}$&  79.0&  97.0& 104.5$^{+   2.0}_{-   4.1}$& 48.75$_{-  1.50}$&  93.6$^{+   0.8}_{-   1.9}$& 283.1$^{+  30.3}_{-  46.3}$&...&...&   0.1\\
ESO$\,$048-017&1.59$\cdot 10^{9}$& 29.52&   2.1$^{ -   0.5}_{+   1.7}$&  66.1&  79.4&  38.2$^{+   0.3}_{-   1.0}$& 48.75&  36.3$^{+   0.2}_{-   0.7}$&  39.2$^{+   3.6}_{-   7.1}$&   8.4&  95.1&   5.1\\
ESO$\,$054-021&3.52$\cdot 10^{9}$& 16.48&   5.1$^{ -   1.3}_{+   4.1}$& 150.8& 111.4&  49.7$^{+   0.4}_{-   1.3}$&102.75&  49.1$^{+   0.3}_{-   1.0}$&  44.0$^{+   2.9}_{-   6.5}$&  12.7& 145.3&   5.6\\
ESO$\,$079-005&1.99$\cdot 10^{9}$& 23.48&   2.3$^{ -   1.2}_{+   2.3}$&  67.3&  92.7&  51.1$^{+   1.2}_{-   2.2}$& 39.75$_{-  1.50}$&  43.6$^{+   0.4}_{-   0.8}$&  32.5$^{+   6.2}_{-   6.1}$&   7.3& 109.4&   4.7\\
ESO$\,$079-007&2.96$\cdot 10^{9}$& 25.22&   1.5$^{ -   0.4}_{+   1.2}$&  46.4&  89.5&  59.7$^{+   0.4}_{-   1.3}$& 38.25&  57.9$^{+   0.3}_{-   1.0}$&  65.1$^{+   5.3}_{-  10.7}$&   5.6&  96.3&   1.9\\
ESO$\,$085-030&6.34$\cdot 10^{9}$& 31.08&   2.3$^{ -   1.1}_{+   2.3}$&  50.9& 116.1&  94.9$^{+   4.7}_{-   7.1}$& 17.25$^{-  1.50}_{+  1.50}$&  68.0$^{+   0.5}_{-   1.1}$& 111.4$^{+  24.7}_{-  23.2}$&   8.2& 126.3&   2.6\\
ESO$\,$085-047&3.31$\cdot 10^{8}$& 16.70&   2.4$^{ -   0.6}_{+   1.9}$&  75.0&  35.8&  21.5$^{+   0.2}_{-   0.6}$& 72.75&  21.5$^{+   0.2}_{-   0.6}$&  11.9$^{+   0.5}_{-   1.1}$&   4.6&  37.4&   2.4\\
ESO$\,$114-007&1.78$\cdot 10^{9}$& 26.10&   1.5$^{ -   0.4}_{+   1.2}$&  41.7&  60.9&  34.2$^{+   0.3}_{-   0.9}$& 35.25$_{-  1.50}$&  33.4$^{+   0.2}_{-   0.6}$&  23.8$^{+   1.6}_{-   3.2}$&   4.8&  68.8&   3.1\\
ESO$\,$120-012&5.90$\cdot 10^{8}$& 14.40&   2.0$^{ -   0.5}_{+   1.6}$&  65.0&  62.2&  32.8$^{+   0.3}_{-   0.9}$& 39.75$_{-  1.50}$&  30.4$^{+   0.1}_{-   0.4}$&  30.8$^{+   1.6}_{-   3.6}$&   3.4&  67.8&   4.3\\
ESO$\,$120-021&1.42$\cdot 10^{8}$& 18.17&   1.9$^{ -   0.5}_{+   1.5}$&  60.0&  65.8&  21.4$^{+   0.2}_{-   0.5}$& 57.75&  21.4$^{+   0.2}_{-   0.5}$&  21.8$^{+   2.3}_{-   4.3}$&   3.5&  42.6&  11.3\\
ESO$\,$145-025&9.68$\cdot 10^{8}$& 22.50&   2.9$^{ -   0.7}_{+   2.3}$&  82.5&  80.4&  31.8$^{+   0.3}_{-   1.0}$& 80.25&  31.8$^{+   0.3}_{-   1.0}$&  20.0$^{+   2.0}_{-   3.8}$&   8.0& 105.5&   7.3\\
ESO$\,$150-005&6.85$\cdot 10^{8}$& 15.15&   3.5$^{ -   0.9}_{+   2.8}$&  97.5&  91.3&  31.6$^{+   0.4}_{-   1.1}$& 95.25&  31.6$^{+   0.4}_{-   1.1}$&  26.0$^{+   2.1}_{-   4.2}$&   5.3& 113.3&   9.9\\
ESO$\,$159-025&1.36$\cdot 10^{8}$& 11.80&   1.8$^{ -   0.5}_{+   1.5}$&  56.3&  71.6&  18.9$^{+   0.2}_{-   0.5}$& 44.25&  17.5$^{+   0.0}_{-   0.2}$&  27.3$^{+   2.0}_{-   3.8}$&   1.9&  62.5&  21.0\\
ESO$\,$187-035&4.68$\cdot 10^{8}$& 26.20&   1.6$^{ -   0.4}_{+   1.3}$&  49.2&  56.4&  24.1$^{+   0.2}_{-   0.5}$& 48.75$_{-  1.50}$&  24.1$^{+   0.2}_{-   0.5}$&  33.4$^{+   3.4}_{-   6.4}$&   5.2&  66.8&   6.0\\
ESO$\,$202-041&2.10$\cdot 10^{8}$& 16.97&   2.6$^{ -   1.3}_{+   2.6}$&  37.9&  55.1&  20.2$^{+   0.5}_{-   0.9}$& 50.25&  19.7$^{+   0.5}_{-   0.9}$&  30.9$^{+   5.4}_{-   5.5}$&   2.3&  70.6&   9.1\\
ESO$\,$234-043&1.32$\cdot 10^{9}$& 26.95&   2.9$^{ -   1.4}_{+   2.9}$&  60.0&  83.3&  40.5$^{+   1.1}_{-   1.9}$& 57.75&  40.5$^{+   1.1}_{-   1.9}$&  44.8$^{+  11.6}_{-  10.6}$&   6.9& 100.0&   4.3\\
ESO$\,$234-049&8.11$\cdot 10^{9}$& 35.99&   1.5$^{ -   0.4}_{+   1.2}$&  46.7&  87.8&  89.0$^{+   2.1}_{-   3.7}$& 35.25$^{+  1.50}$&  86.3$^{+   1.4}_{-   2.6}$&  53.4$^{+  10.3}_{-  10.1}$&...&...&   0.0\\
ESO$\,$236-039&2.90$\cdot 10^{8}$& 22.22&   1.0$^{ -   0.2}_{+   0.8}$&  32.6&  52.4&  25.5$^{+   0.2}_{-   0.4}$& 35.25$^{+  9.00}$&  24.8$^{+   0.1}_{-   0.4}$&  32.0$^{+   2.0}_{-   4.0}$&   2.7&  58.6&   4.7\\
ESO$\,$237-052&3.21$\cdot 10^{9}$& 36.46&   1.5$^{ -   0.4}_{+   1.2}$&  46.0&  89.7&  48.7$^{+   0.6}_{-   1.1}$& 24.75$_{- 19.50}$&  48.2$^{+   0.3}_{-   0.9}$&  40.2$^{+   3.1}_{-   6.4}$&   8.2& 107.5&   3.3\\
ESO$\,$238-018&3.29$\cdot 10^{9}$& 41.04&   1.0$^{ -   0.2}_{+   0.8}$&  32.2&  56.2&  65.0$^{+   0.7}_{-   1.9}$& 23.25&  59.5$^{+   0.3}_{-   0.9}$&  64.5$^{+   1.4}_{-   3.9}$&...&...&...\\
ESO$\,$245-005&1.08$\cdot 10^{8}$&  4.19&   0.5$^{ -   0.2}_{+   0.5}$& 105.0&  47.5&  20.2$^{+   0.1}_{-   0.2}$&102.75&  20.2$^{+   0.1}_{-   0.2}$&...&   1.3& 100.3&   6.1\\
ESO$\,$245-007&      8.6$\cdot 10^{5}$&  0.41&   4.7$^{ -   2.4}_{+   4.7}$& 100.9&...&   6.0$^{+   0.2}_{-   0.3}$&117.75&   5.6$^{+   0.1}_{-   0.2}$&...&...&...&...\\
ESO$\,$249-008&5.30$\cdot 10^{8}$& 17.39&   0.9$^{ -   0.2}_{+   0.7}$&  34.3&  35.6&  50.0$^{+   2.0}_{-   2.7}$& 14.25$^{-  1.50}_{+  3.00}$&  41.0$^{+   0.4}_{-   0.7}$& 150.5$^{+  27.0}_{-  27.1}$&...&...&...\\
ESO$\,$249-026&5.90$\cdot 10^{7}$& 12.93&   1.1$^{ -   0.3}_{+   0.9}$&  45.0&  53.3&  16.5$^{+   0.1}_{-   0.4}$& 42.75&  16.5$^{+   0.1}_{-   0.4}$&  41.9$^{+   4.2}_{-   7.8}$&   1.9&  46.4&  12.6\\
ESO$\,$249-036&3.32$\cdot 10^{8}$&  9.40&   2.3$^{ -   1.2}_{+   2.3}$&  60.0&  46.0&  19.2$^{+   0.6}_{-   1.0}$& 57.75&  19.2$^{+   0.6}_{-   1.0}$&  20.3$^{+   4.4}_{-   4.1}$&   1.4& 119.3&   6.4\\
ESO$\,$285-048&6.78$\cdot 10^{9}$& 33.14&   2.2$^{ -   0.5}_{+   1.7}$&  65.6& 119.8&  69.9$^{+   0.6}_{-   1.9}$& 47.25&  64.6$^{+   0.3}_{-   1.0}$&  55.1$^{+   4.4}_{-   9.1}$&  12.9& 159.7&   3.3\\
ESO$\,$287-037&4.93$\cdot 10^{9}$& 36.15&   1.9$^{ -   0.5}_{+   1.5}$&  59.8& 134.4&  68.7$^{+   0.6}_{-   1.7}$& 42.75$^{+  3.00}$&  63.7$^{+   0.3}_{-   0.9}$&  38.7$^{+   3.1}_{-   6.2}$&  12.1& 181.9&   4.6\\
ESO$\,$288-013&3.20$\cdot 10^{9}$& 35.61&   1.5$^{ -   0.4}_{+   1.2}$&  47.1&  54.2&  57.6$^{+   0.7}_{-   2.0}$& 32.25&  48.1$^{+   0.1}_{-   0.5}$&  27.7$^{+   1.6}_{-   3.3}$&...&...&   0.4\\
ESO$\,$289-026&1.90$\cdot 10^{9}$& 26.46&   3.9$^{ -   1.9}_{+   3.9}$&  70.9&  78.6&  35.5$^{+   0.9}_{-   1.6}$& 80.25&  34.2$^{+   0.6}_{-   1.2}$&  41.8$^{+  10.6}_{-   9.8}$&   8.1&  95.4&   5.7\\
ESO$\,$298-023&4.43$\cdot 10^{8}$& 19.08&   0.9$^{ -   0.2}_{+   0.7}$&  29.8&  92.3&  31.4$^{+   0.2}_{-   0.6}$& 26.25&  30.9$^{+   0.2}_{-   0.5}$&  70.7$^{+   1.6}_{-   4.2}$&   2.1& 103.4&  10.6\\
\hline
\end{tabular}
\end{center}
\end{small}
}}
\end{table}

\clearpage
%
For the barred galaxies with measurements of $Q_{\rm b}$ and $r_{\rm Qb}$,
the following data are listed in Table~\ref{truncated1} (the complete version can be found at www.oulu.fi/astronomy/S4G$\_$BARFORCE/):
\begin{itemize}
\item Column 1: Galaxy identification.
\item Column 2: Bar maximum gravitational torque ($Q_{\rm b}$).
Errors in $Q_{\rm b}$ are associated to the highest and lowest $h_{\rm R}/h_{\rm z}$ values in the relation from \citet[][]{1998MNRAS.299..595D}.
\item Column 3: Bar maximum gravitational torque radius ($ r_{\rm Qb}$).
\item Column 4: Bar maximum gravitational torque correcting for the dilution of the dark halo ($ Q_{\rm b}^{\rm halo-corr}$).
\item Column 5: Bar maximum gravitational torque radius correcting for the dilution of the dark halo ($ r_{\rm Qb}^{\rm halo-corr}$).
\item Column 6: Bar maximum gravitational torque subtracting the effect of the spiral arms ($ Q_{\rm b}^{\rm bar-only}$).
\item Column 7: Bar maximum gravitational torque radius subtracting the effect of the spiral arms ($ sr_{\rm Qb}^{\rm bar-only}$).
\item Column 8: Bar maximum gravitational torque subtracting the effect of the spiral arms and correcting for the effect of the dark halo $\Big((Q_{\rm b}^{\rm bar-only})^{\rm halo-corr}\Big)$.
\item Column 9: Bar maximum gravitational torque radius subtracting the effect of the spiral arms and correcting for the effect of the dark halo $\Big(( r_{\rm Qb}^{\rm bar-only})^{\rm halo-corr}\Big)$.
\item Column 10: Bar maximum gravitational torque using the relation from \citep[][]{2008MNRAS.383..317S} to calculate the disk thickness ($ Q_{\rm b}^{S08}$).
Errors computed assigning a factor 2 thicker or thinner disk for the force calculation.
\item Column 11: Radial force profiles evaluated at the end of the bar $\Big(Q_{\rm T}(r_{\rm bar})\Big)$.
\item Column 12: Halo-corrected radial force profiles evaluated at the end of the bar $\Big(Q_{\rm T}^{\rm halo-corr}(r_{\rm bar})\Big)$.
\end{itemize}
\begin{table}[h]
\scalebox{0.965}{\parbox{\linewidth}{%
\begin{tiny}
\begin{center}
\caption[]{Gravitational torque parameters and radii, with and without spiral arms and halo correction (i $< 65^{\circ}$).}
\label{truncated1}
\begin{tabular}{|c c c c c c c c c c c c|}
\hline
\rm Galaxy & $Q_{\rm b}$ & $r_{\rm Qb}$ & $Q_{\rm b}^{\rm halo-corr}$ & $r_{\rm Qb}^{\rm halo-corr}$ & $Q_{\rm b}^{\rm bar-only}$ & $r_{\rm Qb}^{\rm bar-only}$ &
$(Q_{\rm b}^{\rm bar-only})^{\rm halo-corr}$ & $(r_{\rm Qb}^{\rm bar-only})^{\rm halo-corr}$ & $Q_{\rm b}^{S08}$ & $Q_{\rm T}(r_{\rm bar})$ & $Q_{\rm T}^{\rm halo-corr}(r_{\rm bar})$ \\
 &  & (") &  & (") &  & (") &  & (") & & & \\
 (1)  & (2) & (3) & (4) & (5) & (6) & (7) & (8) & (9) & (10) & (11) & (12)\\
\hline
ESO$\,$012-010&  0.64$^{ +  0.02}_{ -  0.05}$&   8.2$_{ -   1.5}$&  0.52&   3.8&  0.63&   8.2&  0.51&   3.8&  0.68$^{ +  0.05}_{ -  0.06}$&  0.37$^{ +  0.02}_{ -  0.04}$&  0.17 \\
ESO$\,$013-016&  0.52$^{ +  0.02}_{ -  0.05}$&   6.8&  0.47&   6.8&  0.48&   6.8&  0.44&   6.8&  0.58$^{ +  0.05}_{ -  0.06}$&  0.33$^{ +  0.01}_{ -  0.03}$&  0.26 \\
ESO$\,$026-001&  0.58$^{ +  0.03}_{ -  0.06}$&   5.2&  0.51&   5.2&  0.58&   5.2&  0.51&   5.2&  0.55$^{ +  0.07}_{ -  0.06}$&  0.21$^{ +  0.01}_{ -  0.03}$&  0.09 \\
ESO$\,$027-001&  0.25$^{ +  0.02}_{ -  0.04}$&  15.8&...&...&  0.15&   9.8&...&...&  0.23$^{ +  0.05}_{ -  0.04}$&  0.25$^{ +  0.02}_{ -  0.04}$&... \\
ESO$\,$079-007&  0.36$^{ +  0.02}_{ -  0.04}$&   3.8&  0.33&   3.8&  0.36&   3.8&  0.33&   3.8&  0.32$^{ +  0.04}_{ -  0.04}$&  0.20$^{ +  0.01}_{ -  0.02}$&  0.15 \\
ESO$\,$202-041&  0.45$^{ +  0.07}_{ -  0.07}$&  17.2&  0.11&   9.8&  0.32&  11.2&  0.10&   8.2&  0.45$^{ +  0.07}_{ -  0.07}$&  0.44$^{ +  0.07}_{ -  0.07}$&  0.08 \\
ESO$\,$234-043&  0.57$^{ +  0.04}_{ -  0.05}$&   8.2&  0.42&   6.8&  0.54&   6.8&  0.40&   6.8&  0.57$^{ +  0.04}_{ -  0.05}$&  0.40$^{ +  0.03}_{ -  0.04}$&  0.17 \\
ESO$\,$236-039&  0.31$^{ +  0.02}_{ -  0.05}$&   5.2&  0.17&   5.2&  0.31&   5.2&  0.17&   5.2&  0.26$^{ +  0.05}_{ -  0.04}$&  0.16$^{ +  0.01}_{ -  0.02}$&  0.07 \\
ESO$\,$285-048&  0.41$^{ +  0.01}_{ -  0.03}$&   3.8&  0.38&   3.8&  0.40&   3.8&  0.37&   3.8&  0.39$^{ +  0.03}_{ -  0.03}$&  0.27$^{ +  0.02}_{ -  0.04}$&  0.21 \\
ESO$\,$287-037&  0.36$^{ +  0.01}_{ -  0.02}$&   6.8$_{ -   1.5}$&  0.30&   5.2&  0.35&   5.2&  0.29&   5.2&  0.36$^{ +  0.03}_{ -  0.03}$&  0.30$^{ +  0.01}_{ -  0.03}$&  0.17 \\
ESO$\,$289-026&  0.74$^{ +  0.11}_{ -  0.10}$&  18.8$_{ -   1.5}$&  0.55&   3.8&  0.73&  17.2&  0.55&   3.8&  0.74$^{ +  0.11}_{ -  0.10}$&  0.40$^{ +  0.04}_{ -  0.05}$&  0.12 \\
ESO$\,$298-023&  0.43$^{ +  0.02}_{ -  0.03}$&   3.8$_{ -   1.5}$&  0.24&   2.2&  0.41&   2.2&  0.23&   2.2&  0.39$^{ +  0.03}_{ -  0.03}$&...&... \\
ESO$\,$340-017&  0.55$^{ +  0.02}_{ -  0.04}$&   8.2&  0.50&   6.8&  0.55&   8.2&  0.50&   6.8&  0.52$^{ +  0.05}_{ -  0.05}$&  0.19$^{ +  0.01}_{ -  0.01}$&  0.09 \\
ESO$\,$340-042&  0.64$^{ +  0.02}_{ -  0.05}$&   3.8&  0.61&   3.8&  0.62&   3.8&  0.59&   3.8&  0.63$^{ +  0.06}_{ -  0.06}$&  0.38$^{ +  0.02}_{ -  0.05}$&  0.32 \\
ESO$\,$345-046&  0.30$^{ +  0.01}_{ -  0.02}$&   3.8&  0.25&   3.8&  0.20&   3.8&  0.16&   3.8&  0.28$^{ +  0.02}_{ -  0.02}$&  0.27$^{ +  0.01}_{ -  0.02}$&  0.20 \\
ESO$\,$358-020&  0.37$^{ +  0.02}_{ -  0.04}$&   5.2&  0.33&   5.2&  0.36&   5.2&  0.33&   5.2&  0.35$^{ +  0.04}_{ -  0.04}$&  0.12$^{ +  0.01}_{ -  0.01}$&  0.10 \\
ESO$\,$359-031&  0.33$^{ +  0.02}_{ -  0.04}$&   8.2&  0.31&   8.2&  0.31&   8.2&  0.28&   8.2&  0.28$^{ +  0.04}_{ -  0.04}$&  0.24$^{ +  0.01}_{ -  0.02}$&  0.21 \\
ESO$\,$404-003&  0.38$^{ +  0.01}_{ -  0.03}$&   6.8&  0.33&   5.2&  0.32&   5.2&  0.29&   3.8&  0.35$^{ +  0.03}_{ -  0.03}$&  0.37$^{ +  0.01}_{ -  0.03}$&  0.29 \\
ESO$\,$404-012&  0.28$^{ +  0.02}_{ -  0.03}$&   9.8&  0.24&   9.8&  0.22&   9.8&  0.19&   8.2&  0.26$^{ +  0.04}_{ -  0.04}$&  0.27$^{ +  0.02}_{ -  0.03}$&  0.23 \\
ESO$\,$418-008&  0.48$^{ +  0.02}_{ -  0.04}$&   3.8&  0.39&   3.8&  0.45&   3.8&  0.37&   3.8&  0.45$^{ +  0.05}_{ -  0.05}$&  0.29$^{ +  0.01}_{ -  0.03}$&  0.20 \\
ESO$\,$418-009&  0.60$^{ +  0.03}_{ -  0.06}$&   8.2&  0.47&   8.2&  0.49&   6.8&  0.39&   6.8&  0.54$^{ +  0.08}_{ -  0.08}$&  0.43$^{ +  0.02}_{ -  0.04}$&  0.32 \\
ESO$\,$420-009&  0.12$^{ +  0.01}_{ -  0.02}$&   2.2&  0.10&   2.2&  0.13&   2.2&  0.11&   2.2&  0.10$^{ +  0.02}_{ -  0.02}$&  0.07$^{ +  0.01}_{ -  0.02}$&  0.06 \\
ESO$\,$422-005&  0.34$^{ +  0.02}_{ -  0.04}$&   6.8&  0.18&   5.2&  0.33&   6.8&  0.17&   5.2&  0.31$^{ +  0.04}_{ -  0.04}$&...&... \\
ESO$\,$440-046&  0.36$^{ +  0.03}_{ -  0.06}$&   3.8&  0.27&   3.8&  0.39&   3.8&  0.30&   3.8&  0.34$^{ +  0.07}_{ -  0.07}$&  0.15$^{ +  0.02}_{ -  0.03}$&  0.09 \\
ESO$\,$441-017&  0.52$^{ +  0.02}_{ -  0.04}$&   2.2&  0.51&   2.2&  0.49&   2.2&  0.49&   2.2&  0.47$^{ +  0.04}_{ -  0.04}$&  0.25$^{ +  0.01}_{ -  0.01}$&  0.24 \\
ESO$\,$443-069&  0.59$^{ +  0.03}_{ -  0.08}$&   6.8&...&...&  0.55&   6.8&...&...&  0.53$^{ +  0.08}_{ -  0.05}$&  0.40$^{ +  0.02}_{ -  0.04}$&... \\
ESO$\,$443-085&  0.49$^{ +  0.02}_{ -  0.04}$&   6.8&  0.39&   3.8&  0.46&   6.8&  0.37&   3.8&  0.48$^{ +  0.04}_{ -  0.04}$&  0.34$^{ +  0.02}_{ -  0.03}$&  0.18 \\
ESO$\,$445-089&  0.36$^{ +  0.01}_{ -  0.03}$&   3.8$_{ -   1.5}$&  0.35&   3.8&  0.33&   2.2&  0.32&   2.2&  0.37$^{ +  0.03}_{ -  0.03}$&  0.26$^{ +  0.02}_{ -  0.03}$&  0.24 \\
ESO$\,$482-035&  0.54$^{ +  0.03}_{ -  0.04}$&   9.8$^{ +   1.5}$&  0.45&   9.8&  0.51&   9.8&  0.43&   9.8&  0.53$^{ +  0.07}_{ -  0.06}$&  0.32$^{ +  0.02}_{ -  0.03}$&  0.21 \\
ESO$\,$485-021&  0.52$^{ +  0.02}_{ -  0.05}$&   8.2&  0.24&   8.2&  0.51&   8.2&  0.24&   8.2&  0.50$^{ +  0.06}_{ -  0.05}$&  0.38$^{ +  0.02}_{ -  0.04}$&  0.07 \\
ESO$\,$508-007&  0.33$^{ +  0.01}_{ -  0.03}$&   3.8&  0.27&   3.8&  0.29&   3.8&  0.24&   3.8&  0.30$^{ +  0.03}_{ -  0.03}$&  0.23$^{ +  0.01}_{ -  0.02}$&  0.15 \\
ESO$\,$510-059&  0.85$^{ +  0.07}_{ -  0.12}$&   9.8&  0.78&   9.8&  0.83&   9.8&  0.77&   9.8&  0.96$^{ +  0.18}_{ -  0.16}$&  0.43$^{ +  0.03}_{ -  0.06}$&  0.35 \\
\hline
\end{tabular}
\end{center}
\end{tiny}
}}
\end{table}

\clearpage
For the barred galaxies with measured $A_{\rm 2}^{\rm max}$ and $r_{\rm A2}$ (the maximum of the higher order even Fourier component amplitudes could only be identified for a fraction of them),
the following data are listed in Table~\ref{truncated2} (the complete version is provided at www.oulu.fi/astronomy/S4G$\_$BARFORCE/):
\begin{itemize}
\item Column 1: Galaxy identification.
\item Column 2: Bar intrinsic ellipticity ($\epsilon$).
\item Column 3: Bar maximum normalized m=2 Fourier amplitude ($A_{2}^{\rm max}$).
\item Column 4: Bar maximum normalized m=2 Fourier amplitude radius ($r_{\rm A2}$).
\item Column 5: Bar maximum normalized m=4 Fourier amplitude ($A_{4}^{\rm max}$).
\item Column 6: Bar maximum normalized m=4 Fourier amplitude { radius} ($r_{\rm A4}$).
\item Column 7: Bar maximum normalized m=6 Fourier amplitude ($A_{6}^{\rm max}$).
\item Column 8: Bar maximum normalized m=6 Fourier amplitude radius ($r_{\rm A6}$).
\item Column 9: Bar maximum normalized m=8 Fourier amplitude ($A_{8}^{\rm max}$).
\item Column 10: Bar maximum normalized m=8 Fourier amplitude radius ($r_{\rm A8}$).
\item Column 11: Normalized m=2 Fourier amplitude evaluated at the end of the bar $\Big(A_2(r_{\rm bar})\Big)$.
\end{itemize}
\begin{table}[h]
\begin{small}
\begin{center}
\caption[]{Bar intrinsic ellipticities and maximum normalized Fourier amplitudes and radii (m=2,4,6,8) (i $< 65^{\circ}$).}
\label{truncated2}
\begin{tabular}{|c c c c c c c c c c c|}
\hline
Galaxy & $\epsilon$ & $A_{2}^{\rm max}$ & $r_{\rm A2}$ & $A_{4}^{\rm max}$ & $r_{\rm A4}$ & $A_{6}^{\rm max}$ & $r_{\rm A6}$ & $A_{8}^{\rm max}$ & $r_{\rm A8}$ & $A_2(r_{\rm bar})$ \\
  &  &  &  (") &  & (") &  & (") &  & (") & \\
 (1)  & (2) & (3) & (4) & (5) & (6) & (7) & (8) & (9) & (10) & (11) \\
\hline
ESO$\,$012-010&  0.69&  0.48&  15.8&  0.19&  15.8&...&...&...&...&  0.32 \\
ESO$\,$013-016&  0.61&  0.39&   9.8&  0.13&   8.2&  0.10&   9.8&...&...&  0.26 \\
ESO$\,$026-001&  0.60&  0.64&   6.8&  0.28&   6.8&  0.13&   6.8&  0.08&   8.2&  0.24 \\
ESO$\,$027-001&  0.55&  0.49&  14.2&...&...&...&...&...&...&  0.48 \\
ESO$\,$079-007&  0.56&  0.19&   6.8&  0.11&   8.2&...&...&...&...&  0.09 \\
ESO$\,$202-041&  0.59&  0.47&  15.8&  0.25&  15.8&...&...&...&...&  0.39 \\
ESO$\,$234-043&  0.66&  0.50&  14.2&...&...&...&...&...&...&  0.42 \\
ESO$\,$236-039&  0.34&  0.14&   5.2&...&...&...&...&...&...&  0.10 \\
ESO$\,$285-048&  0.59&  0.32&   8.2&  0.09&  12.8&...&...&...&...&  0.27 \\
ESO$\,$287-037&  0.43&  0.29&   9.8&...&...&...&...&...&...&  0.22 \\
ESO$\,$289-026&  0.78&  0.78&  17.2&  0.40&  17.2&  0.21&  18.8&  0.19&  17.2&  0.53 \\
ESO$\,$298-023&...&  0.33&   8.2&...&...&...&...&...&...&... \\
ESO$\,$340-017&  0.48&  0.62&  11.2&...&...&...&...&...&...&  0.49 \\
ESO$\,$340-042&  0.73&  0.53&   8.2&  0.19&   8.2&...&...&...&...&  0.44 \\
ESO$\,$345-046&  0.43&  0.18&   5.2&...&...&...&...&...&...&  0.16 \\
ESO$\,$358-020&  0.49&  0.22&   8.2&...&...&...&...&...&...&  0.15 \\
ESO$\,$359-031&  0.40&  0.19&   9.8&  0.10&  12.8&...&...&...&...&  0.18 \\
ESO$\,$404-003&  0.50&  0.33&   6.8&...&...&...&...&...&...&  0.29 \\
ESO$\,$404-012&  0.61&  0.36&  11.2&  0.13&   9.8&...&...&...&...&  0.36 \\
ESO$\,$418-008&...&  0.25&   6.8&  0.14&   9.8&  0.05&   6.8&...&...&  0.16 \\
ESO$\,$418-009&  0.74&  0.31&   9.8&  0.16&   9.8&...&...&...&...&  0.31 \\
ESO$\,$420-009&  0.38&  0.08&   3.8&  0.02&   3.8&...&...&...&...&  0.06 \\
ESO$\,$422-005&...&  0.26&   9.8&  0.12&   8.2&...&...&...&...&... \\
ESO$\,$440-046&  0.68&  0.17&  11.2&...&...&...&...&...&...&  0.17 \\
ESO$\,$441-017&...&  0.26&   5.2&  0.06&   5.2&...&...&...&...&  0.22 \\
ESO$\,$443-069&  0.67&  0.56&   6.8&  0.29&   8.2&  0.17&   8.2&  0.09&   8.2&  0.30 \\
ESO$\,$443-085&  0.61&  0.31&   9.8&...&...&...&...&...&...&  0.29 \\
ESO$\,$445-089&  0.56&  0.30&   8.2&...&...&...&...&...&...&  0.29 \\
ESO$\,$482-035&  0.68&  0.52&  11.2&  0.22&  12.8&  0.15&  14.2&  0.06&  11.2&  0.40 \\
ESO$\,$485-021&  0.60&  0.44&  14.2&  0.16&  14.2&...&...&...&...&  0.42 \\
ESO$\,$508-007&  0.47&  0.21&   6.8&...&...&...&...&...&...&  0.17 \\
ESO$\,$508-024&  0.42&  0.28&   5.2&  0.07&   6.8&...&...&...&...&  0.15 \\
ESO$\,$510-059&...&  0.78&  11.2&  0.45&   9.8&  0.30&  11.2&  0.24&  11.2&  0.61 \\
ESO$\,$541-004&...&  0.10&  15.8&...&...&...&...&...&...&  0.09 \\
ESO$\,$547-005&  0.43&  0.19&   8.2&...&...&...&...&...&...&  0.14 \\
ESO$\,$548-025&  0.73&  0.61&  12.8&  0.30&  14.2&  0.18&  14.2&  0.08&  17.2&  0.45 \\
ESO$\,$548-032&...&  0.39&   9.8&  0.12&   8.2&  0.10&  11.2&...&...&  0.33 \\
ESO$\,$549-018&  0.45&  0.22&  18.8&...&...&...&...&...&...&  0.19 \\
ESO$\,$572-018&  0.56&  0.78&  11.2&  0.17&  11.2&...&...&...&...&  0.08 \\
\hline
\end{tabular}
\end{center}
\end{small}
\end{table}

\clearpage
\section{Catalogue of measurements}
Figure~\ref{all_plot_summary} summarizes all the measurements carried out in this work for the barred galaxy NGC$\:$4314.
The following plots are presented (for all the galaxies in our sample, an electronic version of these figures and the data used to create them can be found at www.oulu.fi/astronomy/S4G$\_$BARFORCE/):
\begin{itemize}
\item Panel (A): 
3.6~$\mu$m image de-projected to the disk plane (in units of $AB$ mag with an upper and lower magnitude threshold of 25 and 18).
Axes are in units of arcsec.
The blue line traces the visual estimate of the length and position angle of the bar.
The light green ellipse corresponds to the isophote fitting the bar (maximum ellipticity).
The purple and light blue circles (dashed lines) have radii equal
to $r_{\rm A2}$ and $r_{\rm Qb}$, respectively.
The red circumference delimits an area of radius twice the size of the bar: $r_{\rm circle}=\rm 2 \cdot$ max(r$_{\rm vis},\rm r_{\varepsilon})$.
\item Panel (B): 
Fourier-smoothed density with the axisymmetric component (m=0) subtracted.
Contours of equal $Q_{\rm T}$ are overplotted in white.
The dotted lines indicate the regions where the tangential forces change sign, while the red line represents the visual measurement of the bar length and PA.
The outer circle delimits a region of roughly twice the size of the bar (the same as in panel A).
\item Panel (C): 
Gravitational torque map.
In the bottom of the panel the bar shows the colour scale used for the plot.
Axes are in arcsec.
Bar length and ellipticity are traced with a black solid line.
The map is zoomed over the region delimited in panel (A).
The inner dotted circle corresponds to the P4 effective radii of the bulge, in case it is fitted in P4.
\item Panel (D):
$Q_{\rm T}$ radial profiles for three different disk thickness models (the nominal value is the middle one),
with the visual bar length highlighted with a dashed line.
The dotted line indicates $r_{\rm Qb}$.
The radius range is from zero to twice the bar size.
\item Panel (E):
Normalized Fourier amplitude $A_{2}$ radial profile.
The dotted line indicates $r_{\rm A2}$.
The vertical dashed line corresponds to the bar length.
\item Panel (F):
m=2 Fourier intensity phase.
\item Panel (G):
Stellar contribution to the rotation curve for three different disk thickness models.
The radius is given in arcsec.
The vertical dashed line corresponds to the bar length.
The horizontal black line indicates the maximum rotation as deduced from the stellar mass.
The solid orange line corresponds to a straight line with the slope inferred from the linear term of the $m=3$ degree polynomial fit of the inner part of the rotation curve (dashed line).
The dotted vertical orange line delimits the region where this fit is made, which is taken between the galaxy centre and the radius of the maximum rotation within one fourth of $R_{25.5}$.
\item Panel (H):
Summary of measurements for the galaxy with the same notation as in Tables \ref{truncated65}, \ref{truncated1}, and \ref{truncated2}.
\item Panel (I)\footnote{Panels (I) and (J) are available only for the galaxies with usable H{\sc\,i} velocity measurements in the literature.
When the halo component of the RC is not estimated, the bar-only $Q_{\rm T}$ profiles are included in panel (D).}:
Rotation curve decomposition based on the stellar mass mass rotation curve, inferred from the 3.6 $\mu$m photometry.
The amplitude of the URC halo component (isothermal sphere) has been corrected so that the total model matches the H{\sc\,i} maximum velocity at the optical radius (horizontal green line).
The vertical dashed line indicates the length of the bar.
\item Panel (J):
Raw 3.6 $\mu$m and bar-only $Q_{\rm T}$ profiles, with and without halo correction.
The maxima of the force profiles in the bar region and the bar
size are highlighted with vertical lines.
\end{itemize}
\begin{figure}
   \centering 
   \includegraphics[width=0.67\textwidth]{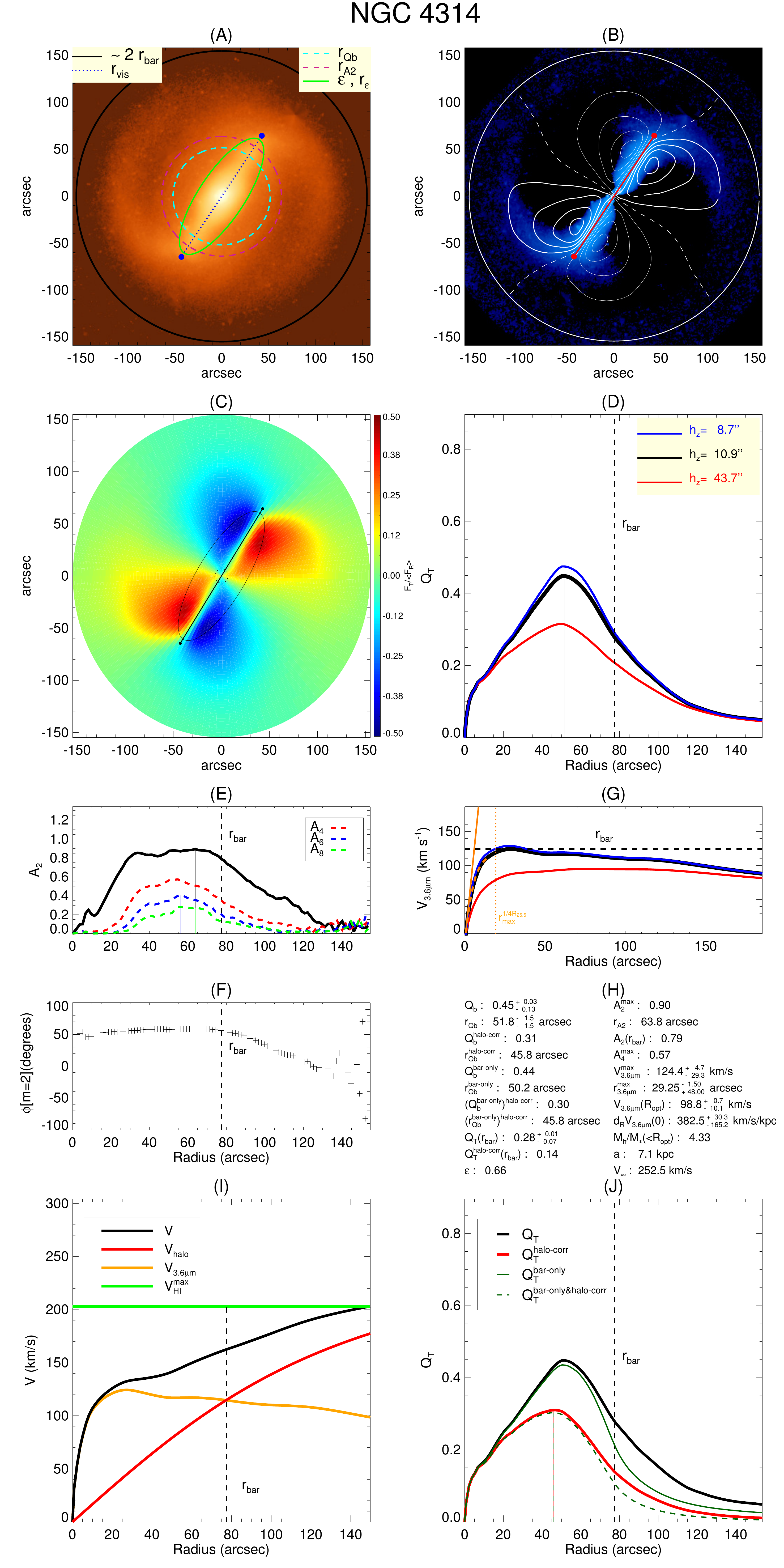}
\caption{
Summary of measurements over the barred galaxy NGC$\:$4314.
}
\label{all_plot_summary}
\end{figure}
\twocolumn
\section{Measurements over P5-mass maps}\label{appendix4}
\begin{figure}
     \centering  
     \begin{tabular}{l c r}
     \includegraphics[width=0.44\textwidth]{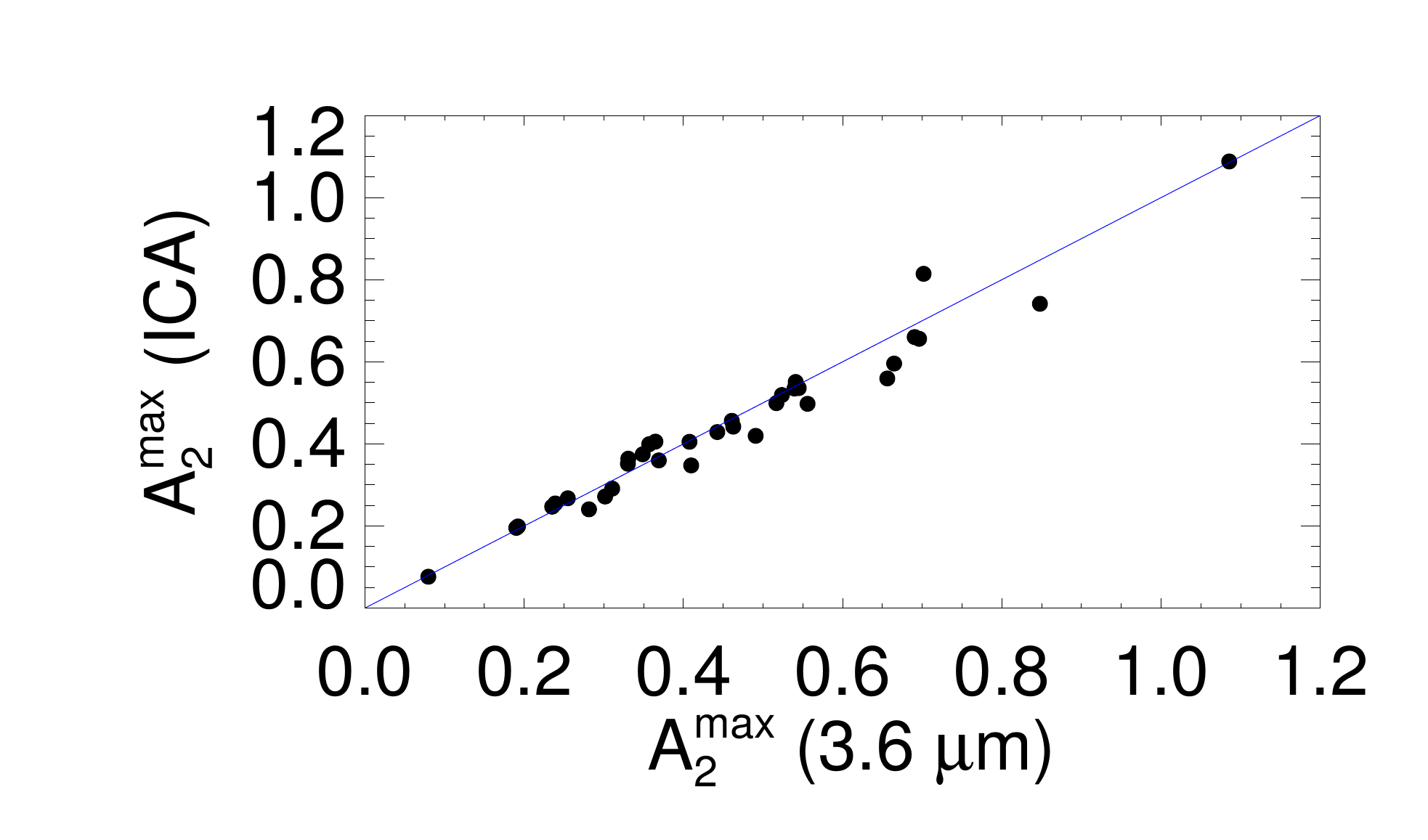}\\[-5ex]
     \includegraphics[width=0.44\textwidth]{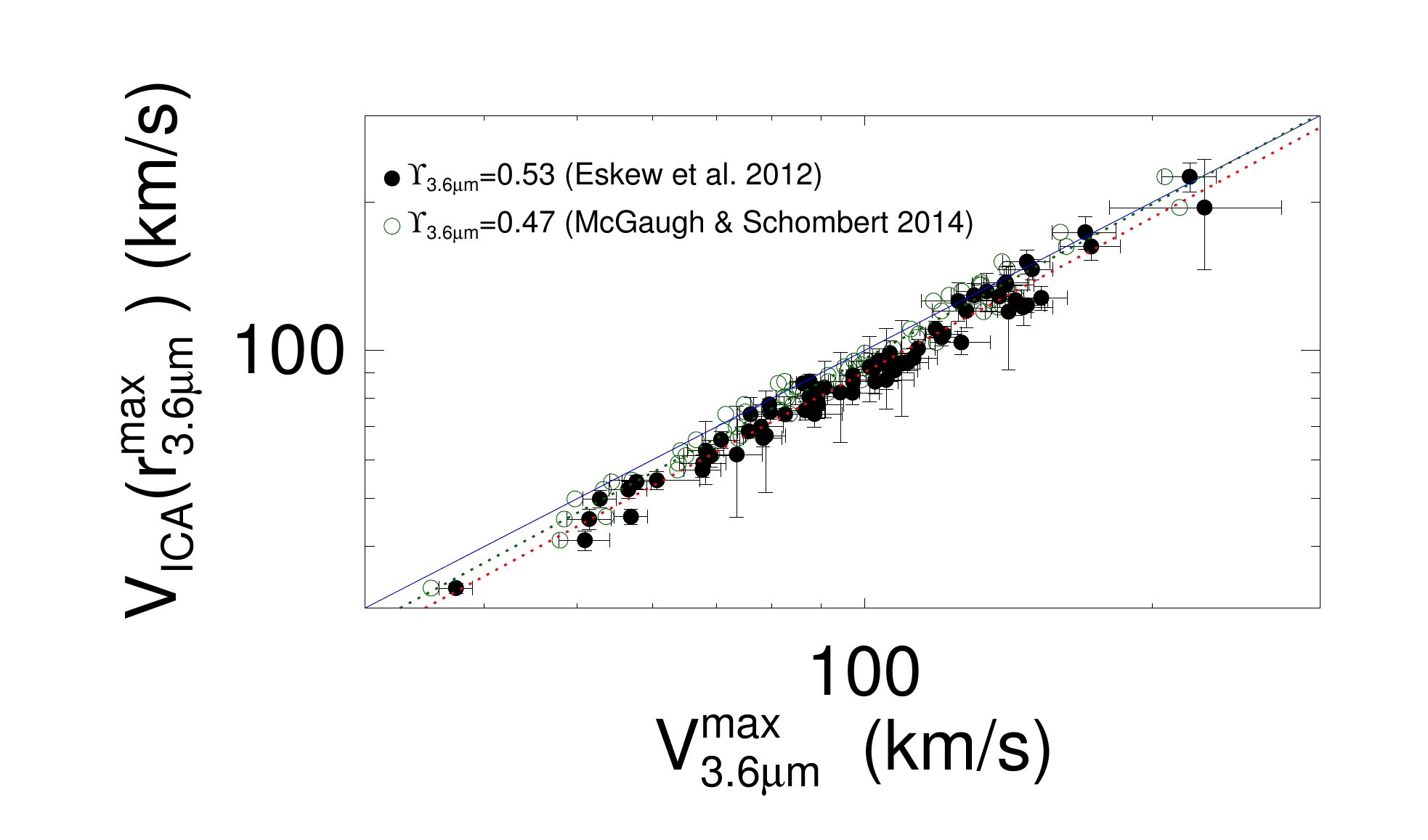}\\[-5ex]
     \includegraphics[width=0.44\textwidth]{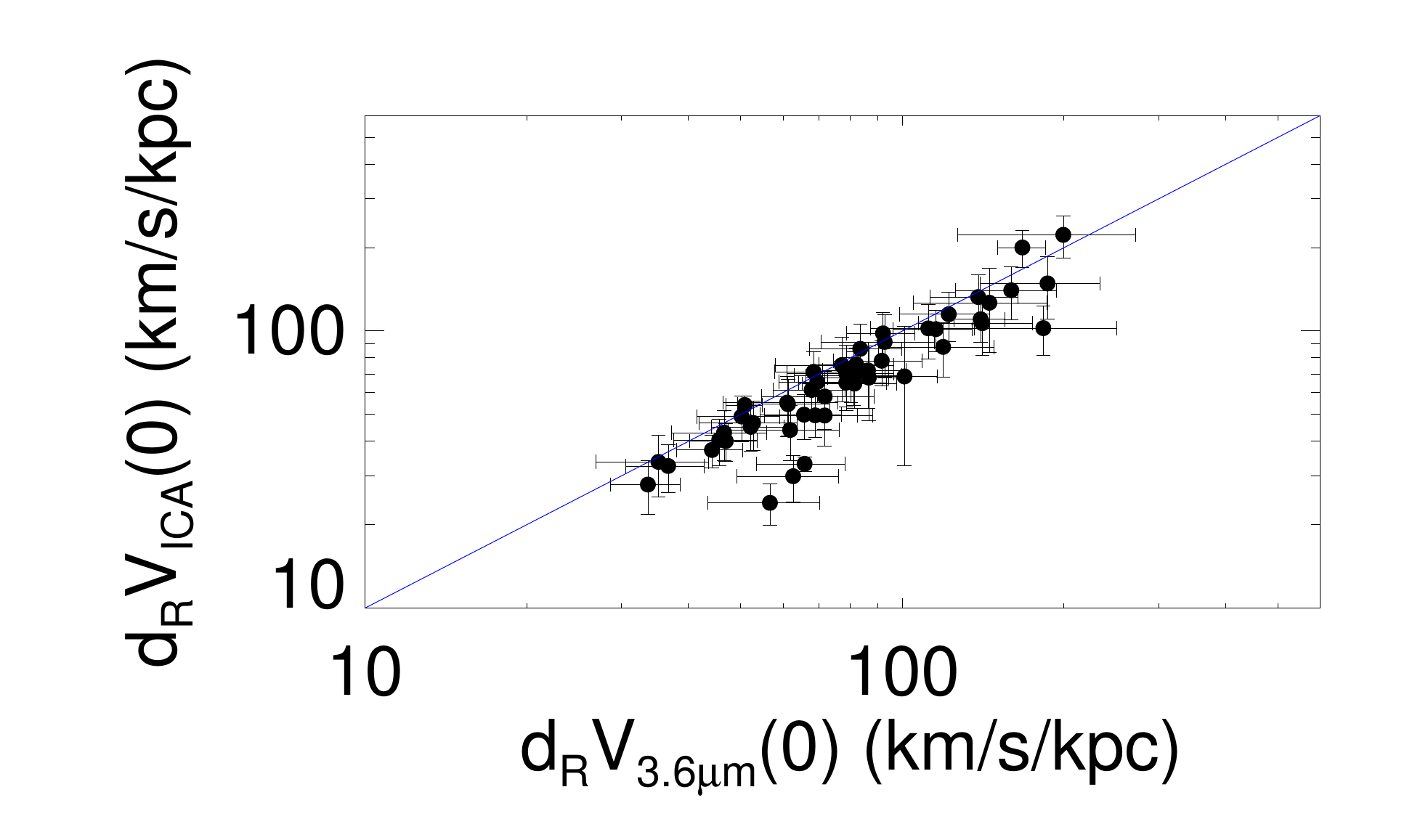}\\
     \end{tabular}
\caption{  
For a sample of S$^{4}$G galaxies, we compare the bar maximum normalized $m=2$ Fourier amplitude (\emph{upper panel}) as measured over 3.6~$\mu$m and ICA-corrected maps.
Likewise, stellar velocities and inner slopes obtained from 3.6~$\mu$m ($M/L=0.53$ from E2012) and mass maps \citep[$M/L=0.6$ from][]{2014ApJ...788..144M} are compared (\emph{central and lower panels}).
Error bars are computed based on boundary disk thicknesses, but the uncertainties coming from the $M/L$ are indeed larger.
A lower mass-to-light ratio at 3.6~$\mu$m, $M/L=0.47$ \citep[][]{2014AJ....148...77M,2015ApJS..219....5Q}, is also tested for the comparison of maximum velocities of the stellar component of the rotation curve
(green empty circles in the central panel). The result from the statistical tests and the number of galaxies in the different subsamples are indicated in Table~\ref{ICA_36_statistics}.
When the uncertainty in $d_{\rm R}V_{3.6\mu \rm m}(0)$ associated with the disk thickness is larger than 75 km s$^{-1}$ kpc$^{-1}$ (upper limit arbitrarily chosen), we exclude the galaxies from the statistics presented in this section.}
\label{parameter_comparisons_mass_a2_rot}
\end{figure}
\begin{figure*}
     %
     \centering  
     %
     \begin{tabular}{l c r}
     \includegraphics[width=0.32\textwidth]{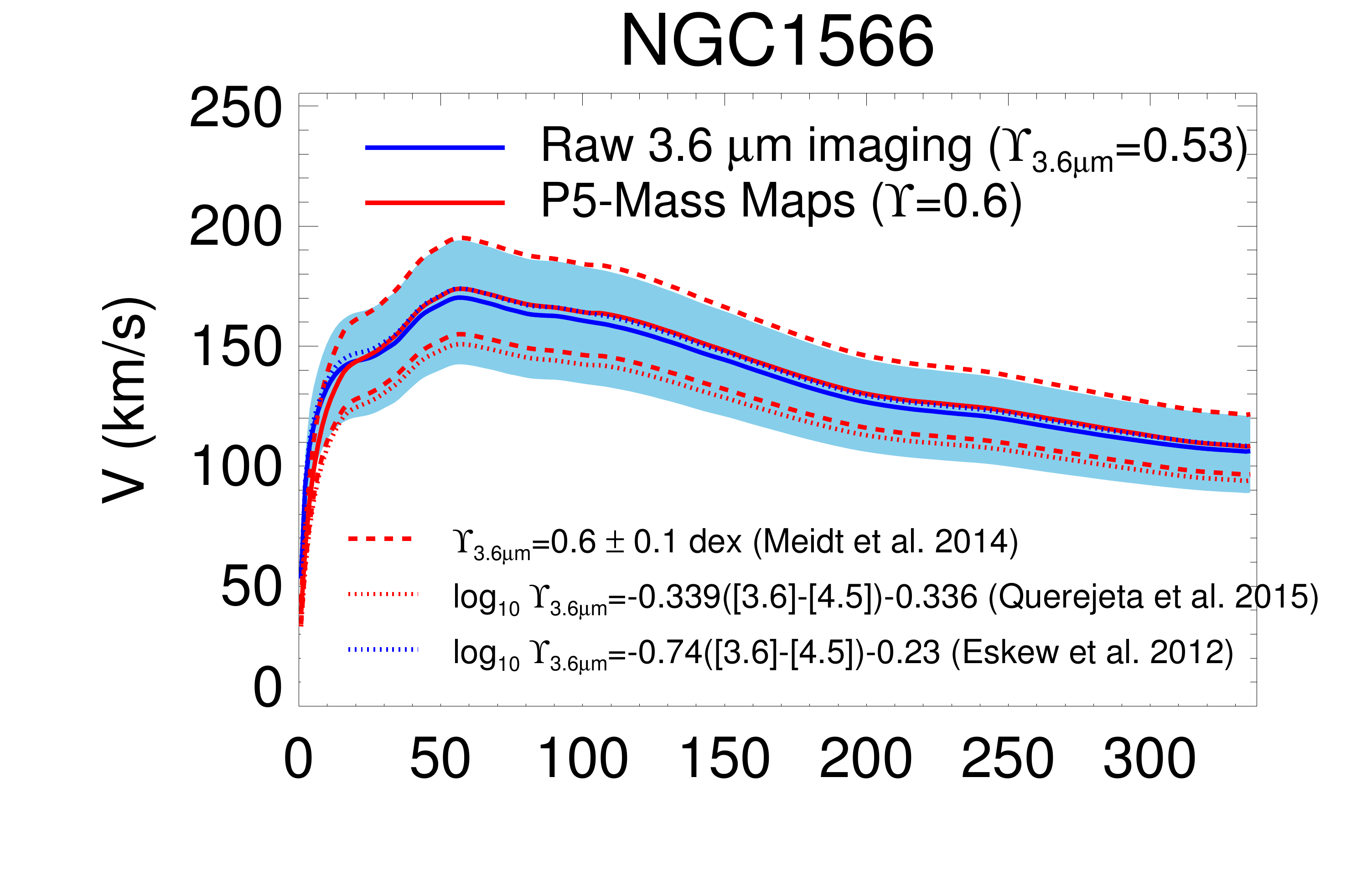}
     \includegraphics[width=0.32\textwidth]{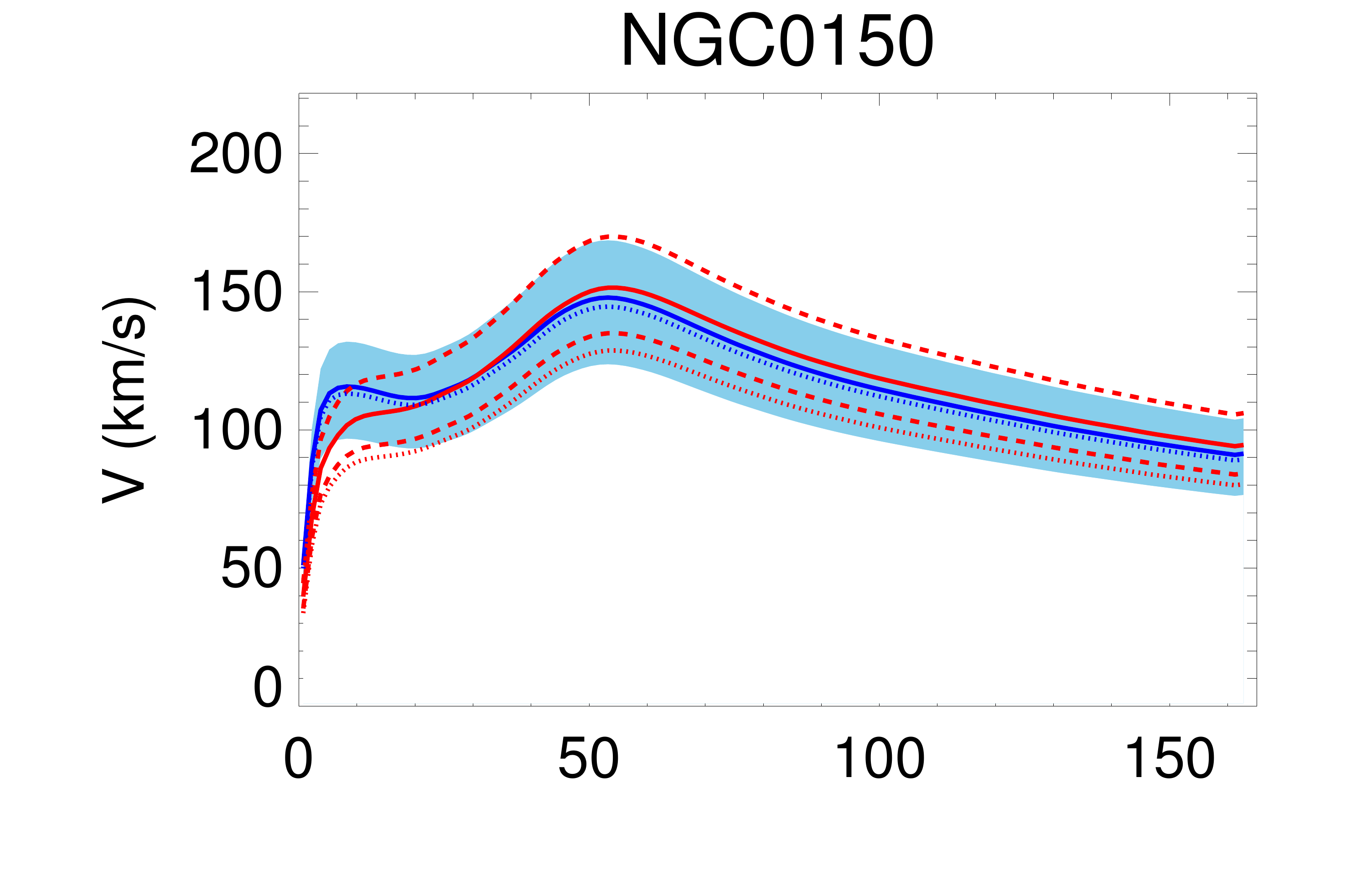}
     \includegraphics[width=0.32\textwidth]{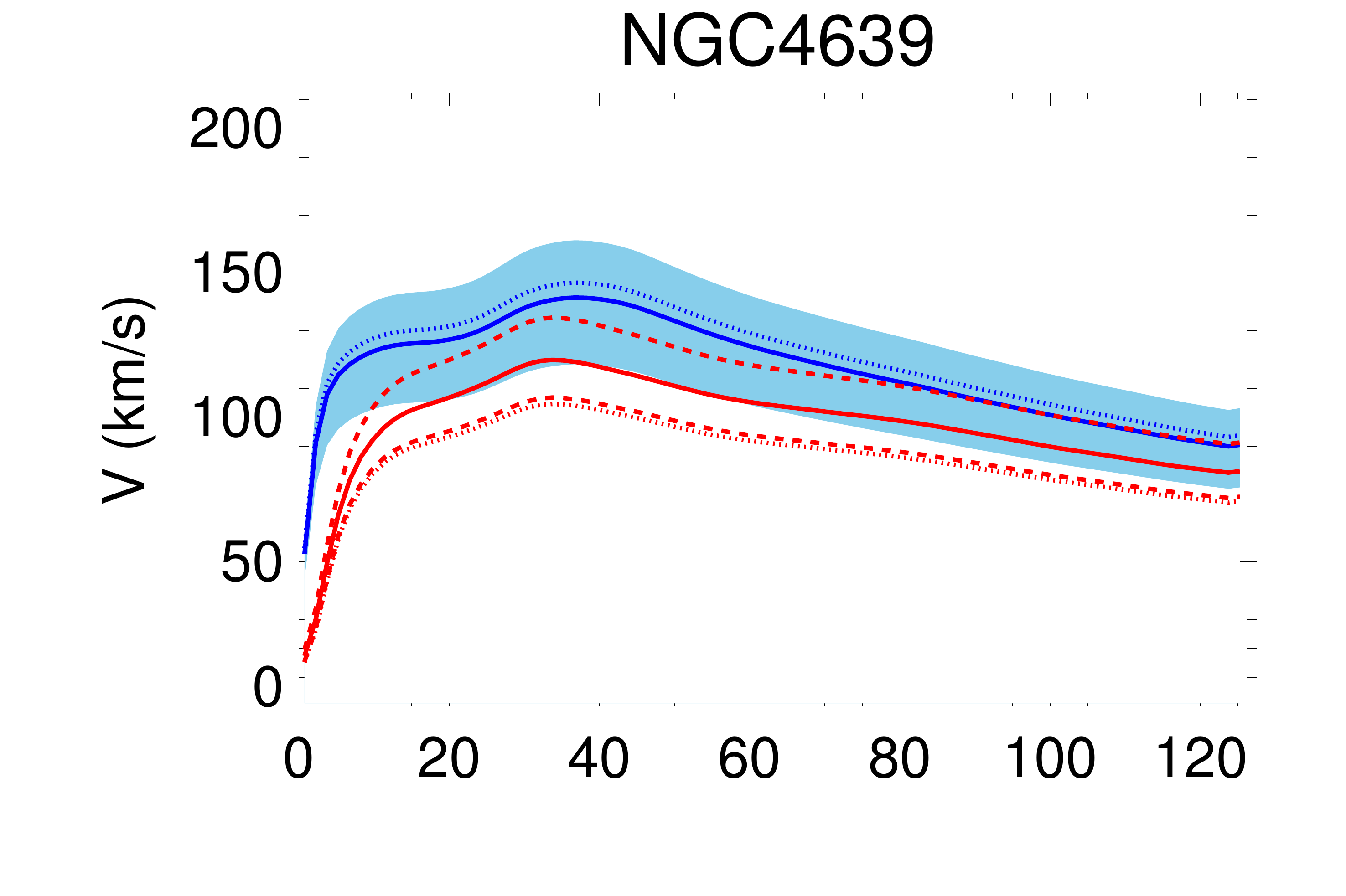}\\[-3ex]
     %
     \includegraphics[width=0.32\textwidth]{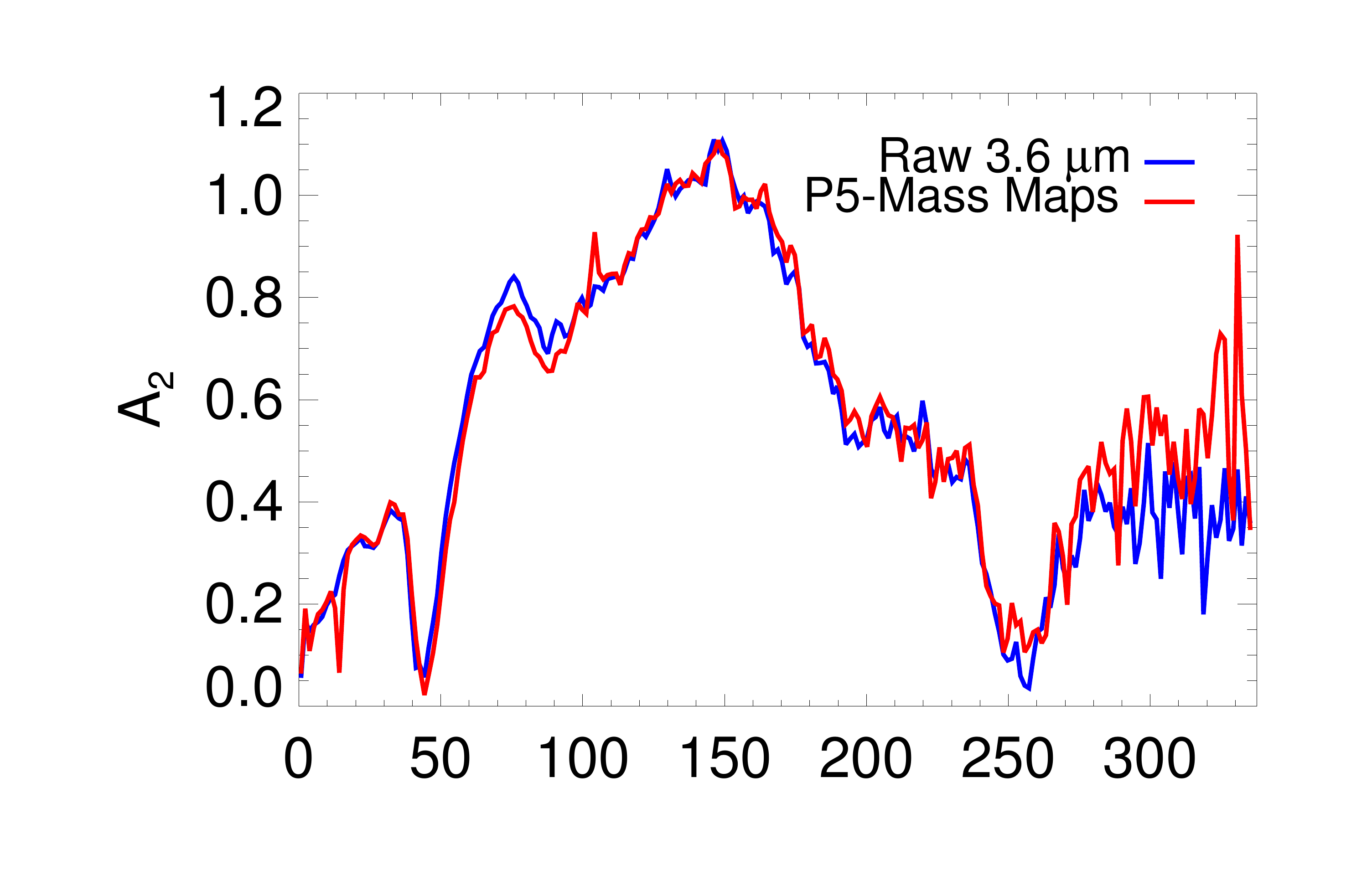}
     \includegraphics[width=0.32\textwidth]{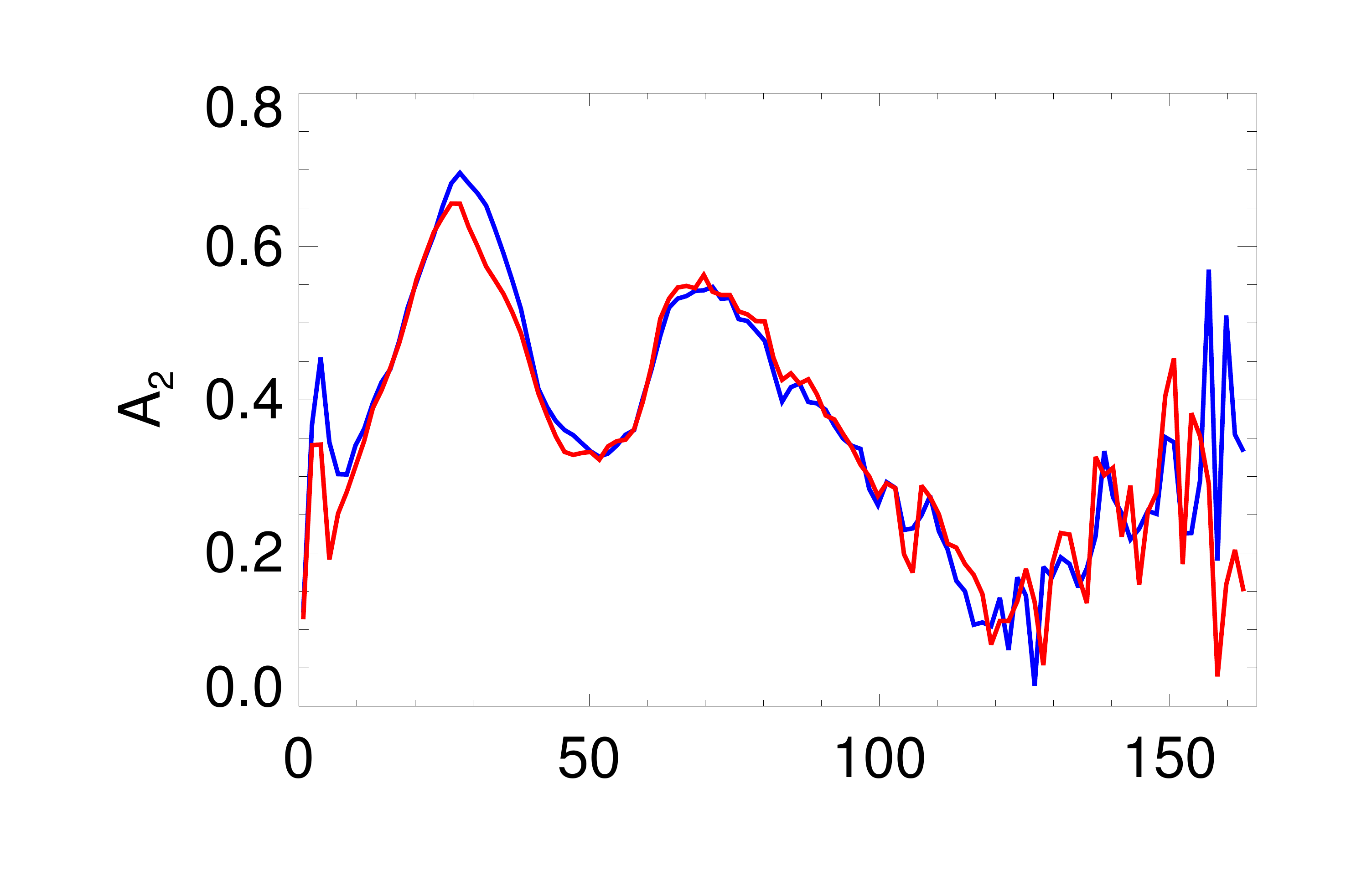}
     \includegraphics[width=0.32\textwidth]{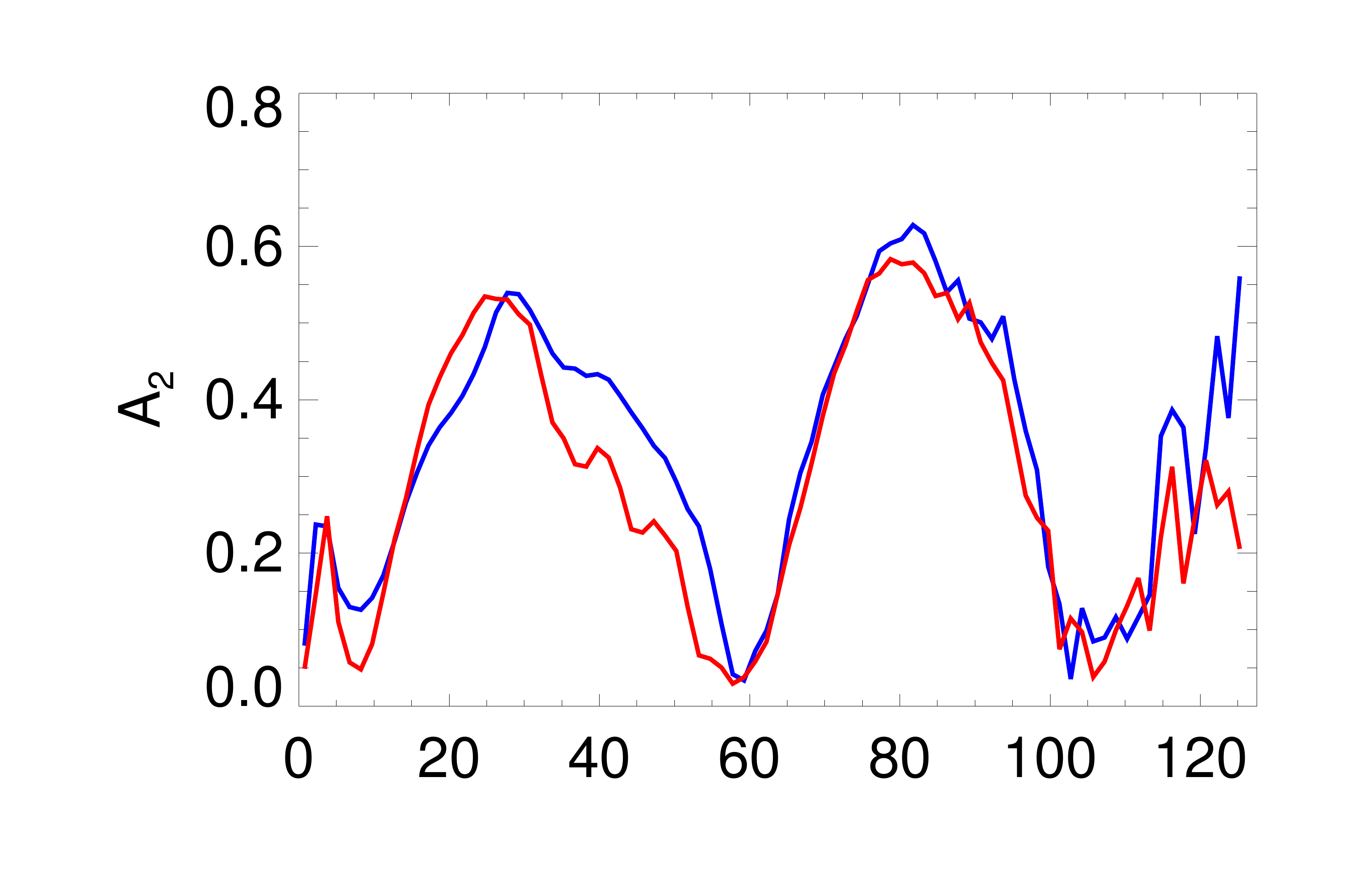}\\[-3ex]
     %
     \includegraphics[width=0.32\textwidth]{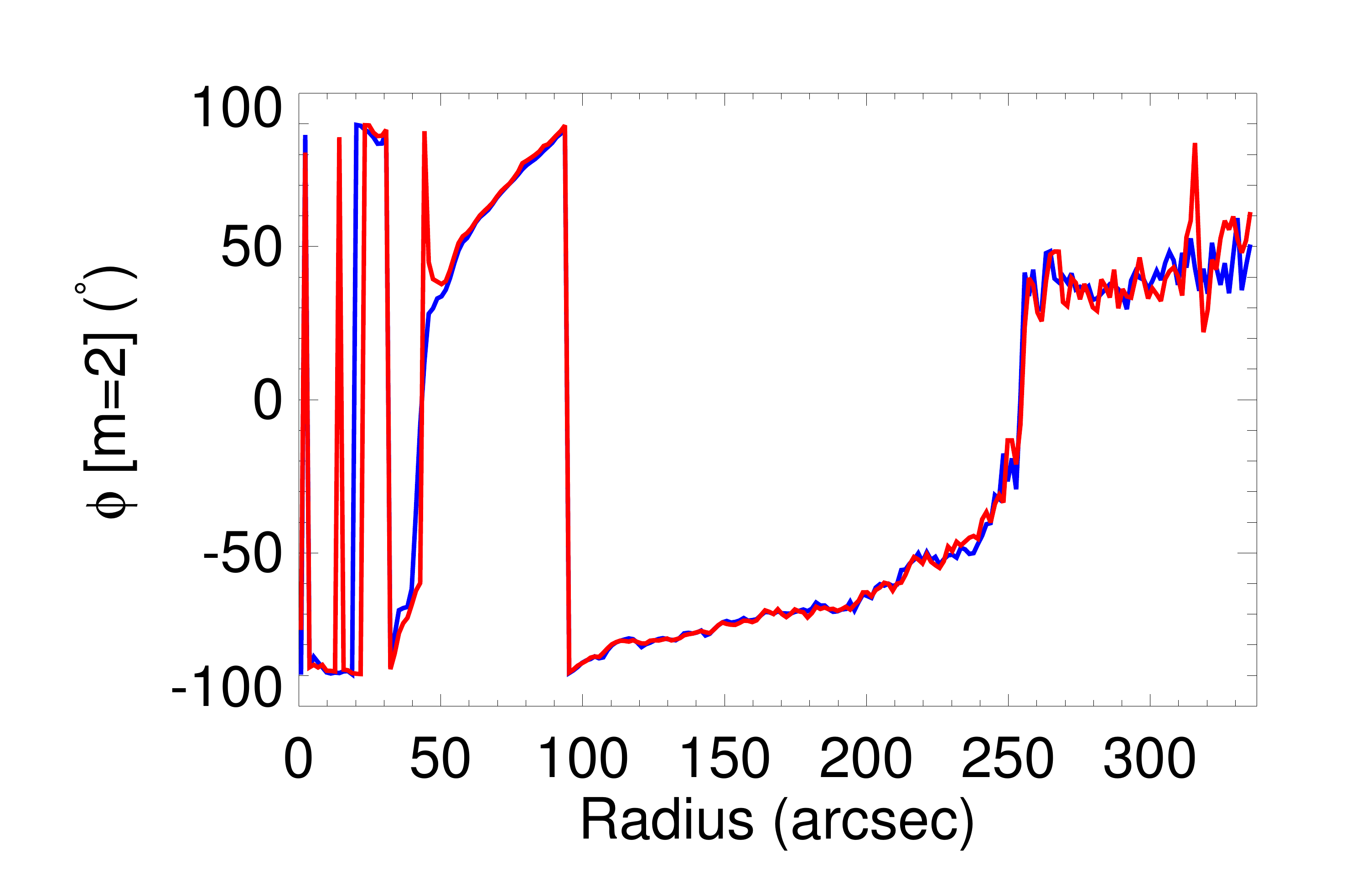}
     \includegraphics[width=0.32\textwidth]{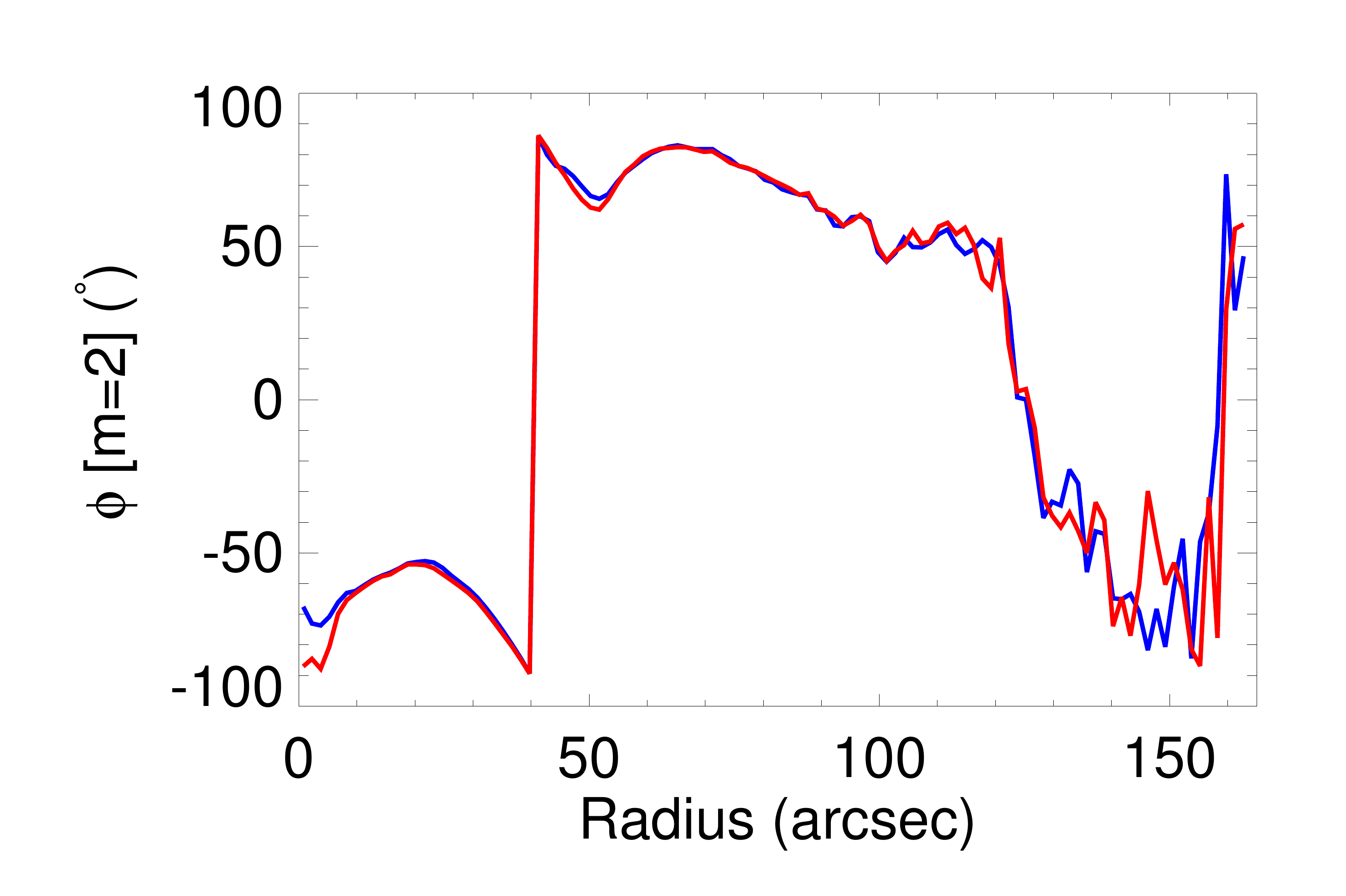}
     \includegraphics[width=0.32\textwidth]{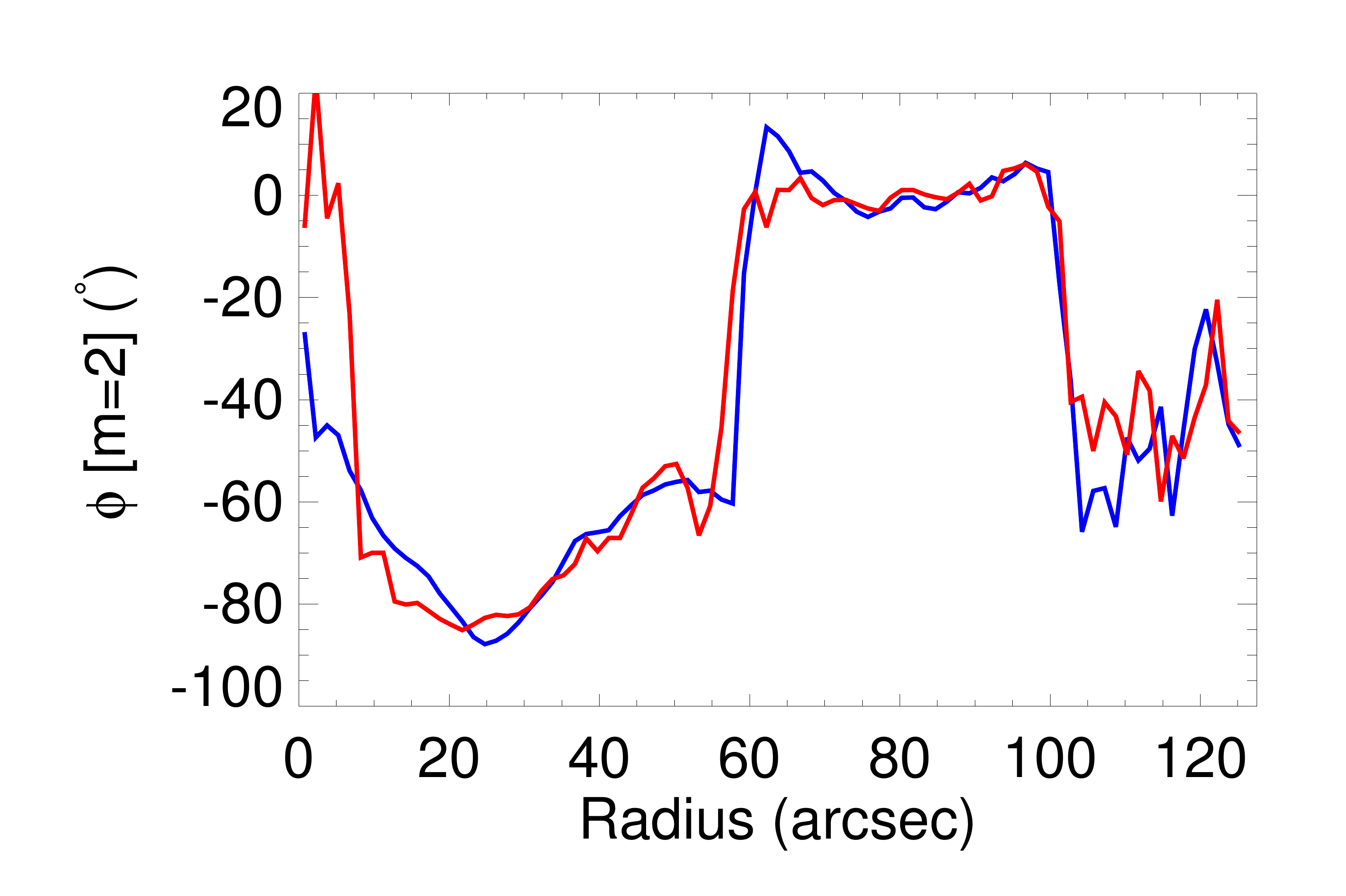}\\
     %
     \end{tabular}
     %
\caption{
Circular velocity (\emph{row 1}), normalized $m=2$ Fourier amplitude (\emph{row 2}) and $m=2$ Fourier phase (\emph{row 3}),
inferred from the 3.6~$\mu$m (blue) and ICA-corrected (red) potentials for the barred galaxies NGC$\,$1566, NGC$\,$0150, and NGC$\,$4639 (\emph{Cols. 1-3}). 
The blue shaded area in the upper panels shows the uncertainty of $V_{3.6\mu \rm m}$ associated to a $30\%$ uncertainty in $\Upsilon_{3.6 \mu \rm m}$, which is assumed to be radially constant.
The red dashed lines correspond to P5-mass maps error bars considering the $0.1$ $dex$ uncertainty deduced in \citet[][]{2012ApJ...744...17M}.
Circular velocities computed as a function of a $[3.6]-[4.5]$ colour-dependent $M/L$ are also shown.
}
\label{rc_massmaps}     
\end{figure*}
\begin{table}
\centering
\begin{tabular}{| c| c| c| c| c|}
\hline
Parameter deviation & Mean &
Median & $|\Delta|$ & $N_{\rm bin}$ \\
\hline
$Q_{\rm b}/Q_{\rm b}^{\rm ICA}$ &  0.93 &  0.89 &  0.03 & 34\\
$A_{\rm 2}^{\rm max}/A_{\rm 2}^{\rm max(ICA)}$ &  1.04 &  1.05 &  0.02 & 34\\
$V_{3.6\mu \rm m}^{\rm max}/V_{\rm ICA}(r_{3.6\mu \rm m}^{\rm max})$ &  1.10 &  1.10 &  9.64 & 71\\
$d_{\rm R}V_{3.6 \mu \rm m}(0)/d_{\rm R}V_{\rm ICA}(0)$ &  1.16 &  1.04 & 11.46 & 51\\
\hline
\end{tabular}
\caption{Comparison of bar strength and circular velocity measurements as obtained from 3.6~$\mu$m images and from ICA corrected mass maps for a subsample of S$^{4}$G galaxies.
The mean and the median of the ratio of the two values are provided, together with the mean of their absolute difference ($|\Delta|$, in the same units as in Figure~\ref{parameter_comparisons_mass_a2_rot}) and the number of galaxies in the bins.}
\label{ICA_36_statistics}
\end{table}

%
%
In Sect.~\ref{contaminants} we quantified the effect of non-stellar contaminants on the bar force,
finding no systematic deviation when comparing $Q_{\rm b}$ values computed from P5-mass maps to those obtained directly from the 3.6~$\mu$m images.
Using the same sample of 34 galaxies with available ICA analysis, 
we repeated a similar test and studied whether non-stellar emission in the bar region can significantly change the Fourier amplitudes of bars ($A_{2}$ and $m=2$ phase). 
The improvement in the obtained $A_{2}^{\rm max}$ values after correcting for the non-stellar contaminants is also small (Fig.~\ref{parameter_comparisons_mass_a2_rot}).
As with $Q_{\rm T}$, noisy $A_{2}$ profiles in 3.6~$\mu$m do not smooth out when using P5-mass maps.

Furthermore, we extended the study to the stellar component of the rotation curves,
deriving the maximum velocity and the inner slope (when possible) from P5 mass maps for a sub-sample of $\sim 70$ galaxies.
In Fig.~\ref{parameter_comparisons_mass_a2_rot} we compare these to the original parameters obtained from raw 3.6 $\mu$m images
(see also Table~\ref{ICA_36_statistics}).
We find that the stellar contributions to the circular velocities are typically overestimated by $\sim 10\%$ when the latter are used.
This error is of the same order as the uncertainty in $V_{3.6 \mu \rm m}$ arising from the disk thickness determination.
In principle, such difference might be expected to arise for spirals, 
given the typically higher star formation rates (thus more emission from hot dust) at the spiral arm(s) region. 
Curiously enough, this systematic overestimation does not seem to be strongly dependent of the stellar mass:
it might naively be expected that the deviation is higher among the faintest systems (where the specific SFR is higher).
In the case of the inner slope of $V_{3.6 \mu \rm m}$, where measurement uncertainties are larger, the mean difference is of the order of $\sim 15\%$.

\citet[][]{2014AJ....148...77M} recently pointed out a lower mass-to-light ratio in 3.6~$\mu$m $\Upsilon_{\rm 3.6\mu m}=0.47$ (with a scatter of 0.1 dex),
which is also compatible with \citet[][]{2015ApJS..219....5Q} and $\sim 15\%$ lower than the $M/L$ assumed in this work. 
In the central panel of Fig.~\ref{parameter_comparisons_mass_a2_rot} we also reassess the comparison between the maximum velocity of the stellar component of the rotation curve derived from contaminant-free maps and from 3.6 $\mu$m after
assigning $\Upsilon_{\rm 3.6\mu m}=0.47$ to the near-IR images, finding a lower mean difference of $\sim 3\%$.

In Fig.~\ref{rc_massmaps} we present the circular velocity, normalized $m=2$ Fourier amplitude, and $m=2$ Fourier phase
obtained from the 3.6 $\mu$m and ICA-corrected potentials for a subset of three barred galaxies.
The force profiles of NGC$\,$1566 and NGC$\,$0150 were shown in Sect.~\ref{contaminants}.
For these galaxies, none of the measurements vary substantially when using P5 products.
The Fourier profiles and circular velocity curve of
NGC\ 4639 are also shown in the same figure.
We have chosen this galaxy in the interest of presenting a case with a noticeable difference in the (stellar) circular velocity curve 
depending on whether one takes 3.6~$\mu$m or the ICA-corrected mass map for the gravitational potential calculation 
(while the Fourier decomposition remains very similar).
Still, the difference in the stellar contribution to the rotation curves between the mass maps and the raw 3.6~$\mu$m images
is of the same order as the uncertainty on $V_{3.6 \mu \rm m}$ that is due to the $M/L$ ratio.
On the other hand, we have assumed a constant $M/L$ for all Hubble types.
However, $M/L$ can also be expressed as a function of the $[3.6]-[4.5]$ colour \citep[][]{2012AJ....143..139E,2015ApJS..219....5Q}.
In Fig.~\ref{rc_massmaps} we also indicate the resulting rotation curves if such a $\Upsilon_{3.6 \mu \rm m}=f([3.6]-[4.5])$ were taken into account, 
obtaining a deviation that is smaller than the error bars.
We also checked that the trends and mean differences discussed in Fig.~\ref{parameter_comparisons_mass_a2_rot} did not change when considering 
a colour-dependent M/L.
\section{Universal rotation curve and the halo correction on the force profiles}\label{URC_appendix}
\begin{figure}
   \centering  
   \begin{tabular}{l r}
   \includegraphics[width=0.5\textwidth]{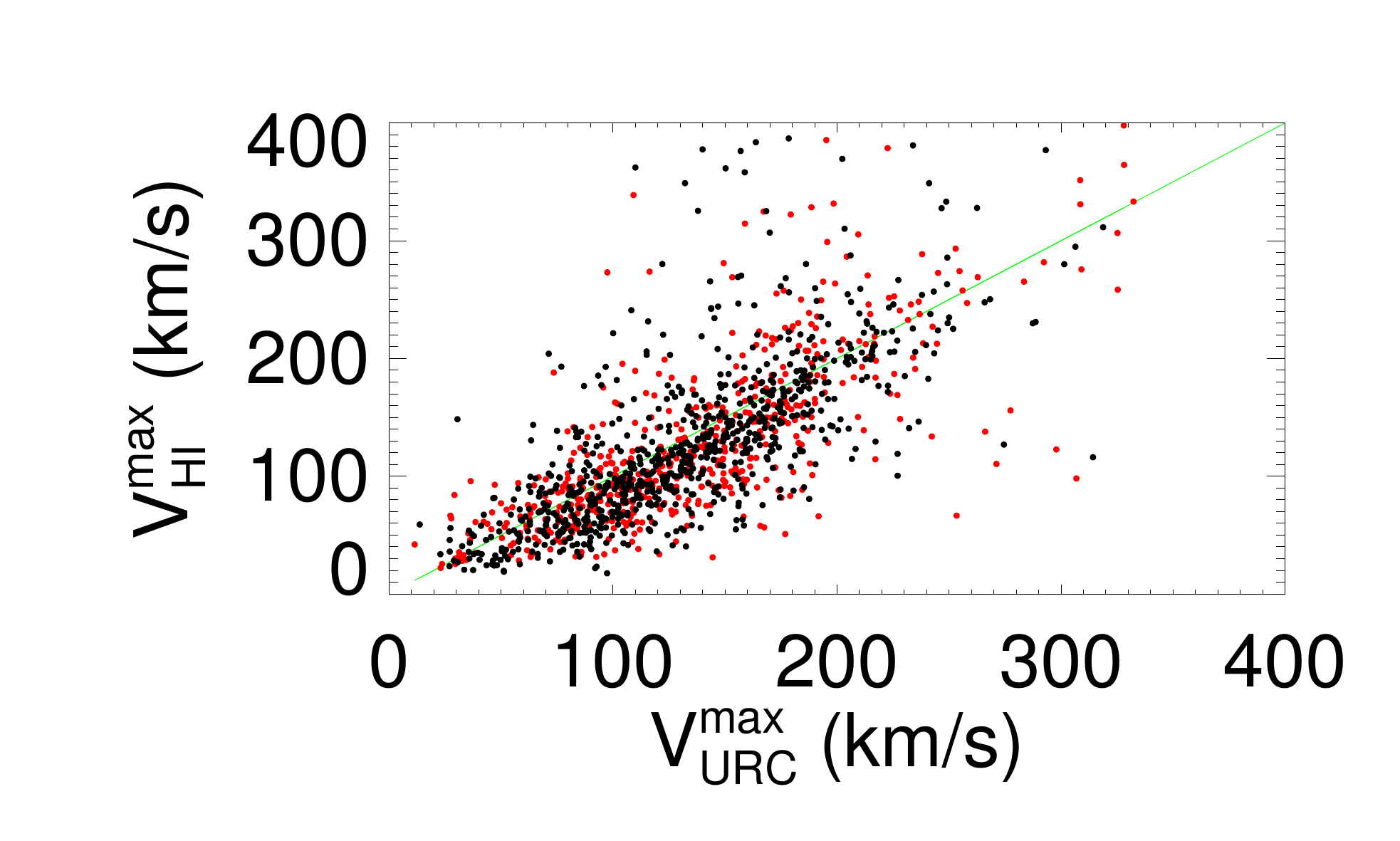}\\[-5ex]
   \includegraphics[width=0.5\textwidth]{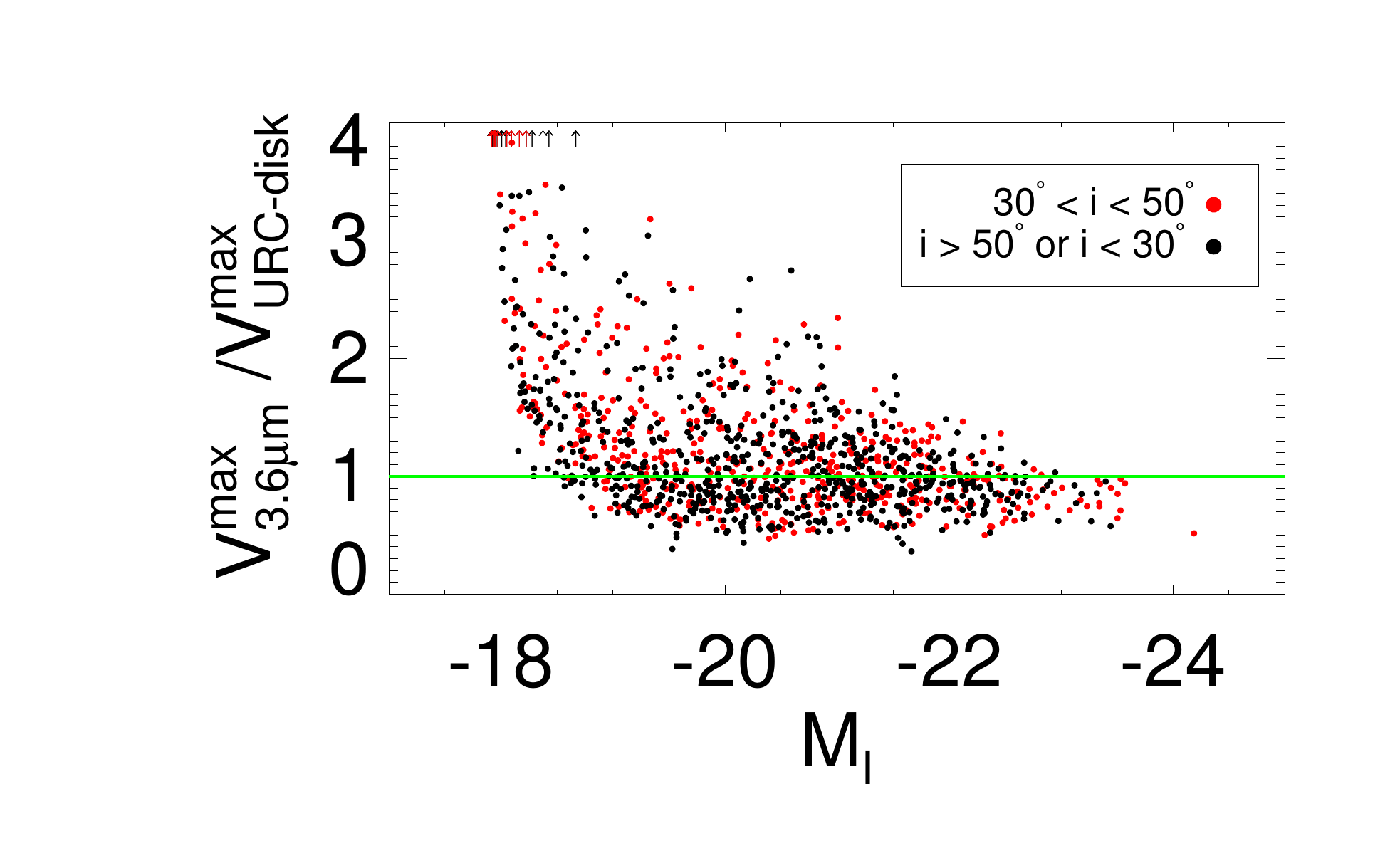}
   \end{tabular}
   \caption{
\emph{First row:}
Maximum H{\sc\,i} circular velocity from the literature (corrected for inclination) vs. maximum of the universal rotation curve.
\emph{Second row:}
Ratio of the maximum of the circular velocity disk component as derived from the URC and from the 3.6~$\mu$m potential vs. absolute corrected $I$-band magnitude.
}
\label{fig_urc_comparison}
\end{figure}
Based on the analysis of the rotation curves of $\sim1100$ spiral galaxies from \citet[][]{1995ApJS...99..501P} and \citet[][]{1992ApJS...81..413M},
PSS concluded that the shape and amplitude of an RCs can be described by a global parameter of the galaxy, such as the luminosity,
and that every RC had a univocal decomposition into a thin exponential disk and a halo component.
Counterexamples to this so-called universal rotation curve (URC), which was first claimed by \citet[][]{1980ApJ...238..471R},
can be found in \citet[][]{1998astro.ph.12013B} or \citet[][]{1997PhDT........13V}, but in spite of them, the URC remains useful for statistical purposes \citep[][]{2000FCPh...21....1B}.
Here we check how well i) the disk velocity amplitudes predicted by the URC fit with those derived from S$^4$G images, and whether
ii) the maximum velocity amplitude agrees with the $V_{\rm HI}^{\rm max}$ from literature.

The URC predicts the rotation velocities at any radius as a function of the $I$-band absolute magnitude \citep[][]{1997ASPC..117..198H}:
\begin{equation}
\begin{split}
V_{\rm URC}(x,M_{\rm I})&=204\cdot10^{-0.12(M_{\rm I}+22)} \cdot \\
\quad  & \cdot \sqrt{V^{2}_{\rm URC-disk}(x,M_{\rm I})+V^{2}_{\rm URC-halo}(x,M_{\rm I})},
\end{split}
\label{urc_tot}
\end{equation}
\begin{equation}
\label{urc_disk}
V^{2}_{\rm URC-disk}(x,M_{\rm I})=\frac{\big(0.72-0.176(M_{\rm I}+22)\big)1.97x^{1.22}}{(x^{2}+0.78^{2})^{1.43}},
\end{equation}
\begin{equation}
\label{urc_halo}
\begin{split}
V^{2}_{\rm URC-halo}(x,M_{\rm I})&=x^{2}[0.28+0.176(M_{\rm I}+22)]\cdot \\
\quad   & \cdot \frac{1+2.25\cdot10^{-0.16(M_{\rm I}+22)}}{x^{2}+2.25\cdot10^{-0.16(M_{\rm I}+22)}},
\end{split}
\end{equation}
with $x=r/R_{\rm opt}$.
We note that Eq.~\ref{urc_disk} approximates the rotation curve of an exponential disk,
while Eq.~\ref{urc_halo} corresponds to the isothermal halo model (see also Eq.~\ref{urc_iso}).

In BLS2004 the halo was modelled based on the URC prediction for the linear dark-to-luminous matter ratio derived by PSS
as a function of the galaxy luminosity in the $B$ band ($L/L_{\ast}$, with $L_{\ast}=10^{10.4}L_{\sun}$ and $L$ corrected for extinction):
\begin{equation}
M_{\rm h}/M_{\ast}|_{\rm URC}(x,L/L_{\ast})=0.4(L/L_{\ast})^{-0.9}x^{3}\frac{1+1.5^2(L/L_{\ast})^{0.4}}{x^{2}+1.5^{2}(L/L_{\ast})^{0.4}}.
\label{frac_URC}
.\end{equation}
From the visible bulge plus disk rotation curve ($V_{\rm d}$) evaluated at the optical radius, BLS2004 determined the dark matter halo rotation amplitude by setting 
$M_{\rm h}/M_{\ast}|_{\rm URC}(x)=V_{\rm URC-halo}(x)^{2}/V_{\rm d}(x)^{2}$ at $x=1$.

Taking advantage of the availability of H{\sc\,i} data for most of the galaxies in the S$^{4}$G sample,
we test how well the predicted velocities (based on the URC) match the observed circular rotation (Fig.~\ref{fig_urc_comparison}).
In the statistical sense, the URC model estimates the RC maximum velocities fairly well.
We also compare the URC disk component with our measured $V_{3.6 \rm \mu m}$ at their peak values, observing disagreement among the faintest galaxies.

In Fig.~\ref{all20_rotcur_decomposition_things} our rotation curve decomposition models are compared with
 $H_{\alpha}$ rotation curves from \citet{2006MNRAS.367..469D} and \citet{2008MNRAS.385..553D}, for a sub-sample of SINGS galaxies \citep[][]{2003PASP..115..928K}.
Likewise, in Fig.~\ref{all20_rotcur_decomposition_block} we study the observed H{\sc\,i} velocity curves from \citet[][]{2008AJ....136.2648D},
taken from THINGS \citep[][]{2008AJ....136.2563W}. 
Altogether, we have ten galaxies (three of which are barred) to compare to our ad hoc RC models. 
In almost all the galaxies, the modelled curve seems to fit the observations fairly well, the outliers being NGC$\,$2841 
(with a rotation curve of similar shape as $V_{3.6 \mu \rm m}$) and the inner part of the rotation curve of the barred galaxy NGC$\,$2903.
For all the systems, we obtain a $\sim 15-20\%$ mean dispersion of our models with respect to the observed rotation curves across the optical disk, 
which could be due to the gas component lacking in our decompositions.

Additionally, in the lower panel of Fig.~\ref{halo-galaxy} we tested the robustness of the halo model and the correction of the forces by
assessing the effect of a measurement error on the used input parameters.
We assumed a 20$\%$ overestimation error on the mass-to-light ratio, the distance to the galaxy, and the observed velocity;
and we tested the effect of reducing by 1 mag the $I$-band total luminosity.
We also checked the effect of matching the observed maximum velocity at $2.2h_{\rm R}$
\citep[radius of the maximum velocity of an exponential disk;][]{1970ApJ...160..811F} instead of at $R_{\rm opt}$.
A possible inaccuracy assuming that the maximum circular rotation takes place at the optical radius was also
considered when testing the overestimation of the observed velocity 
(e.g. the difference between either the absolute maximum or the asymptotic limit of $V_{\rm URC}$ and its value at $R_{\rm opt}$ is always lower than 20$\%$).
None of these variations result in a significant change in the $Q_{\rm b}^{\rm halo-corr}$ value (deviation $<10 \%$).
Moreover, we searched for a possible relation between the disk inclination and $M_{\rm h}/M_{\ast}$,
finding no dependence and thus excluding a possible bias due to observation uncertainties
for highly inclined and face-on galaxies that might affect our halo corrections.
We discarded any major underestimation of the halo amplitude (and thus of the halo correction) that would be due to the assumption of a radially constant mass-to-light
ratio \citep[e.g.][]{2010A&A...521A..82P} because this was one of the factors that determined the nominal $30 \%$ uncertainty
\citep[or even higher, according to][]{2014ApJ...788..144M} on the mass-to-light ratio estimation by E2012, 
and radial colour gradients are not that pronounced in our wavelengths.
\onecolumn
\begin{figure}
\centering
\begin{tabular}{c c c c}
\includegraphics[width=0.24\textwidth]{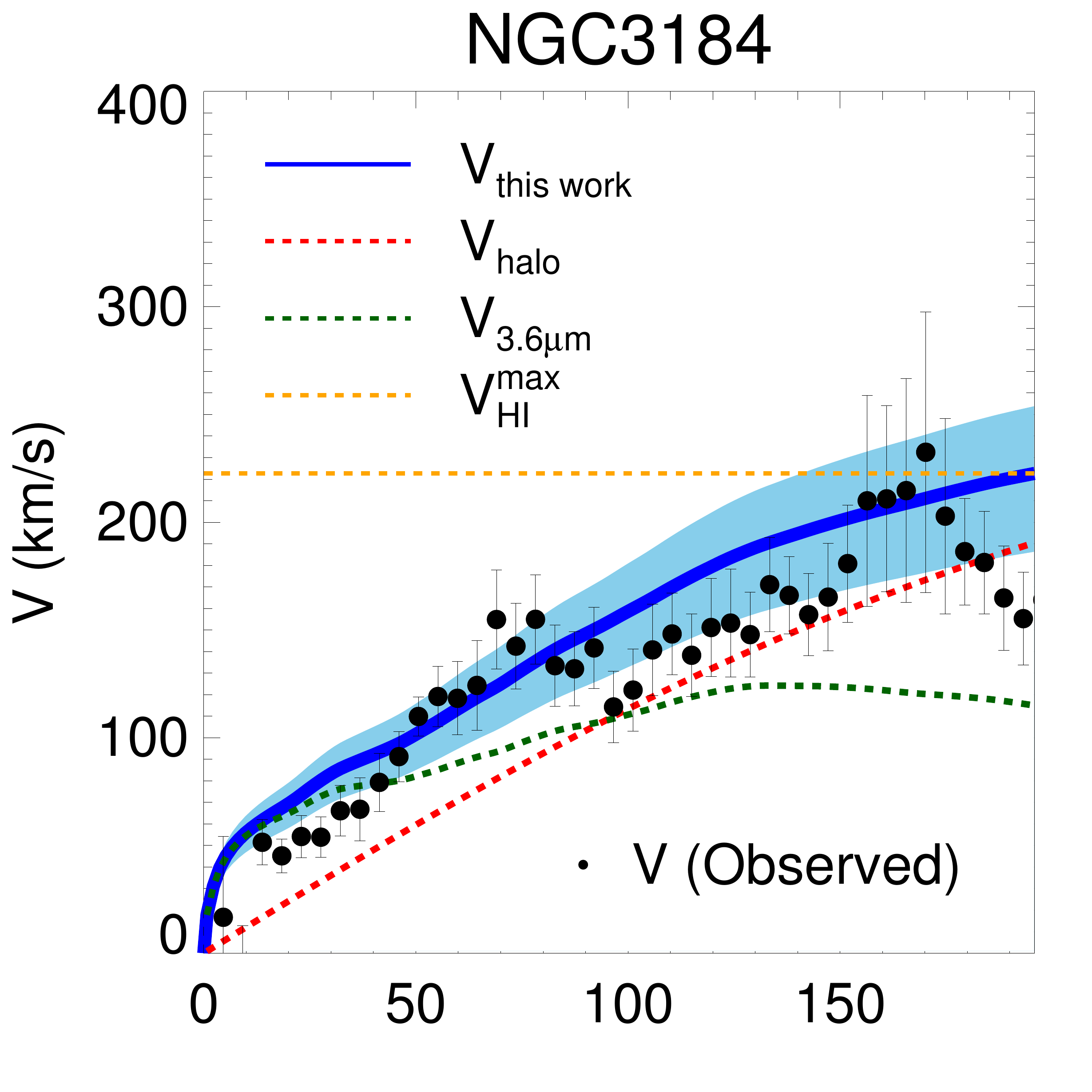}
\includegraphics[width=0.24\textwidth]{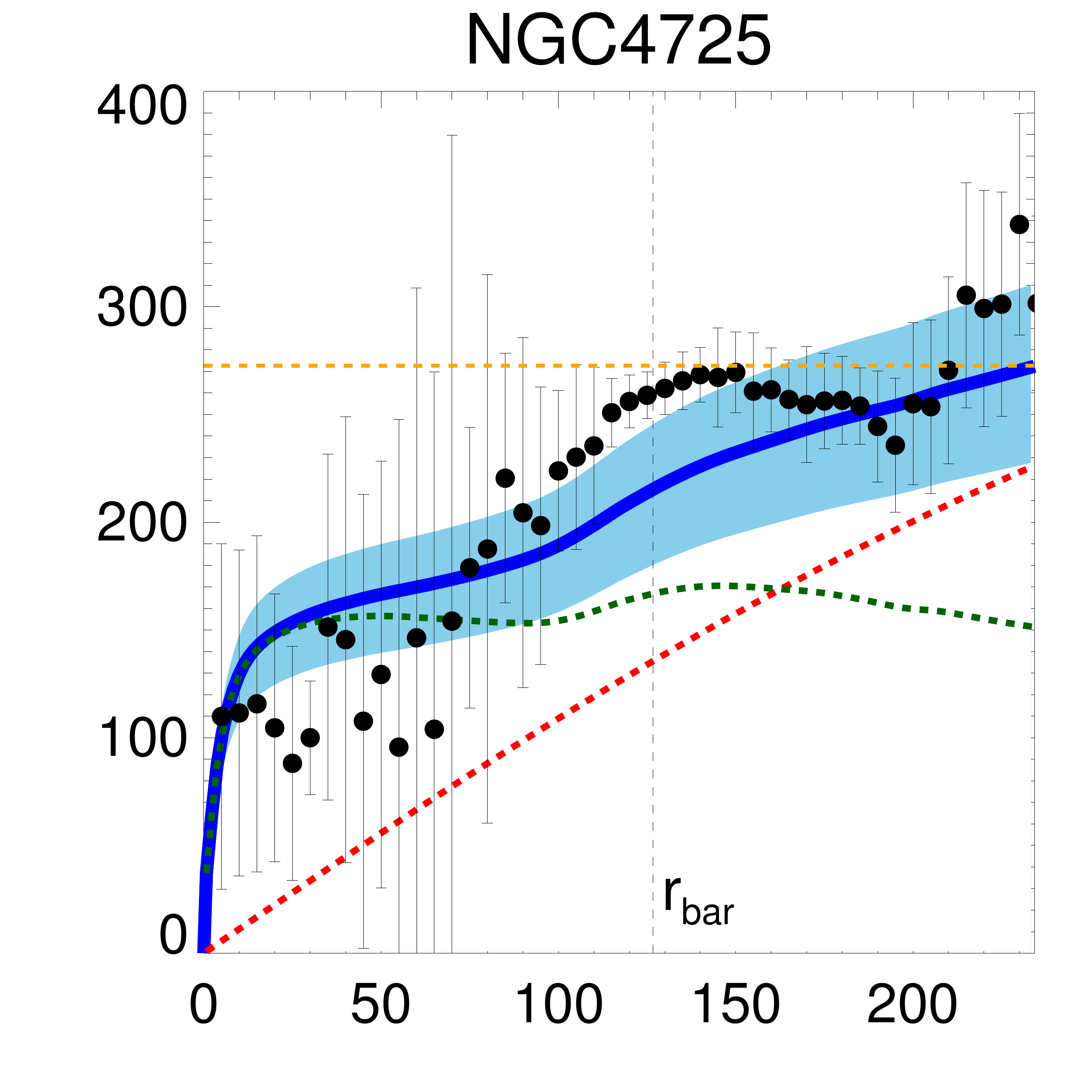}\\
\includegraphics[width=0.24\textwidth]{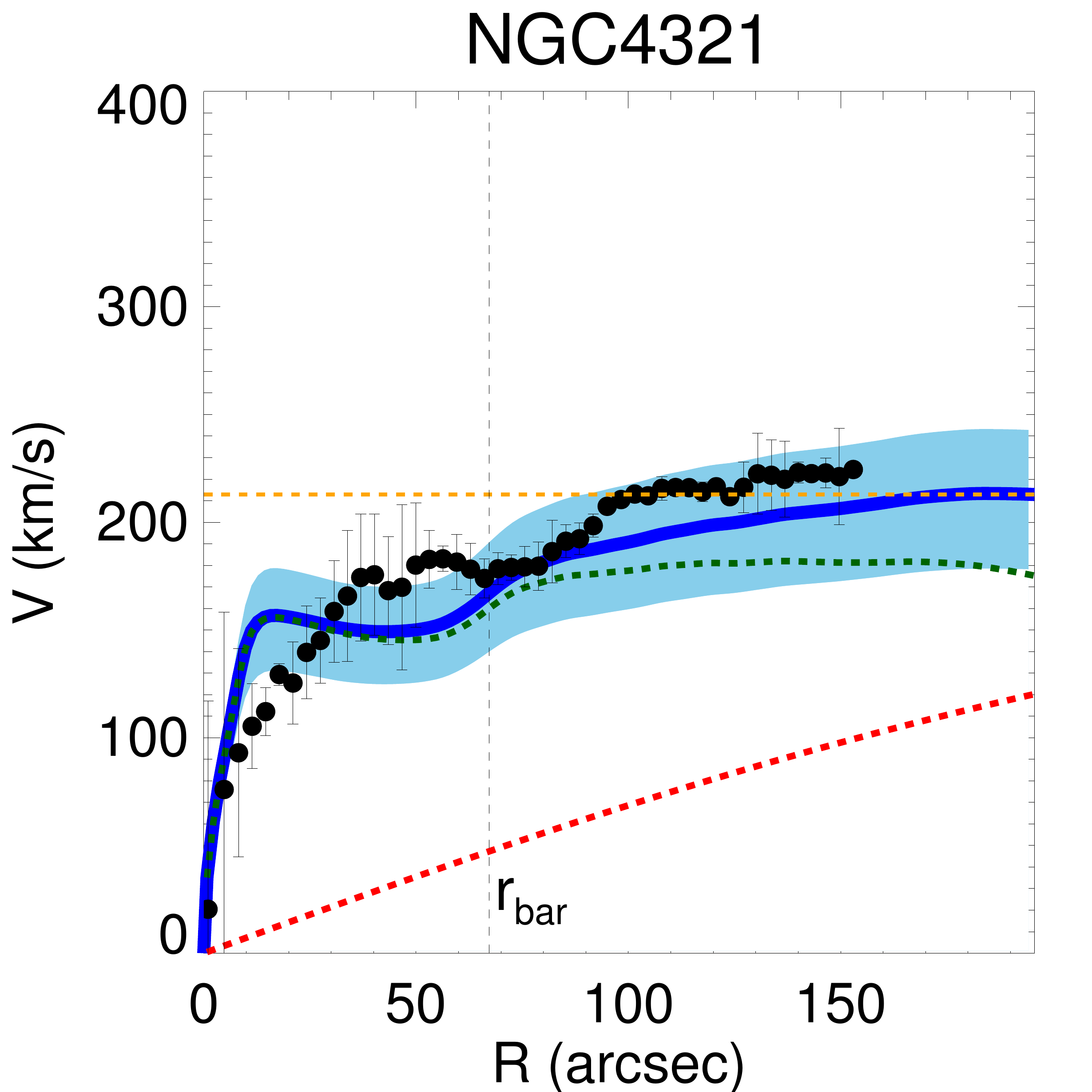}
\includegraphics[width=0.24\textwidth]{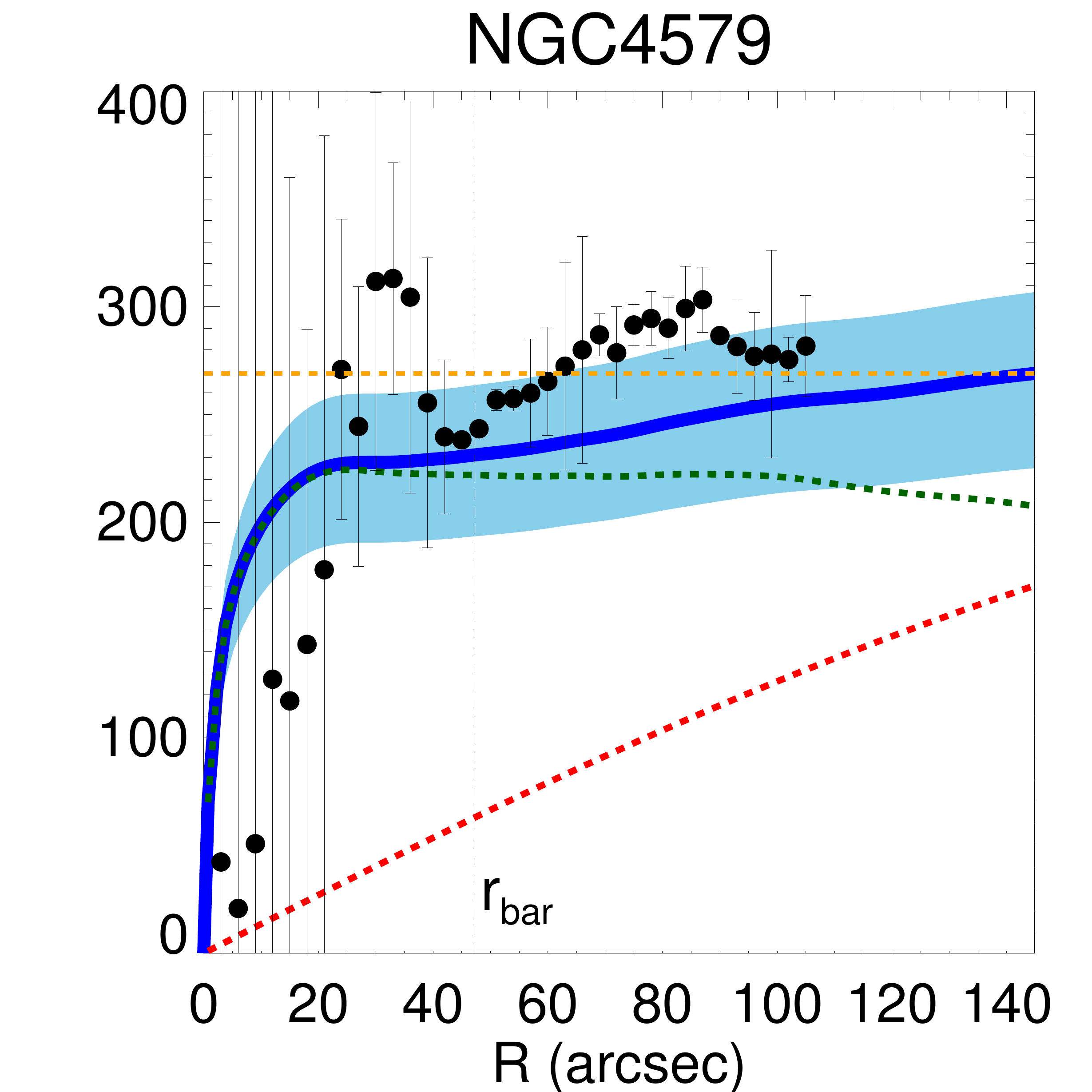}
\end{tabular}
\caption{Rotation curve decomposition models for a sub-sample of 4 galaxies from the SINGS sample (Kennicutt et al. 2003) with disk inclinations lower than 65$^{\circ}$ and the $H{\alpha}$ rotation curves from Daigle et al. (2006) and Dicare et al. (2008).
The RCs have been corrected for inclination using P4 orientations.
The legend in the first pannel shows the meaning of the different lines and points. For every pannel, the x-axis  indicates the galactocentric radius in units of arcsec (limited to the optical radius, $\sim 3.2 h_{\rm R}$) and the y-axis corresponds to the rotation velocity in units of km/s.
The shaded area shows the uncertainty in our rotation curve models due to a 30$\%$ uncertainty in the M/L ratio.
Galaxy RCs are sorted in terms of their mass (increasing from left to right, top to bottom).}
\label{all20_rotcur_decomposition_things}
\end{figure}

\begin{figure}
\centering
\begin{tabular}{c c c c}
\includegraphics[width=0.24\textwidth]{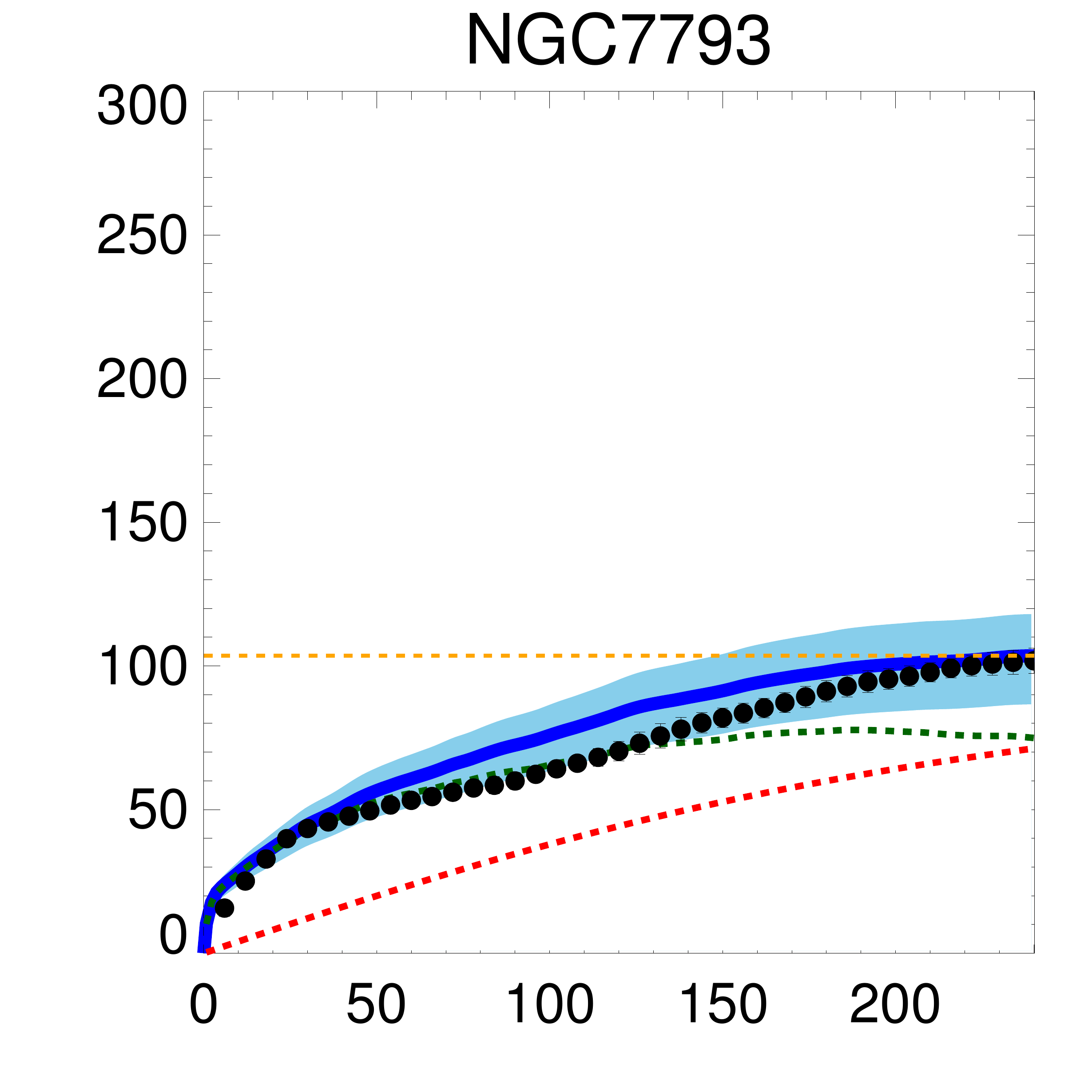}
\includegraphics[width=0.24\textwidth]{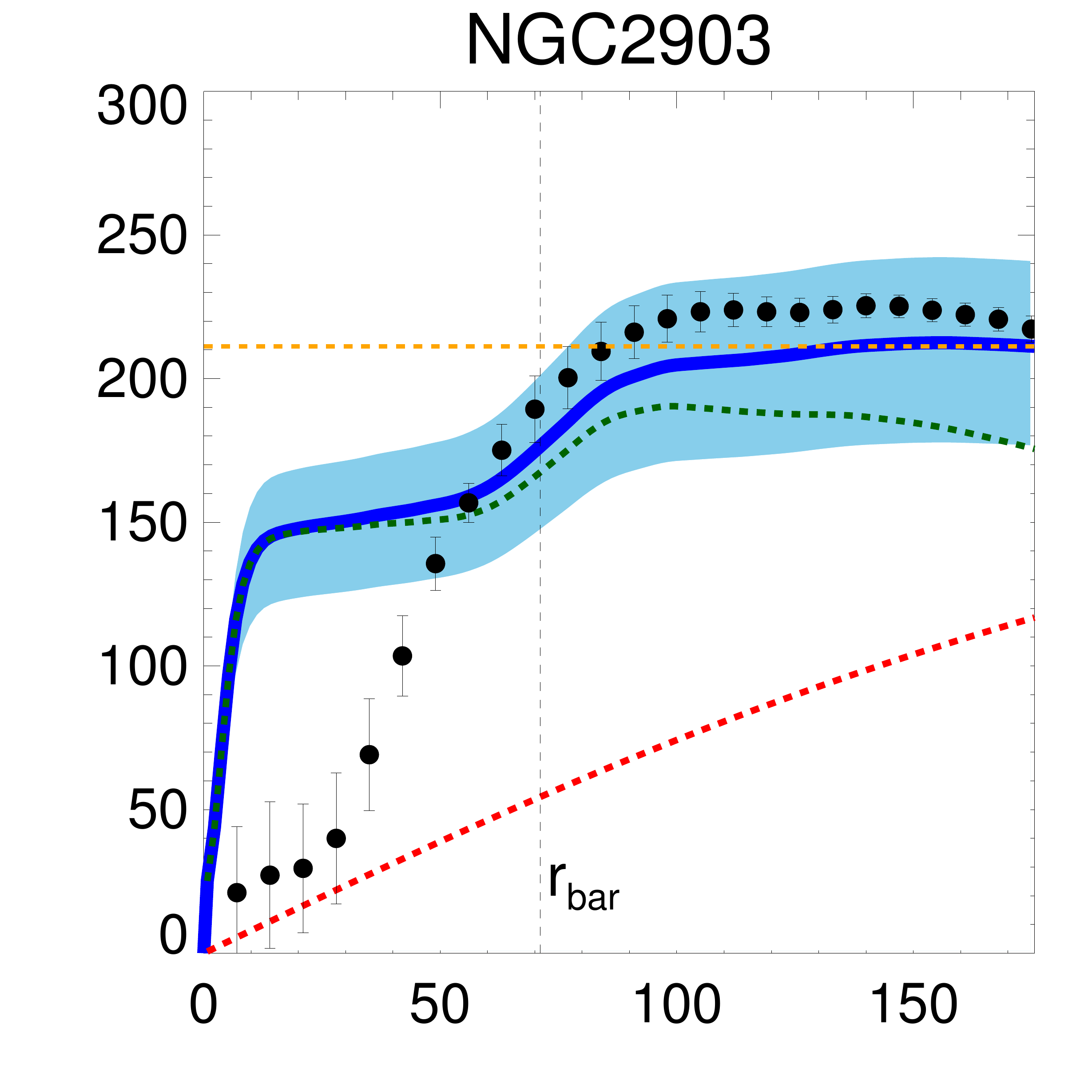}
\includegraphics[width=0.24\textwidth]{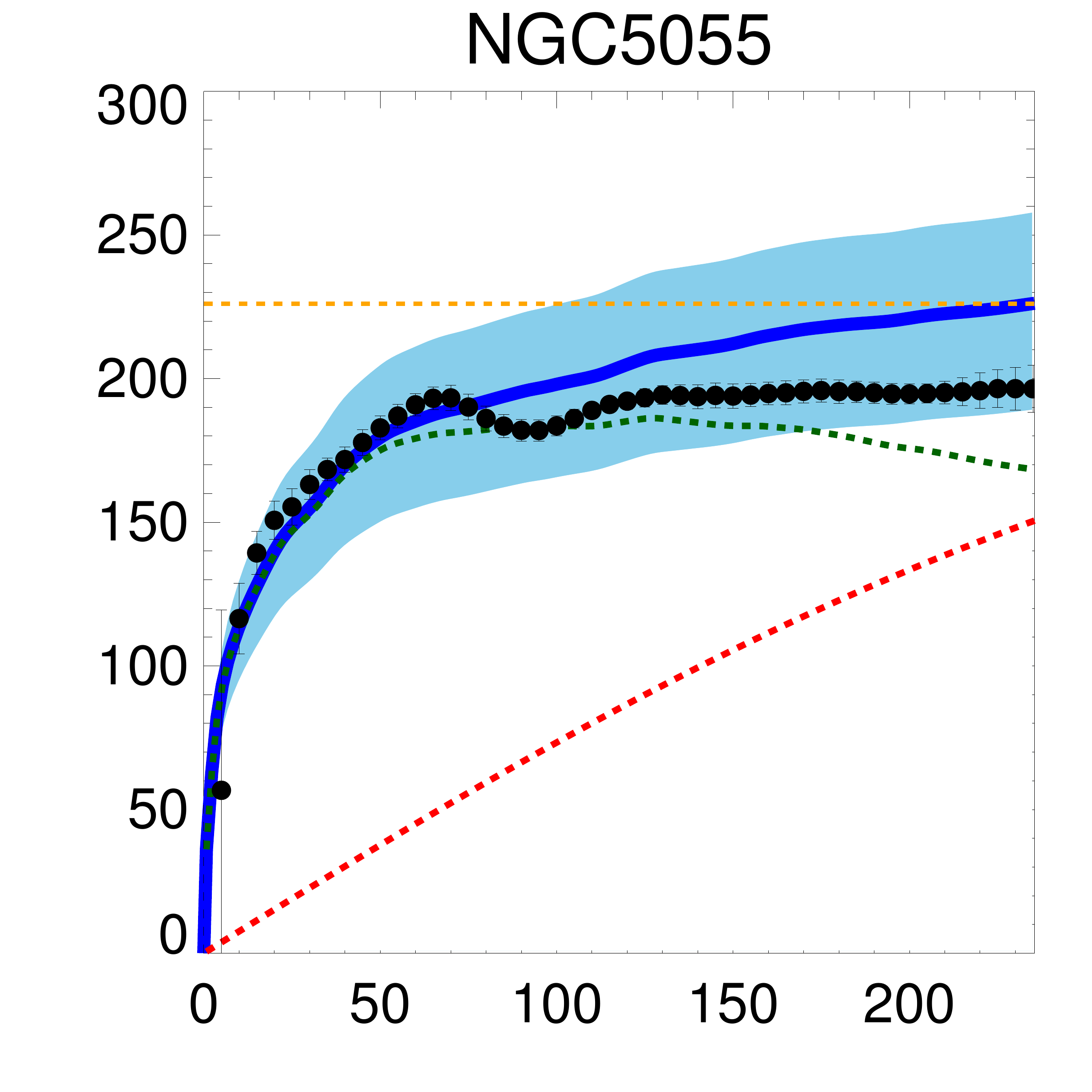}
\includegraphics[width=0.24\textwidth]{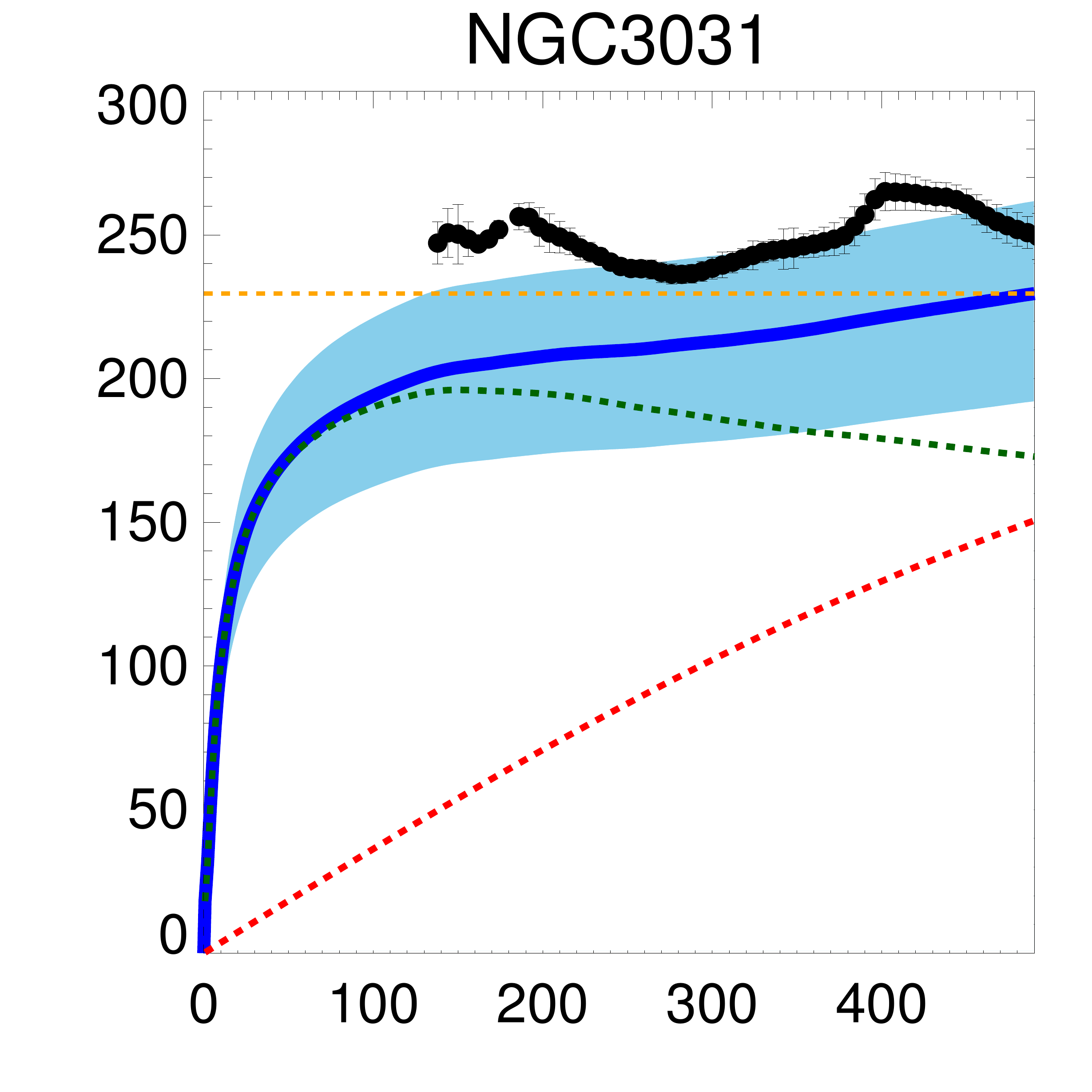}\\
\includegraphics[width=0.24\textwidth]{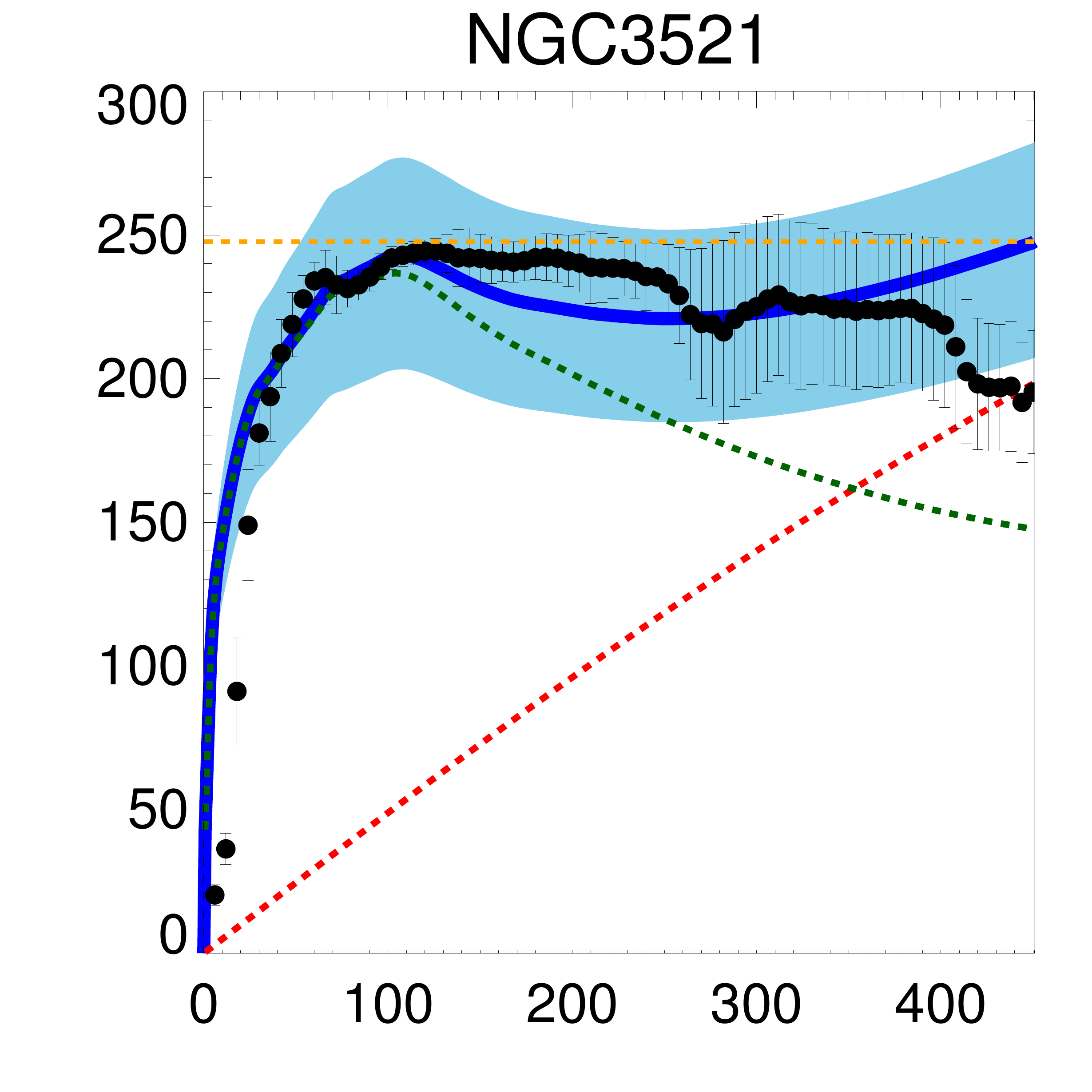}
\includegraphics[width=0.24\textwidth]{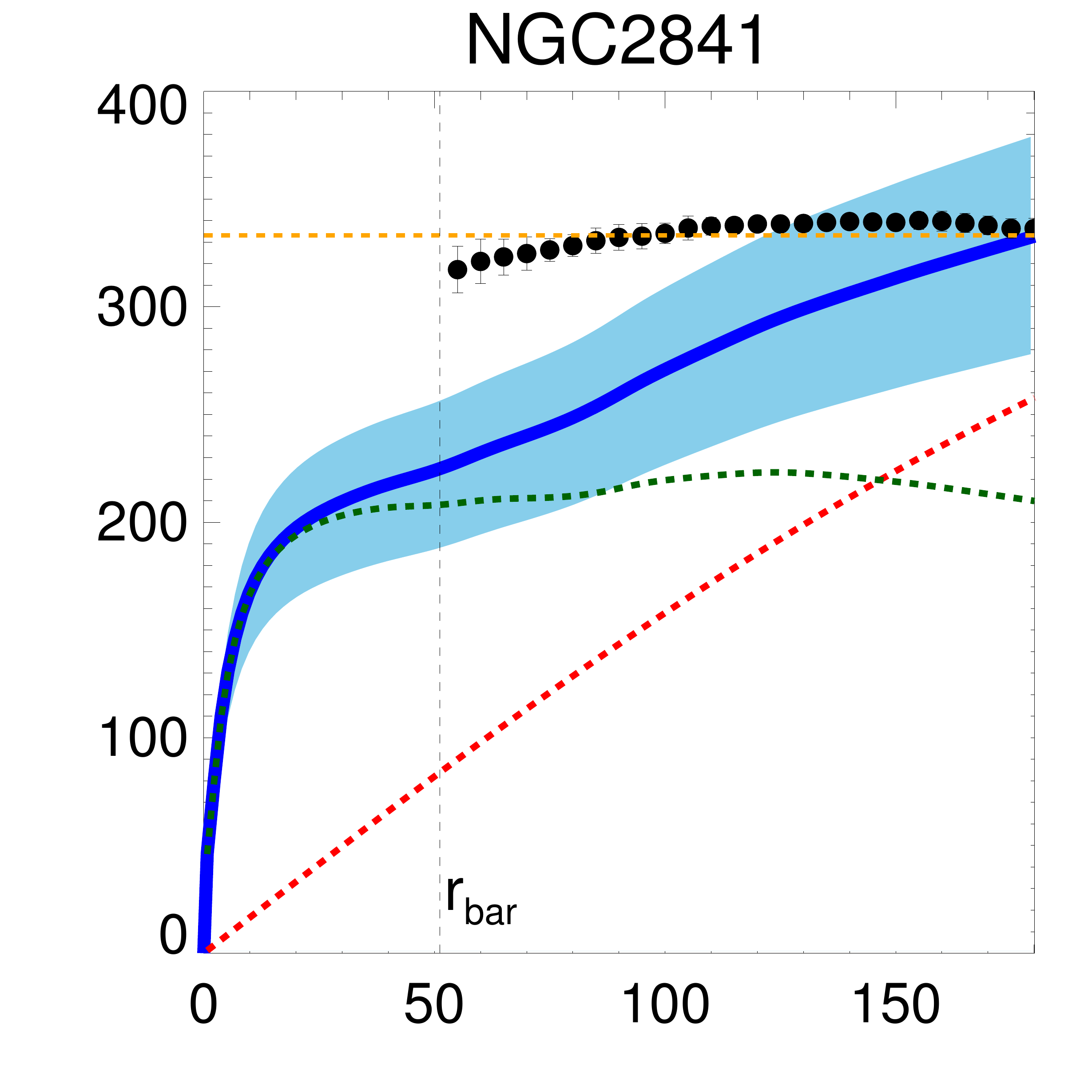}
\end{tabular}
\caption{The same as in the previous figure but using the H{\sc\,i} rotation curves from de Blok et al. (2008)
for a sub-sample of 6 galaxies taken from the THINGS sample (Walter et al. 2008).
}
\label{all20_rotcur_decomposition_block}
\end{figure}

\twocolumn
\section{Halo-to-stellar mass ratio and the bar fraction}\label{DM_appendix}
\begin{figure}
    \begin{tabular}{l r}
    \includegraphics[width=0.43\textwidth]{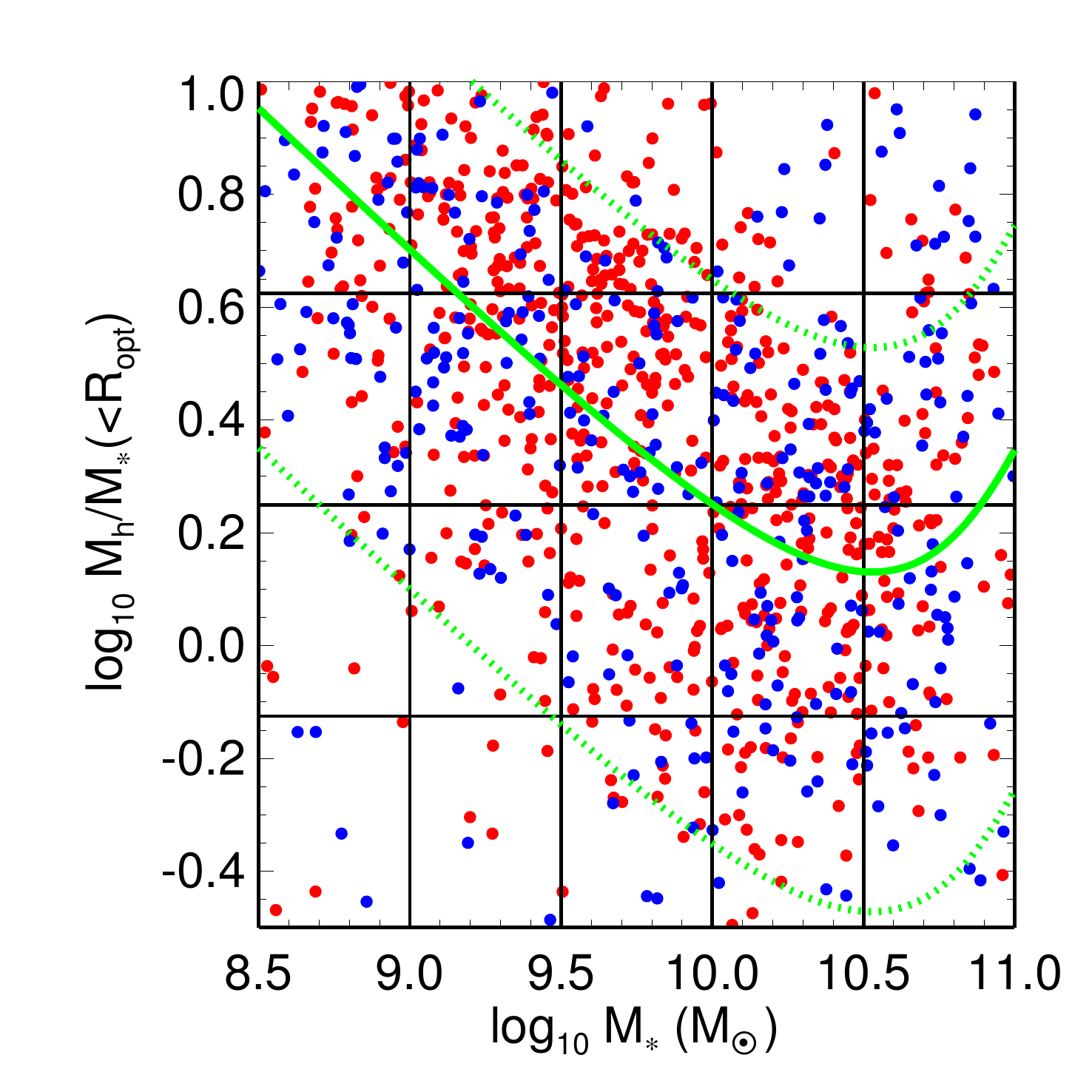}\\
    \includegraphics[width=0.425\textwidth]{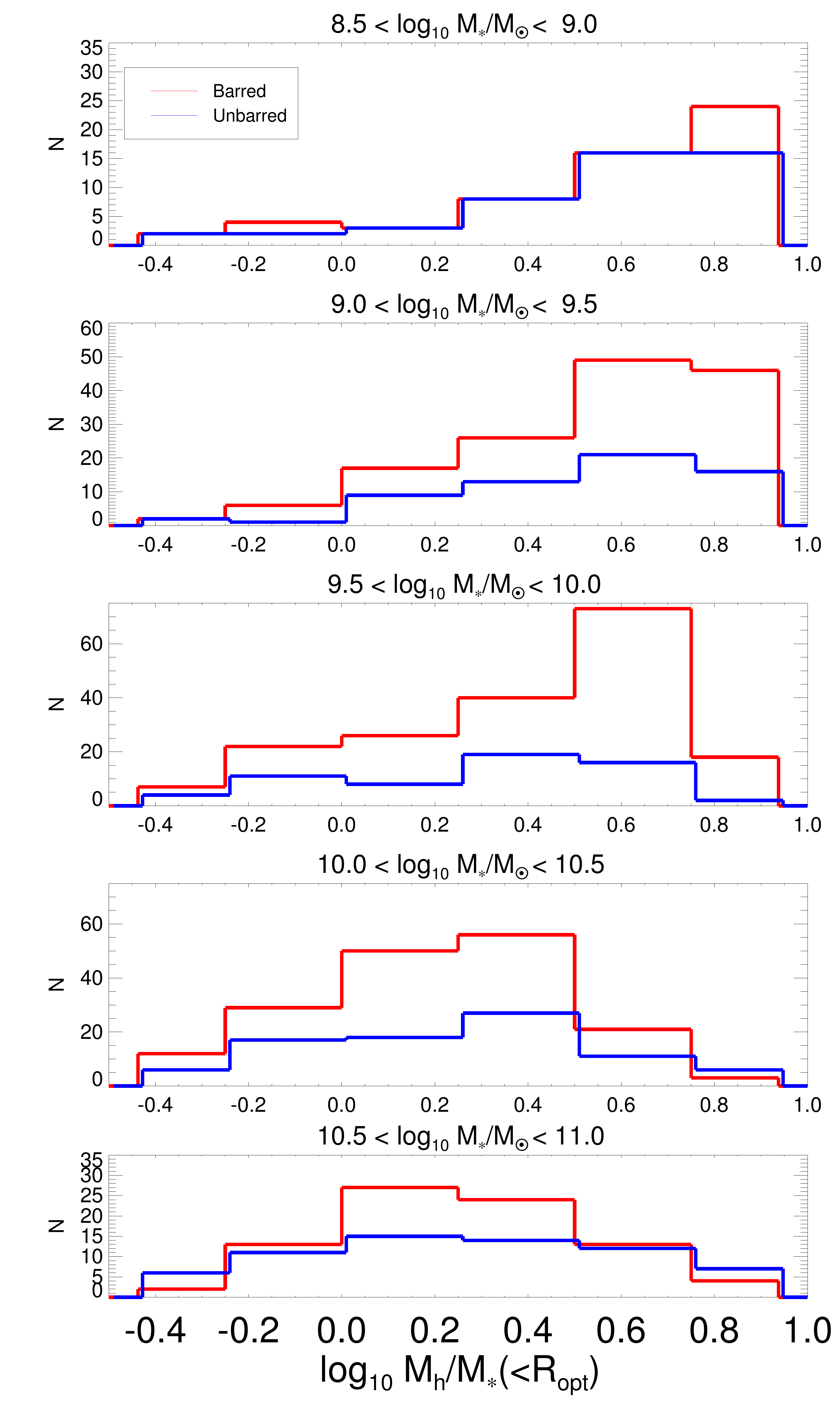}
    \end{tabular}
\caption{
Co-dependence between halo-to-stellar mass ratio, the total stellar mass, and the bar fraction (B2015) are studied by binning the $M_{\rm h}/M_{\ast}$-$M_{\ast}$ space.
In the upper panel, red and blue filled circles correspond to barred and non-barred galaxies, respectively.
For a given stellar mass interval, we display the distribution of barred and non-barred galaxies vs. $M_{\rm h}/M_{\ast}$.
P-values after performing a two-sample K-S test are listed in Table~\ref{k-s-tests}): according to the test, the differences in the distribution of barred and non-barred galaxies are not statistically significant for any of the mass intervals.
The green lines correspond to the total halo-to-stellar ratio from \citet[][]{2010ApJ...710..903M} scaled down by constant factors $\sim 0.04$ (solid line),
$\sim 0.1$ and $\sim 0.01$ (dashed lines).
}
\label{bar_frac_mass_ks}
\end{figure}
In Sect.~\ref{SHMR} we showed a clear dependence between the bar fraction and both the halo-to-stellar mass ratio, evaluated inside the optical radius, and the stellar mass of the host galaxy.  
We also showed that $M_{\rm h}/M_{\ast}(<R_{\rm opt})$ scales inversely with $M_{\ast}$ for masses $M_{\ast}<10^{11}M_{\odot}$.
To probe co-dependences between $M_{\rm h}/M_{\ast}$, $M_{\ast}$ and $f_{\rm bar}$, we studied the distribution of barred and non-barred galaxies \citep[as classified in][]{2015ApJS..217...32B} as a function of $M_{\rm h}/M_{\ast}$
for a given $M_{\ast}$ bin. We observe no significant difference between them (see Fig.~\ref{bar_frac_mass_ks}).
To confirm this result, we performed a two-sample Kolmogorov-Smirnov (see the p-values listed in Table~\ref{k-s-tests}),
finding that the cumulative distribution function of $M_{\rm h}/M_{\ast}$ is similar for barred and non-barred galaxies, given a certain stellar mass range.

We also performed K-S tests for galaxies with relative bar sizes smaller or greater than the median $r_{\rm bar}/h_{\rm R}$ and also for the different bar families
(SAB and S$\underline{\rm A}$B on one hand, and SB and SA$\underline{\rm B}$ on the other),
all of them compared to the family of non-barred galaxies. 
In all cases, the cumulative distribution function of $M_{\rm h}/M_{\ast}$ is not significantly different for barred and non-barred galaxies
(regardless of size, strength, and bar detection criterion), given a certain stellar mass range.
\begin{tiny}
\begin{table*}
\small
\scriptsize
\centering
\begin{tabular}{ c c c c c c}
\hline
\multicolumn{1}{|c}{\centering K-S p-value:} & \multicolumn{1}{|c}{\centering $8.5 < \rm log_{10} (M_{\ast}/M_{\odot}) < 9$} & \multicolumn{1}{|c}{\centering $9 < \rm log_{10} (M_{\ast}/M_{\odot}) < 9.5$} &
\multicolumn{1}{|c}{\centering $9.5 < \rm log_{10} (M_{\ast}/M_{\odot}) < 10$} & \multicolumn{1}{|c}{\centering $10 < \rm log_{10} (M_{\ast}/M_{\odot}) < 10.5$} & \multicolumn{1}{|c|}{\centering $10.5 < \rm log_{10} (M_{\ast}/M_{\odot}) < 11$} \tabularnewline
\hline
Visual& $  0.63$ & $  0.17$ & $  0.01$ & $  0.25$ & $  0.55$\\
Ellipse fitting& $  0.34$ & $  0.02$ & $  0.32$ & $  0.25$ & $  0.11$\\
$\exists$ $A_{2}^{\rm max}$ $\&$ $Q_{\rm b}$& $  0.81$ & $  0.03$ & $  0.15$ & $  0.63$ & $  0.14$\\
\hline
\multicolumn{6}{c}{$r_{\rm bar} > 0.9 h_{\rm R}$}  \vspace{0.1cm} \\
Visual& $  0.36$ & $  0.28$ & $  0.04$ & $  0.16$ & $  0.29$\\
Ellipse fitting& $  0.34$ & $  0.20$ & $  0.47$ & $  0.09$ & $  0.09$\\
$\exists$ $A_{2}^{\rm max}$ $\&$ $Q_{\rm b}$& $  0.68$ & $  1.00$ & $  0.38$ & $  0.19$ & $  0.18$\\
\hline
\multicolumn{6}{c}{$r_{\rm bar} \le 0.9 h_{\rm R}$}  \vspace{0.1cm} \\
Visual& $  0.52$ & $  0.26$ & $  0.01$ & $  0.69$ & $  0.46$\\
Ellipse fitting& $  0.43$ & $  0.03$ & $  0.54$ & $  0.68$ & $  0.14$\\
$\exists$ $A_{2}^{\rm max}$ $\&$ $Q_{\rm b}$& $  0.00$ & $  0.00$ & $  0.08$ & $  0.74$ & $  0.21$\\
\hline
\multicolumn{6}{c}{$SB+SA\underline{B}$}  \vspace{0.1cm} \\
Visual& $  0.85$ & $  0.11$ & $  0.07$ & $  0.11$ & $  0.27$\\
Ellipse fitting& $  0.93$ & $  0.02$ & $  0.42$ & $  0.10$ & $  0.36$\\
$\exists$ $A_{2}^{\rm max}$ $\&$ $Q_{\rm b}$& $  0.99$ & $  0.04$ & $  0.44$ & $  0.29$ & $  0.44$\\
\hline
\multicolumn{6}{c}{$SAB+S\underline{A}B$}  \vspace{0.1cm} \\
Visual& $  0.09$ & $  0.47$ & $  0.00$ & $  0.67$ & $  0.22$\\
Ellipse fitting& $  0.21$ & $  0.09$ & $  0.31$ & $  0.72$ & $  0.01$\\
$\exists$ $A_{2}^{\rm max}$ $\&$ $Q_{\rm b}$& $  0.45$ & $  0.35$ & $  0.01$ & $  0.80$ & $  0.03$\\
\hline
\end{tabular}
\caption{
Significance level of the Kolmogorov-Smirnov statistic between two sub-samples of barred and non-barred galaxies studied as a
function of the stellar-to-halo mass ratio. The sub-samples were drawn from the stellar mass bins shown in the lower panel of Figure~\ref{bar_frac_mass} (the $M_{\ast}$
thresholds are displayed in the first row). K-S tests are also performed for galaxies with relative bar sizes lower or greater than the median $r_{\rm bar}/h_{\rm R}$
and also for the different bar families (SAB and $S\underline{\rm A}$B on one side, and SB and SA$\underline{\rm B}$ on the other), all of them compared to the family of non-barred
galaxies. In general, the distribution of $M_{\rm h}/M_{\ast}$ is not significantly different for barred and non barred galaxies (regardless of size and strength),
given a certain stellar mass range (indicated by p-values $\ge$ 0.01).}
\label{k-s-tests}
\end{table*}
\end{tiny}

\end{appendix}
\end{document}